\documentclass[a4paper,english]{article}
\usepackage[T1]{fontenc}
\usepackage[utf8]{inputenc}
\usepackage{babel}
\usepackage{verbatim}
\usepackage{float}
\usepackage{mathtools}
\usepackage{enumitem}
\usepackage{amsmath}
\usepackage{amsthm}
\usepackage{amssymb}
\usepackage{graphicx}
\usepackage{xargs}[2008/03/08]
\usepackage[pdfusetitle,
 bookmarks=true,bookmarksnumbered=true,bookmarksopen=false,
 breaklinks=false,pdfborder={0 0 1},backref=false,colorlinks=false]
 {hyperref}

\makeatletter

\pdfpageheight\paperheight
\pdfpagewidth\paperwidth

\providecommand{\tabularnewline}{\\}

\numberwithin{equation}{section}
\numberwithin{figure}{section}
\theoremstyle{plain}
\newtheorem{thm}{\protect\theoremname}[section]
\theoremstyle{remark}
\newtheorem{rem}[thm]{\protect\remarkname}
\theoremstyle{plain}
\newtheorem{prop}[thm]{\protect\propositionname}
\ifx\proof\undefined
\newenvironment{proof}[1][\protect\proofname]{\par
	\normalfont\topsep6\p@\@plus6\p@\relax
	\trivlist
	\itemindent\parindent
	\item[\hskip\labelsep\scshape #1]\ignorespaces
}{%
	\endtrivlist\@endpefalse
}
\providecommand{\proofname}{Proof}
\fi
\theoremstyle{plain}
\newtheorem{lem}[thm]{\protect\lemmaname}
\theoremstyle{remark}
\newtheorem{claim}[thm]{\protect\claimname}
\theoremstyle{plain}
\newtheorem{cor}[thm]{\protect\corollaryname}

\usepackage{bbm}
\usepackage{dsfont}
\usepackage{yhmath}
\usepackage{colortbl}

\makeatother

\providecommand{\claimname}{Claim}
\providecommand{\corollaryname}{Corollary}
\providecommand{\lemmaname}{Lemma}
\providecommand{\propositionname}{Proposition}
\providecommand{\remarkname}{Remark}
\providecommand{\theoremname}{Theorem}

\begin{document}
\global\long\def\prob{\mu}%
\newcommandx\w[1][usedefault, addprefix=\global, 1={\Lambda,\lambda}]{\mathsf{w}_{#1}}%
\global\long\def\per{\mathrm{per}}%
\global\long\def\bdd{\mathrm{\square}}%
\global\long\def\eps{\epsilon}%
\global\long\def\Z{\mathbb{Z}}%
\global\long\def\V{\mathbb{V}}%
\global\long\def\E{\mathbb{E}}%
\global\long\def\P{\mathbb{P}}%
\global\long\def\F{\mathbb{F}}%
\global\long\def\R{\mathbb{R}}%
\global\long\def\L{\mathcal{L}}%
\global\long\def\zed{\mathfrak{z}}%
\global\long\def\perimiter{\mathrm{perimiter}}%
\global\long\def\nat{\mathbb{N}}%
\global\long\def\subg{\mathrm{sub}}%
\global\long\def\vol{\mathrm{Area}}%
\global\long\def\width{\mathrm{Width}}%
\global\long\def\height{\mathrm{Height}}%
\global\long\def\len{\mathrm{len}}%
\global\long\def\perim{\mathrm{Perimeter}}%
\global\long\def\rect#1#2{\mathrm{R}_{#1,#2}}%
\global\long\def\frect#1{\mathrm{R}_{#1}}%
\global\long\def\rstr{\,\raisebox{-0.5ex}{\rule{.4pt}{1.7ex}}\,}%
\global\long\def\intr#1{\mathrm{int}(#1)}%
\global\long\def\divides{\vert}%
\global\long\def\ver{\mathrm{ver}}%
\global\long\def\hor{\mathrm{hor}}%

\global\long\def\rectlat{\mathcal{L}}%
\newcommandx\refls[2][usedefault, addprefix=\global, 1=\Lambda, 2=R]{T_{#1}^{#2}}%
\newcommandx\grid[2][usedefault, addprefix=\global, 1=\Lambda, 2=R]{G_{#1}^{#2}}%
\global\long\def\zedc#1#2{\zed_{#1}^{\hspace{-0.1em}#2}}%
\newcommandx\zedd[3][usedefault, addprefix=\global, 1=\Lambda, 2=R, 3=f]{\left\Vert #3\right\Vert _{#2|#1}}%
\newcommandx\zedi[2][usedefault, addprefix=\global, 1=R, 2=f]{\left\Vert #2\right\Vert _{#1}}%
\newcommandx\zeddd[3][usedefault, addprefix=\global, 1=\Lambda, 2=R, 3=f]{\zedc{#1}{#2}\hspace{-0.3em}\left(#3\right)}%

\global\long\def\types{\mathrm{Types}}%
\global\long\def\comp{\mathrm{comp}}%
\global\long\def\null{{}}%
\global\long\def\cov{\mathrm{Cov}}%

\global\long\def\a{\mathfrak{a}}%
\global\long\def\b{\mathfrak{b}}%
\global\long\def\c{\mathfrak{c}}%
\global\long\def\s{\mathfrak{s}}%

\global\long\def\zoz#1{Z_{#1,\mathrm{1D}}^{0}}%
\global\long\def\zop#1{Z_{#1,\mathrm{1D}}^{\per}}%

\global\long\def\trace{\mathrm{Tr}}%
\global\long\def\indic#1{\mathbf{1}_{#1}}%
\global\long\def\phase{\mathrm{Phase}}%
\global\long\def\xre#1#2{X_{#1}#2}%
\global\long\def\peierls#1{\mathfrak{p}_{#1}}%
\global\long\def\pv{\mathfrak{p}}%
\global\long\def\es#1{\overline{#1}}%
\global\long\def\sg#1#2{X_{#1}#2}%

\global\marginparsep=2cm
\title{Columnar order in random packings of $2\hspace{-0.35ex}\times\hspace{-0.35ex}2$
squares on the square lattice}
\author{Daniel Hadas\thanks{Tel Aviv University, Israel \url{danielhadas1@mail.tau.ac.il}}\and Ron
Peled\thanks{Tel Aviv University, Israel \url{peledron@tauex.tau.ac.il}}}
\maketitle
\begin{abstract}
We study random packings of $2\times2$ squares with centers on the
square lattice $\mathbb{Z}^{2}$, in which the probability of a packing
is proportional to $\lambda$ to the number of squares. We prove that
for large $\lambda$, typical packings exhibit columnar order, in
which either the centers of most tiles agree on the parity of their
$x$-coordinate or the centers of most tiles agree on the parity of
their $y$-coordinate. This manifests in the existence of four extremal
and periodic Gibbs measures in which the rotational symmetry of the
lattice is broken while the translational symmetry is only broken
along a single axis. We further quantify the decay of correlations
in these measures, obtaining a slow rate of exponential decay in the
direction of preserved translational symmetry and a fast rate in the
direction of broken translational symmetry. Lastly, we prove that
every periodic Gibbs measure is a mixture of these four measures.

Additionally, our proof introduces an apparently novel extension of
the chessboard estimate, from finite-volume torus measures to all
infinite-volume periodic Gibbs measures.\bigskip{}
\medskip{}
\begin{figure}[H]
\begin{centering}
\hspace*{-0\textwidth}\includegraphics[viewport=0bp 0bp 432bp 290bp,clip,width=0.88\textwidth]{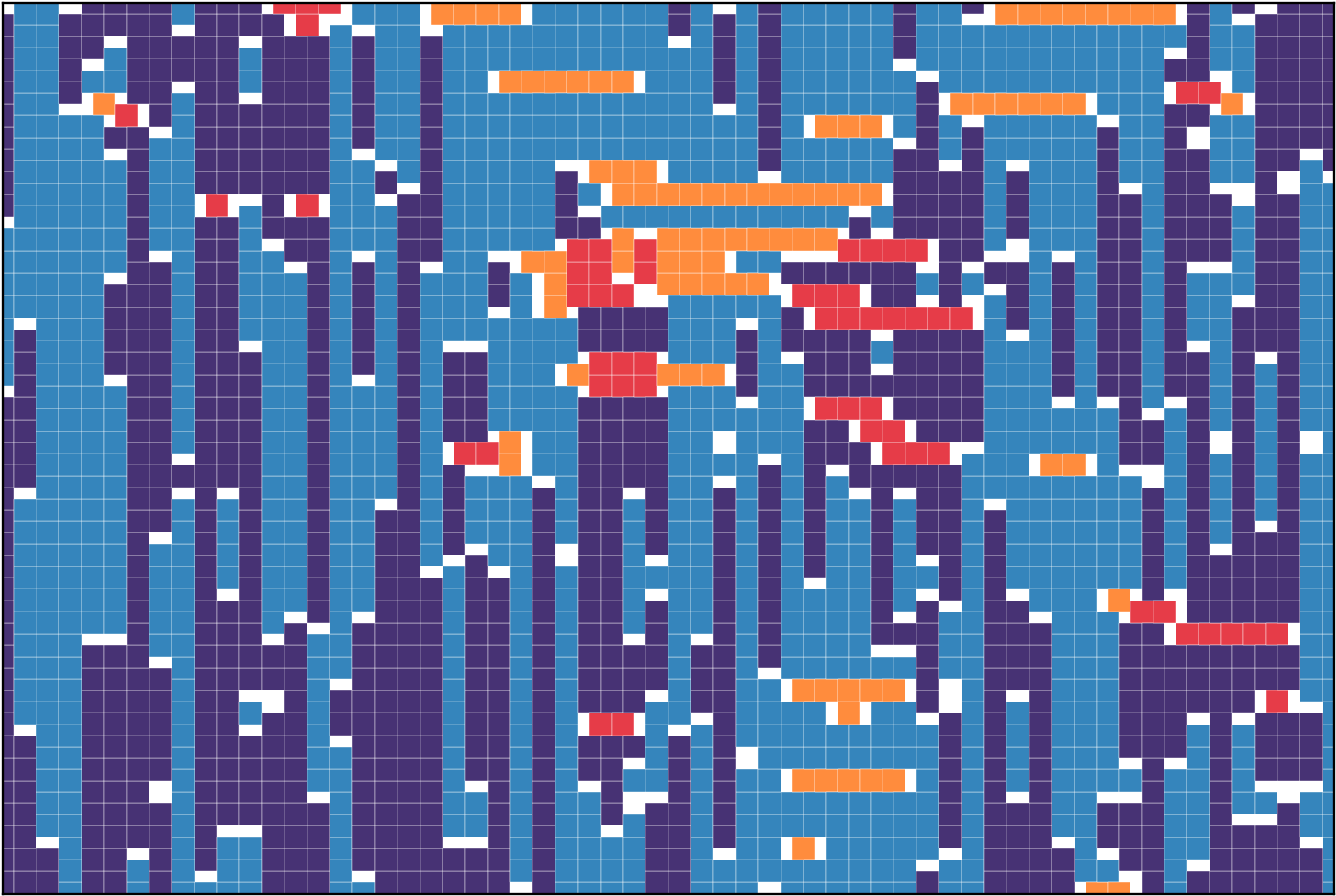}
\par\end{centering}
Columnar order in (a portion of) a high-density packing ($\lambda=130$).
The four colors correspond to the parities of the $x$ and $y$ coordinates
of each tile.
\end{figure}
\end{abstract}
\pagebreak\tableofcontents{}\pagebreak{}

\section{Introduction}

In this paper we study the $2\times2$ hard-square model on the square
lattice $\Z^{2}$. We prove that the model admits \emph{columnar order}
at high fugacity, resulting in multiple Gibbs measures. Moreover,
the set of periodic Gibbs measures is characterized.

\subsection{\label{subsec:the_model}The model}

The square lattice $\mathbb{Z}^{2}$ is embedded in $\mathbb{R}^{2}$
in the natural way. With each vertex $(x,y)\in\Z^{2}$ is associated
a \textbf{tile} $T_{(x,y)}$ that is a closed $2\times2$ square,
with sides parallel to the axes, which is centered at that vertex.
Hard-square configurations are sets of tiles whose interiors are pairwise
disjoint. Precisely, a configuration is represented by a function
$\sigma:\Z^{2}\to\{0,1\}$, with the value $\sigma(v)=1$ indicating
the presence of a tile centered at $v$, so that the \textbf{space
of configurations} is
\[
\Omega\coloneqq\left\{ \sigma\in\{0,1\}^{\Z^{2}}:\text{If }\text{int}(T_{u})\cap\text{int}(T_{v})\neq\emptyset\text{ for \ensuremath{u\neq v}, then }\sigma(u)\sigma(v)=0\right\} 
\]
where $\mathrm{int}(\cdot)$ stands for the interior of a set. Given
a bounded set $\Lambda\subset\R^{2}$ and a configuration $\rho\in\Omega$,
the space of \textbf{configurations with $\rho$-boundary conditions
outside $\Lambda$} is
\[
\Omega_{\Lambda}^{\rho}\coloneqq\left\{ \sigma\in\Omega:\sigma(v)=\rho(v)\text{ on }\Z^{2}\setminus\mathrm{int}(\Lambda)\right\} .
\]

Given additionally a \textbf{fugacity} parameter $\lambda>0$, the
corresponding \textbf{finite-volume} $2\times2$ \textbf{hard-square
model} is the probability measure $\prob_{\Lambda,\lambda}^{\rho}$
on $\Omega_{\Lambda}^{\rho}$ defined by 
\begin{equation}
\prob_{\Lambda,\lambda}^{\rho}(\sigma)\propto\lambda^{\sum_{v\in\mathrm{int}(\Lambda)\cap\mathbb{Z}^{2}}\sigma(v)}\label{eq:intro_measure}
\end{equation}

where we use $\propto$ to denote that the left-hand side is proportional
to the right-hand side. In words, the probability of a configuration
is proportional to $\lambda$ raised to the power of the number of
tiles in $\Lambda$. 

We describe our results in the language of infinite-volume Gibbs measures,
defined via the standard Dobrushin--Lanford--Ruelle prescription.
Precisely, a probability measure $\mu$ on $\Omega$ is an (infinite-volume)\emph{
}\textbf{Gibbs measure} for the $2\times2$ hard-square model at fugacity
$\lambda$ if for every bounded $\Lambda\subset\R^{2}$ the following
holds: Let $\sigma$ be sampled from $\mu$. Conditionally on $\sigma$
restricted to $\Z^{2}\setminus\mathrm{int}(\Lambda)$, the distribution
of $\sigma$ is given by $\mu_{\Lambda,\lambda}^{\rho}$ with $\rho$
being any configuration which coincides with $\sigma$ on $\Z^{2}\setminus\mathrm{int}(\Lambda)$.

Given a sublattice $\mathcal{L}\subset\Z^{2}$ and a probability measure
$\mu$ on $\Omega$, say that $\mu$ is \textbf{$\mathcal{L}$-invariant}
if it is invariant under all translations by vectors from $\mathcal{L}$
and say that $\mu$ is \textbf{$\mathcal{L}$-ergodic} if it is $\mathcal{L}$-invariant
and assigns probability $0$ or $1$ to $\mathcal{L}$-invariant events.
A probability measure on $\Omega$ is called \textbf{periodic} if
it is $\mathcal{L}$-invariant under some \emph{full-rank} sublattice
$\mathcal{L}\subset\Z^{2}$. A probability measure on $\Omega$ is
called \textbf{extremal} if it assigns probability $0$ or $1$ to
all tail events (the tail sigma algebra is the intersection over all
finite $D\subset\mathbb{Z}^{2}$ of the sigma algebra generated by
$\sigma$ restricted to $D^{c}$).

\subsection{\label{subsec:discuss_results}Discussion and results}

Classical methods may be used to show that the $2\times2$ hard-square
model is disordered at \emph{low fugacity}, in the sense that it has
a \emph{unique} Gibbs measure. This follows from either the Dobrushin
uniqueness theorem \cite{dobrushin1968description} or the disagreement
percolation method of van den Berg \cite{cmp/1104252313}, with the
latter method proving uniqueness for $\lambda<p_{c}/(1-p_{c})$ (see
also \cite[Theorem 2.3]{VANDENBERG1994179}), where $p_{c}$ is the
site percolation threshold of the square lattice with nearest and
next-nearest neighbor interactions (i.e., $u,v\in\mathbb{Z}^{2}$
are adjacent if $\|u-v\|_{\infty}=1$, this is dual to the standard
nearest-neighbor site percolation on $\mathbb{Z}^{2}$; numerical
estimates give $p_{c}\approx0.407$ \cite{ziff1992spanning,malarz2005square}).

\begin{figure}[t]
\begin{centering}
\includegraphics[scale=0.5]{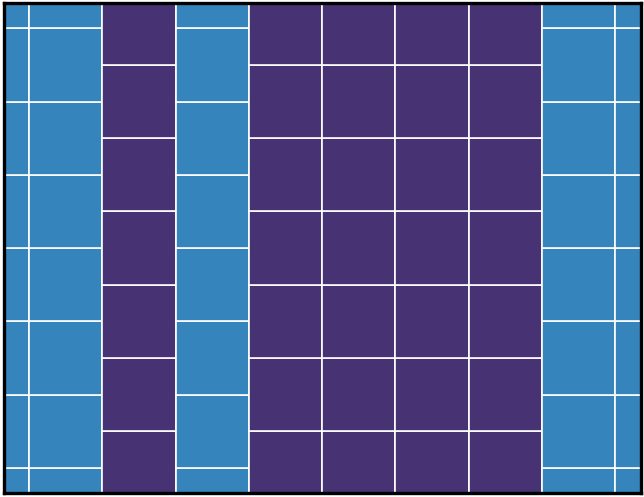}\hfill{}\includegraphics[scale=0.5]{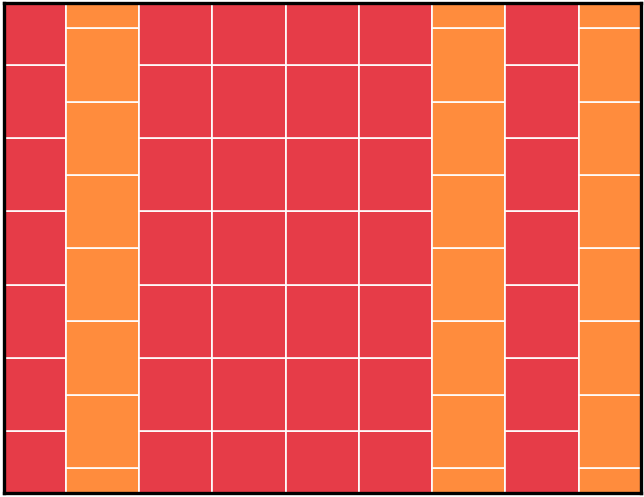}\hfill{}\includegraphics[scale=0.5]{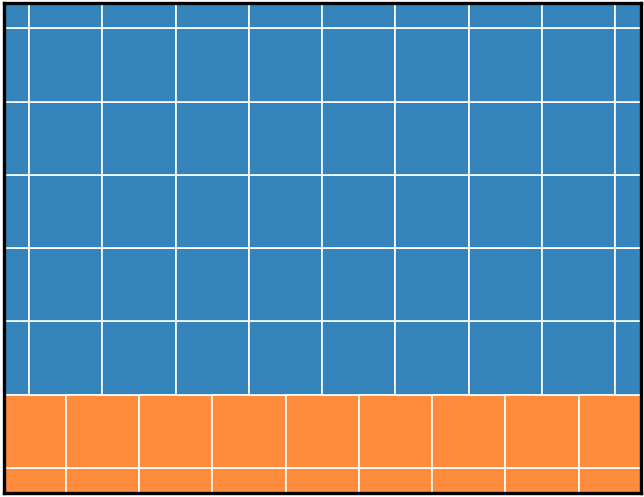}\hfill{}\includegraphics[scale=0.5]{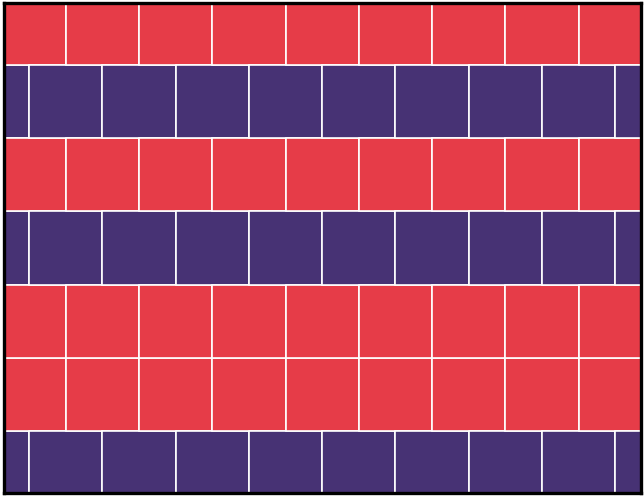}
\par\end{centering}
\caption{\label{fig:fully-packed configurations}Representatives of the four
kinds of fully-packed configurations of the $2\times2$ hard-square
model. The colors of the tiles correspond to the parities of their
$x$ and $y$ coordinates (see Figure \ref{fig:configs} for the precise
correspondence).}
\end{figure}

In this work we study the \emph{high-fugacity} regime of the $2\times2$
hard-square model. To gain intuition, it is instructive to consider
the set of fully-packed configurations; configurations in which the
union of tiles covers the whole of $\mathbb{R}^{2}$. In related models,
such as the nearest-neighbor hard-core model on $\mathbb{Z}^{d}$
(see subsubsection \ref{subsec:The-nearest-neighbor-hard-core}),
one finds that there are a \emph{finite number} of fully-packed configurations,
which are moreover periodic. In some such cases, a Peierls-type argument,
or Pirogov--Sinai theory \cite{pirogov1975phase,pirogov1976phase}
(see also \cite[Chapter 7]{friedli_velenik_2017}), allow to deduce
that typical configurations sampled in the high-fugacity regime behave
as ``small perturbations'' of one of the fully-packed configurations
in the sense that they coincide with this configuration at most places.
In contrast, one easily checks that the $2\times2$ hard-square model
admits a \emph{continuum} (i.e., $2^{\aleph_{0}}$) of fully-packed
configurations, obtained in the following way (see Figure \ref{fig:fully-packed configurations}):
start with the ``square lattice configuration'' $\sigma_{0}\in\Omega$,
which is defined by $\sigma_{0}(x,y)=1$ if and only if both $x$
and $y$ are odd. From $\sigma_{0},$ one can create a continuum of
other fully-packed configurations by ``sliding'' columns of tiles
down by one lattice site; precisely, for each $t:2\mathbb{Z}+1\to\{0,1\}$
one obtains a fully-packed configuration $\sigma^{t}$ by setting
$\sigma^{t}(x,y)=1$ if and only if $x$ is odd and $y\equiv t(x)\bmod2$.
In a similar manner, one can create a continuum of fully-packed configurations
by starting from $\sigma_{0}$ and sliding rows of tiles to the right
by one lattice site. Additional fully-packed configurations may be
generated from the ones described so far by translating a fully-packed
configuration by one lattice site up, or by one lattice site to the
right.

\begin{figure}[t]
\centering{}\includegraphics[width=0.32\textwidth]{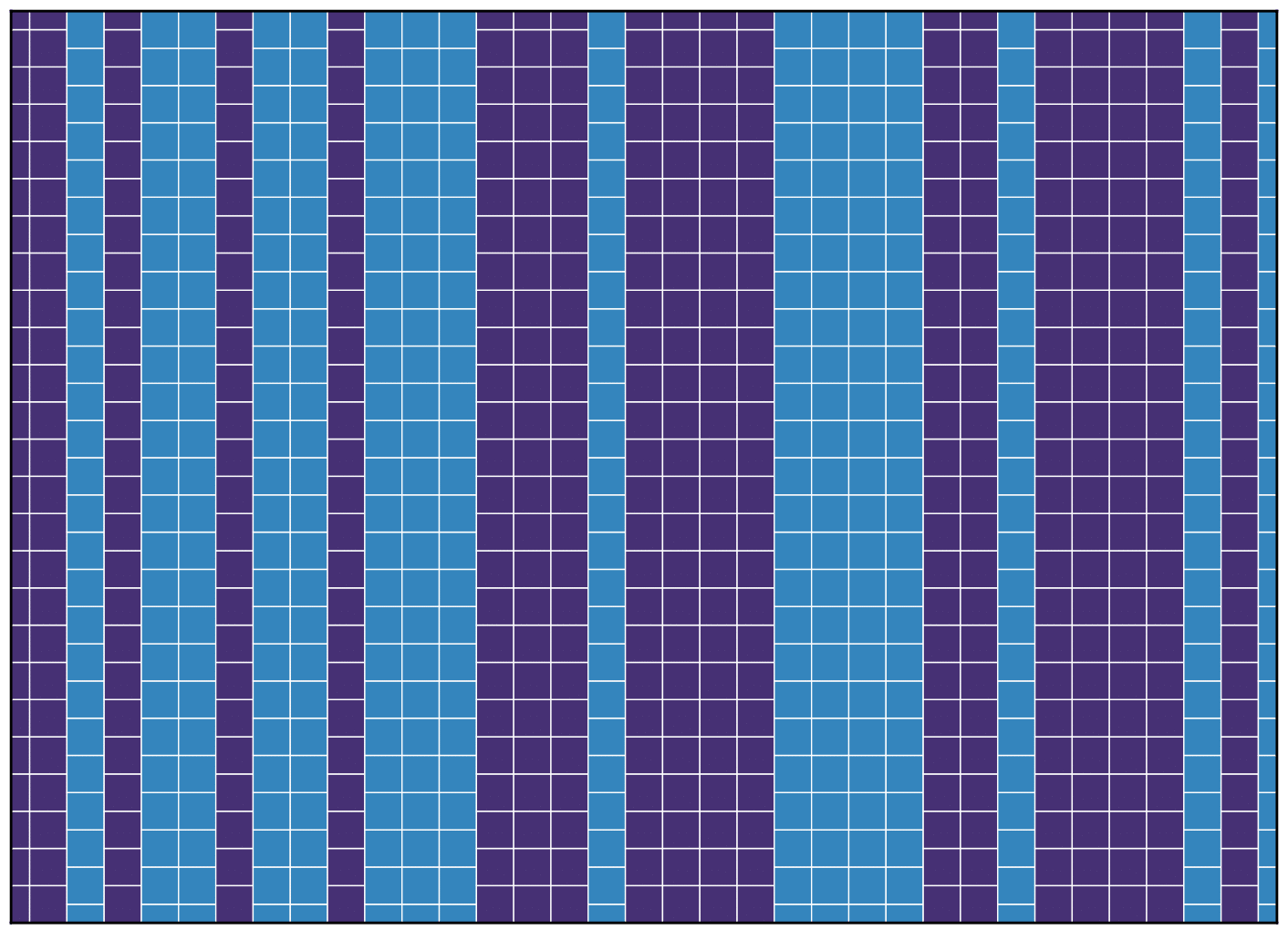}\hfill{}\includegraphics[width=0.32\textwidth]{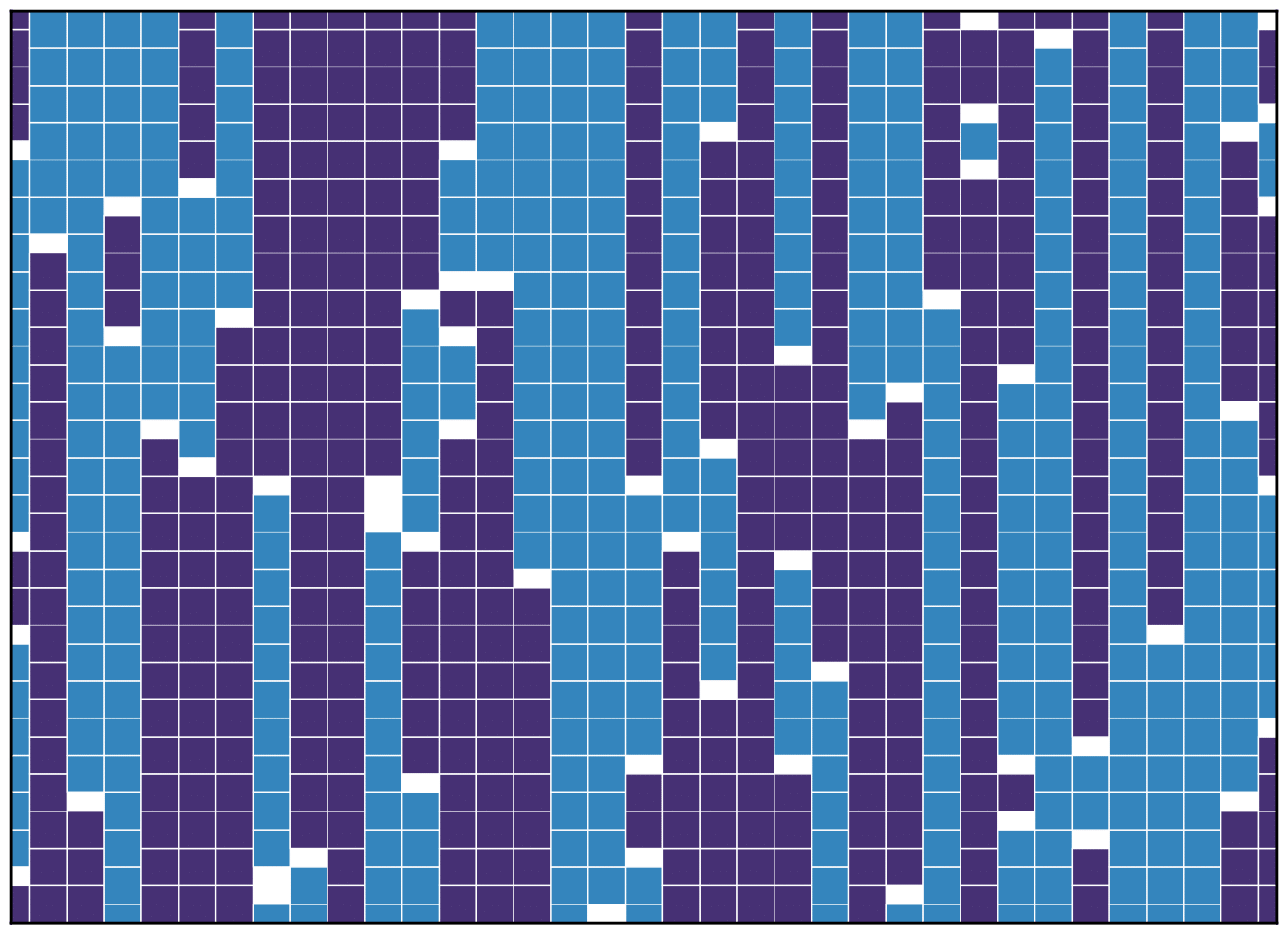}\hfill{}\includegraphics[width=0.32\textwidth]{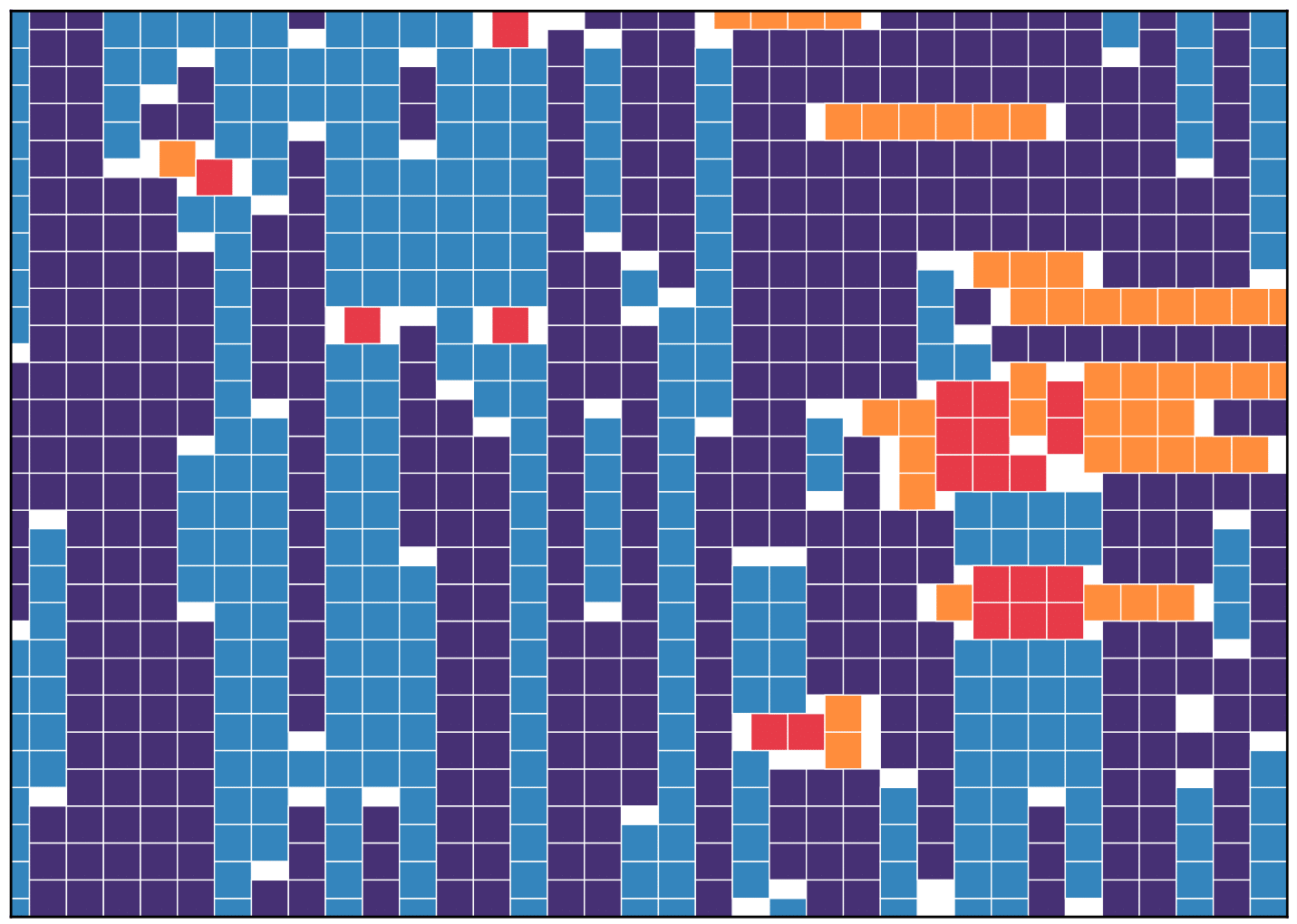}\caption{\label{fig:from fully-packed to high fugacity}The left panel depicts
a fully-packed configuration, arranged in columns. The middle panel
depicts a sample from a union of independent \emph{``}one-dimensional
columnar systems'' at high fugacity (denoted by $\mu_{(\protect\ver,0)}^{\cup\text{1D}}$
in the text). The right panel depicts a sample from the high-fugacity
regime of the $2\times2$ hard-square model. We prove the existence
of a phase for the high-fugacity $2\times2$ hard-square model with
properties resembling a \textquotedblleft small perturbation\textquotedblright{}
of $\mu_{(\protect\ver,0)}^{\cup\text{1D}}$.}
\end{figure}

The existence of this continuum of fully-packed configurations, sometimes
termed the ``sliding phenomenon'', forms an obstacle to Pirogov--Sinai
theory, and is the reason that no rigorous results on the high-fugacity
regime have appeared so far. Fr\"ohlich--Israel--Lieb--Simon \cite[Model 4.4]{frohlichPhaseTransitionsReflection1980}
discussed the $2\times2$ hard-square model and wrote that the ``conventional
wisdom'' is that it has a \emph{unique} Gibbs state in the high-fugacity
regime. Recently, Mazel--Stuhl--Suhov \cite{MSS2019_square} conjectured
that the $2\times2$ hard-square model has a unique Gibbs measure
in the high-fugacity regime, as part of a more general conjecture
on models with the sliding phenomenon. In the physics literature,
while early results \cite{bellemans1967phase,nisbet1974hard} were
inconclusive regarding the existence of a phase transition, modern
studies (see \cite{Fernandes2007_kNN,Nath2016thesis,ramolaColumnarOrderAshkinTeller2015}
and references therein) indicate a \emph{columnar phase} appearing
in the high-fugacity regime of the $2\times2$ hard-square model.
Our work clarifies the situation by proving that the model admits
\emph{multiple} (periodic) Gibbs measures at high fugacity, in agreement
with the modern physics literature and in contrast to the expectations
expressed in \cite{frohlichPhaseTransitionsReflection1980,MSS2019_square}.

Configurations sampled from these Gibbs measures display a similar
columnar (or row) ordering of tiles as the fully-packed configurations
described above; however, the sampled configurations are not simply
perturbations of some fully-packed configuration. Instead, they should
be thought of as a perturbation of a random tiling formed from a ``union
of one-dimensional systems'' in the sense that we now describe (see
Figure \ref{fig:from fully-packed to high fugacity}).

Denote by $\mu_{(\ver,0)}^{\cup\text{1D}}$ the $2\times2$ hard-square
model at fugacity $\lambda$ conditioned so that the $x$-coordinates
of the corners of all tiles are even. By this we mean that in the
columns with odd $x$-coordinate we see a sample from $\mu^{\text{1D}}$,
the unique Gibbs measure of the nearest-neighbor hard-core model on
$\mathbb{Z}$ at fugacity $\lambda$, and these samples are independent
between the different columns (see Figure \ref{fig:from fully-packed to high fugacity},
middle panel). At high fugacity, samples from $\mu^{\text{1D}}$ give
a very dense tiling, with long intervals of fully-packed tiles separated
by a single skip (or, more rarely, multiple skips). The typical length
of the fully-packed intervals is of order $\sqrt{\lambda}$, which
is also the natural length scale at which $\mu^{\text{1D}}$ decorrelates.
Our first theorem proves the existence of a Gibbs measure $\mu_{(\ver,0)}$
for the high-fugacity $2\times2$ hard-square model, which may heuristically
be regarded as a ``small perturbation'' of $\mu_{(\ver,0)}^{\cup\text{1D}}$.
As we discuss after the theorem, this implies the existence of multiple
Gibbs measures at high fugacity.
\begin{thm}
\label{thm:main}There exists $0<\lambda_{0}<\infty$ such that the
$2\times2$ hard-square model at each fugacity $\lambda>\lambda_{0}$
admits a Gibbs measure $\mu_{(\ver,0)}$ satisfying:
\begin{enumerate}
\item \label{item:Invariance-and-extremality}\emph{Invariance and extremality}:
$\mu_{(\ver,0)}$ is $(2\Z\times\Z)$-invariant and extremal. In particular,
$\mu_{(\ver,0)}$ is $(2\Z\times\Z)$-ergodic.
\item \label{item:Columnar-order}\emph{Columnar order}: for all $(x,y)\in\Z^{2}$,
\begin{equation}
\mu_{(\ver,0)}(\sigma(x,y)=1)=\begin{cases}
\Theta(\lambda^{-1}) & x\equiv0\bmod2,\\
\frac{1}{2}-\Theta(\lambda^{-1/2}) & x\equiv1\bmod2,
\end{cases}\label{eq:columnar order}
\end{equation}
where $a=\Theta(b)$ indicates that $ca\le b\le Ca$ for some universal
$C,c>0$.
\item \label{item:Decay-of-correlations}\emph{Decay of correlations}: Let
$f,g:\Omega\to[-1,1]$. Suppose that $f(\sigma)$ depends only on
the restriction of $\sigma$ to a set $A\subset\Z^{2}$ and similarly
$g(\sigma)$ depends only on the restriction of $\sigma$ to $B\subset\Z^{2}$.
Then 
\begin{equation}
\left|\cov_{\mu_{(\ver,0)}}(f,g)\right|\le\sum_{u\in A}\,\text{\ensuremath{\sup_{v\in B}}}\,\alpha(u,v)\label{eq:decay of correlations}
\end{equation}
where $\cov_{\mu}(f,g)$ is the covariance of $f(\sigma)$ and $g(\sigma)$
when $\sigma$ is sampled from $\mu$ and where
\begin{equation}
\alpha\big((x_{1},y_{1}),(x_{2},y_{2})\big):=\min\left\{ Ce^{-c|x_{2}-x_{1}|-c\frac{|y_{2}-y_{1}|}{\sqrt{\lambda}}},\left(\tfrac{C\log\lambda}{\sqrt{\lambda}}\right)^{\mathbf{1}_{x_{1}\neq x_{2}}}\right\} \label{eq:decay of correlations distance}
\end{equation}
for some universal $C,c>0$.
\end{enumerate}
\end{thm}
The theorem thus establishes that at high fugacity, the model admits
a Gibbs measure which is invariant to translations in the vertical
direction and satisfies that tiles preferentially occupy vertices
with odd $x$-coordinate (columnar order). This implies the existence
of at least three other Gibbs measures $\mu_{(\ver,1)}$, $\mu_{(\hor,0)}$
and $\mu_{(\hor,1)}$: The measure $\mu_{(\ver,1)}$ is created by
translating the measure $\mu_{(\ver,0)}$ by one lattice site to the
right. The measures $\mu_{(\hor,0)}$ and $\mu_{(\hor,1)}$ are formed
from $\mu_{(\ver,0)}$ and $\mu_{(\ver,1)}$, respectively, by exchanging
the $x$ and $y$ axes. The four measures are distinct (by (\ref{eq:columnar order})),
with the $\mu_{(\ver,i)}$ being $(2\Z\times\Z)$-invariant and extremal
while the $\mu_{(\hor,i)}$ are $(\Z\times2\Z)$-invariant and extremal.

Due to the columnar order, the measure $\mu_{(\ver,0)}$ breaks the
lattice's $90^{\circ}$ rotational symmetry and also its translational
symmetry in the $x$-coordinate, while preserving the translational
symmetry in the $y$-coordinate. The asymmetric role of the two lattice
directions is further manifested in the decay of correlations property.
It is shown that $\mu_{(\ver,0)}$ exhibits exponential decay of correlations
in the $x$-direction with correlation length of order $1$ (at most),
but the rate of exponential decay shown in the $y$-direction is relatively
slow, with a correlation length of order $\sqrt{\lambda}$ (at most).
The second term in the minimum in (\ref{eq:decay of correlations distance})
additionally shows that already the correlations of events in adjacent
(or nearby) columns are rather small when $\lambda$ is large. See
also Remark \ref{Rem:sharpness of second term in correlation decay}
and subsection \ref{subsec:remarks on 2x2 hard squares} for further
discussion of the correlation decay.

Once the existence of multiple extremal Gibbs measures has been established,
one may wonder whether other extremal Gibbs measures exist, or in
other words, whether the four measures exhaust all the possible ways
in which the model may be ordered. Our second theorem establishes
that there are no other \emph{periodic} and extremal measures.
\begin{thm}
\label{thm:main2}There exists $0<\lambda_{0}<\infty$ such that the
following holds for the hard-square model at each fugacity $\lambda>\lambda_{0}$.
There is a unique Gibbs measure $\mu_{(\ver,0)}$ satisfying the properties
listed in Theorem \ref{thm:main}. Moreover, every periodic Gibbs
measure is a convex combination of $\mu_{(\ver,0)}$, $\mu_{(\ver,1)}$,
$\mu_{(\hor,0)}$ and $\mu_{(\hor,1)}$.
\end{thm}
An overview of our proof appears in Section \ref{subsec:Proof-overview}
below.

Our proof uses the \emph{chessboard estimate}, a consequence of the
reflection positivity enjoyed by the $2\times2$ hard-square model.
Our work introduces an apparently novel extension of the chessboard
estimate which may be of interest also for other models. In its standard
form, the chessboard estimate applies to \emph{finite-volume} Gibbs
measures on a discrete torus. In Proposition \ref{prop:chessboard_infinite}
we show that the estimate may be used directly in \emph{infinite volume},
applying to all (infinite-volume) periodic Gibbs measures of the model.
This extension is especially useful in the proofs in Part \ref{part:Characterization-of-the}
of the paper. We formulate the extension solely for the $2\times2$
hard-square model but the provided proof is applicable to other models.

\subsection{\label{subsec:Background-and-related}Background and related works}

This subsection discusses related literature on hard-core models and
liquid crystals.

\subsubsection{General hard-core models}

A hard-core configuration on a graph $G=(V(G),E(G))$ (called an independent
set in combinatorics) is a function $\sigma:V(G)\to\{0,1\}$ satisfying
that if $\{u,v\}\in E(G)$ then $\sigma(u)\sigma(v)=0$. The hard-core
model on a finite $G$, at fugacity $\lambda$, is the probability
measure $\mu$ on hard-core configurations defined by $\mu(\sigma)\propto\lambda^{\sum_{v\in V(G)}\sigma(v)}$.
The hard-core model further arises as a zero-temperature limit of
the anti-ferromagnetic Ising model with a carefully chosen external
field (see, e.g., \cite[around (62)]{peled2020long} for the construction
on $\Z^{d}$). The definition extends, via the usual prescriptions,
to hard-core measures with given boundary conditions and Gibbs measures
on infinite graphs. A basic challenge in statistical physics is to
characterize the Gibbs measures of the hard-core model on an infinite
graph at different values of the fugacity. As mentioned, general results
\cite{dobrushin1968description,cmp/1104252313} lead to the unicity
of Gibbs measures at sufficiently low fugacities (on bounded-degree
graphs). Thus, the main interest is in other, intermediate and high,
regimes of the fugacity.

\subsubsection{\label{subsec:The-nearest-neighbor-hard-core}The nearest-neighbor
hard-core model on $\protect\Z^{d}$}

A prototypical example is the hard-core model on the lattice $\mathbb{Z}^{d}$
(with nearest-neigbor edges). The model admits exactly two fully-packed
configurations $\sigma^{\text{even}}$ and $\sigma^{\text{odd}}$
(configurations where $\max\{\sigma(u),\sigma(v)\}=1$ whenever $u$
is adjacent to $v$), with $\sigma^{\text{even}}(v)=1$ on the even
sublattice (even sum of coordinates) and $\sigma^{\text{odd}}(v)=1$
on the odd sublattice. A seminal result of Dobrushin \cite{dobrushin1968problem}
proves that at high fugacity the model has two extremal Gibbs measures
$\mu^{\text{even}},\mu^{\text{odd}}$, invariant under parity-preserving
shifts, such that samples of $\mu^{\text{even}}$ ($\mu^{\text{odd}}$)
coincide with $\sigma^{\text{even}}$ ($\sigma^{\text{odd}}$) at
most vertices (the density of coinciding vertices goes to $1$ as
$\lambda\to\infty$). In fact, one can define the measures $\mu^{\text{even}},\mu^{\text{odd}}$
at all fugacities, as suitable infinite-volume limits, and establish
that the model admits multiple Gibbs measures if and only if $\mu^{\text{even}}\neq\mu^{\text{odd}}$.
It is believed that there is a unique transition point $\lambda_{c}(d)$
from a unique to multiple Gibbs measures on $\mathbb{Z}^{d}$, and
that for $\lambda>\lambda_{c}(d)$ all periodic Gibbs measures are
convex combinations of $\mu^{\text{even}},\mu^{\text{odd}}$; however,
these facts are presently unknown. There are examples of graphs on
which there are multiple transitions between uniqueness and multiplicity
of Gibbs measures as $\lambda$ increases \cite{brightwell1999nonmonotonic}.
The fact that $\lambda_{c}(d)\to0$ as $d\to\infty$ (for a suitable
definition of $\lambda_{c}(d)$) was proved by Galvin--Kahn \cite{galvin2004phase},
with the rate of decay improved in \cite{peled2014odd}. It is believed
that $\lambda_{c}(d)$ behaves as $d^{-1+o(1)}$ as $d\to\infty$
but this also remains unknown.

\subsubsection{\label{subsec:monomer_dimer}The monomer-dimer model}

A phenomenon of a different nature was discovered by Heilmann--Lieb
\cite{heilmann1970monomers,heilmann1972theory}, in their study of
the \emph{monomer-dimer model}. Monomer-dimer configurations on a
graph $H$ are subsets of edges with the property that no two edges
share a common endpoint (called matchings in combinatorics). The probability
assigned to a configuration by the monomer-dimer model on a finite
graph is proportional to $\lambda$ raised to the number of edges
in the configuration. The monomer-dimer model is thus equivalent to
the hard-core model on the line graph of $H$ (the graph with vertex
set $E(H)$ and with distinct $e,e'\in E(H)$ adjacent if $e\cap e'\neq\emptyset$).
Heilmann and Lieb made the surprising discovery that the monomer-dimer
model has a unique Gibbs measure on all bounded-degree graphs $H$
at all finite fugacities! An alternative proof was found by van den
Berg \cite{van1999absence}. Thus, this model never exhibits a phase
transition from unique to multiple Gibbs measures (except in the sense
that the limiting model as $\lambda\to\infty$, termed the \emph{dimer
model}, may have multiple Gibbs measures).

\subsubsection{\label{subsec:background_lattice_balls}Euclidean balls with centers
on $\protect\Z^{d}$ and other lattice packings}

One may also study the hard-core model on a modified version of $\mathbb{Z}^{d}$
having a different adjacency structure. The version where each $u\in\mathbb{Z}^{d}$
is adjacent to all $v\in\mathbb{Z}^{d}$ with $0<\|u-v\|_{2}<D$ (WLOG
we assume that $D$ can be realized as the distance between some pair
of points in $\mathbb{Z}^{d}$) has received special attention, motivated
by a continuum version of the hard-core model. Various behaviors have
been predicted for different regimes of $D$ and the fugacity. The
above-mentioned nearest-neighbor hard-core model arises by setting
$D=\sqrt{2},$ while the $2\times2$ hard-square model studied here
is obtained in two dimensions ($d=2$) when $D=2$. Recently, a comprehensive
study of the high-fugacity behavior for the two- and three-dimensional
models with various values of $D$ was undertaken by Mazel--Stuhl--Suhov
\cite{MSS2019_square,mazel2021kepler} (also on the triangular and
hexagonal lattice \cite{MSS2018_trihex}; see also \cite{MSS2020_short}).

In two dimensions, the work \cite{MSS2019_square} characterizes the
periodic hard-core configurations of maximal density for each $D$.
Moreover, Pirogov--Sinai theory is used to show that, with few exceptions,
the extremal Gibbs measures arise as perturbations of (a subset of)
these configurations in a suitable sense. The exceptions are a finite
number of values of $D$ (the list of which was confirmed independently
in \cite{MSS2019_square} and in \cite{krachun2020_MSS_sliding_list})
for which there are \emph{infinitely many} periodic hard-core configurations
of maximal density; these always come about as a result of a ``sliding
instability'' in the configurations (similarly to the sliding phenomenon
described above for the $2\times2$ hard-square model). Pirogov--Sinai
theory does not apply in these exceptional cases and their high-fugacity
behavior remained unclear. It was conjectured in \cite{MSS2019_square}
that in these cases there is a unique Gibbs measure at high fugacities.
As mentioned above, our study clarifies the case of the $2\times2$
hard-square model, refuting the conjecture in this case. Further discussion
is in subsection \ref{subsec:remarks on sliding}.

In three dimensions, the work \cite{mazel2021kepler} studies an infinite
family of values of $D$, discovering a rich set of possibilities
for the corresponding periodic configurations of maximal density and
drawing conclusions on the periodic and extremal Gibbs measures of
the model. The case $D=2$ corresponds to the packing of $2\times2\times2$
cubes with centers on $\Z^{3}$ and exhibits the sliding phenomenon.
It is conjectured in \cite{mazel2021kepler} that sliding leads to
the unicity of periodic and extremal Gibbs measures at high fugacities.
In subsection \ref{subsec:Cubes-and-rods} we discuss our predictions
for the hard-cube model on $\Z^{d}$.

Jauslin--Lebowitz \cite{jauslinCrystallineOrderingLarge2018,jauslinHighFugacityExpansionLee2018}
studied random packings of a (general) tile in $\R^{d}$ and its lattice
translates (i.e., the hard-core model on a discrete periodic graph
in $\R^{d}$). Their work also excludes sliding cases (including the
$2\times2$ hard-square model), by requiring that the specified tile
fully tiles $\R^{d}$ in a finite number of periodic and isometric
ways, and further limiting the ways in which defects may occur in
these periodic tilings. Using Pirogov--Sinai theory, they prove the
existence of high-fugacity extremal and periodic Gibbs states corresponding
to crystalline order according to the possible periodic tilings. They
further establish that the pressure and correlation functions have
expansions in powers of the inverse fugacity with a positive radius
of convergence.

\subsubsection{\label{subsec:disscusion_LC}Liquid crystals}

The term ``columnar order'' that we use originates in the study
of liquid crystals \cite{gennesPhysicsLiquidCrystals1993}. There,
one studies a material composed of molecules in three-dimensional
space and classifies its state according to the symmetries of its
structure. In a \emph{gas} or \emph{liquid} state, the molecules are
disordered in the sense that their distribution retains both the (continuous)
rotational and translational symmetries of $\mathbb{R}^{3}$. On the
other end of the spectrum are \emph{crystal} states, in which the
symmetry group is \emph{discrete}. Liquid crystals are ``intermediate''
states of matter, in which the symmetries of $\mathbb{R}^{3}$ are
partially broken. Three of the main categories of such states are
the \emph{nematic}, in which the rotational symmetry is broken while
the full $\mathbb{R}^{3}$ translational invariance is preserved,\emph{
smectic}, in which the rotational symmetry is broken and also the
translational symmetry is broken along \emph{one axis} (an $\mathbb{R}^{2}$
translational symmetry is retained) and \emph{columnar}, in which
the rotational symmetry is broken and also the translational symmetry
is broken along \emph{two axes} (an $\mathbb{R}$ translational symmetry
is retained). A seminal work in the physics of liquid crystals is
that of Onsager \cite{onsager1949effects}, who considered long, thin,
rod-like molecules in $\mathbb{R}^{3}$ with a pure hard-core interaction
(i.e., molecules are only constrained not to overlap) and predicted
a transition from a disordered to a nematic phase as the density of
the molecules increases.

In two dimensions, we use the term\emph{ nematic} to refer to a model
in which rotational symmetry is broken while translational symmetry
is retained, and the term \emph{columnar} to refer to a model in which
the rotational symmetry is broken while the translational symmetry
is broken only along a single axis. While the terminology of liquid
crystal phases was originally introduced in the continuum, it is also
used for lattice models with a similar meaning, classifying models
in terms of which of the lattice symmetries are broken. 

We are not aware of previous mathematically rigorous proofs of columnar
order in a hard-core model. Nematic order has been given rigorous
proof in several models, including the following: 
\begin{itemize}
\item Heilmann--Lieb \cite{heilmann1979lattice} established rotational
symmetry breaking for several lattice models using reflection positivity
methods and conjectured the absence of translational order.\\
Their models include a dimer model with attractive forces on the square
lattice \cite[Model I]{heilmann1979lattice} for which the conjecture
was recently established by Jauslin--Lieb \cite{jauslin2018nematic}
using Pirogov--Sinai methods, completing the proof of nematic order
(Jauslin--Lieb add that there is little doubt that similar proofs
could be devised for the other models in \cite{heilmann1979lattice}).
Alberici \cite{alberici2016cluster} studied the same model in the
case of non-equal horizontal and vertical dimer activities and proved
the absence of translational order using a cluster expansion.
\item Ioffe--Velenik--Zahradník \cite{ioffe2006entropy} established a
nematic phase for a system of horizontal and vertical rods on a square
lattice having unit width and varying lengths, with hard-core interactions
and specific length-dependent activities; in the case that all lengths
are allowed, they prove their result via an exact mapping to an Ising
model.
\item Disertori--Giuliani \cite{disertori2013nematic} considered rods
of unit width and fixed length $k$ on the square lattice with pure
hard-core interaction and proved that for large $k$, the system has
a nematic phase in an \emph{intermediate} density regime via coarse-graining
to an effective contour model and Pirogov--Sinai methods (see also
subsection \ref{subsec:Cubes-and-rods}).
\item Disertori--Giuliani--Jauslin \cite{disertori2020plate} considered
anisotropic plates with a finite number of allowed orientations in
the \emph{continuum} $\mathbb{R}^{3}$ with pure hard-core interaction
and established a nematic phase for an intermediate density regime.
\end{itemize}
We also mention the work of Bricmont--Kuroda--Lebowitz \cite[Concluding Remarks]{bricmont1984structure}
where, following Ruelle \cite{ruelle1971existence}, rotational-symmetry
breaking is proved in a system of zero-width rods in $\mathbb{R}^{2}$
with finitely many allowed orientations. Nematic order has further
been conjectured in several models, including the following: Abraham--Heilmann
\cite{abraham1980interacting} introduced a three-dimensional model
which extends the two-dimensional model of Heilmann--Lieb \cite[Model I]{heilmann1979lattice},
proved rotational symmetry breaking and conjectured the absence of
translational order. Angelescu--Zagrebnov \cite{angelescuLatticeModelLiquid1982}
and Zagrebnov \cite{zagrebnovLongrangeOrderLatticegas1996} studied
molecules on a lattice with an internal (``spin'') degree of freedom
with \emph{continuous} rotational symmetry. They showed, using a combination
of the infra-red bound and chessboard estimates, that the rotational
symmetry is broken at low temperature, and conjectured that a nematic
phase appears. We also mention the works \cite{biskupOrbitalOrderingTransitionMetal2005,nussinov2004orbital,biskupOrderDisorderOrder2004}
in which models with a continuum of ground states are analyzed, with
the conclusion being that the ground-state degeneracy is (partially)
lifted at low positive temperatures due to the different excitations
available to each degenerate configuration.

\subsection{Proof overview\label{subsec:Proof-overview}}

\subsubsection{Existence of multiple Gibbs measures}

In Part \ref{part:Existence-of-multiple}, the model is shown to admit
several Gibbs measures via an involved Peierls-type argument. The
keys to the argument are a ``coarse-grained identification'' of
vertically ordered and horizontally ordered regions, and a proof that
the resulting interfaces are rare. We proceed to describe these points,
starting with some key concepts that we introduce.

\begin{figure}[t]
\centering{}\includegraphics{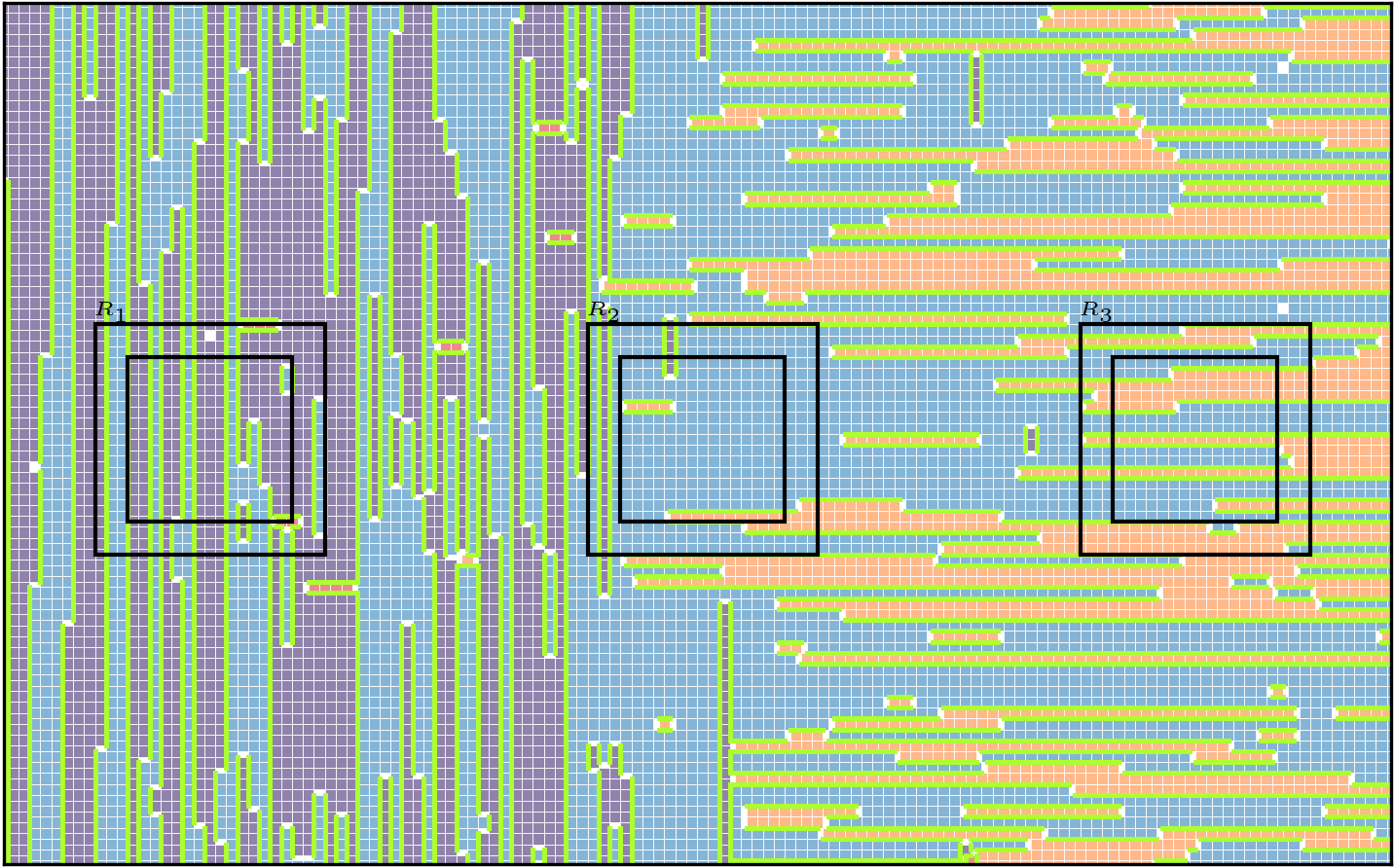}\caption{\label{fig:sticks}Sticks (green lines) in a configuration. On the
left there is an abundance of vertical sticks while on the right there
is an abundance of horizontal sticks. The interface region is not
crossed by long sticks (of either orientation), a feature which we
rely upon in order to prove that interface regions are rare.\protect \\
The rectangles $R_{1},R_{2},R_{3}$ are drawn with their concentric
$R^{-}$ rectangles (with $N=7$). The rectangles $R_{1}$ and $R_{3}$
are properly divided by vertical and horizontal sticks, respectively,
while $R_{2}$ is not properly divided.}
\end{figure}

\textbf{Sticks:} Tiles are classified to four types according to the
parities of the coordinates of their bottom left corner. A stick edge
of a configuration is an edge of $\mathbb{Z}^{2}$ which lies on the
boundary of two tiles of different type. A stick is a maximal path
of stick edges. It is easily seen that sticks are either horizontal
or vertical segments and that sticks of different orientation cannot
meet. The idea is that long vertical (horizontal) sticks should be
abundant in regions of vertical (horizontal) columnar order, while
the interfaces between differently ordered regions are not crossed
by long sticks (see Figure \ref{fig:sticks}).

\textbf{Properly-divided rectangles:} To classify regions into vertically
and horizontally ordered we consider the crossing of rectangles by
sticks. We call a rectangle $R$ divided, if there is a stick crossing
$R$ in the horizontal or vertical direction; importantly, a rectangle
cannot be divided \emph{both} horizontally and vertically. For technical
reasons, we further define $R$ to be properly divided if it is divided
by a stick which also divides $R^{-}$, a rectangle concentric to
$R$ having slightly smaller dimensions; specifically, we fix a large
integer $N$ (independent of $\lambda$), suppose the dimensions of
$R$ are integers divisible by $N$ and let the dimensions of $R^{-}$
be $\frac{N-2}{N}\width(R)\times\frac{N-2}{N}\height(R)$ (see Figure
\ref{fig:dividing}). Rectangles are thus classified into properly
divided vertically, properly divided horizontally or not properly
divided. The concept of properly-divided rectangles is only used with
squares when proving the existence of multiple Gibbs measures. Rectangular
$R$ are used in the proofs of our other results.

\textbf{Identification of interfaces:} Let $b(\lambda)$ be an (integer)
length scale, later chosen to satisfy (\ref{eq:mesoscopic length scale})
below. We consider the square grid $b(\lambda)\Z\times b(\lambda)\Z$,
and associate with each of its points a square of side length $Nb(\lambda)$
with its bottom-left corner at that point. These grid squares are
thus partially overlapping, with the amount of overlap chosen to ensure
the following property: two squares associated to neighboring positions
on the grid cannot be properly divided in distinct directions. Thus,
if grid squares are properly divided vertically in one region and
horizontally in another region then necessarily these regions are
separated by a ``contour'' of (points associated to) grid squares
that are \emph{not} properly divided.

\textbf{Multiple Gibbs measures from a Peierls-type argument:} Let
$\mu$ be a periodic Gibbs measure (at least one such measure exists
by compactness arguments). Our main technical lemma, Lemma \ref{lem:main},
implies that for suitably chosen $b(\lambda)$ and $N$, in samples
from $\mu$, long contours of grid squares that are not properly divided
are highly unlikely to occur at any given position (see next paragraph).
A union bound over contours then shows that, $\mu$-almost surely,
there is either an infinite connected component of grid squares that
are properly divided horizontally or an infinite connected component
of grid squares that are properly divided vertically, but not both.
Due to the lattice's $90^{\circ}$ rotational symmetry, this implies
the existence of at least two periodic Gibbs measures.

\textbf{The basic estimate: }Lemma \ref{lem:main} implies, for large
constant $N$, that rectangles~$R$ whose width and height are at
most $c_{0}\lambda^{1/2}$ satisfy 
\begin{equation}
\mathbb{\mu}(R\text{ is not properly divided)\ensuremath{\le e^{-c_{0}\vol(R)\lambda^{-1/2}}}}\label{eq:main technical bound}
\end{equation}

where $\mathbb{\mu}$ stands for any periodic Gibbs measure and $c_{0}>0$
is a small universal constant. The lemma moreover shows that this
estimate is multiplicative in the sense that the probability that
$n$ disjoint rectangles of the same dimensions (technically required
to have their corners on a shift of the grid $\width(R)\Z\times\height(R)\Z$)
are all not properly divided is at most the RHS of (\ref{eq:main technical bound})
raised to the power $n$. To use the estimate (\ref{eq:main technical bound})
in our Peierls-type argument we choose $b(\lambda)$ to be a \emph{mesoscopic}
length scale, satisfying
\begin{equation}
C\lambda^{1/4}<Nb(\lambda)<c_{0}\lambda^{1/2}\label{eq:mesoscopic length scale}
\end{equation}
for a large universal constant $C>0$. The fugacity $\lambda$ is
required to be large in order for this interval to be non-empty.
\begin{rem}
We offer some motivation for the form of the bound (\ref{eq:main technical bound}).
Recall that typical ordered vertical regions heuristically behave
as small perturbations of a ``union of 1D vertical systems'', i.e.,
samples from the measure $\mu_{(\ver,0)}^{\cup\text{1D}}$ (Figure
\ref{fig:from fully-packed to high fugacity}, middle panel) or its
shift to the right by one lattice site. Suppose we wanted to prove
(\ref{eq:main technical bound}) and its multiplicative version for
$\mu_{(\ver,0)}^{\cup\text{1D}}$. Let us show that there is a lower
bound on the probability which matches (the multiplicative version
of) (\ref{eq:main technical bound}) up to the constant in the exponent.
Indeed, Let $R_{1},...,R_{n}$ be rectangles of equal dimensions positioned
exactly one below the other (the top edge of $R_{i}$ is the bottom
edge of $R_{i-1}$). One way in which none of the $R_{i}$ will be
properly divided is if the configuration restricted to $\cup R_{i}$
is the ``square lattice configuration'' (defined by $\sigma(x,y)=\mathbf{1}_{x,y\equiv1\bmod2}$).
What is the probability of this event under the measure $\mu_{(\ver,0)}^{\cup\text{1D}}$?
Say that a face of $\Z^{2}$ is a vacancy if it is uncovered by the
tiles of the configuration. The event occurs when there are no vacancies
in $\cup R_{i}$ and all columns in the 1D systems ``enter $\cup R_{i}$
with the same phase''. Since vacancy pairs (as vacancies necessarily
come in pairs) are distributed roughly as a Poisson process with intensity
proportional to $\lambda^{-1/2}$ we conclude that the $\mu_{(\ver,0)}^{\cup\text{1D}}$
probability of the event is roughly $2^{-\frac{1}{2}\width(R_{1})}e^{-c'n\vol(R_{1})\lambda^{-1/2}}$
for some $c'>0$ (the first factor accounts for the phases). In comparison,
the upper bound resulting from (\ref{eq:main technical bound}) is
$e^{-c_{0}n\vol(R_{1})\lambda^{-1/2}}$, which has the same form as
$n\to\infty$.
\end{rem}
\textbf{Applying the chessboard estimate: }The first step in the proof
of the bound (\ref{eq:main technical bound}) is to apply the chessboard
estimate. This step is described with the following notation: Fix
a rectangle $\Lambda\subset\Z^{2}$ which we view as the domain. For
a configuration $\sigma\in\Omega$, define its weight in $\Lambda$
to be 
\begin{equation}
\w(\sigma):=\lambda^{-\frac{1}{4}\#\{\text{vacancies of \ensuremath{\sigma} in \ensuremath{\Lambda}}\}}.\label{eq:intro_weight_per}
\end{equation}
For a rectangle $S$ denote $\L^{S}:=\width(S)\Z\times\height(S)\Z$.
Given an event $E$ we write $Z_{\Lambda,\lambda}^{\per}(E)$ for
the sum of $\w(\sigma)$ over all $\sigma\in E$ with $\sigma$ periodic
with respect to $\Lambda$ (i.e. invariant to translations by elements
of $\L^{\Lambda}$). We also set $Z_{\Lambda,\lambda}^{\per}:=Z_{\Lambda,\lambda}^{\per}(\Omega)$.
Informally, the weight (\ref{eq:intro_weight_per}) is proportional
to the analog of (\ref{eq:intro_measure}) in periodic boundary conditions
(see also around (\ref{eq:weight of a configuration})).

The proof of (\ref{eq:main technical bound}) and its multiplicative
extension is reduced, via the chessboard estimate (or rather, its
infinite-volume extension in Section \ref{subsec:Infinite-volume chessboard}),
to showing that
\begin{equation}
\frac{Z_{\Lambda,\lambda}^{\per}(E_{R,R^{-}})}{Z_{\Lambda,\lambda}^{\per}}\le\left(e^{-c_{0}\vol(R)\lambda^{-1/2}}\right)^{\frac{\vol(\Lambda)}{\vol(R)}}\label{eq:chessboard bound}
\end{equation}
where $E_{R,R^{-}}$ is the ``disseminated version'' of the event
that $R$ is not properly divided, i.e., the event that all the translates
of $R$ by elements of $\L^{R}$ are not properly divided. The chessboard
estimate requires that the dimensions of $\Lambda$ are even integer
multiples of the corresponding dimensions of $R$.

The bound (\ref{eq:chessboard bound}) is proved via an upper bound
on the numerator and a lower bound on the denominator. We proceed
to describe these two bounds.

\textbf{Lower bound via 1D systems:} The lower bound on $Z_{\Lambda,\lambda}^{\per}$
is easy. It is obtained by restricting to configurations corresponding
to the ``union of 1D systems'' (i.e., all tiles have the same horizontal
parity), for which an explicit solution of a linear recurrence relation
(Proposition \ref{prop:1d periodic}) yields for some $c_{1}>0$ that
\begin{equation}
Z_{\Lambda,\lambda}^{\per}\ge e^{c_{1}\lambda^{-1/2}\vol(\Lambda)}.\label{eq:overview_1d}
\end{equation}

\textbf{Upper bound for the disseminated event:} The difficult part
lies in obtaining an upper bound on $Z_{\Lambda,\lambda}^{\per}(E_{R,R^{-}})$.
By several steps of simplification (see next paragraph), we pass to
bounding instead the value of $Z_{\Lambda,\lambda}^{1}(E_{M})$ which
is defined as follows. The event $E_{M}$ is the event that all sticks
are of length at most $M$, and for an event $E$, the value of $Z_{\Lambda,\lambda}^{1}(E)$
is the sum of weights of configurations in $E\cap\Omega_{\Lambda}^{\rho}$
(as defined in subsection \ref{subsec:the_model}) for the fully-packed
boundary condition $\rho(x,y)=\mathbf{1}_{x,y\equiv1\bmod2}$. The
value of $M$ is chosen to be $c_{2}\lambda^{1/2}$.

\textbf{Simplifications steps:} First, it is shown that the effect
of boundary conditions is insignificant for the purposes of (\ref{eq:chessboard bound})
(asymptotically as $\Lambda\uparrow\Z^{2}$), thus one may estimate
$Z_{\Lambda,\lambda}^{1}(E_{R,R^{-}})$ instead of $Z_{\Lambda,\lambda}^{\per}(E_{R,R^{-}})$.
Second, we introduce an event $E_{M,A}$, that requires the sticks
whose extension to a line crosses a translate of $R^{-}$ by an element
of $\L^{R}$ to have length at most $M$ (the notation $E_{M,A}$
is used for consistency with the main text; there $A$ is a set parameterizing
the event while here we define the event as a special case resulting
from a specific choice of $A$). One checks that 
\begin{equation}
E_{R,R^{-}}\subset E_{M,A}\text{ when }M\ge\max\{2\width(R),2\height(R)\}\label{eq:overview_Mcond}
\end{equation}
(this holds for the $R$ of (\ref{eq:main technical bound}) by choosing
$c_{0}<c_{2}/2$). Third and lastly, it remains to bound from above
$Z_{\Lambda,\lambda}^{1}(E_{M,A})$ in terms of $Z_{\Lambda,\lambda}^{1}(E_{M})$.
The proof of this bound is quite technical, however, the essential
idea is simply that the additional constraints imposed by $E_{M}$
on top of those of $E_{M,A}$ concern only a small fraction (approximately
$4/N$) of the area of $\Lambda$. Namely, they concern the complement
of the union of translates of $R^{-}$ by elements of $\L^{R}$.

Eventually, for the chosen value of $M$, the simplification steps
result in the following bound for large $\Lambda$:
\begin{equation}
Z_{\Lambda,\lambda}^{\per}(E_{R,R^{-}})\le e^{c_{3}\lambda^{-1/2}\vol(\Lambda)}Z_{\Lambda,\lambda}^{1}(E_{M}),\label{eq:overview_simplification}
\end{equation}
where for a fixed $c_{2}$, we have $c_{3}\to0$ as $N\to\infty$.

\textbf{Configuration without long sticks:} We are left with the more
essential task of bounding $Z_{\Lambda,\lambda}^{1}(E_{M})$. This
is achieved by our key Lemma \ref{lem:counting_argument}, which shows
that for some large $C_{4}>0$, if $M<\lambda^{1/2}/C_{4}$ then
\begin{equation}
Z_{\Lambda,\lambda}^{1}(E_{M})\le\left(1+\frac{C_{4}M}{\lambda}\right)^{^{\vol(\Lambda)}}.\label{eq:overview_short sticks}
\end{equation}

The proof is via direct combinatorial counting arguments. A first
observation is that sticks must end at vacant faces. Therefore, configurations
contain ``connected components'' composed of sticks and vacancies
together. We consider all possibilities for such connected components,
up to translations. Then, the proof of (\ref{eq:overview_short sticks})
reduces to suitably bounding the sum $\sum_{H}\lambda^{-\frac{1}{4}v_{H}}$
where $H$ ranges over all those possibilities in which all sticks
have length at most $M$ and $v_{H}$ denotes the number of vacancies
in $H$. To this end, we define $k_{H}$, a quantity satisfying that
$k_{H}-2$ is the number of ``degrees of freedom'' one has for extending
and contracting the sticks of $H$ (this is generally less than the
number of sticks in $H$ since following the sticks in a cycle of
$H$ must lead back to the starting point). Geometric considerations
lead to the bound $v_{H}\ge\max\{2(k_{H}-1),4\}$. This supplies the
necessary control for the requisite bound on the above sum.

\textbf{Conclusion of the basic estimate:} As mentioned above, the
basic estimate (\ref{eq:main technical bound}) follows from (\ref{eq:chessboard bound}).
The latter bound, on the probability of $E_{R,R^{-}}$, then follows
by combining (\ref{eq:overview_1d}), (\ref{eq:overview_simplification})
and (\ref{eq:overview_short sticks}), under the assumption that 
\[
c_{1}-c_{3}-C_{4}c_{2}>c_{0}>0.
\]
For (\ref{eq:overview_simplification}), we also require that $c_{0}<c_{2}/2$
so that the condition of (\ref{eq:overview_Mcond}) holds.

To satisfy this, the constants are chosen in the following order:
First, $c_{1}$ and $C_{4}$ are fixed. Then we choose $c_{2}=c_{1}/(2C_{4})$,
and subsequently choose $N$ so that $c_{3}<c_{1}/2$. This allows
to choose $c_{0}$ satisfying the two inequalities of the previous
paragraph.

\subsubsection{Fine properties of Gibbs measures and the characterization of periodic
Gibbs measures}

In Part \ref{part:Characterization-of-the}, we prove the fine properties
and characterization results stated in Theorem \ref{thm:main} and
Theorem \ref{thm:main2}. We briefly describe our proofs here.

Let $\L$ be a sufficiently sparse full-rank lattice $\L$. The previous
arguments imply that every $\L$-ergodic Gibbs measure satisfies either
that most long sticks are vertical (``$\ver$'' measure) or most
long sticks are horizontal (``$\hor$'' measure). The next step
is to refine this classification, proving that $\L$-ergodic Gibbs
measures come in exactly one of four ``phases'' $(\ver,0),(\ver,1),(\hor,0),(\hor,1)$
according to the orientation of most long sticks and their parity
(the parity of a vertical stick is the parity of its $x$-coordinate
and the parity of a horizontal stick is the parity of its $y$-coordinate). 

First, we use an inductive procedure on length scales, employing a
Peierls-type argument driven by the basic estimate (\ref{eq:main technical bound})
in each step, showing that for ``$\ver$'' measures, even very thin
rectangles, of dimensions $C\times c\sqrt{\lambda}$ for a large universal
$C>0$ and small universal $c>0$, are typically properly divided
vertically. Via an additional Peierls-type argument, this allows the
further classification into the phases $(\ver,0)$ and $(\ver,$1)
by noting that if two long vertical sticks of opposite parities are
near each other, then there is a long rectangle bounded between them
which must contain an atypically large number of vacancies (tail bounds
for the number of vacancies are readily obtained by the infinite-volume
version of the chessboard estimate). Classification of ``$\hor$''
measures is analogous.

The fact that there exist exactly four $\L$-ergodic measures, which
are furthermore extremal, along with quantitative decay of correlation
estimates and precise invariance properties, is achieved by the method
of disagreement percolation \cite{cmp/1104252313,VANDENBERG1994179}
(see Theorem \ref{thm:disagreement percolation}): Let $\mu,\mu'$
be Gibbs measures and let $\sigma,\sigma'$ be two \emph{independent}
samples from $\mu$ and $\mu'$, respectively. If, almost surely,
there is no infinite path where $\sigma,\sigma'$ disagree then $\mu=\mu'$
and this common measure is extremal. Moreover, decay of correlation
estimates are obtained by bounding the probability that disagreement
paths connect distant vertices. This reduces our task to that of controlling
long disagreement paths between independent samples from $\L$-ergodic
measures of the same phase. A key fact is that in a one-dimensional
system, disagreement paths must terminate at the first vacancy (in
any of the two configurations). This allows to control the length
of disagreement paths in regions where both configurations have their
tiles arranged in columns (or rows) of the same parity. Other regions
are rare, with suitable quantitative control, by the assumption that
the measures have the same phase and by the thin rectangle crossing
results of the previous paragraph.

Lastly, the proof of the estimate $\mu_{(\ver,0)}(\sigma(x,y)=1)=O(\lambda^{-1})$
for even $x$ (part of the columnar order property (\ref{eq:columnar order}))
relies on the fact that between two vertical sticks of even parity,
if there is some tile with an even $x$-coordinate of its center then
there are at least four vacancies in its row between the two sticks.
The other parts of (\ref{eq:columnar order}) are relatively simple
consequences of the previously-obtained information (Section \ref{subsec:tile_probabilities}).

\subsection{Reader's guide}

The fundamental task of proving that the $2\times2$ hard-square
model admits multiple Gibbs measures at high fugacity is achieved
in Part \ref{part:Existence-of-multiple}. Section \ref{sec:preliminaries}
contains the basic definitions used throughout and some simple estimates
on the effect of boundary conditions. Section \ref{sec:Chessboard-estimates}
establishes reflection positivity of the $2\times2$ hard-square model,
presents the chessboard estimate in finite volume and extends its
applicability to infinite volume. Section \ref{sec:main_lemma} introduces
the notion of sticks and proves that mesoscopic rectangles are typically
divided by sticks. This fact is then used in Section \ref{sec:multiple_gibbs}
to derive the existence of multiple Gibbs measures via a Peierls-type
argument.

Part \ref{part:Characterization-of-the} is devoted to proving the
existence of a Gibbs measure with the properties stated in Theorem
\ref{thm:main}, and proving the characterization of periodic Gibbs
measures stated in Theorem \ref{thm:main2}. Section \ref{sec:Peierls-type-arguments}
sets up a convenient framework for Peierls-type arguments. Classification
of periodic-ergodic Gibbs measures into four phases is established
in Section \ref{sec:Four-phases}. The disagreement percolation method
is introduced in Section \ref{sec:decay}, where it is used to prove
extremality of the periodic-ergodic Gibbs measures and to bound their
rate of correlation decay, as well as to characterize the periodic
Gibbs measures. Lastly, columnar order, as well as additional correlation
decay estimates, are established in Section \ref{sec:Columnar-order}.

Part \ref{part:Discussion} is devoted to further discussion and open
questions.

\subsection{Acknowledgments}

This research was supported in part by Israel Science Foundation grants
861/15 and 1971/19 and by the European Research Council starting grant
678520 (LocalOrder) and consolidator grant 101002733 (Transitions).
We are grateful to Izabella Stuhl and Yuri Suhov for initial discussions
on lattice packings. We thank Omer Angel, Thomas Bohman, Nishant Chandgotia,
Aernout van Enter, Ohad Feldheim and Yinon Spinka for helpful comments
and discussions. We also thank an annonymous referee for their careful
reading and thoughtful comments.\pagebreak{}

\part{\label{part:Existence-of-multiple}Existence of multiple Gibbs measures}

In this part we prove the existence of multiple Gibbs measures for
the $2\times2$ hard-square model on the square lattice $\Z^{2}$
at high fugacity (Corollary \ref{cor:multiple Gibbs measures}). The
results of this part will further be instrumental in the second part,
where we prove the more refined results stated in the introduction.

\section{Preliminaries\label{sec:preliminaries}}

\subsection{Basic definitions\label{subsec:Basic-definitions}}

In this section we provide some of the basic definitions used throughout
the paper. The definitions from the introduction are repeated, sometimes
in a different (but equivalent) formulation.

\textbf{Elementary objects:} We use the convention $\mathbb{N}:=\{1,2,\ldots\}$.
We use the standard coordinate system in $\mathbb{R}^{2}$\marginpar{$\nat$,$\R$}
where the $x$ axis points to the right and the $y$ axis points upwards.

In this paper, the term rectangle refers to an axis parallel rectangle
with corners in integer coordinates, formally viewed as \emph{closed}
set. For integers $x,y\in\Z$ and positive integers $K,L\in\nat$,
define $\rect{K\times L}{(x,y)}\coloneqq[x,x+K]\times[y,y+L]\subset\R^{2}$\marginpar{$\rect{K\times L}{(x,y)}$,

$\width$,$\height$,

$\perim$,$\vol$,

$\partial R$}, that is, the rectangle with its bottom left corner at $(x,y)$ and
with side lengths $K$ and $L$. For a rectangle $R=\rect{K\times L}{(x,y)}$
denote $\width(R)=K$, $\height(R)=L$, $\perim(R)=2K+2L$ and $\vol(R)=KL$
(but note that $R$ contains $(K+1)(L+1)$ points of $\Z^{2}$). When
$K,L,x,y$ are even we say that $R$ is an \textbf{even rectangle}.
We also use the shorthand $\frect{K\times L}\coloneqq R_{K\times L,(0,0)}$.
The boundary of $R$ as a set in $\R^{2}$ is denoted $\partial R$.

For\marginpar{$f\rstr_{C}$, $\mathbf{1}_{A}$} sets $A,B,C$ and a
function $f:A\to B$ we denote by $f\rstr_{C}$ the restriction of
$f$ to $A\cap C$. We denote the indicator function of a set $A$
by $\mathbf{1}_{A}$.

We\marginpar{$\V,\E_{\square},\mathbb{F},\E_{\boxtimes}$} consider
two graphs having the vertex set $\mathbb{V}\coloneqq\mathbb{Z}^{2}$
(we will use these two notations interchangeably). The \textbf{nearest
neighbors graph} is $(\V,\E_{\square})$ where $\E_{\square}=\{uv:u,v\in\V,\left\Vert v-u\right\Vert _{1}=1\}$
is the set of edges connecting nearest neighbors in $\Z^{2}$. This
is a planar graph, and we consider its edges to be embedded in $\R^{2}$
as line segments. Therefore its \textbf{faces} may be thought of as
$1\times1$ squares. We define $\mathbb{F}\coloneqq\{R_{1\times1,v}:v\in\V\}$.
We also consider the \textbf{nearest and next-to-nearest neighbors
graph} $(\V,\E_{\boxtimes})$ where $\E_{\boxtimes}=\{uv:u,v\in\V,\left\Vert v-u\right\Vert _{\infty}=1\}$.
When discussing vertices in $\V$ we use terms such as ``$\square$-adjacent''
and ``$\boxtimes$-connected component'' with the obvious meaning.

With\marginpar{$T_{(x,y)}$} each vertex $(x,y)$ is associated a
tile that is a $2\times2$ square centered at that vertex: $T_{(x,y)}\coloneqq R_{2\times2,(x-1,y-1)}$.
We define the \textbf{parity} of a tile centered at $(x,y)$ to be
$(x-1\bmod2,y-1\bmod2)$, so there are $4$ possible parities for
a tile. The first component of the parity of a tile is termed its
\textbf{horizontal parity} while the second component is termed its
\textbf{vertical parity}.

\textbf{Configuration spaces:}\marginpar{$\sigma,\Omega$\\
$\intr{\cdot}$} We think of hard square configurations as sets of tiles whose interiors
are pairwise disjoint. Formally, a configuration is represented by
a function $\sigma:\V\to\{0,1\}$, with the value $1$ corresponding
to centers of tiles, so that the space of configurations is
\[
\Omega\coloneqq\{\sigma\in\{0,1\}^{\V}:\sigma(u)=\sigma(v)=1\implies\text{int}(T_{v})\cap\text{int}(T_{u})=\emptyset\}
\]
where $\mathrm{int}(\cdot)$ stands for the interior of a set. The
space $\Omega$ is equipped with the standard Borel measurable structure
(induced by the product topology).

Let\marginpar{$\Lambda$} be $\Lambda$ a rectangle. We will work
with several restricted sets of configurations, corresponding to different
choices of boundary conditions for $\Lambda$:
\begin{itemize}
\item Define the set of $\Lambda$-periodic configurations as\marginpar{$\Omega_{\Lambda}^{\per}$}
\begin{align*}
\Omega_{\Lambda}^{\per}\coloneqq\{\sigma\in\Omega:\forall v\in\V,\,\sigma\left(v\right) & =\\
\sigma\left(v+(\width(\Lambda),0)\right) & =\sigma\left(v+(0,\height(\Lambda))\right)\}.
\end{align*}
We emphasize that only the dimensions of $\Lambda$ enter into the
definition of $\Omega_{\Lambda}^{\per}$. To stress this point, we
will often use the notation $\Lambda=\frect{K\times L}$ (omitting
the corner position) when working with periodic boundary conditions.
\item Let $\rho\in\Omega$, and define the set of configurations with $\rho$-boundary
conditions outside $\Lambda$ to be\marginpar{$\Omega_{\Lambda}^{\rho}$}
\begin{equation}
\Omega_{\Lambda}^{\rho}\coloneqq\{\sigma\in\Omega:\forall v\in\V\setminus\mathrm{\intr{\Lambda}},\,\rho(v)=\sigma(v)\}.\label{eq:Omega Lambda rho}
\end{equation}
\item We\marginpar{$\Omega_{\Lambda}^{0},\Omega_{\Lambda}^{1}$} give names
to two special cases of this definition. We write $\Omega_{\Lambda}^{0}$
for the case that $\rho$ is identically $0$ and call $\Omega_{\Lambda}^{0}$
the set of configurations with \textbf{free boundary conditions} outside
$\Lambda$. When $\Lambda$ is an \emph{even} rectangle, we also write
$\Omega_{\Lambda}^{1}$ for the case that $\rho(x,y)=\mathbf{1}_{x,y=1\bmod2}$
and call $\Omega_{\Lambda}^{1}$ the set of configurations with \textbf{fully-packed
boundary conditions} outside $\Lambda$.\textbf{ }We point out that
for an even rectangle $\Lambda$, free and fully-packed boundary conditions,
realized by $\Omega_{\Lambda}^{0}$ and $\Omega_{\Lambda}^{1}$, induce
the same set of feasible configurations in $\Lambda$ (and thus the
same partition function and measure, according to the definitions
below), but the distinction between them will be convenient in Section
\ref{sec:main_lemma}.
\begin{figure}[t]
\includegraphics{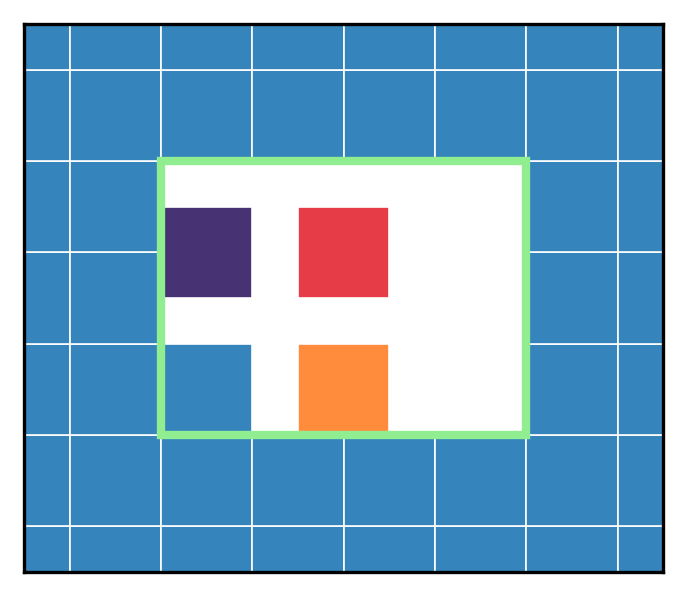}\hspace*{\fill}\includegraphics{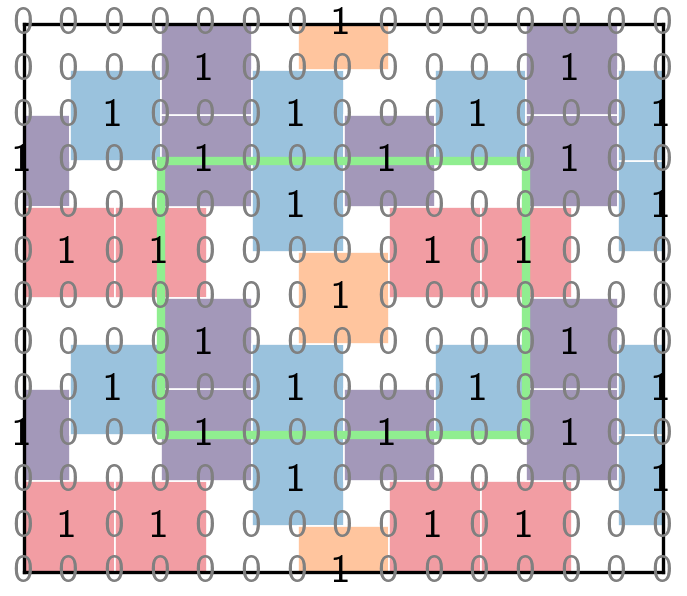}

\caption{\label{fig:configs}Two configurations in $\Omega$. The boundary
of the even rectangle $\Lambda=\protect\rect{8\times6}{(0,0)}$ is
shown in green.\protect \\
On the left: a configuration in $\sigma\in\Omega_{\Lambda}^{1}$.
The tiles inside $\Lambda$ are $T_{(1,1)},T_{(1,4)},T_{(4,1)},T_{(4,4)}$,
of parities $(0,0),(0,1),(1,0),(1,1)$ respectively, and colored blue,
orange, deep blue, red, respectively. Throughout the paper, we color
tiles according to their parities in this way.\protect \\
On the right: a configuration $\sigma\in\Omega_{\Lambda}^{\protect\per}$.
The values of $\sigma$ appear centered on points of $\protect\Z^{2}$,
and the corresponding tiles are in the background. }

\end{figure}
\end{itemize}
For a set $\Lambda$ and a measurable function $f:\Omega\to\R$, say
that $f$ is $\Lambda$-local if
\begin{equation}
\sigma\rstr_{\Lambda}=\sigma'\rstr_{\Lambda}\implies f(\sigma)=f(\sigma'),\quad\forall\sigma,\sigma'\in\Omega.\label{eq:local definition}
\end{equation}
An event $E$ is called $\Lambda$-local if $\indic E$ is $\Lambda$-local.

\textbf{Measures:} Given a configuration $\sigma\in\Omega$, and a
face $f\in\mathbb{F}$. We say that $f$ is a \textbf{vacancy} (or
that $f$ is vacant), if it is not contained in a tile of $\sigma$.
Otherwise we say that it is \textbf{occupied}.

Let $\lambda>0$\marginpar{$\lambda$} denote \textbf{fugacity}. Informally,
in the hard squares model, the probability of a configuration is proportional
to $\lambda^{n}$ where $n$ is the number of tiles. Equivalently,
it is proportional to $\lambda^{-v/4}$ where $v$ is the number of
vacancies. Formally, for a rectangle $\Lambda$, we define the weight
of a configuration according to the number of vacant faces as follows:\marginpar{$\w(\sigma)$}
\begin{equation}
\w(\sigma)\coloneqq\lambda^{-\frac{1}{4}\#\{f\in\mathbb{F}:f\subset\Lambda\text{ and }f\text{ is vacant in \ensuremath{\sigma}}\}}.\label{eq:weight of a configuration}
\end{equation}

For $^{*}$ denoting either $^{\per}$ or $^{\rho}$, define the \textbf{hard-square
Gibbs measure $\prob_{\Lambda,\lambda}^{*}$,} at fugacity $\lambda$
in the finite volume $\Lambda$ with boundary conditions $^{*}$,\textbf{
}as the measure on $\Omega_{\Lambda}^{*}$ assigning probability\marginpar{$\prob_{\Lambda,\lambda}^{\per}$, $\prob_{\Lambda,\lambda}^{\rho}$}
\[
\prob_{\Lambda,\lambda}^{*}(\sigma)=\frac{\w(\sigma)}{Z_{\Lambda,\lambda}^{*}}
\]
to each configuration $\sigma$, where $Z_{\Lambda,\lambda}^{*}\coloneqq\sum_{\sigma\in\Omega_{\Lambda}^{*}}\w(\sigma)$
is the partition function. The assumption that $\Lambda$ is a rectangle
ensures that this coincides with the definitions (\ref{eq:intro_measure})
and (\ref{eq:intro_weight_per}) given in the introduction. It is
convenient to further define the weight of an event $E\subset\Omega$
under the boundary conditions $^{*}$ to be\marginpar{$Z_{\Lambda,\lambda}^{*}(E)$}
\begin{equation}
Z_{\Lambda,\lambda}^{*}(E):=\sum_{\sigma\in E\cap\Omega_{\Lambda}^{*}}\w(\sigma),\label{eq:weight of an event}
\end{equation}

so that $Z_{\Lambda,\lambda}^{*}=Z_{\Lambda,\lambda}^{*}(\Omega)$
and $\prob_{\Lambda,\lambda}^{*}(E)=Z_{\Lambda,\lambda}^{*}(E)/Z_{\Lambda,\lambda}^{*}$.
Throughout the text $\lambda$ will denote the fugacity and we will
often omit it from the notation.

A measure $\prob$ on $\Omega$ with the natural sigma-algebra, is
said to be an \textbf{infinite volume Gibbs measure} if for every
rectangle $\Lambda$ and measurable function $f$, it holds almost
surely for $\sigma$ sampled from $\mu$ that
\begin{equation}
\mu(f\,|\,\sigma\rstr{}_{\V\setminus\intr{\Lambda}})=\mu_{\lambda,\Lambda}^{\sigma}(f).\label{eq:DLR}
\end{equation}
Note that while we only use rectangular domains in this definition,
the resulting infinite volume Gibbs measures are the same as the ones
defined after (\ref{eq:intro_measure}) in the introduction.

\textbf{Transformations:} Let $\eta:\V\to\V$, (usually $\eta$ will
be a restriction of an isometry of $\R^{2}$). For $\sigma\in\Omega$,
define $\eta\sigma\coloneqq\sigma\circ\eta$. For a function $f:\Omega\to\R$,
define $\eta f$ by 
\begin{equation}
\eta f(\sigma)=f(\eta\sigma)=f(\sigma\circ\eta).\label{eq:func_shift}
\end{equation}
Analogously, for an event $E\subset\Omega$ define $\eta E\coloneqq\{\sigma\in\Omega:\eta\sigma\in E\}$.
For a measure $\mu$ of $\Omega$, define $\eta\mu$ by $\eta\mu(E)=\mu(\eta E)$.

For\marginpar{$\eta_{v}$} $v\in\V$, define $\eta_{v}:\V\to\V$ by
$\eta_{v}(u)=u+v$.

\textbf{Constant notation:} Many of our claims introduce constants
in phrasings such as ``There is $c>0$ such that ...''. For clarity,
when referring to such constants at later parts of the argument we
add the number of the claim as a subscript (e.g.~$c_{4.4}$).

\subsection{Comparison of boundary conditions}

Fix a fugacity parameter $\lambda>0$. The goal of this section is
to bound the effects of boundary conditions on the expectation value
of observables. The approach is standard, though some care is needed
due to the hard constraints inherent in the model. We present two
bounds of this type, applicable in slightly different settings.
\begin{prop}
\label{prop:boundary_fixing}Let $\Lambda$ be a rectangle and let
$\rho\in\Omega$. Define $m^{\rho,\Lambda}:\Omega\to\Omega_{\Lambda}^{\rho}$
\marginpar{$m^{\rho,\Lambda}$}as follows: $m^{\rho,\Lambda}(\sigma)$
be obtained from $\sigma$ by first setting $\sigma(v)$ to $\rho(v)$
for every $v\in\V\setminus\mathrm{\intr{\Lambda}}$, and then removing
any tile that has its center in $\intr{\Lambda}$ and overlaps with
another tile.

Let $E\subset\Omega$. Then
\[
Z_{\Lambda}^{*}(E)\le C(\lambda)^{\perim(\Lambda)}Z_{\Lambda}^{\rho}(m^{\rho,\Lambda}(E))
\]
where $C(\lambda)$ depends only on $\lambda$ and $*$ stands for
either a configuration in $\Omega$ or for the symbol $\per$.
\end{prop}
\begin{proof}
For $\sigma'\in m^{\rho,\Lambda}(E)$ it holds that 
\begin{equation}
\#\{\sigma\in E\cap\Omega_{\Lambda}^{*}:m^{\rho,\Lambda}(\sigma)=\sigma'\}\le2^{\frac{3}{2}\perim(\Lambda)}.\label{eq:bound on preimage}
\end{equation}
Indeed, if $*$ is a configuration in $\Omega$ then all $\sigma$
in this set are identical except on the points of $\intr{\Lambda}\cap\V$
which are adjacent to $\partial\Lambda$. If $*=\per$, configurations
in this set coincide on all points of $\intr{\Lambda}\cap\V$ which
are not adjacent to $\partial\Lambda$, which implies (\ref{eq:bound on preimage})
when taking into account the periodicity constraint.

Additionally, for any $\sigma\in\Omega$,
\[
\w(\sigma)\le\max\{\lambda,\lambda^{-1}\}^{\frac{1}{4}\perim(\Lambda)}\w(m^{\rho,\Lambda}(\sigma)).
\]

Thus
\begin{align*}
Z_{\Lambda}^{*}(E) & =\sum_{\sigma\in E\cap\Omega_{\Lambda}^{*}}\w(\sigma)=\sum_{\sigma'\in m^{\rho,\Lambda}(E)}\sum_{\substack{\sigma\in E\cap\Omega_{\Lambda}^{*}\\
m^{\rho,\Lambda}(\sigma)=\sigma'
}
}\w(\sigma)\\
 & \le\sum_{\sigma'\in m^{\rho,\Lambda}(E)}2^{\frac{3}{2}\perim(\Lambda)}\max\{\lambda,\lambda^{-1}\}^{\frac{1}{4}\perim(\Lambda)}\w(\sigma')\\
 & =C(\lambda)^{\perim(\Lambda)}Z_{\Lambda}^{\rho}(m^{\rho,\Lambda}(E))
\end{align*}
where the last equality is since $m^{\rho,\Lambda}(E)\subset\Omega_{\Lambda}^{\rho}.$
\end{proof}
\begin{prop}
\label{prop:boundary}Let $\Lambda'\subset\Lambda$ be rectangles
and further assume that the Euclidean distance from $\Lambda'$ to
$\R^{2}\setminus\Lambda$ is at least $2$. Let $f:\Omega\to[0,\infty)$
be a $\Lambda'$-local function. Then there is $C(\lambda)$ (depending
only on $\lambda$) such that
\[
\mu_{\Lambda'}(f)\le C(\lambda)^{\perim(\Lambda)}\mu_{\Lambda}(f)
\]
where $\mu_{\Lambda}$ may stand for $\mu_{\Lambda}^{\per}$, $\mu_{\Lambda}^{\rho}$
for some $\rho\in\Omega$, or an infinite-volume Gibbs measure $\mu$,
and, independently, $\mu_{\Lambda'}$ may stand for $\mu_{\Lambda'}^{\per}$,
$\mu_{\Lambda'}^{\rho'}$ for some $\rho'\in\Omega$, or an infinite-volume
Gibbs measure $\mu'$.
\end{prop}
\begin{proof}
By the DLR condition (\ref{eq:DLR}), we have $\mu(f)=\mu(\mu_{\Lambda}^{\sigma}(f))$.
Similarly, $\mu_{\Lambda}^{\per}=\mu_{\Lambda}^{\per}(\mu_{\Lambda}^{\sigma}(f))$.
The analogous equalities hold with $\Lambda'$ instead of $\Lambda$.
Therefore it suffices to prove for every $\rho,\rho'\in\Omega$ that
\begin{equation}
\mu_{\Lambda'}^{\rho'}(f)\le C(\lambda)^{\perim(\Lambda)}\mu_{\Lambda}^{\rho}(f).\label{eq:boundary comparison inequality}
\end{equation}

Let $E$ be the event $\{\sigma:\sigma\rstr_{\partial\Lambda'}=\rho'\rstr_{\partial\Lambda'}\}$.
The fact that $f$ is $\Lambda'$-local implies that

\[
\mu_{\Lambda'}^{\rho'}(f)=\mu_{\Lambda}^{\rho}(f\,|\,E).
\]

We conclude that

\[
\mu_{\Lambda'}^{\rho'}(f)=\frac{\mu_{\Lambda}^{\rho}(f\cdot\indic E)}{\mu_{\Lambda}^{\rho}(E)}\le\frac{\mu_{\Lambda}^{\rho}(f)}{\mu_{\Lambda}^{\rho}(E)}
\]

and thus (\ref{eq:boundary comparison inequality}) will follow from
showing that $\mu_{\Lambda}^{\rho}(E)\ge C(\lambda)^{-\perim(\Lambda)}$.
The latter inequality is a simple consequence of the fact that the
Euclidean distance from $\Lambda'$ to $\R^{2}\setminus\Lambda$ is
at least $2$ (proved similarly to Proposition \ref{prop:boundary_fixing}).
\end{proof}

\section{Chessboard estimates\label{sec:Chessboard-estimates}}

In this section we state the chessboard estimate for the $2\times2$
hard-square model, after setting up the necessary definitions. We
do not give a proof of the chessboard estimate and refer to \cite{frohlichPhaseTransitionsReflection1978},
\cite{shlosman1986method}, \cite{biskup2009reflection}, \cite[Chapter 10]{friedli_velenik_2017},
\cite[Section 2.7.1]{PeledSpinka2019}, \cite{hadas2022chessboard}
for pedagogical references.

Additionally we prove a version of the chessboard estimate applicable
to periodic infinite-volume Gibbs measures.

We use the following definitions for a rectangle $R=\rect{K\times L}{(x_{0},y_{0})}$: 
\begin{itemize}
\item Define the grid of $R$ and its origin-shifted version:\marginpar{$\grid[\null],\mathcal{\rectlat}^{R}$}
\begin{equation}
\grid[\null]\coloneqq(x_{0}+K\Z)\times(y_{0}+L\Z)\quad\text{and}\quad\mathcal{\rectlat}^{R}:=K\Z\times L\Z.\label{eq:G^R and L^R}
\end{equation}
\item Let $\refls[\null]$\marginpar{$\refls[\null]$} denote the group
generated by reflections of $\R^{2}$ through the horizontal and vertical
lines that intersect with $\grid[\null]$. Precisely, $\refls[\null]$
is the set of $\tau:\R^{2}\to\R^{2}$ satisfying, for some $m,n\in\Z$,
that
\begin{align*}
\text{either} &  &  & \tau(x,y)_{1}=2(x_{0}+mK)-x &  & \text{or} &  & \tau(x,y)_{1}=x+2mK, & \text{and}\\
\text{either} &  &  & \tau(x,y)_{2}=2(y_{0}+nL)-y &  & \text{or} &  & \tau(x,y)_{2}=y+2nL.
\end{align*}
\item Importantly\marginpar{$\tau_{R,v}$}, for each $v\in\grid[\null]$,
there is a unique isometry in $\refls[\null]$ which maps $R$ to
$\rect{K\times L}v$; we denote this isometry by $\tau_{R,v}$.
\item For $f:\Omega\to\R$ and $\tau\in\refls[\null]$ recall that $\tau f$
is defined by (\ref{eq:func_shift}) where $\tau$ is implicitly restricted
to $\Z^{2}$.
\item Recall that an $R$-local function (or event) is defined by (\ref{eq:local definition}).
It will be essential that $R$ is \emph{closed}.
\end{itemize}

\subsection{Finite volume}

Throughout this subsection we fix a rectangle $\Lambda$ and derive
properties of the measure $\mu_{\Lambda}^{\per}$. We remind the reader
that $\mu_{\Lambda}^{\per}$ is supported on $\Lambda$-periodic configurations
(i.e., $\sigma\in\Omega_{\Lambda}^{\per}$) and though we stick to
our convention of regarding configurations as defined on the infinite
lattice $\Z^{2}$, the reader should keep in mind that configurations
$\sigma\in\Omega_{\Lambda}^{\per}$ are naturally defined on the torus
$\Z^{2}/\mathcal{\rectlat}^{\Lambda}$.

We start in Section \ref{subsec:Reflection-positivity} by showing
that $\mu_{\Lambda}^{\per}$ has reflection positivity with respect
to reflection lines passing through vertices in $\V$. This is a standard
consequence of the fact that the model has only nearest-neighbor and
next-nearest-neighbor interactions (i.e., interactions involving only
the $8$ nearest vertices). We continue in Section \ref{subsec:The-chessboard-seminorm}
to derive the chessboard estimate for $\mu_{\Lambda}^{\per}$ ---
a standard consequence of reflection positivity. We give the name
chessboard seminorm (see (\ref{eq:chessboard norm def})) to the function
$\zedd[][][\cdot]$ which appears on the right-hand side of the chessboard
estimate (see also \cite[equation (5.47)]{biskup2009reflection})
and derive some of its basic properties.

\subsubsection{\label{subsec:Reflection-positivity}Reflection positivity}

In this subsubsection we establish the basic reflection positivity
property of the $2\times2$ hard-square model.
\begin{lem}
[Reflection positivity]\label{lemma:reflection_positivity} Let
$R=\rect{K\times L}{(x_{0},y_{0})}$ be a rectangle and let $f$ be
an $R$-local function. Then
\[
\mu_{\Lambda}^{\per}\left(f\cdot\tau f\right)\ge0
\]
when either $\Lambda=\frect{2K\times L}$ and $\tau$ is the reflection
$\tau_{R,(x_{0}+K,y_{0})}$, or $\Lambda=\frect{K\times2L}$ and $\tau$
is the reflection $\tau_{R,(x_{0},y_{0}+L)}$.
\end{lem}
\begin{proof}
We prove only the first case as the second one is analogous. Denote
$R_{1}\coloneqq\rect{K\times L}{(x_{0}+K,y_{0})}$ and observe that
$\tau f$ is an $R_{1}$-local function. Let $\mathcal{F}$ be the
sigma algebra generated by the $\sigma(x,y)$ with $x\equiv x_{0}\pmod K$.
The fact that the model has only nearest- and next-nearest-neighbor
interactions, which are also symmetric, implies that $f$ and $\tau f$
are independent and identically distributed under $\mu_{\Lambda}^{\per}$
conditioned on $\mathcal{\mathcal{F}}$. Thus,
\[
\mu_{\Lambda}^{\per}\left(f\cdot\tau f\,|\,\mathcal{F}\right)=\left(\mu_{\Lambda}^{\per}\left(f\,|\,\mathcal{F}\right)\right)^{2}\ge0
\]

and taking expectations of both sides concludes the proof.
\end{proof}

\subsubsection{\label{subsec:The-chessboard-seminorm}The chessboard seminorm and
the chessboard estimate}

In this subsubsection we discuss the chessboard estimate for the measure
$\mu_{\Lambda}^{\per}$.

We say that $R$\textbf{ is a block of} $\Lambda$ when $R$ and $\Lambda$
are rectangles satisfying that $2\width(R)$ divides $\width(\Lambda)$
and $2\height(R)$ divides $\height(\Lambda)$. In this case we make
the following definitions, that depend on $R$ and on the dimensions
of~$\Lambda$:
\begin{itemize}
\item Set\marginpar{$\refls$} $\refls:=\refls[\null]/\mathcal{L}^{\Lambda}$,
i.e., the quotient of the group $\refls[\null]$ by the group of translations
by vectors of $\mathcal{L}^{\Lambda}$. Our assumption that $R$ is
a block of $\Lambda$ implies that the latter group is indeed a subgroup
of the former. Note that $\#\refls=\frac{\vol(\Lambda)}{\vol(R)}$.
\item We observe that while an element $\tau\in\refls$ is formally an equivalence
class of isometries, it may also be thought of as a single isometry
of the torus $\Z^{2}/\rectlat^{\Lambda}$. Thus, $\sigma\circ\tau$
is well defined for $\sigma\in\Omega_{\Lambda}^{\per}$, which allows,
given $f:\Omega\to\R$, to further define $\tau f:\Omega_{\Lambda}^{\per}\to\R$
(via (\ref{eq:func_shift})).
\item For an $R$-local function $f:\Omega\to\R$ define\marginpar{$\zedd[][][\cdot]$}
\begin{align}
\zedd[][][f] & \coloneqq\left[\mu_{\Lambda}^{\per}\left(\prod_{\tau\in\refls}\tau f\right)\right]^{1/\#\refls}.\label{eq:chessboard norm def}
\end{align}
Note that the expectation in this definition is necessarily non-negative
by reflection positivity (Lemma \ref{lemma:reflection_positivity}),
so that $\zedd[][][f]$ is well defined and satisfies 
\begin{equation}
\zedd[][][f]\ge0.\label{eq:non-negativity of zeta}
\end{equation}
Note also that $\zedd[][][\cdot]$ further depends on $\lambda$ but,
for brevity, we omit this dependence from the notation.

For an $R$-local event $E$ we write $\zedd[][][E]:=\zedd[][][\mathbf{1}_{E}]$.

We call $\zedd[][][\cdot]$ the $(R,\Lambda)$-\textbf{chessboard
seminorm.} The name `seminorm' is justified by Proposition \ref{prop:seminorm}
below. The notation $\zedd[][][\cdot]$, as an alternative to the
$\zed$ notation used in \cite[equation (5.47)]{biskup2009reflection},
is chosen to better remind the reader of the seminorm properties.
\end{itemize}
\begin{figure}[H]
\begin{centering}
\includegraphics{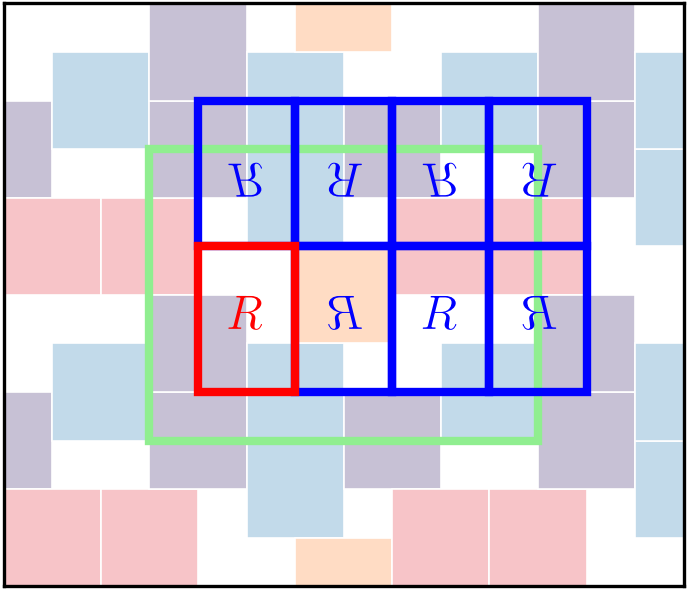}\caption{In the background the rectangle $\Lambda=\protect\rect{8\times6}{(0,0)}$
is shown in green, with a configuration in $\Omega_{\Lambda}^{\protect\per}$.
The red rectangle $R=\protect\rect{2\times3}{(1,1)}$ is a block of
$\Lambda$. The red and blue rectangles are mappings of $R$ by $8$
elements of $\protect\refls[\protect\null]$ that form a representative
set of $\protect\refls$.}
\par\end{centering}
\end{figure}

\begin{prop}
[Chessboard estimate]\label{prop:chessboard} Let $R$ be a block
of $\Lambda$, let $A\subset\refls$, and let $(f_{\tau})_{\tau\in A}$
be $R$-local functions. Then
\[
\mu_{\Lambda}^{\per}\left(\prod_{\tau\in A}\tau f_{\tau}\right)\le\prod_{\tau\in A}\zedd[][][f_{\tau}].
\]
\end{prop}
\begin{proof}
This is a standard consequence of reflection positivity (Lemma \ref{lemma:reflection_positivity});
see \cite[Theorem 10.11 and Remark 10.15]{friedli_velenik_2017} or
\cite[Theorem 5.8]{biskup2009reflection}. In both references, the
proof is given for the case of reflection positivity ``through edges/bonds''
and it is remarked that an analogous result holds for reflection positivity
``through vertices/sites'' (as in our case).
\end{proof}
We proceed to note several basic properties of the chessboard seminorm
$\zedd[][][\cdot]$. The first two properties justify the name seminorm
while the last two properties imply that $\zedd[][][\cdot]$ restricted
to $R$-local events is an outer measure (as in \cite[Lemma 5.9]{biskup2009reflection}.
Countable subadditivity follows from additivity as there are only
finitely many $R$-local events).
\begin{prop}
[positive homogeneity, triangle inequality and monotonicity]\label{prop:seminorm}The
mapping $f\mapsto\zedd$, where $f$ ranges over $R$-local functions,
satisfies the following properties:
\end{prop}
\begin{enumerate}
\item Homogeneity: $\zedd[][][\alpha f]=\left|\alpha\right|\zedd$ for $\alpha\in\R$.
In particular, $\zedd\ge0$.
\item Triangle inequality: $\zedd[][][f_{0}+f_{1}]\le\zedd[][][f_{0}]+\zedd[][][f_{1}]$.
\item Monotonicity: $\zedd[][][g]\ge\zedd[][][f]$ whenever $g\ge f\ge0.$
\end{enumerate}
\begin{proof}
Since $\#\refls$ is even, we have $\prod_{\tau\in\refls}\tau(\alpha f)=\left|\alpha\right|^{\#\refls[\Lambda]}\prod_{\tau\in\refls}\tau f$.
It follows from the definition of $\zedd[][][\cdot]$ that $\zedd[][][\alpha f]=\left|\alpha\right|\zedd$.

The triangle inequality follows from
\begin{align*}
\left(\zedd[][][f_{0}+f_{1}]\right) & ^{\#\refls}=\\
\text{(by definition)} & =\mu_{\Lambda}^{\per}\left(\prod_{\tau\in\refls}\tau(f_{0}+f_{1})\right)\\
\text{(expanding brackets)} & =\sum_{r:\refls\to\{0,1\}}\mu_{\Lambda}^{\per}\left(\prod_{\tau\in\refls}\tau f_{r(\tau)}\right)\\
\text{(by the chessboard estimate)} & \le\sum_{r:\refls\to\{0,1\}}\prod_{\tau\in\refls}\zedd[][][f_{r(\tau)}]\\
\text{(factorizing)} & =\left(\zedd[][][f_{0}]+\zedd[][][f_{1}]\right)^{\#\refls}.
\end{align*}

Monotonicity follows from the definition of $\zedd$ by the monotonicity
of $\mu_{\Lambda}^{\per}$.
\end{proof}
Lastly, we note a simple relation between the chessboard seminorms
of rectangles with different dimensions.
\begin{lem}
[``Recursive chessboard estimate'']\label{lem:chessboard_recursive}
Let $R$ and $S$ be blocks of $\Lambda$ and assume that the corners
of $S$ are in $\grid[\null]$. Let $A\subset\refls[\null]$ be such
that $\cup_{\tau\in A}\tau R\subset S$. For each $\tau\in A$, let
$f_{\tau}$ be an $R$-local function. Then
\[
\zedd[][S][\prod_{\tau\in A}\tau f_{\tau}(\sigma)]\le\prod_{\tau\in A}\zedd[][][f_{\tau}].
\]
\end{lem}
\begin{proof}
Denote $g:=\prod_{\tau\in A}\tau f_{\tau}(\sigma)$, so that $g$
is an $S$-local function. Our assumptions imply that $\refls[][S]\subset\refls$.
By the definition of $g$,
\begin{equation}
\prod_{\iota\in\refls[][S]}\iota g=\prod_{\iota\in\refls[][S]}\prod_{\tau\in A}\iota\tau f_{\tau}\quad\text{on \ensuremath{\Omega_{\Lambda}^{\per}}}.\label{eq:distinct reflections}
\end{equation}
Our assumption that $\cup_{\tau\in A}\tau R\subset S$ shows that
each choice of $\iota\in\refls[][S]$ and $\tau\in A$ gives a distinct
element $\iota\tau\in\refls$. Therefore, by the chessboard estimate,
\begin{equation}
\mu_{\Lambda}^{\per}\left(\prod_{\iota\in\refls[][S]}\prod_{\tau\in A}\iota\tau f_{\tau}\right)\le\prod_{\iota\in\refls[][S]}\prod_{\tau\in A}\zedd[][][f_{\tau}]=\left(\prod_{\tau\in A}\zedd[][][f_{\tau}]\right)^{\#\refls[][S]}.\label{eq:chessboard with distinct reflections}
\end{equation}
Substituting (\ref{eq:distinct reflections}) in the LHS of (\ref{eq:chessboard with distinct reflections}),
we get that, by the definition (\ref{eq:chessboard norm def}),
\[
\zedd[][S][g]\le\prod_{\tau\in A}\zedd[][][f_{\tau}].\qedhere
\]
\end{proof}

\subsection{Infinite volume\label{subsec:Infinite-volume chessboard}}

Recall that we call an (infinite-volume) Gibbs measure periodic if
it is invariant under translations by some full-rank sublattice of
$\Z^{2}$. Our goal in this section is to provide a version of the
chessboard estimate applicable to periodic Gibbs measures (Proposition
\ref{prop:chessboard_infinite} below). This will be used in later
sections to apply a Peierls-type argument directly in infinite volume.

We have not seen the chessboard estimate formulated directly in infinite
volume before, though we mention that a different approach was used
by Biskup--Koteck\'y \cite{biskup2006forbidden} in order to apply
a Peierls argument driven by a chessboard estimate to periodic (infinite-volume)
Gibbs measures. 

\subsubsection{The chessboard seminorm and its basic properties}

We begin by defining a ``limit of $\zedd[][][\cdot]$ as $\Lambda\to\infty$''.
Let $R$ be a rectangle and let $f$ be an $R$-local function. Define\marginpar{$\zedi[][\cdot]$}
\begin{equation}
\zedi\coloneqq\limsup_{n\to\infty}\zedd[\frect{n!\times n!}]\label{eq:infinite-volume zeta}
\end{equation}
noting that $R$ is a block of $\frect{n!\times n!}$ for almost all
$n$.
\begin{rem}
In fact $\lim_{(m,n)\to(\infty,\infty)}\zedd[\frect{2m\width(R)\times2n\height(R)}]$
exists, but the proof of this fact is complicated by the ``boundary
overlaps between blocks'' in the definition of $\zedd[\Lambda]$.
To avoid proving this fact, we have chosen the somewhat arbitrary
definition above.
\end{rem}
The basic properties of $\zedd[\Lambda][][\cdot]$ transfer directly
to the limiting definition (\ref{eq:infinite-volume zeta}).
\begin{prop}
[positive homogeneity, triangle inequality and monotonicity for infinite
volume]\label{prop:seminorm_limit}The mapping $f\mapsto\zedi$,
where $f$ ranges over $R$-local functions, satisfies the properties
stated in Proposition \ref{prop:seminorm}.
\end{prop}
\begin{proof}
The properties follow from Proposition \ref{prop:seminorm} and the
subadditivity of the limit superior.
\end{proof}
Lemma \ref{lem:chessboard_recursive} also admits an immediate extension.
\begin{lem}
[``Recursive chessboard estimate'' for infinite volume]\label{lem:inf_recursive_chessboard}Let
$R$ and $S$ be rectangles and assume that the corners of $S$ are
in $\grid[\null]$. Let $A\subset\refls[\null]$ be such that $\cup_{\tau\in A}\tau R\subset S$.
For each $\tau\in A$, let $f_{\tau}$ be an $R$-local function.
Then
\[
\zedi[S][\prod_{\tau\in A}\tau f_{\tau}]\le\prod_{\tau\in A}\zedi[][f_{\tau}].
\]
\end{lem}
\begin{proof}
The inequality follows from Lemma \ref{lem:chessboard_recursive}
using the definition (\ref{eq:infinite-volume zeta}) (making use
of the fact that definition (\ref{eq:infinite-volume zeta}) involves
a limsup rather than a liminf).
\end{proof}

\subsubsection{The chessboard estimate}

This subsubsection is devoted to the proof of the following statement.
\begin{prop}
[Chessboard estimate for infinite volume]\label{prop:chessboard_infinite}
Let $R$ be a rectangle. Let $A\subset\refls[\null]$ be finite and
let $(f_{\tau})_{\tau\in A}$ be $R$-local functions. Then
\[
\mu\left(\prod_{\tau\in A}\tau f_{\tau}\right)\le\prod_{\tau\in A}\zedi[][f_{\tau}]
\]
for all periodic Gibbs measures $\mu$. 
\end{prop}
The following lemma is the main tool in the proof.
\begin{lem}
\label{lem:zed_for_infinite}Let $R$ be a rectangle and let $f$
be an $R$-local function. Then
\begin{equation}
\mu(f)\le\zedi\label{eq:zedd for periodic measures}
\end{equation}
for all periodic Gibbs measures $\mu$.
\end{lem}
The proof of the lemma relies on the following two auxiliary claims.
The first claim is a weak form of Proposition \ref{prop:chessboard_infinite}
which follows from the finite-volume chessboard estimate and a comparison
of boundary conditions.
\begin{claim}
\label{claim:weak infinite-volume chessboard}Let $R$ be a rectangle
and let $g$ be a nonnegative $R$-local function. Let $(A_{n})_{n\ge1}$
be finite subsets of $\refls[\null]$ satisfying 
\begin{equation}
\frac{\text{\text{diam}}(\cup_{\tau\in A_{n}}\tau R)}{\#A_{n}}\xrightarrow[n\to\infty]{\substack{}
}0\label{eq:Folner}
\end{equation}
where we denote the diameter of subsets of $\R^{2}$ by $\text{diam(\ensuremath{\cdot})}$.
Then
\[
\limsup_{n\to\infty}\sqrt[\#A_{n}]{\mu\left(\prod_{\tau\in A_{n}}\tau g\right)}\le\zedi[][g]
\]

for all Gibbs measures $\mu$.
\end{claim}
\begin{proof}
Fix $n\ge1$ large. Denote $\Lambda_{m}:=\rect{m!\times m!}{(-m!/2,-m!/2)}$
for $m\ge2$.

Proposition \ref{prop:boundary} implies that, for sufficiently large
$m$,
\[
\mu\left(\prod_{\tau\in A_{n}}\tau g\right)\le C_{\ref{prop:boundary}}(\lambda)^{5\text{diam}(\cup_{\tau\in A_{n}}\tau R)}\mu_{\Lambda_{m}}^{\per}\left(\prod_{\tau\in A_{n}}\tau g\right).
\]
To see this, let $\Lambda'$ be the smallest rectangle containing
$\cup_{\tau\in A_{n}}\tau R$ and let $\Lambda\subset\Lambda_{m}$
be the smallest rectangle for which the Euclidean distance from $\Lambda'$
to $\R^{2}\setminus\Lambda$ is at least $2$ (noting that $\perim(\Lambda)\le5\text{diam}(\cup_{\tau\in A_{n}}\tau R)$).
Then use the domain Markov property to write $\mu_{\Lambda_{m}}^{\per}\left(\prod_{\tau\in A_{n}}\tau g\right)=\mu_{\Lambda_{m}}^{\per}\left(\mu_{\Lambda}^{\sigma}\left(\prod_{\tau\in A_{n}}\tau g\right)\right)$
and apply Proposition \ref{prop:boundary} with $\mu_{\Lambda'}=\mu$
and $\mu_{\Lambda}=\mu_{\Lambda}^{\rho}$ for all possible $\rho$.

For sufficiently large $m$, it holds that $R$ is a block of $\Lambda_{m}$
and $\cup_{\tau\in A_{n}}\tau R\subset\Lambda_{m}$, so that the elements
of $A_{n}$ belong to distinct equivalence classes in $\refls[\Lambda_{m}]$.
Therefore, by the finite-volume chessboard estimate of Proposition
\ref{prop:chessboard},
\[
\mu_{\Lambda_{m}}^{\per}\left(\prod_{\tau\in A_{n}}\tau g\right)\le\zedd[\Lambda_{m}][][g]^{\#A_{n}}
\]

Combining the last two displays shows that

\[
\sqrt[\uproot{3}{\scriptstyle \#A_{n}}]{\mu\left(\prod_{\tau\in A_{n}}\tau g\right)}\le C(\lambda)^{\frac{\text{\text{diam}}(\cup_{\tau\in A_{n}}\tau R)}{\#A_{n}}}\zedd[\Lambda_{m}][][g]
\]

The claim follows by taking limits superior, first as $m\to\infty$
and then as $n\to\infty$, and using our assumption (\ref{eq:Folner})
and the definition (\ref{eq:infinite-volume zeta}) of $\zedi[][g]$.
\end{proof}
The second claim is a simple application of Taylor's theorem.

\begin{claim}
\label{claim:ergodic theorem application}Let $M\in\mathbb{N}$. Let
$S_{n}\subset M\Z^{2}$ be finite and let $A_{n}:=\{\tau_{s}:s\in S_{n}\}$,
where $\tau_{s}:\R^{2}\to\R^{2}$ is the shift $\tau_{s}(u):=u+s$.
Let $0<\eps<1/2$ and let $g:\Omega\to[1-\eps,1+\eps]$ be a measurable
function. Then
\begin{equation}
\mu\left(\sqrt[\uproot{5}{\scriptstyle \#A_{n}}]{\prod_{\tau\in A_{n}}\tau g}\right)=\mu(g)+O(\eps^{2})\label{eq:geometric average inequality}
\end{equation}

for every $M\Z^{2}$-invariant $\mu$. Here $O(\eps^{2})$ denotes
an expression whose absolute value is at most $C\eps^{2}$ for a universal
constant $C>0$.
\end{claim}
\begin{proof}
Set $g=1+\eps f$, so that $\left|f\right|\le1$. Then
\begin{align*}
\mu\left(\sqrt[\uproot{4}{\scriptstyle \#A_{n}}]{\prod_{\tau\in A_{n}}\tau g}\right) & =\mu\left(\exp\left(\frac{1}{\#A_{n}}\sum_{\tau\in A_{n}}\log(1+\eps\tau f)\right)\right)=\\
(\text{by Taylor's theorem}) & =\mu\left(\exp\left(\frac{1}{\#A_{n}}\sum_{\tau\in A_{n}}\left(\eps\tau f+O(\eps^{2})\right)\right)\right)\\
 & =\mu\left(\exp\left(\eps\left(\frac{1}{\#A_{n}}\sum_{\tau\in A_{n}}\tau f\right)+O(\eps^{2})\right)\right)\\
(\text{by Taylor's theorem}) & =\mu\left(1+\eps\left(\frac{1}{\#A_{n}}\sum_{\tau\in A_{n}}\tau f\right)+O(\eps^{2})\right)\\
 & =1+\frac{\eps}{\#A_{n}}\sum_{\tau\in A_{n}}\mu(\tau f)+O(\eps^{2})\\
(\text{since \ensuremath{\mu} is \ensuremath{M\Z^{2}} invariant}) & =1+\eps\mu(f)+O(\eps^{2})\\
 & =\mu(g)+O(\eps^{2}).\qedhere
\end{align*}
\end{proof}
We now deduce Lemma \ref{lem:zed_for_infinite}.
\begin{proof}
[Proof of Lemma \ref{lem:zed_for_infinite}]Let $\mathcal{L}\subset\Z^{2}$
be a full-rank sublattice and let $\mu$ be an $\mathcal{L}$-invariant
Gibbs measure. Recall the definition of $\mathcal{L}^{R}$ from (\ref{eq:G^R and L^R})
and note that $\mathcal{L}\cap2\mathcal{L}^{R}$ is also a full-rank
sublattice. Let $M\in\mathbb{N}$ be such that $M\Z^{2}\subset\mathcal{L}\cap2\mathcal{L}^{R}.$
For $n\in\mathbb{N}$, set $S_{n}:=\{M,2M,...,Mn\}^{2}\subset M\Z^{2}$
and let $A_{n}$ be as in Claim \ref{claim:ergodic theorem application}.
Observe that $A_{n}\subset\refls[\null]$ as $S_{n}\subset2\mathcal{L}^{R}$.
For $0<\eps<\frac{1}{2\max|f|}$, set $g=1+\eps f$ and observe that
\begin{multline*}
1+\eps\zedi[][f]\overset{{\scriptscriptstyle (\text{i})}}{\ge}\zedi[][g]\overset{{\scriptscriptstyle (\text{ii})}}{\ge}\limsup_{n\to\infty}\sqrt[{\scriptstyle n^{2}}]{\mu\left(\prod_{\tau\in A_{n}}\tau g\right)}\overset{{\scriptscriptstyle (\text{iii})}}{\ge}\limsup_{n\to\infty}\mu\left(\sqrt[\uproot{2}{\scriptstyle n^{2}}]{\prod_{\tau\in A_{n}}\tau g}\right)\\
\overset{{\scriptscriptstyle (\text{iv})}}{=}\mu(g)+O((\eps\max|f|)^{2})=1+\eps\mu(f)+O(\eps^{2}(\max|f|)^{2}),
\end{multline*}

where (i) follows from subadditivity of $\zedi[][\cdot]$ (Proposition
\ref{prop:seminorm_limit}), (ii) follows from Claim \ref{claim:weak infinite-volume chessboard},
noting that $A_{n}$ satisfies (\ref{eq:Folner}), (iii) follows from
Jensen's inequality and (iv) follows from Claim \ref{claim:ergodic theorem application},
noting that $\mu$ is $M\Z^{2}$-invariant. The lemma follows by taking
$\eps$ to zero.
\end{proof}
Finally, we turn to the proof of the infinite-volume chessboard estimate.
\begin{proof}
[Proof of Proposition \ref{prop:chessboard_infinite}]Recall the
definition of $\grid[\null][R]$ from (\ref{eq:G^R and L^R}). Let
$S$ be a rectangle whose corners lie in $\grid[\null][R]$, and contains
$\bigcup_{\tau\in A}\tau R$. Observe that $\prod_{\tau\in A}\tau f_{\tau}(\sigma)$
is an $S$-local function. Applying Lemma \ref{lem:zed_for_infinite}
and Lemma \ref{lem:inf_recursive_chessboard}, 
\[
\mu\left(\prod_{\tau\in A}\tau f_{\tau}(\sigma)\right)\le\zedi[S][\prod_{\tau\in A}\tau f_{\tau}(\sigma)]\le\prod_{\tau\in A}\zedi[][f_{\tau}].\qedhere
\]
\end{proof}

\section{\label{sec:main_lemma}Mesoscopic rectangles are divided by sticks}

The goal of this work is to establish a form of columnar, or row,
order for the $2\times2$ hard-square model at high fugacity. Recall
from the introduction (see Figure \ref{fig:from fully-packed to high fugacity})
that in a vertically ordered state of this type, the tiles organize
in columns of width $2$ which are non-interacting at most places.
I.e., the system may be thought of as a perturbation of a product
system in which each column follows a one-dimensional hard-square
model. It is instructive to note that one-dimensional systems at high
fugacity $\lambda$ consist of segments of fully-packed tiles whose
lengths are typically of order $\sqrt{\lambda}$ --- a mesoscopic
length scale which will be important in our arguments.

Motivated by this description, we introduce the notion of sticks.
Informally, a vertical (horizontal) stick is the line separating two
finite columns (rows) in which the tiles are fully packed but have
a different vertical (horizontal) offset. Columnar order leads to
an abundance of vertical sticks of mesoscopic length (length of order
$\sqrt{\lambda}$) while row order similarly leads to an abundance
of horizontal sticks of mesoscopic length. Importantly, vertical and
horizontal sticks cannot meet. Using this fact, it will be shown that,
in a suitable sense, the interface between regions of columnar and
row order is characterized by the presence of mesoscopic rectangles
which are not divided by a stick.

The goal of this section is to prove Lemma \ref{lem:main} below,
which roughly states that mesoscopic rectangles are divided by a stick
with high probability. By the loose term ``mesoscopic rectangle''
we mean a rectangle whose side lengths are small compared to $\sqrt{\lambda}$
but whose area is large compared to $\sqrt{\lambda}$. Moreover, this
probabilistic estimate applies multiplicatively to collections of
disjoint mesoscopic rectangles using the chessboard estimate of Proposition
\ref{prop:chessboard_infinite}. This will lead, in Section \ref{sec:multiple_gibbs},
to the existence of multiple Gibbs measures (one with a predominance
of vertical sticks and another with a predominance of horizontal sticks)
through a Peierls-type argument.

We proceed with the formal definitions and results.

\textbf{Sticks:} Let $\sigma\in\Omega$, and let the following definitions
depend on $\sigma$. Recall from Section \ref{subsec:Basic-definitions}
that the parity of a tile centered at $(x,y)$ is $(x-1\bmod2,y-1\bmod2)$.
An edge of $\E_{\square}$ that bounds two faces in $\mathbb{F}$
that are respectively contained in two tiles of $\sigma$ having \emph{distinct}
parities, is called a \textbf{stick edge} (for $\sigma$) --- each
stick edge is naturally vertical or horizontal. A \textbf{stick} is
a maximal path of stick edges (possibly infinite in one or both directions).

A case analysis shows that a vertical stick edge may never meet a
horizontal stick edge at a vertex. Thus a stick may be viewed as a
vertical or horizontal segment in $\R^{2}$. Note also that sticks
are pairwise disjoint.

In later sections we will classify sticks according to their orientation
and parity. We say that a stick is of type $\boldsymbol{(\ver,0)}$,
and call it ``a $(\ver,0)$ stick'', if it is vertical and passes
through points with even $x$-coordinate. Equivalently, a stick is
of type $(\ver,0)$ if it bounds only on tiles with horizontal parity
$0$. We define analogously the types $\boldsymbol{(\ver,1)},$$\boldsymbol{(\hor,0)}$
and $\boldsymbol{(\hor,1)}$.

Let $R=\rect{K\times L}{(x,y)}$ be a rectangle and consider a vertical
segment whose endpoints are $(x_{1},y_{1})$ and $(x_{1},y_{2})$
with $y_{1}<y_{2}$. We say that the segment \textbf{vertically divides}
$R$ if $y_{1}\le y\le y+L\le y_{2}$ and $x<x_{1}<x+K$. We make
an analogous definition for horizontal segments. A segment is said
to \textbf{divide} $R$ if either it is a vertical segment dividing
$R$ vertically or it is a horizontal segment dividing $R$ horizontally.

Note that with these definitions, the event that a stick divides $R$
is $R$-local. This may be seen using the fact that if an edge $e\in\E_{\square}$
is such that its image in $\R^{2}$ is contained in $R$ but not contained
in $\partial R$, then the event that $e$ is a stick edge is $R$-local.

Our goal in this section is to prove the following probabilistic bound
on the prevalence of dividing sticks in mesoscopic rectangles. Recall
the infinite-volume chessboard seminorm $\zedi[][\cdot]$ defined
in (\ref{eq:infinite-volume zeta}).
\begin{lem}
\label{lem:main}There is $c>0$ such that if rectangles $S\subset R$
satisfy 
\begin{gather}
\frac{1}{c}\le\width(R),\height(R)\le c\lambda^{1/2},\label{eq:dim_abs}\\
\width(S)\ge(1-c)\width(R)\text{\ensuremath{\quad}and\ensuremath{\quad}}\height(S)\ge(1-c)\height(R),\label{eq:dim_rat}
\end{gather}
then
\begin{equation}
\zedi[R][\text{no stick divides both \ensuremath{R} and \ensuremath{S}}]\le e^{-c\vol(R)\lambda^{-1/2}}.\label{eq:main_lemma}
\end{equation}
\end{lem}
\begin{figure}[H]
\includegraphics[width=1\textwidth]{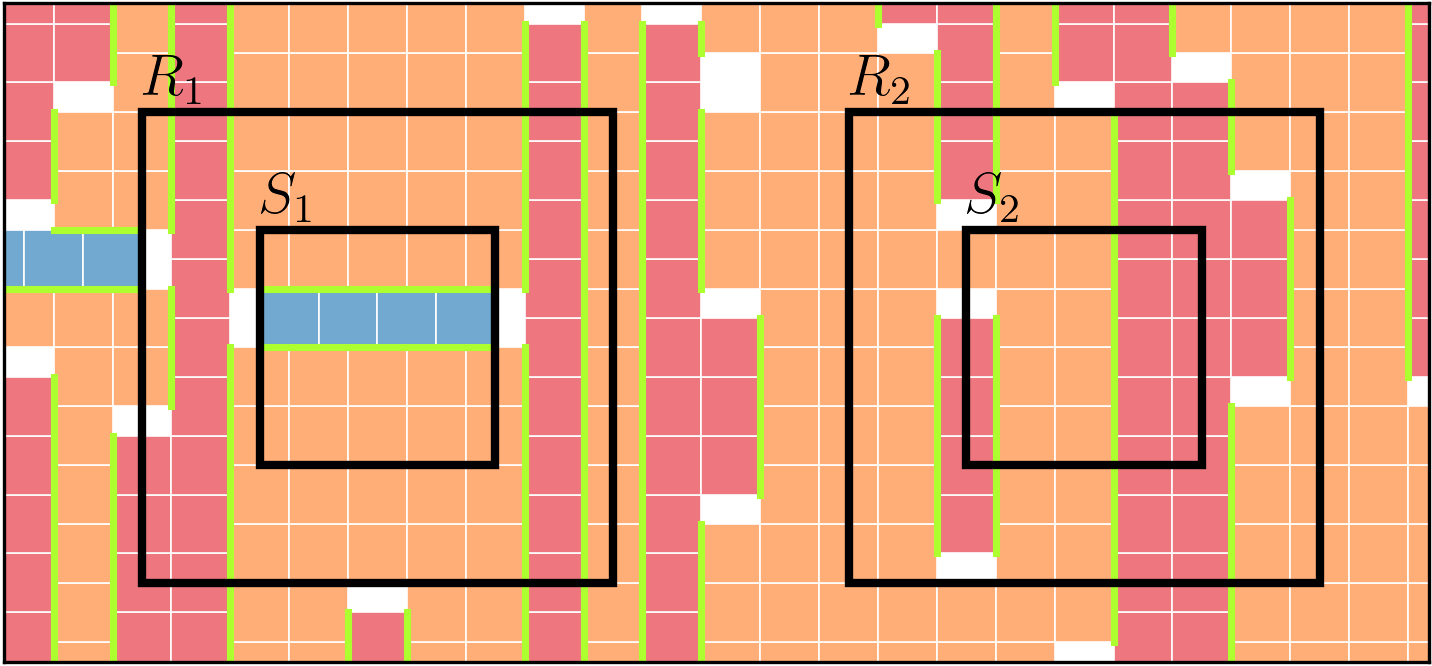}

\caption{\label{fig:dividing}The sticks of the configuration are highlighted
in green. No stick divides both $R_{1}$ and $S_{1}$ although each
of them is divided by a stick. A stick divides both $R_{2}$ and $S_{2}$.\protect \\
In the terminology of Section \ref{sec:multiple_gibbs}, if $R_{1}=\protect\rect{16\times16}{(0,0)}$
and $N=4$ then $S_{1}=R_{1}^{-}$, $S_{2}=R_{2}^{-}$ and $R_{2}$
is properly divided by a $(\protect\ver,1)$ stick while $R_{1}$
is not properly divided. In symbols, $(0,0)\protect\notin\Psi^{4\times4}$
and $(6,0)\in\Psi_{(\protect\ver,1)}^{4\times4}$.}
\end{figure}

\subsection{One-dimensional systems}

The proof of the probability estimate (\ref{eq:main_lemma}) involves
giving an upper bound on the total weight of configurations (mostly)
without long sticks and comparing it with a lower bound on the the
total weight of all configurations (with suitable boundary conditions).
The first task will be handled in the subsequent sections whereas
here we focus on the simpler second task. Having in mind that high-fugacity
systems are expected to order in a columnar fashion (as we aim to
prove in this paper), it is natural to obtain a lower bound for the
two-dimensional system via lower bounds for one-dimensional systems
(which should be thought of as single columns of tiles in the two-dimensional
system). We proceed to develop such bounds.

It is simplest to define the one-dimensional model as the restriction
of the two-dimensional model to a rectangle of width 2 (the width
of a single tile). Thus we define the partition function of a one-dimensional
system of size $L$ with free boundary conditions by\marginpar{$\zoz L$}
\[
\zoz L\coloneqq Z_{\frect{2\times L}}^{0}
\]
and the partition function of a one-dimensional system of size $L$
with periodic boundary conditions by\marginpar{$\zop L$}
\[
\zop L\coloneqq Z_{\frect{2\times L}}^{\per}(E),
\]
where $E$ is the event that all tiles have even horizontal parity,
and using the notation (\ref{eq:weight of an event}) for the weight
of an event. In these definitions we again follow our convention of
omitting the fugacity parameter $\lambda$ from the notation.

The next proposition provides a lower bound for the partition function
of periodic one-dimensional systems.
\begin{prop}
\label{prop:1d periodic}$\zop L\ge\left(1+\frac{1}{2}\lambda^{-1/2}\right)^{L}$
for all $\lambda>0$ and all even $L\ge0$.
\end{prop}
\begin{proof}
Let $A$ be the set of configurations in $\Omega_{\frect{2\times L}}^{\per}$
where all tiles have even horizontal parity. There is a one-to-one
correspondence between configurations in $A$ and the set $B$ of
sequences $r\in\{0,1\}^{\{0,1,...,L\}}$ satisfying that $r_{0}=r_{L}$
and $r_{i}r_{i+1}=0$ for $0\le i<L$; the correspondence is defined
by $\left(r(\sigma)\right)_{i}=\sigma(1,i)$.

Recalling the formula (\ref{eq:weight of a configuration}) for the
weight of a configuration, we note the identity
\[
\w[\frect{2\times L},\lambda](\sigma)=\lambda^{-\frac{1}{2}\sum_{i=0}^{L-1}(1-r(\sigma)_{i})(1-r(\sigma)_{i+1})}.
\]
This follows by observing that the vacancies in $\sigma$ necessarily
come in horizontally-adjacent pairs, and that such pairs correspond
in $r(\sigma)$ to pairs of consecutive $0$ values. Thus
\[
\zop L=\sum_{\sigma\in A}\w[\frect{2\times L},\lambda](\sigma)=\sum_{r\in B}\lambda^{-\frac{1}{2}\sum_{i=0}^{L-1}(1-r_{i})(1-r_{i+1})}=\trace\left(\begin{pmatrix}\lambda^{-1/2} & 1\\
1 & 0
\end{pmatrix}^{L}\right),
\]
where the $0$ in the matrix corresponds to the restriction of not
having consecutive $1$ values in $r$. The eigenvalues of $\begin{pmatrix}\lambda^{-1/2} & 1\\
1 & 0
\end{pmatrix}$ are
\begin{align*}
\gamma_{\pm} & =\frac{\lambda^{-1/2}\pm\sqrt{\lambda^{-1}+4}}{2},
\end{align*}
whence, for even $L$,
\[
\zop L=\gamma_{+}^{L}+\gamma_{-}^{L}\ge\gamma_{+}^{L}\ge\left(1+\frac{1}{2}\lambda^{-1/2}\right)^{L}.\qedhere
\]
\end{proof}
The above one-dimensional bound implies the following lower bound
for the partition function of two-dimensional systems.
\begin{cor}
\label{cor:1d_bound}For each $c<\frac{1}{4}$, there is $\lambda_{0}$
such that for all $\lambda>\lambda_{0}$ and all even rectangles $\Lambda$,
\[
Z_{\Lambda,\lambda}^{\per}\ge e^{c\lambda^{-1/2}\vol(\Lambda)}.
\]
\end{cor}
\begin{proof}
The total weight of configurations in $\Omega_{\Lambda,\lambda}^{\per}$
with all tiles having even horizontal parity (as in the center panel
of Figure \ref{fig:from fully-packed to high fugacity}) is $\left(\zop{\height(\Lambda)}\right)^{\width(\Lambda)/2}$.
Thus, by Proposition \ref{prop:1d periodic},
\[
Z_{\Lambda,\lambda}^{\per}\ge\left(\zop{\height(\Lambda)}\right)^{\width(\Lambda)/2}\ge\left(1+\frac{1}{2}\lambda^{-1/2}\right)^{\vol(\Lambda)/2}
\]
from which the corollary follows.
\end{proof}
We prove also a lower bound on $\zoz L$ that will be used in Proposition
\ref{prop:shminiot}. This bound is useful in particular when $L$
has the same order of magnitude as $\lambda^{1/2}$.
\begin{prop}
\label{prop:1d_square_bound}$\zoz L\ge1+\frac{L^{2}}{8\lambda}$
for all $\lambda>0$ and all even $L\ge0$.
\end{prop}
\begin{proof}
Since $L$ is even, there is a configuration in $\Omega_{\frect{2\times L}}^{0}$
in which $\frect{2\times L}$ is fully packed with tiles, and this
configuration has weight $1$. We consider also configurations having
one tile less than the fully-packed configuration. Each such configuration
has weight $\lambda^{-1}$, and one checks that the number of such
configurations is exactly $\frac{L/2\cdot(L/2+1)}{2}$. This shows
that $\zoz L\ge1+\frac{L^{2}+2L}{8\lambda}$ from which the proposition
follows.
\end{proof}
We will make extensive use of the following simple corollary in Part
\ref{part:Characterization-of-the} (it will not be used in Part \ref{part:Existence-of-multiple}).
\begin{cor}
\label{cor:single_vacancy_Z}Let $f\in\mathbb{F}$ be a face. Then:
\begin{enumerate}
\item $\zedi[f][\text{\ensuremath{f} is vacant}]\le\lambda^{-1/4}$ and
\item for each $c<1/4$ and sufficiently large $\lambda$, $\zedi[f][\text{\ensuremath{f} is occupied}]\le1-c\lambda^{-1/2}$.
\end{enumerate}
\end{cor}
\begin{proof}
Consider an even rectangle $\Lambda$, and a face $f$. Then $f$
is a block of $\Lambda$. The empty configuration $0\in\Omega_{\Lambda}^{\per}$
has weight $\w(0)=\lambda^{-\frac{1}{4}\vol(\Lambda)}$. A fully packed
configuration has weight $1$, whence $Z_{\Lambda}^{\per}\ge1$. This
gives
\begin{align*}
\zedd[][f][\text{\ensuremath{f} is vacant}] & =\left(\mu_{\Lambda}^{\per}(\text{all faces of \ensuremath{\Lambda} are vacant})\right)^{1/\vol(\Lambda)}\\
 & =\left(\frac{\w(0)}{Z_{\Lambda}^{\per}}\right)^{1/\vol(\Lambda)}\le\lambda^{-1/4}.
\end{align*}
For the second item, note that a fully packed configuration in $\Omega_{\Lambda}^{\per}$
is either composed of fully packed columns or of fully packed rows
(see Figure \ref{fig:fully-packed configurations}). In the case of
columns, say, there is a global choice of parity for the horizontal
offset of the columns and, for each column, two possibilities to choose
its vertical offset. Thus 
\begin{equation}
Z_{\Lambda}^{\per}(\text{\ensuremath{\Lambda} is fully packed})\le2\cdot2^{\width(\Lambda)}+2\cdot2^{\height(\Lambda)}.\label{eq:fully_packed_Z}
\end{equation}
Let $c<1/4$, and choose some $c<c_{\ref{cor:1d_bound}}<\frac{1}{4}$
. Then
\begin{align*}
\zedd[][f][\text{\ensuremath{f} occupied}] & =\left(\mu_{\Lambda}^{\per}(\text{all faces of \ensuremath{\Lambda} are occupied})\right)^{1/\vol(\Lambda)}\\
\left(\text{by (\ref{eq:fully_packed_Z})}\right) & \le\left(\frac{2\cdot2^{\width(\Lambda)}+2\cdot2^{\height(\Lambda)}}{Z_{\Lambda}^{\per}}\right)^{1/\vol(\Lambda)}\\
\left(\text{Corollary \ref{cor:1d_bound}}\right) & \le\left(\frac{2\cdot2^{\width(\Lambda)}+2\cdot2^{\height(\Lambda)}}{e^{c_{\ref{cor:1d_bound}}\lambda^{-1/2}\vol(\Lambda)}}\right)^{1/\vol(\Lambda)}\\
 & \xrightarrow{\frac{\perim(\Lambda)}{\vol(\Lambda)}\ensuremath{\to0}}e^{-c_{\ref{cor:1d_bound}}\lambda^{-1/2}}\\
\left(\substack{\text{for sufficiently large \ensuremath{\lambda},}\\
\text{since \ensuremath{c<c_{\ref{cor:1d_bound}}}}
}
\right) & \le1-c\lambda^{-1/2}.
\end{align*}
The bounds on $\zedi[f][\cdot]$ now follow using definition (\ref{eq:infinite-volume zeta}).
\end{proof}

\subsection{\label{sec:graph-counting-argument}Configurations without long sticks}

For $M\ge1$, denote by $E_{M}\subset\Omega$ \marginpar{$E_{M}$}the
set of configurations in which all sticks are of length at most $M$.
For an even rectangle $\Lambda$, consider $E_{M}\cap\Omega_{\Lambda}^{1}$,
the set of configurations \emph{with fully-packed boundary conditions}
in $E_{M}$. This subsection is devoted to proving the following ``weighted
counting lemma'', bounding the total weight of all such configurations.
\begin{lem}
\label{lem:counting_argument} There exists $C>0$ such that for every
$\lambda>0$, $M\ge1$ and even rectangle $\Lambda$, if $M<\lambda^{1/2}/C$
then
\[
Z_{\Lambda}^{1}(E_{M})\le\left(1+\frac{CM}{\lambda}\right)^{^{\vol(\Lambda)}}.
\]
\begin{rem}
The bound of the lemma is sharp up to the value of the constant $C$,
at least when $M\le\max\{\width(\Lambda),\height(\Lambda)\}$. Let
us sketch how a matching lower bound may be obtained. Observe that
$E_{M}$ contains the set of configurations 
\[
\tilde{E}_{M}\coloneqq\{\sigma\in\Omega:\,\,\forall(x,y)\in\Z^{2},\,\,(2\divides x\,\text{or}\,M\divides y)\implies\sigma(x,y)=0\}.
\]
Indeed for $\sigma\in\tilde{E}_{M}$ all stick edges are vertical,
and no stick intersects a line of the form $y=y_{0}$ with $M$ dividing
$y_{0}$, limiting the length of vertical sticks to be at most $M$.
Assume for simplicity that $\Lambda=\rect{K\times L}{(0,0)}$ where
$M$ divides $L$. Then configurations in $\Omega_{\Lambda}^{0}\cap\tilde{E}_{M}$
are sums of the form $\sum_{i=0}^{K/2-1}\sum_{j=0}^{L/M-1}\sigma_{i,j}$
where $\sigma_{i,j}\in\Omega_{\rect{2\times M}{(2i,Mj)}}^{0}$. This,
together with Proposition \ref{prop:1d_square_bound} gives $Z_{\Lambda}^{1}(\tilde{E}_{M})=Z_{\Lambda}^{0}(\tilde{E}_{M})=(\zoz M)^{\frac{KL}{2M}}\ge(1+\frac{cM^{2}}{\lambda})^{\vol(\Lambda)/M}$,
which matches the upper bound of the lemma, up to the value of $C$,
since $M<\lambda^{1/2}/C$.
\end{rem}
\end{lem}

\subsubsection{Components}

Recall the definition of a stick edge and further define a \textbf{vacancy
edge} as an edge in $\E_{\square}$ that bounds a vacant face. A \textbf{regular
edge} is defined as one that is neither a stick edge nor a vacancy
edge.

We define a \textbf{marked graph}, as a directed graph where each
edge is marked as either horizontal or vertical, and also marked as
either a vacancy edge, a stick edge or a regular edge. Formally, it
is a triplet $\left(V,E,f\right)$ where $\left(V,E\right)$ is a
directed graph and $f:E\to\{\text{``horizontal''},\text{``vertical''}\}\times\{\text{``vacancy''},\text{``stick'',}\text{``regular''}\}$.

For a configuration $\sigma$ its \textbf{configuration graph} $G_{\sigma}$\marginpar{$G_{\sigma}$}
is defined to be a marked graph, that is obtained as follows. We direct
each edge of $(\mathbb{\V},\E_{\square})$ either upwards or to the
right, and mark it as horizontal or vertical, in accordance with our
standard embedding of $(\mathbb{\V},\E_{\square})$ in the plane.
Then we mark each edge with the information of whether it is a stick,
vacancy or regular edge in $\sigma$. Finally, we remove the regular
edges (while keeping all vertices). We note for later use that every
vertex in a configuration graph is either isolated, an internal vertex
of a stick (in which case it has degree exactly $2$) or is incident
to a vacancy in $\sigma$, and these cases are mutually exclusive.
In particular, there are no vertices of degree exactly one.

Define $\mathcal{H}$ \marginpar{$\mathcal{H}$}to be the family of
abstract marked graphs, that may appear as finite connected components
of a configuration graph. We emphasize that $\mathcal{H}$ includes
the \emph{trivial }graph, having a single vertex and no edges. The
word ``abstract'' is used to signify that two elements of $\mathcal{H}$
are considered equal if they are isomorphic as marked directed graphs,
which means that an isomorphism must preserve the directions and markings
of the edges. Formally we write:
\[
\mathcal{H}\coloneqq\left\{ H:\substack{\text{there exist \ensuremath{\sigma\in\Omega} such that \ensuremath{H} is}\\
\text{ a finite connected component of \ensuremath{G_{\sigma}}}
}
\right\} .
\]

Each $H\in\mathcal{H}$ can be realized as a connected component of
some $G_{\sigma}$, by definition, and each such realization yields
an emedding of $H$ in $\R^{2}$. By the \textbf{image} of such an
embedding we mean the set in $\R^{2}$ formed by the union of all
of its vertices and edges. For a non-trivial $H$, this is the same
as the union of all edges.
\begin{prop}
\label{prop:rigidity}Let $H\in\mathcal{H}$ be non-trivial. Suppose
$H$ appears as a connected component of both $G_{\sigma_{1}}$ and
$G_{\sigma_{2}}$, for some $\sigma_{1},\sigma_{2}\in\Omega$. Then
the two resulting embeddings of $H$, as well as the vacancies and
tiles of $\sigma_{1}$and $\sigma_{2}$ bounding on the images of
these embeddings, are the same up to a global translation.
\end{prop}
\begin{proof}
The embeddings are the same up to a global translation since $H$
is connected and the vector in $\R^{2}$ pointing from the head of
an edge in $G_{\sigma}$ to its tail is uniquely determined by its
marking as ``horizontal'' or ``vertical''. To show that the vacancies
and tiles bounding on the images of the embeddings are the same up
to the global translation, let us fix a $G_{\sigma}$ for which $H$
is a connected component and a face $f$ bounding on an edge $e$
of the resulting embedding, and explain how the information in the
embedding uniquely determines whether the face is a vacancy or part
of a tile in $\sigma$ and in the latter case, the parity of the tile. 

It is simple to see that $f$ is a vacancy in $\sigma$ if and only
if it is surrounded by vacancy edges (all of which are necessarily
in $H$). Thus suppose, without loss of generality, that $e$ is horizontal,
that $f$ is the face directly above it, and that $f$ is in a tile.
Denote the other other three edges in $\E_{\square}$ incident to
the right end of $e$ by $e_{1},e_{2},e_{3}$, in clockwise order.
Consider the first of these edges that appears in $G_{\sigma}$. If
it is $e_{1}$ or $e_{3}$, a case analysis shows that the tile covering
$f$ has its center above the left end of $e$. If it is $e_{2}$,
then there is also a tile directly above $e_{2}$, and the parity
of this tile is the same as that of the tile directly above $e$ (they
may or may not be the same tile). By an inductive argument (as $H$
is finite), we may assume that the tile directly above $e_{2}$ is
already known.
\end{proof}

\subsubsection{The partition function of $\mathcal{H}_{M}$}

For $M\ge1$, define $\mathcal{H}_{M}\subset\mathcal{H}$ \marginpar{$\mathcal{H}_{M}$}by
\[
\mathcal{H}_{M}\coloneqq\left\{ H\in\mathcal{H}:\substack{\text{paths of stick edges in \ensuremath{H}}\\
\text{ have length at most \ensuremath{M}}
}
\right\} =\left\{ H:\substack{\text{there exist \ensuremath{\sigma\in E_{M}} such that \ensuremath{H} is}\\
\text{ a finite connected component of \ensuremath{G_{\sigma}}}
}
\right\} .
\]

Consider $G_{\sigma}$ for some $\sigma\in\Omega$. Note that the
four bounding edges of a vacancy are necessarily in the same component
of $G_{\sigma}$; we then say that the vacancy \textbf{belongs} to
the component of its bounding edges. For a finite component $H$ of
$G_{\sigma}$, denote by $v_{H}$\marginpar{$v_{H}$} the number of
vacancies that belong to it (See Figure \ref{fig:components}). By
Proposition \ref{prop:rigidity}, we may define $v_{H}$ for an abstract
$H\in\mathcal{H}$, without mention of $\sigma$. We define the \textbf{weight}
of $H$ to be $\lambda^{-v_{H}/4}$. We will deduce Lemma \ref{lem:counting_argument}
from the following bound on the total weight of $\mathcal{H}_{M}$.
\begin{lem}
\label{lem:graph counting}There exists $C>0$ such that for every
$\lambda>0$ and $M\ge1$, if $M<\lambda^{1/2}/C$ then 
\[
\sum_{H\in\mathcal{H}_{M}}\lambda^{-v_{H}/4}\le1+\frac{CM}{\lambda}.
\]
\end{lem}
We again remark that the bound is sharp, up to the value of $C$,
since $\mathcal{H}_{M}$ contains the trivial graph, and also all
marked graphs with two vertical sticks of equal length $1\le k\le M$
bounded by a pair of vacancies at both ends.

Before proving the lemma, let us explain how it implies the main result
of this subsection.
\begin{proof}
[Proof of Lemma~\ref{lem:counting_argument}]Fix an even rectangle
$\Lambda$, $\lambda>0$ and $M\ge1$. For each $H\in\mathcal{H}_{M}$
we designate one vertex as the root, with the designation being arbitrary
except for the requirement that if $H$ is non-trivial then the root
is not the head of a directed edge. It is possible to satisfy this
requirement since $H$ cannot have a directed cycle (as edges are
directed right/up).

For every $\sigma\in E_{M}\cap\Omega_{\Lambda}^{1}$, define a function
$f_{\sigma}:\mathbb{V}\to\mathcal{H}_{M}$ as follows. For $v\in\mathbb{V}$,
if $v$ happens to be the root of a non-trivial component $H$ of
$G_{\sigma}$ then $f_{\sigma}(v)=H$. Otherwise $f_{\sigma}(v)$
is the trivial graph. We will rely on the fact that $f_{\sigma}$
determines $\sigma$, which is implied by Proposition \ref{prop:rigidity}.

Since outside of $\Lambda$ the configurations in $\Omega_{\Lambda}^{1}$
are fully-packed with tiles of the same parity, $f_{\sigma}$ must
assign the trivial graph to any $v$ outside of $\mathbb{V}\cap\Lambda$.
Taking into account the requirement on the root we see that $f_{\sigma}$
is in fact constant outside of the set $V$, defined to be the set
of lower left corners of faces inside $\Lambda$ (vertices at the
right or top boundaries of $\Lambda$ which are in non-trivial components
of $G_{\sigma}$ are necessarily heads of directed edges since their
degree in $G_{\sigma}$ must be at least two, while their out degree
is at most one).

Now
\begin{align*}
Z_{\Lambda}^{1}(E_{M}) & =\sum_{\sigma\in E_{M}\cap\Omega_{\Lambda}^{1}}\w(\sigma)\\
 & \overset{{\scriptscriptstyle (1)}}{\le}\sum_{f:V\to\mathcal{H}_{M}}\prod_{v\in V}\lambda^{-v_{f(v)}/4}\\
 & \overset{{\scriptscriptstyle }}{=}\left(\sum_{H\in\mathcal{H}_{M}}\lambda^{-v_{H}/4}\right)^{\#V}\overset{{\scriptscriptstyle (2)}}{\le}\left(1+\frac{CM}{\lambda}\right)^{\vol(\Lambda)},
\end{align*}
since:\\
(1) The mapping from $\sigma\mapsto f_{\sigma}\rstr_{V}$ is injective
and $\w(\sigma)=\prod_{v\in\mathbb{V}}\lambda^{-v_{f_{\sigma}(v)}/4}$.\\
(2) Lemma \ref{lem:graph counting} and $\#V=\vol(\Lambda)$.
\end{proof}
The remainder of the subsection is devoted to the proof of Lemma \ref{lem:graph counting}.

\subsubsection{Lower bounds on $v_{H}$}

We proceed to obtain lower bounds for the number of vacancies belonging
to a non-trivial $H\in\mathcal{H}$. We first prove a simple lower
bound, showing that $v_{H}\ge4$, and then prove a more involved lower
bound in terms of the number of vertical and horizontal sub-components
of $H$ (as defined below).
\begin{prop}
\label{prop:v-greater-than-4}If $H\in\mathcal{H}$ is non-trivial
then $v_{H}\ge4$.
\end{prop}
\begin{proof}
Consider $H$ as a component of $G_{\sigma}$ for some $\sigma$.
Since the end of a stick is necessarily a corner of a vacancy, at
least one vacancy must belong to $H$. Among the leftmost vacancies
of $H$, the topmost one must differ from the bottommost one, since
otherwise there is a unique leftmost vacancy $f$, and this is not
possible. Indeed, assuming this by contradiction, a case analysis
shows that from the edge bounding on the left of $f$, must extend
a vertical stick, either from the bottom or from the top vertex of
the edge, and this stick must have a vacancy at its other end, contradicting
that $f$ is the unique leftmost vacancy.

Likewise there is no unique rightmost vacancy, no unique topmost vacancy,
and no unique bottommost vacancy. Therefore the topmost rightmost
vacancy, the topmost leftmost vacancy, the bottommost leftmost vacancy
and the bottommost rightmost vacancy are all distinct from each other.
\end{proof}
Let $H$ be a marked graph. Define $H_{\ver}$\marginpar{$H_{\ver},H_{\hor}$}
as the graph obtained from $H$ by removing its horizontal stick edges
(while keeping all vertices and all vacancy edges). We define the
\textbf{vertical sub-components} of $H$ to be the connected components
of $H_{\ver}$ that are not trivial graphs (equivalently, the ones
that have at least one edge), and denote their cardinality by $k_{H,\ver}$\marginpar{$k_{H,\ver},k_{H,\hor}$}.
We note for later reference that for non-trivial $H$, if a vertex
$v$ is isolated in $H_{\ver}$ (i.e., its connected component is
trivial) then it necessarily was an internal vertex of a horizontal
stick in $H$ (using that $H$ has no vertex of degree one). We define
$H_{\hor}$, \textbf{horizontal sub-components} of $H$ and $k_{H,\hor}$
analogously. Then\marginpar{$k_{H}$} define $k_{H}=k_{H,\ver}+k_{H,\hor}$
(See Figure \ref{fig:components}).

Consider some $H\in\mathcal{H}$ as a component of $G_{\sigma}$ for
some $\sigma$. The bounding edges of a vacancy all belong to a single
vertical sub-component of $H$, and the same is true for a horizontal
one. Therefore we may say that the vacancy \textbf{belongs} to a single
vertical sub-component and a single horizontal sub-component.
\begin{lem}
\label{lem:two_vac}Let $H$ be a finite component of $G_{\sigma}$
for some $\sigma\in\Omega$. Let $A$ be a vertical sub-component
of $H$ and let $B$ be a horizontal sub-component $H$. Suppose that
$A$ and $B$ share a vertex. Then there are at least $2$ vacancies
of $\sigma$ that belong to both $A$ and $B$.
\end{lem}
\begin{proof}
The shared vertex must be incident to a vacancy edge for $\sigma$,
otherwise the vertex is incident to both a horizontal and a vertical
stick edge, and this is not possible. Therefore $A$ shares at least
one vacancy with $B$; denote it by $f$. If $f$ shares an edge with
another vacant face, we are done since the two vacancies must belong
to both $A$ and $B$. Therefore we assume that $f$ is an \textbf{isolated
vacancy}.

Adjacent to $f$ must be four tiles. We may assume WLOG (by applying
reflections and translations) that they are arranged as in Figure
\ref{fig:crossing}. Additionally, since $H$ is finite, we may assume
WLOG (by modifying $\sigma$ away from $H$) that $\sigma\in\Omega_{\Lambda}^{1}$
for some even rectangle $\Lambda$.

Consider the union $U$ of all the tiles whose parity has an odd vertical
component; this includes the deep blue and red tiles in the figure
(recall the color convention introduced in Figure \ref{fig:configs}).
The boundary of $U$ is the image of a subgraph $I$ of $(\V,\E_{\square})$
(i.e. it is a union of points in $\Z^{2}$ and segments of length
$1$ connecting some of them). We observe that $U$ is bounded, by
the definition of $\Omega_{\Lambda}^{1}$, and that necessarily $I\subset(G_{\sigma})_{\ver}$. 

As the image of $I$ is a boundary of a region in the plane, all the
degrees of $I$ are necessarily even. Therefore, any path in $I$
whose internal vertices have degree $2$ in $I$, may be extended
to a (simple) cycle in $I$. Let $C^{\ver}$ be the extension of the
pink path in Figure \ref{fig:crossing} to a cycle in $I$. Then $C^{\ver}$
is a subgraph of $A$: this is since $C^{\ver}\subset I\subset(G_{\sigma})_{\ver}$
and $C^{\ver}$ is connected and intersects $A$ which is a connected
component of $(G_{\sigma})_{\ver}$.

\begin{figure}[t]
\centering{}\includegraphics[width=0.2\paperwidth]{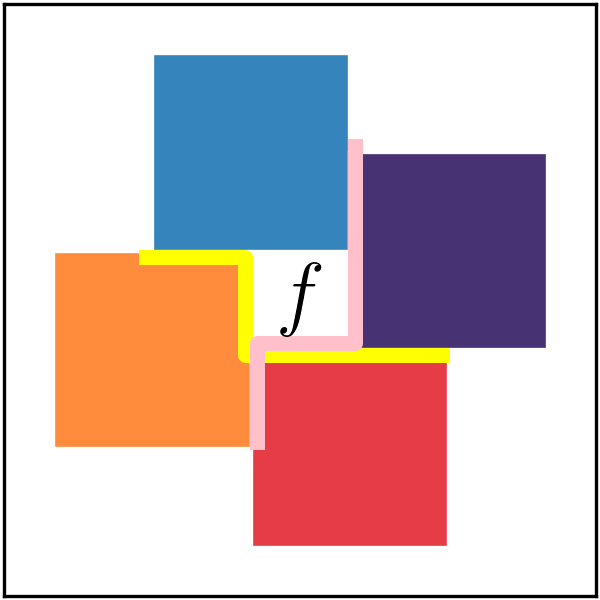}\caption{\label{fig:crossing}The pink and yellow paths intersect near the
vacant face $f$. They extend to cycles $C^{\protect\ver}$ and $C^{\protect\hor}$,
which must intersect near an additional vacancy. For better visibility,
the paths are slightly offset.}
\end{figure}

Repeating the analogous steps with the yellow path in Figure \ref{fig:crossing}
(whose image lies in the boundary of the union of tiles whose parity
has an odd \emph{horizontal} component) gives rise to a cycle $C^{\hor}$
which is a subgraph of $B$.

The cycles $C^{\ver}$ and $C^{\hor}$ must intersect at a shared
vertex of $A$ and $B$ that is not a corner of $f$. This is since
two cycles in the plane that intersect transversally at a point, must
intersect at an additional point. The new shared vertex is necessarily
a corner of a vacancy (as explained in the beginning of the proof),
which is necessarily distinct from $f$. Thus there are at least two
vacancies belonging to both $A$ and $B$.
\end{proof}
We proceed to deduce a lower bound for $v_{H}$ which improves upon
that of Proposition \ref{prop:v-greater-than-4} when $k_{H}>3$.
\begin{cor}
\label{cor:v-greater-than-2k-2}If $H\in\mathcal{H}$ is non-trivial
then $v_{H}\ge2(k_{H}-1)$.
\end{cor}
\begin{proof}
Consider $H$ as a component of $G_{\sigma}$ for some $\sigma$.

We construct an auxiliary bipartite graph, whose vertices are the
horizontal and the vertical sub-components of $H$. A horizontal sub-component
is adjacent to a vertical one, if they share at least one vertex.

Lemma \ref{lem:two_vac} implies that two components are adjacent
in the auxiliary graph iff there are at least \emph{two} vacancies
that belong to both components. This shows that $v_{H}$ is at least
twice the number of edges in the auxiliary graph (since each vacancy
belongs to a unique vertical and a unique horizontal sub-component).

By definition, $k_{H}$ is the number of vertices in the auxiliary
graph. Note that the auxiliary graph is connected since $H$ is connected.
Therefore the number of edges in the auxiliary graph is at least $k_{H}-1$.
Together with the previous paragraph, this gives the lemma.
\end{proof}
\begin{figure}[t]
(i)~\includegraphics[height=0.3\textwidth]{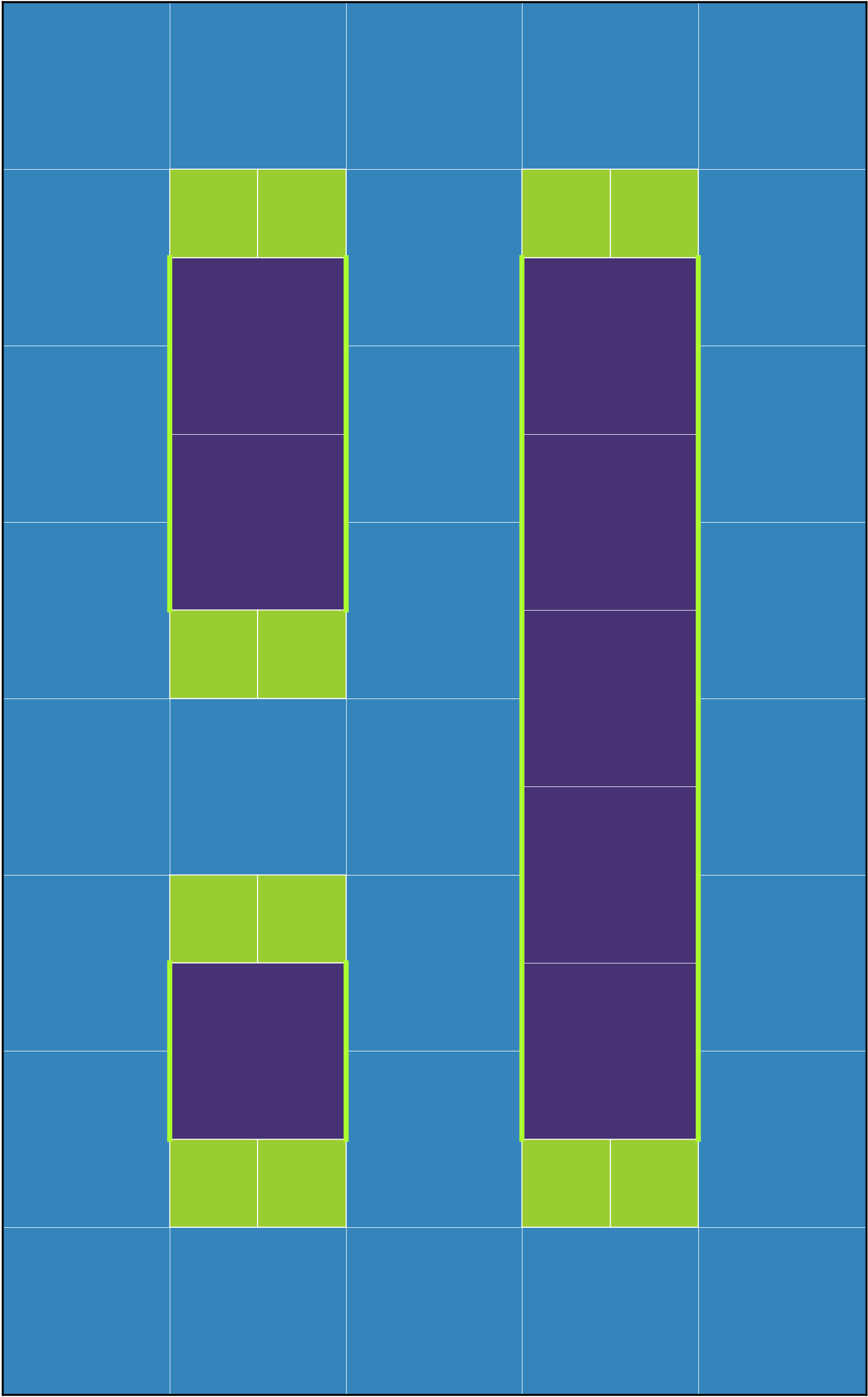}\hspace*{\fill}(ii)~\includegraphics[height=0.3\textwidth]{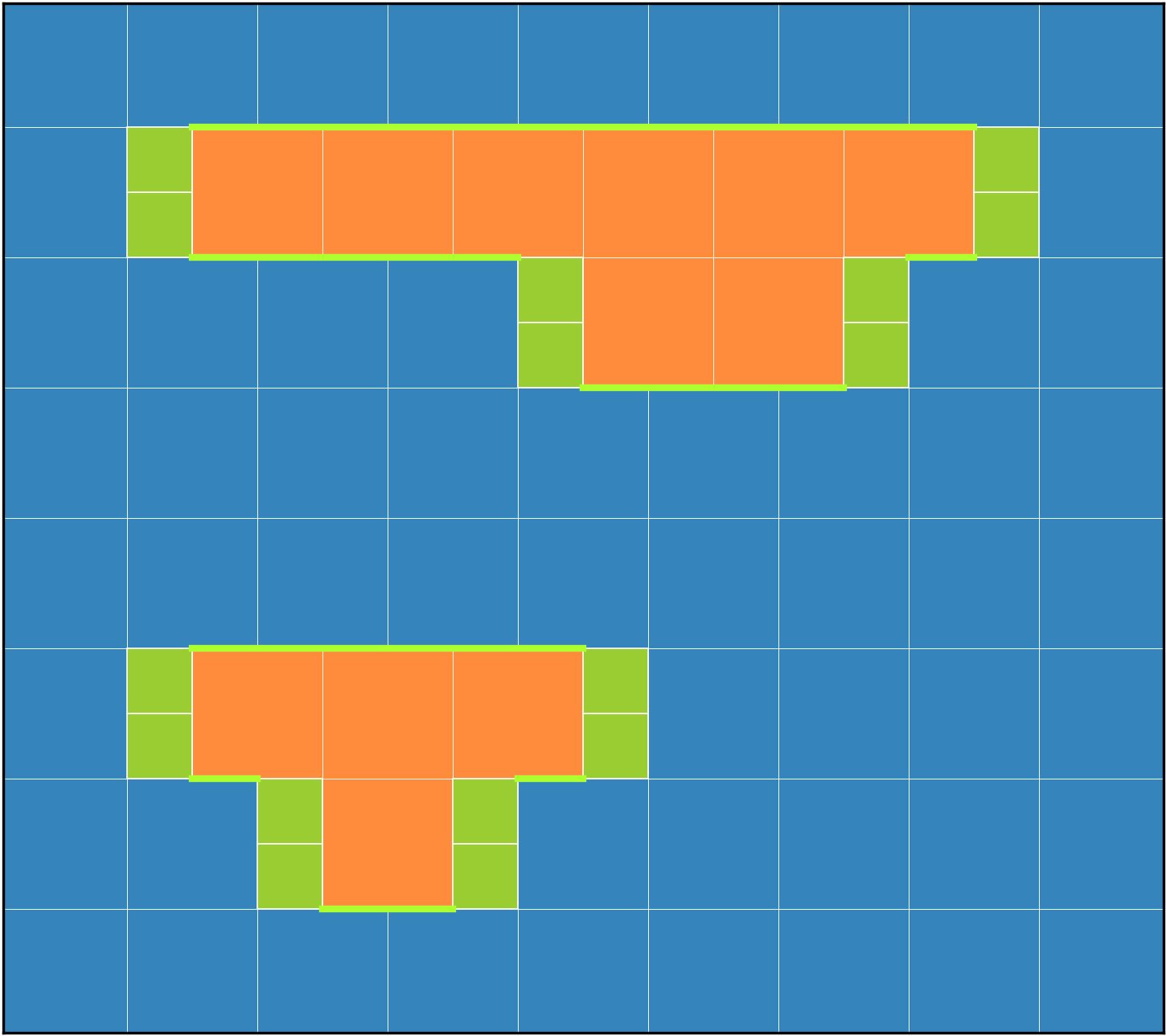}\hspace*{\fill}(iii)~\includegraphics[height=0.3\textwidth]{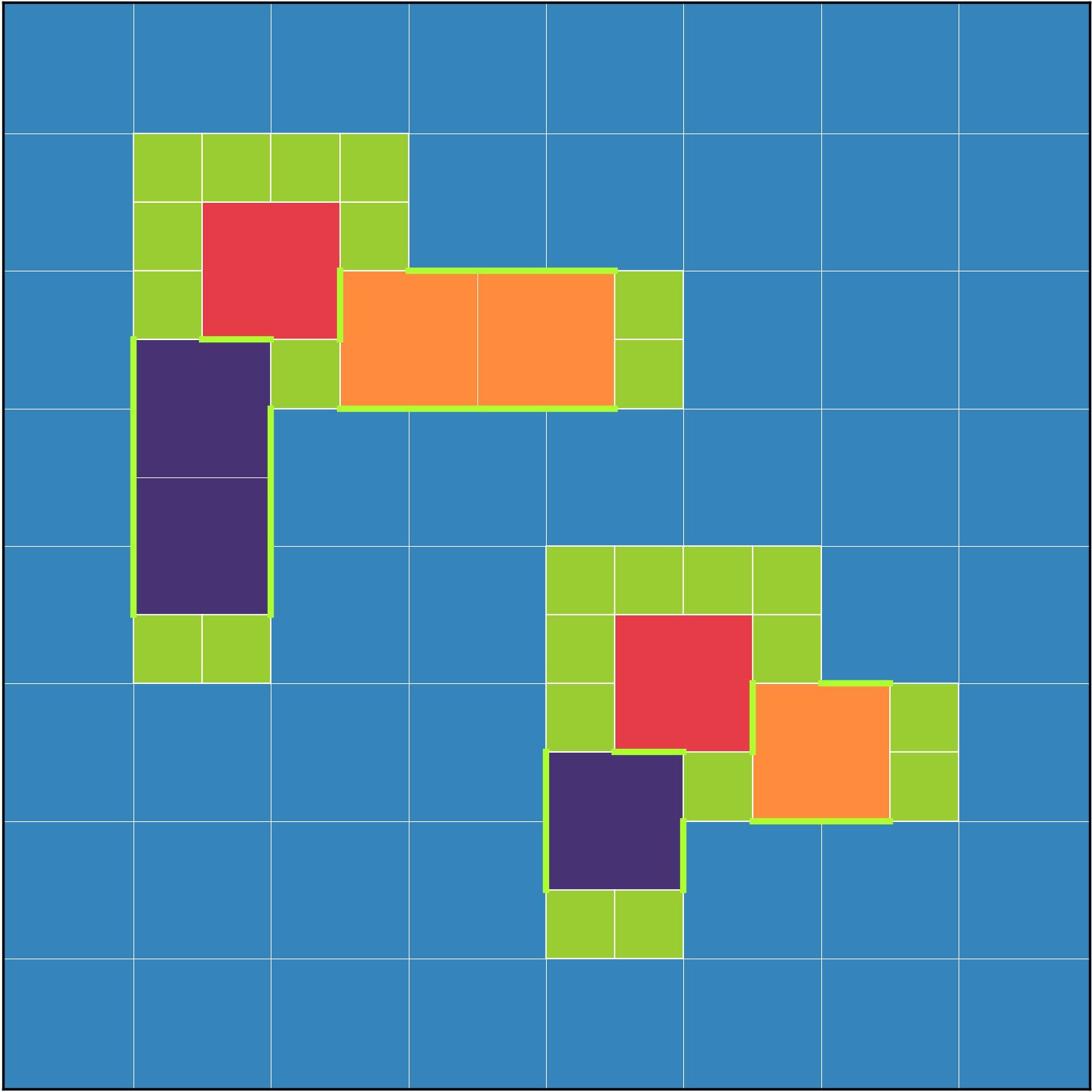}

\caption{\label{fig:components}Each image depicts several components with
the same compressed version, with sticks and vacancies colored green.
Below are the number of vacancies and number of vertical and horizontal
sub-components for a single component in each image:\protect \\
(i)~$v_{H}=4,\quad k_{H}=3,\quad k_{H,\protect\ver}=1,\quad k_{H,\protect\hor}=2$,\protect \\
(ii)~$v_{H}=8,\quad k_{H}=5,\quad k_{H,\protect\ver}=4,\quad k_{H,\protect\hor}=1,$\protect \\
(iii)~$v_{H}=12,\quad k_{H}=4,\quad k_{H,\protect\ver}=2,\quad k_{H,\protect\hor}=2.$}
\end{figure}

\subsubsection{Compressed graphs}

For a marked graph $H$, which is either in $\mathcal{H}$ or is a
vertical or horizontal sub-component of a graph in $\mathcal{H}$,
define its \textbf{compressed version $\comp(H)$} \marginpar{\textbf{$\comp(H)$}}as
follows: every stick (that is, a maximal path of stick edges) is replaced
with a single directed edge pointing from the beginning of the path
to its end, removing all the original internal vertices (noting that
such internal vertices necessarily have degree $2$ in $H$). The
new edge is marked ``stick'' and also ``vertical'' or ``horizontal''
in accordance with the stick that it replaced.

The idea here is that $\comp(I)=\comp(H)$ if $I$ and $H$ are ``the
same up to extending and contracting sticks'' (See Figure \ref{fig:components}).
Lemma \ref{lem:M^k-2} roughly follows from the fact that $k_{H}-2$
is the number of ``degrees of freedom'' in choosing an $I$ with
$\comp(I)=\comp(H)$.
\begin{prop}
\label{prop:compressed_connected_components}Let $H\in\mathcal{H}$
be non-trivial. There are exactly $k_{H,\ver}$ connected components
in $\comp(H)_{\ver}$, all of which are non-trivial. The analogous
claim holds for horizontal sub-components so that, in particular,
$k_{\comp(H)}=k_{H}$.
\end{prop}
\begin{proof}
Recall that, since $H$ is non-trivial, the only trivial connected
components in $H_{\ver}$ arise from internal vertices of horizontal
sticks in $H$. This implies that all connected components of $\comp(H)_{\ver}$
are non-trivial. It remains to check that for each vertex $v\in\comp(H)$,
the connected component of $v$ in $\comp(H)_{\ver}$ necessarily
equals the compressed version of the connected component of $v$ in
$H_{\ver}$. The analogous claims hold for horizontal sub-components.
\end{proof}
\begin{lem}
\label{lem:M^k-2}Let $M\ge1$ and let $H\in\mathcal{H}$ be non-trivial.
Then
\[
\#\{I\in\mathcal{H}_{M}:\comp(I)=\comp(H)\}\le M^{k_{H}-2}.
\]
\end{lem}
\begin{proof}
Given $I\in\mathcal{H}$ we may assign lengths to the stick edges
of $\text{comp}(I)$, such that each is assigned the length of the
path that it replaced. Then $\text{comp}(I)$ together with these
lengths contains sufficient information to reconstruct $I$.

Fix $M\ge1$ and $H\in\mathcal{H}$. An assignment of lengths to the
stick edges of $\text{comp}(H)$ is termed\textbf{ valid }if it arises
from some $I\in\mathcal{H}$ with $\comp(I)=\comp(H)$. If, additionally,
all the assigned lengths are at most $M$, then we say the assignment
is \textbf{$M$-valid}. Thus, the lemma will follow from proving that
the number of $M$-valid length assignments to $\text{comp}(H)$ is
at most $M^{k_{H}-2}$. We will show that
\begin{equation}
\begin{aligned}\text{} & \text{there are at most \ensuremath{M^{k_{H,\ver}-1}} possibilities for the restriction of an}\\
\text{} & \text{\ensuremath{M}-valid length assignment to the horizontal stick edges of \ensuremath{\text{comp}(H)}.}
\end{aligned}
\label{eq:restriction to horizontal edges}
\end{equation}

This, together with the analogous statement for the restriction to
the vertical stick edges will imply the lemma (recalling that $k_{H}=k_{H,\ver}+k_{H,\hor}$).

We first make the following observation. Suppose $\comp(H)$ is endowed
with a valid length assignment, and assign length $1$ to all the
vacancy edges of $\comp(H)$. Consider a closed walk on $\comp(H)$.
Then the sum of signed lengths of horizontal edges in the walk (where
edges that are walked in the opposite direction are counted with a
minus sign) necessarily equals zero, since it represents the total
horizontal movement for a closed walk on a component of $G_{\sigma}$
for some $\sigma$.

Now consider a maximal spanning forest of $\comp(H)_{\ver}$, having
$k_{H,\ver}$ components by Proposition \ref{prop:compressed_connected_components}.
As a spanning forest of $\text{comp}(H)$, it may be extended to a
spanning tree of $\text{comp}(H)$, by adding $k_{H,\ver}-1$ edges
of $\text{comp}(H)$. Denote the set of added edges by $E$. There
are at most $M^{k_{H,\ver}-1}$ possibilities for the restriction
to $E$ of an $M$-valid length assignment, as the assigned length
of each edge is in $\{1,\dots,\left\lfloor M\right\rfloor \}$.

We now prove that the restriction to the horizontal edges of $\text{comp}(H)$
of a valid length assignment is determined by its restriction to $E$,
which proves (\ref{eq:restriction to horizontal edges}) and thus
finishes the proof of the lemma. By construction, the only horizontal
edges in the spanning tree are vacancy edges and the edges in $E$.
Thus, a length assignment to the edges in $E$ determines the length
of all horizontal edges in the spanning tree. Lastly, any horizontal
edge in $\text{comp}(H)$ which is not in the tree, necessarily closes
a cycle with the edges in the tree and thus its length is determined
by the lengths of the horizontal edges in the tree by the observation
above.
\end{proof}

\subsubsection{Proof of Lemma \ref{lem:graph counting}}

For $v\ge4$, denote 
\[
\mathcal{H}'_{v}\coloneqq\{\comp(H):H\in\mathcal{H}\text{, and }v_{H}=v\}.
\]
We claim that $\#\mathcal{H}'_{v}$ grows at most exponentially in
$v$, say 
\begin{equation}
\#\mathcal{H}'_{v}\le C_{1}^{v}.\label{eq:planar_growth}
\end{equation}
This follows from the following two facts: The number of (unlabeled,
simple) planar graphs on $v$ vertices grows at most exponentially
with $v$ \cite{turan1984succinct}, which implies the same for marked
planar graphs (as the number of edges of a planar graph is at most
a constant times its number of vertices). The number of vertices in
$\comp(H)$, for a non-trivial $H\in\mathcal{H}$, is at most $4v_{H}$
(since, in the realization of $H$ as a component of some $G_{\sigma}$,
every vertex which is not an internal vertex of a stick is incident
to a vacancy).

Let $C:=\max\{4C_{1}^{2},2C_{1}^{4}\}.$ Fix some $\lambda>0$ and
$M\ge1$ satisfying $M<\lambda^{1/2}/C$, so that in particular, $C_{1}\lambda^{-1/4}M^{1/2}<1/2$.
Then
\begin{align*}
\sum_{H\in\mathcal{H}_{M}}\lambda^{-v_{H}/4} & =\\
\text{\ensuremath{\left(\substack{\text{by Proposition \ref{prop:v-greater-than-4},}\\
\text{and summing over possibilities}\\
\text{for the compressed version of \ensuremath{H}}
}
\right)}} & =1+\sum_{v\ge4}\sum_{H'\in\mathcal{H}'_{v}}\sum_{\substack{H\in\mathcal{H}_{M}\\
\comp(H)=H'
}
}\lambda^{-v/4}\\
\left(\substack{\text{by Lemma \ref{lem:M^k-2}}\\
\text{and Proposition \ref{prop:compressed_connected_components}}
}
\right) & \le1+\sum_{v\ge4}\sum_{H'\in\mathcal{H}'_{v}}\lambda^{-v/4}M^{k_{H'}-2}\\
\left(\substack{\text{by Proposition \ref{prop:compressed_connected_components},}\\
\text{Corollary \ref{cor:v-greater-than-2k-2} and \ensuremath{M\ge1}}
}
\right) & \le1+\sum_{v\ge4}\sum_{H'\in\mathcal{H}'_{v}}\lambda^{-v/4}M^{v/2-1}\\
\left(\text{by inequality (\ref{eq:planar_growth})}\right) & \le1+\sum_{v\ge4}C_{1}^{v}\lambda^{-v/4}M^{v/2-1}\\
\left(\text{rearrangment}\right) & \le1+M^{-1}\sum_{v\ge4}(C_{1}\lambda^{-1/4}M^{1/2})^{v}\\
\text{\ensuremath{\left(\substack{\text{using \ensuremath{C_{1}\lambda^{-1/4}M^{1/2}<1/2}}\\
 \text{and summing a geometric series} 
}
 \right)}} & \le1+2C_{1}^{4}\lambda^{-1}M\le1+\frac{CM}{\lambda}.\tag*{{\qed}}
\end{align*}

\subsection{Configurations mostly without long sticks}

Recall that $E_{M}$ is the event that all sticks have length at most
$M$, and that the weight of $E_{M}$ under fully-packed boundary
conditions was estimated in Lemma \ref{lem:counting_argument}. The
proof of our main lemma, Lemma \ref{lem:main}, requires an extension
of Lemma \ref{lem:counting_argument} in which the weight of a larger
event is estimated. The larger event is parameterized by a collection
of horizontal and vertical line segments and consists of configurations
in which all sticks are of length at most $M$, \emph{except} maybe
the sticks contained in one of the segments of the collection. We
proceed to describe this extension.

Let $M\ge1$ and let $A$ be a collection \marginpar{$A$}of vertical
and horizontal line segments of the form $\{x=x_{0},y_{0}\le y\le y_{1}\}$
or $\{y=y_{0},x_{0}\le x\le x_{1}\}$ with $x_{0},y_{0},x_{1},y_{1}\in\Z$.
Define $E_{M,A}$ \marginpar{$E_{M,A}$}to be the event that every
stick whose length is longer than $M$ is fully contained in one of
the segments in $A$. Denote by $\len(A)$ \marginpar{$\len(A)$}the
total length of the segments in $A$.
\begin{prop}
\label{prop:shminiot}There exists $C\ge1$ such that the following
holds for all fugacities $\lambda>0$. Let $\Lambda$ be an even rectangle.
Let $M>C$ and let $A$ be a collection of line segments as above.
Then
\begin{equation}
Z_{\Lambda}^{1}(E_{M,A})\le\exp\left(\frac{C\lambda}{M^{3}}\len(A)\right)Z_{\Lambda}^{1}(E_{M}).\label{eq:stick_dilution-1}
\end{equation}
\end{prop}
\begin{proof}
Fix $M$, $A$ and $\Lambda$ as above. For the span of this proof,
we say that a stick or segment is \textbf{long }if its length is more
than $M$.

We assume WLOG that all segments in $A$ are contained in $\Lambda$,
as replacing each segment in $A$ by its intersection with $\Lambda$
leaves the set $\Omega_{\Lambda}^{1}\cap E_{M,A}$ unaltered. Similarly,
assume WLOG that no segment in $A$ lies at distance exactly $1$
from an edge of $\Lambda$ parallel to it (as a stick contained in
such a segment implies that a tile is centered on a point in $\partial\Lambda$,
violating the boundary conditions). We also assume WLOG that all line
segments in $A$ are long, as the event $E_{M,A}$ is invariant to
the removal of segments from $A$ whose length is at most $M$.

Choose a collection of vertical and horizontal line segments $I_{1},\dots,I_{N}$
\marginpar{$I_{1},\dots,I_{N}$}of length $\lceil\frac{1}{2}M\rceil$
whose endpoints are on $\Z^{2}$ and whose union equals the union
of the segments in $A$, in such a way that
\[
N\le\text{\ensuremath{\frac{3\len(A)}{M}}}.
\]

One may check that if a long stick is contained in a segment of $A$
then necessarily the stick contains one of $I_{1},\dots,I_{N}$.

For each $1\le i\le N$, let $D_{i}$ \marginpar{$D_{i}$}be the event
that $I_{i}$ is not contained in a long stick.

For $0\le k\le N$, set $F_{k}:=E_{M,A}\cap\bigcap_{i=1}^{k}D_{i}$,
\marginpar{$F_{i}$}so that $F_{0}=E_{M,A}$ and $F_{N}=E_{M}$. Observe
that
\begin{equation}
\frac{Z_{\Lambda}^{1}(E_{M,A})}{Z_{\Lambda}^{1}(E_{M})}=\prod_{i=1}^{N}\frac{Z_{\Lambda}^{1}(F_{i-1})}{Z_{\Lambda}^{1}(F_{i})}.\label{eq:bound by ratios}
\end{equation}
We will show that for a sufficiently large universal constant $C$,
for each $1\le i\le N$ and $M>C$, it holds that
\begin{equation}
\frac{Z_{\Lambda}^{1}(F_{i-1})}{Z_{\Lambda}^{1}(F_{i})}\le1+\frac{C\lambda}{M^{2}}.\label{eq:ratio of Zs}
\end{equation}
This suffices for the proposition, as substituting this bound into
(\ref{eq:bound by ratios}) implies (\ref{eq:stick_dilution-1}) (with
a larger $C$) by using the bound on $N$.

Fix $1\le i\le N$. We will define a mapping $m:(F_{i-1}\setminus F_{i})\to2^{F_{i-1}}$
\marginpar{$m$}and show that it satisfies
\begin{equation}
m(\sigma_{1})\cap m(\sigma_{2})=\emptyset\text{\ensuremath{\quad}for distinct \ensuremath{\sigma_{1},\sigma_{2}\in F_{i-1}\setminus F_{i}},}\label{eq:dstinct images}
\end{equation}
and, for $M>C$,
\begin{equation}
Z_{\Lambda}^{1}(m(\sigma))\ge\left(1+\frac{M^{2}}{C\lambda}\right)\w[\Lambda](\sigma)\text{\ensuremath{\quad}for \ensuremath{\sigma\in(F_{i-1}\setminus F_{i})\cap\Omega_{\Lambda}^{1}}.}\label{eq:weight increase}
\end{equation}

The existence of $m$ with these properties implies that 
\[
Z_{\Lambda}^{1}(F_{i-1})\ge\left(1+\frac{M^{2}}{C\lambda}\right)Z_{\Lambda}^{1}(F_{i-1}\setminus F_{i})
\]
by summing over (\ref{eq:weight increase}). The last display and
the fact that $F_{i}\subset F_{i-1}$ imply (\ref{eq:ratio of Zs}).

We proceed to define the mapping $m$. Assume WLOG that $I_{i}$ is
vertical. Choose \marginpar{$J$} $J$ to be a rectangle contained
in $\Lambda$ with $\width(J)=2$ and with one of its vertical sides
coinciding with $I_{i}$ (this is possible by the first two assumptions
made at the beginning of the proof). Define $m(\sigma):=\Omega_{J}^{\sigma}$,
where we recall from (\ref{eq:Omega Lambda rho}) that $\Omega_{J}^{\sigma}$
is the set of configurations which agree with $\sigma$ on all tiles
which are not fully contained in $J$. To show that $m$ is correctly
defined we must prove that $\Omega_{J}^{\sigma}\subset F_{i-1}$,
which we shall do in the last two paragraphs of the proof.

To show (\ref{eq:dstinct images}), let $\sigma_{1},\sigma_{2}\in F_{i-1}\setminus F_{i}$,
and assume that $\Omega_{J}^{\sigma_{1}}\cap\Omega_{J}^{\sigma_{2}}\neq\emptyset$.
Then $\sigma_{1}\rstr_{\intr J^{c}}=\sigma_{2}\rstr_{\intr J^{c}}$.
In particular, $\sigma_{1}$ agrees with $\sigma_{2}$ on the tiles
bounding on $I_{i}$ on the side opposite to $J$. Thus, using the
fact that $I_{i}$ is contained in a stick of both $\sigma_{1}$ and
$\sigma_{2}$ (since $\sigma_{1},\sigma_{2}\notin D_{i})$ we deduce
that $\sigma_{1}\rstr_{\intr J}=\sigma_{2}\rstr_{\intr J}$, whence
$\sigma_{1}=\sigma_{2}$.

We now show (\ref{eq:weight increase}). Fix $\sigma\in(F_{i-1}\setminus F_{i})\cap\Omega_{\Lambda}^{1}$.
Let $\sigma_{0}$ be the configuration obtained from $\sigma$ by
removing all tiles fully contained in $J$ and denote the union of
these tiles by\marginpar{$J_{\sigma}$} $J_{\sigma}$. Since $\sigma\notin D_{i}$,
the rectangle $J$ is fully covered by tiles of $\sigma$, whence
$J_{\sigma}$ is a rectangle of width $2$ and its height is even
and satisfies $\lceil\frac{1}{2}M\rceil-2\le\height(J_{\sigma})\le\lceil\frac{1}{2}M\rceil$.
We then observe that
\begin{equation}
m(\sigma)=\Omega_{J}^{\sigma}=\Omega_{J_{\sigma}}^{\sigma}=\sigma_{0}+\Omega_{J_{\sigma}}^{0}\coloneqq\{\sigma_{0}+\tilde{\sigma}:\tilde{\sigma}\in\Omega_{J_{\sigma}}^{0}\}.\label{eq:alternative_configurations}
\end{equation}
We note that $\w[\Lambda](\sigma_{0}+\tilde{\sigma})=\w[\Lambda](\sigma)\w[J_{\sigma}](\tilde{\sigma})$
for all $\tilde{\sigma}\in\Omega_{J_{\sigma}}^{0}$; also, $m(\sigma)\subset\Omega_{\Lambda}^{1}$
(since $\sigma\in\Omega_{\Lambda}^{1}$ and $J\subset\Lambda$) and
thus
\[
Z_{\Lambda}^{1}(m(\sigma))=Z_{\Lambda}^{0}(J_{\sigma})\w[\Lambda](\sigma).
\]
Recalling that $Z_{\Lambda}^{0}(J_{\sigma})=\zoz{\height(J_{\sigma})}$
and that $\height(J_{\sigma})$ is even and at least $\frac{1}{2}M-2$,
we obtain (\ref{eq:weight increase}) from the one-dimensional estimate
of Proposition \ref{prop:1d_square_bound} by choosing $C$ large
enough and using the hypothesis $M>C$.

It remains to prove that $\Omega_{J}^{\sigma}\subset F_{i-1}$ for
each $\sigma\in F_{i-1}\setminus F_{i}$. Fix $\sigma\in F_{i-1}\setminus F_{i}$
and some $\sigma'\in m(\sigma)=\Omega_{J_{\sigma}}^{\sigma}$. We
have that $\sigma\in F_{i-1}$, which means that each long stick of
$\sigma$ is contained in a segment of $A$, and does not contain
any of $I_{1},\dots,I_{i-1}$. We will show that each long stick of
$\sigma'$ is contained in a stick of $\sigma$. Thus each long stick
of $\sigma'$ is contained in a segment of $A$, and does not contain
any of $I_{1},\dots,I_{i-1}$, whence $\sigma'\in F_{i-1}$.

We now show that each long stick of $\sigma'$ is contained in a stick
of $\sigma$. Indeed, every horizontal stick edge in $\sigma'$ is
a stick edge in $\sigma$. Every vertical stick edge in $\sigma'$
that is not a stick edge in $\sigma$ bounds on a tile contained in
$J_{\sigma}$ which does not appear in $\sigma$. As $\height(J_{\sigma})$
is even, $J_{\sigma}$ contains pairs of horizontally-adjacent vacancies
of $\sigma'$ above and below that tile. This implies that such vertical
stick edges are part of sticks whose length is less than $\height(J_{\sigma})\le\lceil\frac{1}{2}M\rceil\le M$.
\end{proof}

\subsection{Proof of the main lemma}
\begin{proof}
[Proof of Lemma \ref{lem:main}]Let $c>0$ be a constant sufficiently
small to satisfy some assumptions that will follow. Let $S\subset R$
be rectangles satisfying the hypotheses (\ref{eq:dim_abs}) and (\ref{eq:dim_rat})
of the lemma. Note in particular that we may assume the fugacity $\lambda$
to be large by taking $c$ sufficiently small in (\ref{eq:dim_abs}).

Denote by $f$ the indicator of the ($R$-local) event that no stick
divides both $R$ and $S$. Fix an arbitrary even rectangle $\Lambda$
for which $R$ is a block. Recalling (\ref{eq:infinite-volume zeta}),
the definition of $\zedi$, we aim to bound 
\[
\zedd[][][f]=\mu_{\Lambda}^{\per}\left(\prod_{\tau\in\refls}\tau f\right)^{\frac{\vol(R)}{\vol(\Lambda)}}
\]
and take the limit as $\width(\Lambda),\height(\Lambda)\to\infty$.
We assume for convenience that $R$ and $\Lambda$ have their bottom
left corner at the origin. Denote by $E_{R,S}$ \marginpar{$E_{R,S}$}the
event that for all $\tau\in\refls[\null][R]$, there is no stick dividing
both $\tau R$ and $\tau S$, and observe that on $\Omega_{\Lambda}^{\per}$,
the function $\prod_{\tau\in\refls}\tau f$ is the indicator of $E_{R,S}$.

Recall Proposition \ref{prop:boundary_fixing} and the function $m^{\rho,\Lambda}$
defined there. We use this proposition to bound $\mu_{\Lambda}^{\per}(E_{R,S})$.
Fix $\rho\in\Omega$ to be the fully-packed configuration $\rho(x,y)=\mathbf{1}_{x,y=1\bmod2}$.
By the proposition,
\[
Z_{\Lambda}^{\per}(E_{R,S})\le C_{\ref{prop:boundary_fixing}}(\lambda)^{\perim(\Lambda)}Z_{\Lambda}^{1}(m^{\rho,\Lambda}(E)).
\]

Aiming to apply the bound of Proposition \ref{prop:shminiot}, we
will choose $M$ and $A$ so that
\begin{equation}
m^{\rho,\Lambda}(E_{R,S})\subset E_{M,A}.\label{eq:m containment}
\end{equation}
We postpone the choice of $M$ and assume for now 
\begin{equation}
M\ge2\max\{\width(R),\height(R)\},\label{eq:main_M_req}
\end{equation}
and let $A$ be the set of integer translates of the sides of $\Lambda$
which are both contained in $\Lambda$ and disjoint from $\bigcup_{\tau\in\refls[\null]}\intr{\tau S}$.

Let us check that (\ref{eq:m containment}) holds. Indeed let $\sigma\in E_{R,S}$
and denote $\sigma'=m^{\rho,\Lambda}(\sigma)$. To show that $\sigma'\in E_{M,A}$,
we consider a stick $s'$ of $\sigma'$ with length more than $M$,
and show that it is contained in a segment of $A$. Note that $s'\subset\Lambda$
by the choice of $\rho$. Assume WLOG that $s'$ is vertical and consider
its extension to a horizontal translation of a vertical side of $\Lambda$.
If this extension is a side of $\Lambda$ then we are done, as it
is an element of $A$. Otherwise, by the choice of $\rho$ and using
that $\Lambda$ is an even rectangle, every edge of $s'$ is also
a stick edge in $\sigma$, and thus $s'$ is contained in a stick
$s$ of $\sigma$. The stick $s$ is of length at least $2\height(R)$
and thus must divide some rectangle $\tau R$ or lie on its boundary.
But, as $\sigma\in E_{R,S}$, it cannot divide $\tau S$. This implies
that the extension of $s'$ is disjoint from $\bigcup_{\tau\in\refls[\null]}\intr{\tau S}$
and thus an element of $A$. We conclude that (\ref{eq:m containment})
holds.

So far, we have
\begin{align*}
\zedd[\Lambda]^{\frac{\vol(\Lambda)}{\vol(R)}} & =\mu_{\Lambda}^{\per}\left(E_{R,S}\right)=\frac{Z_{\Lambda}^{\per}(E_{R,S})}{Z_{\Lambda}^{\per}}\le\frac{C_{\ref{prop:boundary_fixing}}(\lambda)^{\perim(\Lambda)}Z_{\Lambda}^{1}(E_{M,A})}{Z_{\Lambda}^{\per}}.
\end{align*}

Putting $\Lambda=\frect{n!\times n!}$ and taking the limit $n\to\infty$
we learn that 
\begin{equation}
\zedi=\limsup_{n\to\infty}\zedd[\frect{n!\times n!}]\le\limsup_{\width(\Lambda),\height(\Lambda)\to\infty}\left(\frac{Z_{\Lambda}^{1}(E_{M,A})}{Z_{\Lambda}^{\per}}\right)^{\frac{\vol(R)}{\vol(\Lambda)}}\label{eq:main_limit}
\end{equation}
and we are left with bounding the RHS.

Combining the main results of the three previous subsections, respectively:
Corollary \ref{cor:1d_bound}, Lemma \ref{lem:counting_argument}
and Proposition \ref{prop:shminiot} we have
\begin{equation}
\frac{Z_{\Lambda}^{1}(E_{M,A})}{Z_{\Lambda}^{\per}}\le\frac{\exp\left(\frac{C_{\ref{prop:shminiot}}\lambda}{M^{3}}\len(A)\right)\left(1+\frac{C_{\ref{lem:counting_argument}}M}{\lambda}\right)^{^{\vol(\Lambda)}}}{\exp\left(c_{\ref{cor:1d_bound}}\lambda^{-1/2}\vol(\Lambda)\right)}\label{eq:main_rat_1}
\end{equation}
given that we fix some $0<c_{\ref{cor:1d_bound}}<1/4$, say $c_{\ref{cor:1d_bound}}=1/8$,
and make the assumptions that $\lambda$ is sufficiently large and
that
\begin{equation}
C_{\ref{prop:shminiot}}<M<\frac{\lambda^{1/2}}{C_{\ref{lem:counting_argument}}}.\label{eq:M requisite upper bound}
\end{equation}

Let us bound $\len(A)$. The number of vertical segments in $A$ is
\[
\width(\Lambda)+1-[\width(S)-1]\frac{\width(\Lambda)}{\width(R)},
\]
thus their total length is
\begin{align*}
 & \vol(\Lambda)\left(1-\frac{\width(S)}{\width(R)}+\frac{1}{\width(R)}+\frac{1}{\width(\Lambda)}\right)\\
 & \le\vol(\Lambda)\left(1-\left(1-c\right)+\frac{1}{1/c}+\frac{1}{2/c}\right)\\
 & \le3c\vol(\Lambda)
\end{align*}
using the hypotheses (\ref{eq:dim_rat}) and (\ref{eq:dim_abs}) and
the fact that $R$ is a block of $\Lambda$. A similar bound holds
for the horizontal segments, and thus
\[
\len(A)\le6c\vol(\Lambda).
\]

We set
\[
M=\frac{c_{\ref{cor:1d_bound}}}{2C_{\ref{lem:counting_argument}}}\lambda^{1/2}
\]
and require $c$ to be sufficiently small for (\ref{eq:dim_abs})
to imply (\ref{eq:main_M_req}), and $\lambda$ to be sufficiently
large to imply (\ref{eq:M requisite upper bound}). Substitute the
two last displays into (\ref{eq:main_rat_1}) to obtain
\begin{align*}
\left(\frac{Z_{\Lambda}^{1}(E_{M,A})}{Z_{\Lambda}^{\per}}\right)^{\frac{\vol(R)}{\vol(\Lambda)}} & \negmedspace\le\exp\left(\left[6cC_{\ref{prop:shminiot}}\frac{\lambda^{3/2}}{M^{3}}+C_{\ref{lem:counting_argument}}M\lambda^{-1/2}-c_{\ref{cor:1d_bound}}\right]\vol(R)\lambda^{-1/2}\right)\\
 & \le\exp\left(\left[6cC_{\ref{prop:shminiot}}\frac{\lambda^{3/2}}{M^{3}}-c_{\ref{cor:1d_bound}}/2\right]\vol(R)\lambda^{-1/2}\right)\\
 & \le\exp\left(-c\vol(R)\lambda^{-1/2}\right),
\end{align*}

where $c$ is chosen sufficiently small for the last inequality. Combining
this with (\ref{eq:main_limit}) we get the lemma.
\end{proof}

\section{Existence of multiple Gibbs measures \label{sec:multiple_gibbs}}

In this section we prove Theorem \ref{thm:dichotomy} which shows,
for periodic Gibbs measures with sufficiently large fugacity, that
almost surely exactly one of two symmetric invariant events holds.
Corollary \ref{cor:multiple Gibbs measures} concludes from this the
non-uniqueness of Gibbs measures.

For the rest of the paper, fix some integer\marginpar{$N$} $N>2$
such that $\frac{2}{N}\le c_{\ref{lem:main}}$ (where $c_{\ref{lem:main}}$
is the constant from the statement of Lemma \ref{lem:main}). For
a rectangle $R$ with dimensions divisible by $N$, denote by\marginpar{$R^{-}$}
$R^{-}$ the rectangle sharing its center with $R$ and having ($\width(R^{-}),\height(R^{-}))=\frac{N-2}{N}(\width(R),\height(R))$.
Observe that our choice of $N$ ensures that the assumption (\ref{eq:dim_rat})
of Lemma \ref{lem:main} is satisfied when $S=R^{-}.$ We say that
a stick \textbf{divides $R$ properly} if it divides both $R$ and
$R^{-}$ (see Figure \ref{fig:dividing}).

For $K,L\in\nat$ and a configuration $\sigma\in\Omega$, define a
set\marginpar{$\Psi_{(\ver,0)}^{K\times L}$} $\Psi_{(\ver,0)}^{K\times L}(\sigma)\subset\V$
as follows: for $(x,y)\in\V$, set $R=\rect{KN\times LN}{(xK,yL)}$
and say that $(x,y)\in\Psi_{(\ver,0)}^{K\times L}(\sigma)$ if $R$
is divided properly by some $(\ver,0)$ stick of $\sigma$. We make
three analogous definitions by putting $(\ver,1)$, $(\hor,0)$ or
$(\hor,1)$ instead of $(\ver,0)$ in the definition above. Also\marginpar{$\Psi_{\ver}^{K\times L}$\smallskip{}

$\Psi^{K\times L}$} define 
\begin{align*}
\Psi_{\ver}^{K\times L}(\sigma) & :=\Psi_{(\ver,0)}^{K\times L}(\sigma)\cup\Psi_{(\ver,1)}^{K\times L}(\sigma),\\
\Psi_{\hor}^{K\times L}(\sigma) & :=\Psi_{(\hor,0)}^{K\times L}(\sigma)\cup\Psi_{(\hor,1)}^{K\times L}(\sigma),\\
\Psi^{K\times L}(\sigma) & :=\Psi_{\ver}^{K\times L}(\sigma)\cup\Psi_{\hor}^{K\times L}(\sigma).
\end{align*}

The following lemma is key to our use of the Peierls argument. It
shows that regions with long vertical sticks must be separated from
regions with long horizontal sticks.
\begin{lem}
\label{lem:stick_collision}Let $K,L\in\nat$ and $\sigma\in\Omega$.
If $u\in\Psi_{\ver}^{K\times L}(\sigma)$ and $v\in\Psi_{\hor}^{K\times L}(\sigma)$
then $u,v$ are not neighbors in $(\V,\E_{\square})$.
\end{lem}
\begin{proof}
Assume $u\in\Psi_{\ver}^{K\times L}(\sigma)$ and $v\in\Psi_{\hor}^{K\times L}(\sigma)$
and assume by contradiction that $u,v$ are neighbors in $(\V,\E_{\square})$.
The situation has enough symmetry that we may assume WLOG that $u=(0,0)$
and $v=(0,1)$. Then $(0,0)\in\Psi_{\ver}^{K\times L}(\sigma)$ implies
that $R=\rect{KN\times LN}{(0,0)}$ is divided by a vertical stick.
Also, $(0,1)\in\Psi_{\hor}^{K\times L}(\sigma)$ implies that a horizontal
stick divides $\rect{KN\times LN}{(0,L)}$ and $\rect{K(N-2)\times L(N-2)}{(K,2L)}$.
This horizontal stick must also divide $R$, and thus intersects the
vertical stick. As sticks cannot intersect, this is a contradiction.
\end{proof}
In the next theorem, as well as in many of our later uses, the discussion
focuses on a Gibbs measure $\mu$ and the notation $\Psi_{\ver}^{K\times L}$
(and its relatives) is used without explicit mention of $\sigma$.
In these cases it is understood that $\sigma$ is randomly sampled
from $\mu$.
\begin{thm}
\label{thm:dichotomy}There are $C,c>0$ such that the following holds.
Let $b\in\nat$ satisfy $C\lambda^{1/4}<b<c\lambda^{1/2}$. Let $\mu$
be a periodic Gibbs measure. Then
\begin{equation}
\mu(\text{exactly one of \ensuremath{\Psi_{\ver}^{b\times b}} and \ensuremath{\Psi_{\hor}^{b\times b}} has an infinite \ensuremath{\square}-component})=1.\label{eq:exactly one infinite component}
\end{equation}
\end{thm}
\begin{proof}
We will prove that for each $C_{0}>0$, under the hypotheses of the
theorem, for each \marginpar{$A$}finite $A\subset\V$ it holds that
\begin{equation}
\mu\left(A\cap\Psi^{b\times b}=\emptyset\right)\le e^{-C_{0}\#A}.\label{eq:Peierls condition}
\end{equation}

Taking $C_{0}$ large enough, by the Peierls argument, (\ref{eq:Peierls condition})
implies that $\Psi^{b\times b}$ almost surely contains a unique infinite
$\square$-component $I$ (we assume here familiarity with the Peierls
argument. However, Lemma \ref{lem:peierls_inf_comp} below provides
a proof). The infinite component $I$ must be contained in either
$\Psi_{\ver}^{b\times b}$ or $\Psi_{\hor}^{b\times b}$ since their
union is $\Psi^{b\times b}$, and for $u\in\Psi_{\ver}^{b\times b},v\in\Psi_{\hor}^{b\times b}$
it cannot be that $(u,v)\in\E_{\square}$ by Lemma \ref{lem:stick_collision}.
Thus it remains to prove (\ref{eq:Peierls condition}).

Choose $c=c_{\ref{lem:main}}/N$ and let $C>0$ be a constant, sufficiently
large to satisfy some assumptions that will follow. Let $b\in\nat$,
$\lambda>0$ satisfy $C\lambda^{1/4}<b<c\lambda^{1/2}$, and let $\mu$
be a periodic Gibbs measure for the fugacity $\lambda$. Fix a finite
$A\subset\V$.

Choose\marginpar{$v,A'$} $v\in\{0,\dots,N-1\}^{2}$ such that $A'\coloneqq(v+N\Z^{2})\cap A$
satisfies $\#A'\ge\#A/N^{2}$. Set $R=\rect{bN\times bN}{bv}$. Lemma
\ref{lem:main} is applicable to $R$ and $S=R^{-}$. Indeed (\ref{eq:dim_rat})
holds by the choice of $N$, the right part of (\ref{eq:dim_abs})
holds by the choice of $c$, and the left part of (\ref{eq:dim_abs})
then holds for sufficiently large $C$ (as $b>C\lambda^{1/4}>C^{2}/c$).
Thus the lemma yields
\[
\zedi[R][f]\le e^{-c_{\ref{lem:main}}\vol(R)\lambda^{-1/2}}=e^{-c_{\ref{lem:main}}(bN)^{2}\lambda^{-1/2}}
\]

where $f$ is the indicator of the event that $R$ is not divided
properly. The definition of $A'$ implies that for each $u\in A'$,
the indicator of the event that $u\notin\Phi^{b\times b}$ is of the
form $\tau f$, where $\tau\in\refls[\null][R]$ is distinct for each
$u$. By the infinite volume chessboard estimate (Proposition \ref{prop:chessboard_infinite}),
this says that
\begin{align*}
\mu\left(A\cap\Psi^{b\times b}=\emptyset\right) & \le\mu\left(A'\cap\Psi^{b\times b}=\emptyset\right)\le(\zedi[R][f])^{\#A'}\\
 & \le e^{-c_{\ref{lem:main}}(bN)^{2}\lambda^{-1/2}\#A'}\le e^{-c_{\ref{lem:main}}b^{2}\lambda^{-1/2}\#A}.
\end{align*}

Inequality (\ref{eq:Peierls condition}) now follows from the assumption
that $b>C\lambda^{1/4}$ when $C$ is sufficiently large.
\end{proof}
\begin{cor}
\label{cor:multiple Gibbs measures}There exists $\lambda_{0}$ such
that for all $\lambda>\lambda_{0}$ there are at least two periodic
Gibbs measures.
\end{cor}
\begin{proof}
By compactness, there is a subsequence of $\left(\mu_{R_{L\times L,(-L/2,-L/2)}}^{\per}\right)_{L\in2\nat}$
which converges in distribution to a Gibbs measure $\mu$. The periodic
boundary conditions ensure that $\mu$ is $\Z^{2}$-invariant.

Let $\lambda$ be sufficiently large so that we may choose $b\in\nat$
satisfying $C_{\ref{thm:dichotomy}}\lambda^{1/4}<b<c_{\ref{thm:dichotomy}}\lambda^{1/2}$.
Let $E_{\ver}$ be the event that $\Psi_{\ver}^{b\times b}$ has an
infinite $\square$-component and define $E_{\hor}$ analogously.
Theorem \ref{thm:dichotomy} shows that 
\begin{equation}
\mu(\text{exactly one of \ensuremath{E_{\ver}} or \ensuremath{E_{\hor}} occurs})=1.\label{eq:exactly one infinite component for mu}
\end{equation}

Assume without loss of generality that $\mu(E_{\ver})>0$. Write $\mu_{\ver}$
for the measure $\mu$ conditioned on $E_{\ver}$. The fact that $E_{\ver}$
is a $b\Z^{2}$-invariant event implies that $\mu_{\ver}$ is a Gibbs
measure, and is $b\Z^{2}$-invariant.

Define $\mu_{\hor}\coloneqq\tau\mu_{\ver}$ where $\tau$ is the reflection
defined by $\tau(x,y)=(y,x)$. Then $\mu_{\ver},\mu_{\hor}$ are $b\Z^{2}$-invariant
Gibbs measures and they are distinct since $\mu_{\hor}(E_{\hor})=1$
while $\mu_{\ver}(E_{\hor})=0$ by (\ref{eq:exactly one infinite component for mu}).
\end{proof}
\pagebreak{}

\part{\label{part:Characterization-of-the}Characterization of the periodic
Gibbs measures}

In this part we show that at high fugacity the set of periodic Gibbs
measures is the convex hull of exactly four periodic and extremal
Gibbs measures, proving Theorem \ref{thm:main2}. We also investigate
some properties of the extreme measures leading to a proof of Theorem
\ref{thm:main}.

\section{Peierls-type arguments and strongly percolating sets\label{sec:Peierls-type-arguments}}

In this section we introduce the non-standard terminology of $\eps$-strongly-percolating
sets. We use it to state and prove the Peierls argument and some related
propositions. The reason for doing so is that this terminology will
allow us in later sections to easily apply the Peierls argument repeatedly,
and on grids with different spacing.

\subsection{Definitions}

Let $B$ be a random set (with respect to a measure $\mathbb{P}$)\marginpar{$\mathbb{P}$},
and let $\eps\ge0$. Say that $B$ is \textbf{$\eps$-rare} if for
every finite set $A$,
\[
\mathbb{P}(A\subset B)\le\eps^{\#A}.
\]

For the rest of the paper, \marginpar{$\eps_{0}$}fix
\begin{equation}
\eps_{0}=1/21.\label{eq:eps_0 def}
\end{equation}
For a random set $\Psi\subset\V$ say that $\Psi$ is\textbf{ $\eps$-strongly
percolating}, if either $\eps\ge\eps_{0}$ or there is an $\eps$-rare
set $B$ (on the same probability space) such that $\Psi$ almost
surely contains an infinite $\square$-component of $\V\backslash B$.
If $\Psi$ is $\eps$-strongly percolating for some $0<\eps<\eps_{0}$,
we say that $\Psi$ is strongly percolating (if $\eps\ge\eps_{0}$,
the statement that $\Psi$ is $\eps$-strongly percolating is vacuous).
We denote\marginpar{$\pv(\cdot)$}
\[
\pv_{\mathbb{P}}(\Psi)=\inf\{\eps\ge0:\text{\ensuremath{\Psi} is \ensuremath{\eps}-strongly-percolating}\}
\]

and usually omit $\mathbb{P}$ from the notation.

To get a feeling for the above definitions, note the following two
points: An $\eps$-strongly percolating set is a random set, which
necessarily contains an infinite $\square$-component if $\eps<\eps_{0}$
(while nothing is guaranteed if $\eps\ge\eps_{0}$). In addition,
the smaller $\eps$ is, the ``larger'' the $\eps$-strongly percolating
set is, in the sense that the condition of being $\epsilon$-strongly
percolating becomes stricter as $\eps$ decreases. With this in mind,
$\pv_{\mathbb{P}}(\Psi)$ is a ``measure of the size of $\Psi$'',
with smaller values corresponding to larger size.

For sets $U,V,B\subset\V$, we say for a set $B$ that it $\square$-\textbf{separates}
$U$ from $V$ if there is no $\square$-path starting in $U$, ending
in $V$, and contained in $\V\setminus B$.

The following additional definitions come into play near the end of
the section. Recall from subsection \ref{subsec:Basic-definitions}
that a translation by a vector $v$ is denoted by $\eta_{v}$. For
an event $E\subset\Omega$, define\marginpar{$\es E$} the random
set $\es E:\Omega\to2^{\V}$ by
\[
\es E(\sigma)\coloneqq\{v\in\V:\sigma\in\eta_{v}E\}.
\]
Intuitively, $E$ should be thought of as a local property of the
configuration and $\es E$ is the random set of positions where this
local property holds.

For\marginpar{$\sg{K\times L}{}$} a set (or a random set) $\Psi$,
define
\[
\sg{K\times L}{\Psi}\coloneqq\{(x,y)\in\V:(Kx,Ly)\in\Psi\}.
\]
In other words, $\sg{K\times L}{\Psi}$ is formed by restricting $\Psi$
to the grid $K\mathbb{Z}\times L\mathbb{Z}$ and then rescaling so
that this grid becomes $\V$.

\subsection{Reformulation of the Peierls argument}

In this subsection we formulate and prove several Peierls-type results
using the terminology of strongly-percolating sets.

The following lemma is a standard fact on the connectivity of separating
sets in $\V$. We provide a proof for completeness, following the
ideas of Timár \cite{timar2013boundary}, as we do not have a reference
for the precise result that we need. The lemma is similar to a special
case of \cite[Theorem 3]{timar2013boundary}.
\begin{prop}
\label{prop:seperating_is_connected}(Connectivity of minimal separators)
Let $U,V\subset\V$ be two $\boxtimes$-connected sets and let $B\subset\V$
be a minimal (with respect to inclusion) set that $\square$-separates
$U$ from $V$. Then $B$ is $\boxtimes$-connected.
\end{prop}
\begin{proof}
Introduce two auxiliary vertices $u,v$ and consider the graph
\[
G_{U,V}=\left(\V\cup\{u,v\},\medspace\E_{\square}\cup\{uw:w\in U\}\cup\{vw:w\in V\}\cup\E_{\boxtimes}[U]\cup\E_{\boxtimes}[V]\right).
\]
where $E[A]$ stands for the set of edges in $E$ with both endpoints
in $A$. Note that $B$ is a minimal set separating $u$ from $v$
in $G_{U,V}$.

Define $\mathcal{C}$ to be the set of cycles consisting of the $4$-cycles
in $(\V,\E_{\square})$ and all the triangles in $G_{U,V}$. Let us
show that $\mathcal{C}$ generates the cycle space of $G_{U,V}$ (the
set of spanning subgraphs with even degrees, viewed as a vector space
over the two-element finite field). Let $C$ be an element of the
cycle space of $G_{U,V}$. We show how to add to it cycles from $\mathcal{C}$
to obtain the empty graph. Whenever $\deg_{C}(u)\neq0$, pick two
neighbors of $u$ in $C$, $u_{1},u_{2}\in U$. Since $U$ is $\boxtimes$-connected,
there is a path in $\E_{\boxtimes}[U]$ from $u_{1}$ to $u_{2}$.
By adding, for each edge $e$ in the path, the triangle incident to
$e$ and $u$, we decreased $\deg_{C}(u)$ by $2$ without altering
$\deg_{C}(v)$. Repeating this process, and its analog for $v$, we
can make sure that $C$ has no edges incident to $u$ or $v$. Then,
we may add to $C$ triangles from $(\V,\E_{\boxtimes})$ until $C\subset\E_{\square}$.
It is known that the cycle space of $(\V,\E_{\square})$ is generated
by its $4$-cycles. Thus we have shown that $\mathcal{C}$ generates
the cycle space of $G_{U,V}$.

Our goal is to show that $B$ is $\boxtimes$-connected. Thus it suffices
to take an arbitrary partition, $B=B_{1}\uplus B_{2}$, and find an
edge $w_{1}w_{2}\in\E_{\boxtimes}$ with $w_{1}\in B_{1}$ and $w_{2}\in B_{2}$.

Consider the set $T_{0}$ of edges incident to $B$ in $G_{U,V}$.
The set $T_{0}$ separates $u$ from $v$, thus let us choose a subset
$T\subset T_{0}$ which is a minimal set of edges separating $u$
from $v$. By the minimality of $B$, every vertex $w\in B$ is an
endpoint of an edge in $T$. 

If an edge in $T$ is incident to both $B_{1}$ and $B_{2}$, then
it is in $\E_{\boxtimes}$ and we are done. Otherwise, let $T_{1},T_{2}$
be the sets of edges in $T$ incident to $B_{1},B_{2}$ respectively;
they are both non-empty and form a partition $T=T_{1}\uplus T_{2}$.
Thus by \cite[Lemma 1]{timar2013boundary}, there is a cycle $C\in\mathcal{C}$,
that contains edges $e_{1},e_{2}$ from $T_{1}$ and $T_{2}$ respectively.
The edges $e_{1},e_{2}$ are respectively incident to vertices $w_{1}\in B_{1},w_{2}\in B_{2}$.
In particular $w_{1},w_{2}\in\V\cap C$, and considering the case
that $C$ is a triangle and the case that $C$ is a $4$-cycle in
$\E_{\square}$, it is obvious that $w_{1}w_{2}\in\E_{\boxtimes}$.
\end{proof}
\begin{lem}
[Peierls argument]\label{lem:peierls_inf_comp}If $\eps<\eps_{0}$
and $B$ is an $\eps$-rare set, then $\V\backslash B$ almost surely
has a unique infinite $\square$-component $I$. Moreover, each $\boxtimes$-component
of $\V\backslash I$ is finite.
\end{lem}
\begin{proof}
Let $B$ be an $\eps$-rare set and assume $\eps<\eps_{0}$. We first
show that $\V\backslash B$ almost surely has an infinite $\square$-component.
Denote $C_{n}=\{-n,-n+1,\dots,n\}^{2}$. Consider the random set $D_{n}$
of all points from which a $\square$-path to $C_{n}$ exists in $\V\setminus B$.
Whenever this set is finite, there exists $m>n$ and a finite set
$B'\subset B$ minimal among the sets separating $C_{n}$ from $\V\setminus C_{m}$.
Such $B'$ is $\boxtimes$-connected by Proposition \ref{prop:seperating_is_connected},
and must contain points $(-s,0)$,$(t,0)$ for some $s$,$t$ between
$n$ and $m+1$. Thus $B$ must contain a $\boxtimes$-path of length
at least $n+s$ starting at $(-s,0)$ for some $s\ge n$. Using that
$B$ is $\eps$-rare and summing over such paths gives
\[
\mathbb{P}(\text{\ensuremath{D_{n}} is finite})\le\sum_{s=n}^{\infty}8\cdot7^{n+s-1}\eps^{n+s+1}\xrightarrow{n\to\infty}0
\]
using that $\eps<\eps_{0}\le1/7$. Thus almost surely $D_{n}$ is
infinite for some $n$. Whenever $D_{n}$ is infinite, there is some
point in $C_{n}$ that is contained in an infinite $\square$-connected
component of $\V\setminus B$. Thus $\V\setminus B$ almost surely
has an infinite $\square$-component.

Let $I$ be such an infinite $\square$-component of $\V\setminus B$.
Whenever $\V\setminus I$ has an infinite $\ensuremath{\boxtimes}$-component
$J$, by Proposition \ref{prop:seperating_is_connected} there is
a $\boxtimes$-connected set $B'\subset B$ that $\square$-separates
$I$ from $J$, and $B'$ is infinite (it is easy to see that a set
separating two infinite sets of vertices in $(\V,\E_{\square})$ must
be infinite). But the probability that $B$ contains an infinite $\boxtimes$-connected
set is $0$ (this is since $B$ is $\eps$-rare for $\eps<1/7$).
Thus $\V\setminus I$ has no infinite $\boxtimes$-component.
\end{proof}
The following is an immediate corollary of the definition of $\pv$
and Lemma \ref{lem:peierls_inf_comp}.
\begin{cor}
\label{cor:complement of eps rare}Let $\eps\ge0$. Let $B$ be an
$\eps$-rare set. Then $\V\setminus B$ is $\eps$-strongly percolating,
and in particular $\pv(\V\setminus B)\le\eps$.
\end{cor}
The next lemma strengthens the Peierls-type result of Lemma \ref{lem:peierls_inf_comp}
by bounding the probability that in the complement of a strongly percolating
set, the connected component of a given point has a large diameter.
\begin{lem}
[Quantitative Peierls argument]\label{lem:two_point_peierls}Let
$u\in\V$, let $d\ge0$ and let $\Psi\subset\V$ be a random set.
Let $E$ be the event that there exists a $\boxtimes$-path in $\V\setminus\Psi$
starting at $u$ and ending at some point in $\{v\in\V\colon\left\Vert v-u\right\Vert _{\infty}\ge d\}$.
Then 
\[
\mathbb{P}(E)\le\left(\frac{\pv(\Psi)}{\eps_{0}}\right)^{1+d}.
\]
\end{lem}
\begin{proof}
It suffices to show $\mathbb{P}(E)\le\left(\eps/\eps_{0}\right)^{1+d}$
for each $\eps>\pv(\Psi)$. If $\eps\ge\eps_{0}$ there is nothing
to prove. Otherwise, let $B$ be an $\eps$-rare set such that $\Psi$
contains $I$, the infinite $\square$-component of $\V\backslash B$
(it exists by Lemma \ref{lem:peierls_inf_comp}).

When $u\in\V\setminus\Psi$, consider $C$, the $\boxtimes$-connected
component of $\V\setminus\Psi$ containing $u$. Since $\pv(\Psi)<\eps_{0}$,
it holds almost surely that $C$ is finite, and $C$ is $\square$-separated
from $I$ by $B$ (since $B$ $\square$-separates $I$ from its complement).
Let $B'$ be some minimal subset of $B$ that $\square$-separates
$C$ from $I$. Then by Proposition \ref{prop:seperating_is_connected},
$B'$ is $\boxtimes$-connected.

Denote $u=(x_{u},y_{u})$. Denote by $E_{1}$ the event that $u\in\V\setminus\Psi$
and a point $v=(x_{v},y_{v})\in C$ with $x_{v}-x_{u}\ge d$ exists.
Fix an outcome $\sigma\in E_{1}$. Consider the straight infinite
$\square$-paths extending from $u$ to the left and from $v$ to
the right respectively. These paths intersect $I$ by Lemma \ref{lem:peierls_inf_comp},
thus since $B'$ separates $C$ from $I$, there must be $s,t\ge0$
such that $(x_{u}-s,y_{u}),(x_{v}+t,y,v)\in B'$. Since $B'$ is $\boxtimes$-connected
it contains a $\boxtimes$-path connecting these two points, which
is of length at least $s+d$. Summing over the possibilities for $s$
and over $\boxtimes$-paths of length $s+d$ starting at $(x_{u}-s,y_{u})$,
and using the fact that $B'\subset B$ and $B$ is $\eps$-rare, it
follows that
\begin{align*}
4\mathbb{P}(E_{1}) & \le4\sum_{s=0}^{\infty}8\cdot7^{d+s-1}\eps^{d+s+1}\le\frac{32}{49}\sum_{s=0}^{\infty}(7\eps)^{d+s+1}\\
 & =\frac{32}{49(1-7\eps)}(7\eps)^{d+1}\le\left(\frac{\eps}{\eps_{0}}\right)^{d+1}
\end{align*}

where we have used $\eps<\eps_{0}$. Define $E_{2},E_{3},E_{4}$ in
the same way as $E_{1}$ except that the condition $x_{v}-x_{u}\ge d$
is replaced by $x_{v}-x_{u}\le-d$, $y_{v}-y_{u}\ge d$, or $y_{v}-y_{u}\le-d$
respectively. The bound on $\mathbb{P}(E_{1})$ applies analogously
for $\mathbb{P}(E_{2})$, $\mathbb{P}(E_{3})$, and $\mathbb{P}(E_{4})$.
Since $E=E_{1}\cup E_{2}\cup E_{3}\cup E_{4}$, the lemma follows
from the bound.
\end{proof}

\subsection{Relations between random sets}

In this section we show that the properties of being rare and strongly
percolating are maintained under various set operations, provided
the $\eps$ parameter in these properties is sufficiently small.
\begin{lem}
[Union of rare sets] \label{lem:rare_union} Suppose $B_{i}$ is
$\eps_{i}$-rare for $i\in\{1,\dots,k\}$. Then $\bigcup_{i=1}^{k}B_{i}$
is $\left(k\max_{i=1,\dots,k}\eps_{i}^{1/k}\right)$-rare.%
\end{lem}
\begin{proof}
Denote $B=\text{\ensuremath{\bigcup_{i=1}^{k}B_{i}}}$. Let $A\subset\V$.
For a function $f:A\to\{1,\dots,k\}$, there is some $i$ such that
$\#f^{-1}(i)\ge\frac{\#A}{k}$, thus the probability that $v\in B_{f(v)}$
for each $v\in A$ is at most $\eps_{i}^{\#A/k}$. By a union bound
over all $f:A\to\{1,\dots,k\}$, the probability that $A\subset B$
is at most
\[
\eps\coloneqq(k\max_{i\in\{1,\dots,k\}}\eps_{i}^{1/k})^{\#A},
\]
thus $B$ is $\eps$-rare.
\end{proof}
\begin{lem}
[Intersection of strongly percolating sets]\label{lem:perc_intersection}
\[
\pv\left(\bigcap_{i=1}^{k}\Psi_{i}\right)\le k\sqrt[\uproot{3}k]{\max_{i=1,\dots,k}\pv(\Psi_{i})}\quad\text{for every random \ensuremath{\Psi_{1}},\ensuremath{\ldots},\ensuremath{\Psi_{k}\subset\V}.}
\]

\end{lem}
\begin{proof}
We may assume $k\ge1$ since $\pv(\V)=0$. Let $\eps_{i}>\pv(\Psi_{i})$
for $i\in\{1,\dots,k\}$ and set $\eps\coloneqq k\max_{i\in\{1,\dots,k\}}\eps_{i}^{1/k}$.
It suffices to prove that $\bigcap_{i=1}^{k}\Psi_{i}$ is $\eps$-strongly-percolating.
We may assume that $\eps<\eps_{0}$, otherwise there is nothing to
prove. In particular, since $\eps_{0}<1$, for each $i\in\{1,\dots,k\}$
it holds that $\eps_{i}<\eps_{0}$. So by the definition of strong
percolation we may fix random sets $B_{i}$ and $I_{i}$ such that
$B_{i}$ is $\eps_{i}$-rare, $I_{i}$ is an infinite $\square$-component
of $\V\backslash B_{i}$, and $I_{i}\subset\Psi_{i}$. Denote $B=\text{\ensuremath{\bigcup_{i=1}^{k}B_{i}}}$.
By Lemma \ref{lem:rare_union}, $B$ is $\eps$-rare. Thus by Lemma
\ref{lem:peierls_inf_comp} there is a set $I$ which is an infinite
$\square$-component of $\V\backslash B$, such that all the $\boxtimes$-components
of $\V\backslash I$ are finite. For each $i\in\{1,\dots,k\}$ it
holds that $I\subset\V\backslash B_{i}$, and since it is infinite
and $\boxtimes$-connected, it must be that $I\subset I_{i}$. Thus
$I\subset\bigcap_{i=1}^{k}I_{i}\subset\bigcap_{i=1}^{k}\Psi_{i}$.
\end{proof}
The next lemma considers ``strong percolation of events on sub-grids
of $\V$''. The following is an intuitive description of items \ref{enu:lump}
and \ref{enu:unlump} therein: We associate two random sets to an
event $E$, thought of as percolation processes: Given $k,K,l,L\in\nat$
satisfying $k\divides K$ and $l\divides L$, we have the percolation
$\Psi\coloneqq\sg{k\times l}{\es E}$ corresponding to the $k\mathbb{Z}\times\ell\mathbb{Z}$
grid. In addition, we define a block percolation $\Psi'$ corresponding
to the $K\Z\times L\Z$ grid in which a point is open if all the points
of $\Psi$ in the corresponding block of the $k\Z\times\ell\Z$ are
open. We show that sufficiently strong percolation of $\Psi$ implies
strong percolation of $\Psi'$, and vice versa.
\begin{lem}
\label{lem:grid_change}Let $E\subset\Omega$ be an event. Let $k,K,l,L\in\nat$
where $k\divides K$ and $l\divides L$. Denote 
\[
H\coloneqq\{\eta_{(x,y)}:(x,y)\in(k\mathbb{Z}\times\ell\Z)\cap\left([0,K)\times[0,L)\right)\}
\]
 and $r\coloneqq\frac{KL}{kl}=\#H$. Denote 
\[
\Psi\coloneqq\sg{k\times l}{\es E},\qquad\Psi_{\eta}'\coloneqq\sg{K\times L}{\es{\eta E}}\;\;\text{for each \ensuremath{\eta\in}H,}\qquad\Psi'\coloneqq\bigcap_{\eta\in H}\Psi_{\eta}'.
\]
Then:
\begin{enumerate}
\item \label{enu:lump}${\displaystyle \pv(\Psi')\le r\pv(\Psi)}.$
\item \label{enu:unlump}${\displaystyle \pv(\Psi)\le\sqrt[r]{\pv(\Psi')}}$.
\item \label{enu:partition}${\displaystyle \pv(\Psi)\le\sqrt[r]{r\sqrt[\uproot{2}r]{\max_{\eta\in H}\left\{ \pv(\Psi_{\eta}')\right\} }}}.$
\item \label{enu:partition_chessboard}If $E\subset\Omega$ is $R$-local
for $R=\rect{K\times L}{(0,0)}$, and $E$ is invariant to reflections
through the vertical and horizontal lines passing through the center
of $R$, then for each periodic Gibbs measure $\mu$,
\begin{equation}
\pv_{\mu}(\Psi)\le\sqrt[r]{{\zedi[][E^{c}]}}.\label{eq:chess_pv}
\end{equation}
\end{enumerate}
\end{lem}
\begin{proof}
Define $f:\V\to\V$ by $f(x,y)=\left(\left\lfloor \frac{xk}{K}\right\rfloor ,\left\lfloor \frac{yl}{L}\right\rfloor \right)$
and note that 
\begin{equation}
\Psi'=\{v\in\V:f^{-1}(v)\subset\Psi\}.\label{eq:lump_as_fiber}
\end{equation}

Proof of item \ref{enu:lump}: Assume WLOG that $r\pv(\Psi)<\eps_{0}.$
Let $\eps$ satisfy $\pv(\Psi)<\eps$ and $r\eps<\eps_{0}$. Choose
random sets $B,I$ such that $B$ is $\eps$-rare, $I\subset\Psi$,
and $I$ is a unique (by Lemma \ref{lem:peierls_inf_comp}) infinite
$\square$-component of $\V\setminus B$. Then, by a union bound,
the set $B'\coloneqq f(B)$ is $r\eps$-rare. Thus by Lemma \ref{lem:peierls_inf_comp},
$\V\setminus B'$ has an infinite $\square$-component $I'$. It is
easily seen that $f^{-1}(I')$ is infinite, $\square$-connected and
disjoint from $f^{-1}(B')$. Since $B\subset f^{-1}(B')$, it holds
that $f^{-1}(I')$ is disjoint from $B$ and thus $f^{-1}(I')\subset I$,
as $I$ is the unique infinite $\square$-component of $\V\setminus B$.
For each $v\in I'$, it holds that $f^{-1}(v)\subset I\subset\Psi$.
Thus by (\ref{eq:lump_as_fiber}), we have $I'\subset\Psi'$. The
existence of $B',I'$ as above shows that $\pv(\Psi')\le r\eps$.

Proof of item \ref{enu:unlump}: Assume WLOG that $\sqrt[r]{\pv(\Psi')}<\eps_{0}.$
Let $\eps$ satisfy $\pv(\Psi')<\eps$ and $\sqrt[r]{\eps}<\eps_{0}$.
Choose random sets $B',I'$ such that $B'$ is $\eps$-rare, $I'\subset\Psi'$,
and $I'$ is an infinite $\square$-component of $\V\setminus B'$.
This easily implies that $f^{-1}(I')$ is an infinite $\square$-connected
component of $\V\setminus f^{-1}(B')$. Since $B'$ is $\eps$-rare,
$f^{-1}(B')$ is $\sqrt[r]{\eps}$-rare. It remains to note that $f^{-1}(I')\subset f^{-1}(\Psi')\subset\Psi$
by (\ref{eq:lump_as_fiber}). The existence of $f^{-1}(B'),f^{-1}(I')$
as above shows that $\pv(\Psi')\le\sqrt[r]{\eps}$.

Proof of item \ref{enu:partition}: By Lemma \ref{lem:perc_intersection},
$\pv(\Psi')\le r\sqrt[r]{\max\left\{ \pv(\Psi_{\eta}'):\eta\in H\right\} }$.
The result follows by item \ref{enu:unlump}.

Proof of item \ref{enu:partition_chessboard}: Define $B\coloneqq\sg{k\times l}{\es{E^{c}}}$.
Denote $\eps=\zedi[][E^{c}]$. We will show that $B$ is $\sqrt[r]{\eps}$-rare.
This suffices by Corollary \ref{cor:complement of eps rare} since
$\Psi=\V\setminus B$.

Let $A\subset\V$ be finite. Choose $\eta=\eta_{(x_{0},y_{0})}\in H$
such that for $G\coloneqq\left(\frac{K\Z+x_{0}}{k}\times\frac{L\Z+y_{0}}{l}\right)$
and $A'\coloneqq A\cap G$ it holds that $\#A'\ge\#A/r$.

By the hypotheses of the current item, for each $v\in K\Z\times L\Z=G_{R}$,
it holds that $\tau_{R,v}E^{c}=\eta_{v}E^{c}$ (recall $G_{R}$ and
$\tau_{R,v}$ from Section \ref{sec:Chessboard-estimates}). Let $\sigma$
be sampled from $\eta\mu$. Then the chessboard estimate for infinite
volume (Lemma \ref{prop:chessboard_infinite}), implies that
\begin{equation}
\text{\ensuremath{\sg{K\times L}{\es{E^{c}}(\sigma)}=\{(x,y)\in\V:\eta\sigma\in\tau_{R,(Kx,Ly)}E^{c}\}} is \ensuremath{\eps}-rare.}\label{eq:chessboard_subgrid}
\end{equation}
By the one-to-one mapping $m:\V\to G$, $m(x,y)=(\frac{Kx+x_{0}}{k},\frac{Ly+y_{0}}{l})$
it follows that $G\cap B$ is $\eps$-rare. Thus
\[
\mu(A\subset B)\le\mu(A'\subset G\cap B)\le\eps^{\#A'}\le\left(\sqrt[r]{\eps}\right)^{\#A}.\qedhere
\]
\end{proof}

\subsection{Splitting strongly percolating sets}

The next proposition shows that, in an ergodic setting, when a strongly
percolating set is split into two \emph{separated} random sets, then
one of the two resulting sets is itself strongly percolating (in particular,
this set contains an infinite $\square$-component with probabilty
one).
\begin{prop}
\label{prop:peierls_two}Let $\mu$ be an $\L$-ergodic measure on
$\Omega$ for some lattice $\L\subset K\Z\times L\Z$. Let $E,F\subset\Omega$
be events. Assume that
\[
E\cap\eta_{(K,0)}F=E\cap\eta_{(0,L)}F=\eta_{(K,0)}E\cap F=\eta_{(0,L)}E\cap F=\emptyset.
\]
Then
\[
\min\{\pv_{\mu}(\sg{K\times L}{\es E}),\pv_{\mu}(\sg{K\times L}{\es F})\}\le\pv_{\mu}(\sg{K\times L}{\es{E\cup F}}).
\]
\end{prop}
\begin{proof}
Denote $\Psi_{E}\coloneqq\sg{K\times L}{\es E},\Psi_{F}\coloneqq\sg{K\times L}{\es F},\Psi\coloneqq\sg{K\times L}{\es{E\cup F}}=\Psi_{E}\cup\Psi_{F}$,
and $\eps\coloneqq\pv_{\mu}(\sg{K\times L}{\es{E\cup F}})$. If $\eps\ge\eps_{0}$
there is nothing to prove. Otherwise, there is a random $\eps$-rare
set $B$ such that $\Psi$ almost surely contains an infinite $\square$-component
$I$ of $\V\setminus B$. By Lemma \ref{lem:peierls_inf_comp}, the
set $\Psi$ almost surely has a unique infinite component, denote
it by $I'$. The random set $I'$ is defined from $\Psi$ up to measure
$0$ (without dependence on $B$ and $I$) and satisfies $I\subset I'$.

Thus the events $\{I'\subset\Psi_{E}\}$ and $\{I'\subset\Psi_{F}\}$
are properly defined and $\L$-invariant up to measure $0$. By the
$\L$-ergodicity they each have probability $0$ or $1$. The condition
of the lemma ensures that no element of $\Psi_{E}$ is $\square$-adjacent
to an element of $\Psi_{F}$, and thus each component of $\Psi$ is
contained in either $\Psi_{E}$ or $\Psi_{F}$. This holds in particular
for the component $I'$, thus the union of the two events above holds
almost surely, and as they are $0$-$1$ events one of them holds
almost surely. Thus one of the events $\{I\subset\Psi_{E}\}$ and
$\{I\subset\Psi_{F}\}$ holds almost surely, and this implies by definition
that either $\Psi_{E}$ or $\Psi_{F}$ is $\eps$-strongly-percolating.
\end{proof}

\section{\label{sec:Four-phases}Four phases}

In this section, we improve upon the result of Section \ref{sec:multiple_gibbs}.
On the intuitive level, there it was shown that mesoscopic sticks
in a configuration are either mostly vertical or mostly horizontal.
Here we extend this, by showing that the offset of mesoscopic sticks
(horizontal offset for vertical sticks, and vice versa) is either
mostly even or mostly odd. We also get better quantitative control
over the ``density'' of the sticks. We choose a length scale $\b$
comparable to $\lambda^{1/2}$, and take $\a$ to be a sufficiently
large universal constant. We show that when sticks are mostly vertical
and with even offset, most rectangles of dimensions $N\a\times N\b$
will be divided by a vertical stick of even offset (recall that $N$
was defined as a universal constant).

The above is stated formally in Theorem \ref{thm:ergodic_tetrachotomy},
which is the main result of the section. This theorem gives quantitative
results that will be used in later sections to prove the main theorems
stated in the introduction. Additionally, it already implies an extension
of Corollary \ref{cor:multiple Gibbs measures}: that for all sufficiently
large $\lambda$, there is a set of four affinely independent periodic
Gibbs measures.

The theorems are stated for ergodic Gibbs measures, sometimes requiring
ergodicity with respect to a very sparse lattice such as $\b!\Z^{2}$.
All these theorems may be seen to have implications for any periodic
Gibbs measure using the ergodic decomposition theorem (as will be
done later, in the proof of item \ref{enu:unique_decompostion} of
Lemma \ref{lem:unique Gibbs measure}). Also note that we could have
replaced $\b!\Z^{2}$ by any lattice $\L\subset\b!\Z^{2}$.

Recall from Section \ref{sec:multiple_gibbs} the definitions of $N$
and $\Psi^{K\times L}$, $\Psi_{\ver}^{K\times L}$, $\Psi_{\hor}^{K\times L}$,$\Psi_{(\ver,0)}^{K\times L}$,
$\Psi_{(\ver,1)}^{K\times L}$, $\Psi_{(\hor,0)}^{K\times L}$, $\Psi_{(\hor,1)}^{K\times L}$.
Recall also $\eps_{0}$ from (\ref{eq:eps_0 def}). Define\marginpar{$D^{K\times L}$}
\[
D^{K\times L}\coloneqq\{\sigma\in\Omega:(0,0)\in\Psi^{K\times L}(\sigma)\}.
\]
Equivalently, $D^{K\times L}$ is the event that the rectangles $R=\rect{NK\times NL}{(0,0)}$
and $R^{-}=\rect{(N-2)K\times(N-2)L}{(K,L)}$ are both divided by
a stick. Then for each $(x,y)\in\V$, the event $\eta_{(Kx,Ly)}D^{K\times L}$
holds iff $(x,y)\in\Psi^{K\times L}$, thus $\Psi^{K\times L}=\sg{K\times L}{\es{D^{K\times L}}}$.
Similarly define\marginpar{$D_{\pi}^{K\times L}$} 
\[
D_{\pi}^{K\times L}=\{\sigma\in\Omega:(0,0)\in\Psi_{\pi}^{K\times L}(\sigma)\}
\]
 for $\pi\in\{\ver,\hor,(\ver,0),(\ver,1),(\hor,0),(\hor,1)\}$ and
note that statements analogous to those made for $D^{K\times L}$
hold for $D_{\pi}^{K\times L}$, assuming that $K$ and $L$ are even.

For the rest of the paper, fix\marginpar{$\b$} 
\begin{equation}
\b=\b(\lambda):=2\left\lfloor \frac{c_{\ref{lem:main}}\lambda^{1/2}}{2N}\right\rfloor .\label{eq:b def}
\end{equation}

\begin{thm}
\label{thm:ergodic_tetrachotomy}There is $c>0$ such that for each
sufficiently large $\a\in2\nat$ the following holds:

For all sufficiently large $\lambda$ and every $\b!\Z^{2}$-ergodic
Gibbs measure, exactly one of $\Psi_{(\ver,0)}^{\a\times\b}$,$\Psi_{(\ver,1)}^{\a\times\b}$,$\Psi_{(\hor,0)}^{\b\times\a}$,
and $\Psi_{(\hor,1)}^{\b\times\a}$ is $e^{-c\a}$-strongly-percolating,
while each of the others almost surely has only finite $\boxtimes$-components.
\end{thm}
Let \marginpar{$\mathcal{P}$} 
\[
\mathcal{P}:=\{(\ver,0),(\ver,1),(\hor,0),(\hor,1)\}.
\]
Following Theorem \ref{thm:ergodic_tetrachotomy}, for a $\b!\Z^{2}$-ergodic
Gibbs measure $\mu$, we write\marginpar{$\phase$} $\phase(\mu)=\pi$
for the element $\pi\in\mathcal{P}$ corresponding to the set which
percolates. Formally, the notation $\phase$ also depends on a choice
of $\a$, and only makes sense when $\lambda$ is chosen large as
a function of $\a$, but we omit explicit mention of this in the notation
as we will use $\phase$ in situations where $\a$ and $\lambda$
will be suitably fixed in advance. The following statement explains
how $\phase$ transforms under isometries $\tau$ of $\Z^{2}$. Recall
from Section \ref{subsec:Basic-definitions} that we write $\tau\mu$
for the push-forward of $\mu$ under $\tau$ and that $\eta_{(x,y)}$
denotes a translation by the vector $(x,y)$.
\begin{prop}
\label{prop:isometries and phase}Let $\a\in2\nat$ be sufficiently
large. Let $\lambda$ be sufficiently large (as a function of $\a$).
Let $\mu$ be a $\b!\Z^{2}$-ergodic Gibbs measure. Then
\end{prop}
\begin{enumerate}
\item \label{enu:ph_trans_x}$\phase(\mu)=(\ver,j)$ if and only if $\phase(\eta_{(1,0)}\mu)=(\ver,1-j)$
for $j\in\{0,1\}$.
\item \label{enu:ph_trans_y}If $\phase(\mu)\in\{(\ver,0),(\ver,1)\}$ then
$\phase(\eta_{(0,1)}\mu)=\phase(\mu)$.
\item \label{enu:ph_ref_diag}Let $\tau:\Z^{2}\to\Z^{2}$ be the reflection
$\tau(x,y)=(y,x).$ Then $\phase(\mu)=(\ver,j)$ if and only if $\phase(\tau\mu)=(\hor,j)$,
for $j\in\{0,1\}$.
\item \label{enu:ph_ref_vert}Let $\tau:\Z^{2}\to\Z^{2}$ be the reflection
$\tau(x,y)=(-x,y).$ Then $\phase(\mu)=\phase(\tau\mu)$.
\end{enumerate}
The rest of the section is devoted to the proof of the theorem and
proposition above.

\subsection{Proof of Theorem \ref{thm:ergodic_tetrachotomy}}

We divide the proof of Theorem \ref{thm:ergodic_tetrachotomy} into
parts, corresponding to the three items of the lemma below.

The first item is similar to Theorem \ref{thm:dichotomy}. Unlike
Theorem \ref{thm:dichotomy}, the item concerns an ergodic measure
rather than just a periodic one, it gives a quantitative result, and
it concerns rectangles rather than just squares (i.e. $\a$ does not
necessarily equal $\b$). The proof is similar, except for the use
of the terminology that was introduced in Section \ref{sec:Peierls-type-arguments},
and an additional step where the ergodicity is used to draw a stronger
conclusion.

The second item concerns the ``density'' of the sticks. It shows
that thin rectangles aligned with the the preferred direction of sticks
are usually divided \emph{in this direction}.

The last item implies Theorem \ref{thm:ergodic_tetrachotomy} directly.

We point out that while the first two items do not refer to the fugacity
explicitly, they contain an implicit requirement that $\lambda$ be
large in the assumption that $\a_{0}\le\b$.
\begin{lem}
\label{lem:chotomies}There exist $c,\a_{0}>0$ such that for every
$\lambda>0$, each $\a\in2\nat$ satisfying $\a_{0}\le\a\le\b$, and
every $\b!\Z^{2}$-ergodic Gibbs measure, the following hold:
\begin{enumerate}
\item \label{enu:ergodic_dichotomy_quant}One of $\Psi_{\ver}^{\a\times\b}$
and $\Psi_{\hor}^{\a\times\b}$ is $e^{-c\a}$-strongly-percolating.
\item \label{enu:ergodic_dichotomy_thin}One of $\Psi_{\ver}^{\a\times\b}$
and $\Psi_{\hor}^{\b\times\a}$ is $e^{-c\a}$-strongly-percolating.
\item \label{enu:ergodic_tetrachotomy}There is a universal $\lambda_{0}(\a)$
such that for $\lambda>\lambda_{0}(\a)$, exactly one of\\
 $\Psi_{(\ver,0)}^{\a\times\b}$,$\Psi_{(\ver,1)}^{\a\times\b}$,$\Psi_{(\hor,0)}^{\b\times\a}$,
and $\Psi_{(\hor,1)}^{\b\times\a}$ is $e^{-c\a}$-strongly-percolating
while each of the others almost surely has only finite $\boxtimes$-components.
\end{enumerate}
\end{lem}
\begin{proof}
Let $c,\a_{0}$ be universal constants with $c$ sufficiently small
and $\a_{0}$ sufficiently large to satisfy some assumptions that
will follow. Let $\lambda,\a$ and a Gibbs measure be as above. By
slightly decreasing $c$, for each item, instead of showing that one
of the sets $\Psi$ is $e^{-c\a}$-strongly percolating, it suffices
to prove that one of the random sets $\Psi$ satisfies $\pv_{\mu}(\Psi)\le e^{-c\a}$.\renewcommand{\qedsymbol}{}
\end{proof}
\begin{proof}
[Proof of item \ref{enu:ergodic_dichotomy_quant}] Note that $D^{\a\times\b}$
is $R$-local for $R=\rect{\a N\times\b N}{(0,0)}$, thus by item
\ref{enu:partition_chessboard} of Lemma \ref{lem:grid_change}
\[
\pv(\Psi^{\a\times\b})=\pv(\sg{\a\times\b}{\es{D^{\a\times\b}}})\le\sqrt[N^{2}]{\zedi[R][\Omega\setminus D^{\a\times\b}]}.
\]
Lemma \ref{lem:main} is applicable to $R$ and $S=R^{-}$. Indeed
(\ref{eq:dim_rat}) holds by the choice of $N$ (at the beginning
of Section \ref{sec:multiple_gibbs}), the right part of (\ref{eq:dim_abs})
holds by the definition of $\b$, and the left part of (\ref{eq:dim_abs})
then holds by assuming $\a_{0}\ge\frac{1}{Nc_{\ref{lem:main}}}$.
Thus the lemma yields
\begin{equation}
\zedi[R][\Omega\setminus D^{\a\times\b}]\le e^{-c_{\ref{lem:main}}\vol(R)\lambda^{-1/2}}=e^{-c_{\ref{lem:main}}\a\b N^{2}\lambda^{-1/2}}\le e^{-2c_{\ref{lem:main}}^{2}N\a}\label{eq:strong_perc_eps_1}
\end{equation}
where the last inequality is obtained by requiring $\a_{0}\ge2$,
and noting that by $\a_{0}\le\b$ and (\ref{eq:b def}) it holds that
that $\b\le2c_{\ref{lem:main}}\lambda^{1/2}N$. Combining the two
last displays, and taking sufficiently small $c$, it follows that
\begin{equation}
\pv(\Psi^{\a\times\b})\le e^{-c\a}.\label{eq:dev_pv}
\end{equation}

Note that by Lemma \ref{lem:stick_collision} and the assumption $\a\in2\nat$,
the conditions of Proposition \ref{prop:peierls_two} are satisfied
for
\[
K=\a,L=\b,E=D_{\ver}^{\a\times\b},F=D_{\hor}^{\a\times\b}.
\]

Thus
\[
\min\{\pv(\Psi_{\ver}^{\a\times\b}),\pv(\Psi_{\hor}^{\a\times\b}\}\le\pv(\Psi^{\a\times\b})\le e^{-c\a}.\qedhere
\]
\end{proof}
\begin{proof}
[Proof of item \ref{enu:ergodic_dichotomy_thin}]Denote $\eps=e^{-c\a_{0}}/\eps_{0}$.
We keep $c$ as in item \ref{enu:ergodic_dichotomy_quant} and possibly
increase $\a_{0}$ so that $3\eps<1$. 

Item \ref{enu:ergodic_dichotomy_quant} implies that one of $\Psi_{\ver}^{\b\times\b}$
and $\Psi_{\hor}^{\b\times\b}$ is $e^{-\b c}$-strongly-percolating
for $\mu$. Assume WLOG that $\Psi_{\ver}^{\b\times\b}$ is. We prove
by a decreasing induction that $\pv(\Psi_{\ver}^{\a\times\b})\le e^{-\a c}$
for each $\a\in2\nat$ satisfying $\a_{0}\le\a\le\b$.

Let $\a\in\nat$ satisfy $\a_{0}\le\a<\b$ and assume the induction
hypothesis, that $\pv(\Psi_{\ver}^{(\a+2)\times\b})\le e^{-c(\a+2)}$.
By item \ref{enu:ergodic_dichotomy_quant}, one of $\Psi_{\ver}^{\a\times\b}$
and $\Psi_{\hor}^{\a\times\b}$ is $e^{-c\a}$-strongly-percolating.
Assume by contradiction that $\Psi_{\hor}^{\a\times\b}$ is. Thus
by Lemma \ref{lem:two_point_peierls} (with $d=0$) and the choice
of $\eps$, 
\[
\max\left\{ \,\mu(\Omega\setminus D_{\ver}^{(\a+2)\times\b}),\,\mu(\Omega\setminus D_{\hor}^{\a\times\b}),\,\mu(\Omega\setminus\eta_{(2\a,0)}D_{\hor}^{\a\times\b})\,\right\} \le\eps.
\]
By the assumption $3\eps<1$, the event $D_{\ver}^{(\a+2)\times\b}\cup D_{\hor}^{\a\times\b}\cup\eta_{(2\a,0)}D_{\hor}^{\a\times\b}$
holds with positive probability. That is, there is an outcome $\sigma$
for which $\rect{N\a\times N\b}{(0,0)}$ and $\rect{N\a\times N\b}{(2\a,0)}$
are divided horizontally, and $\rect{(N\a+2N)\times N\b}{(0,0)}$
is divided vertically. By assuming $\a_{0}\ge N$, the union of the
two former rectangles contains the latter one, while all their vertical
dimensions are the same. This implies an intersection of sticks, which
is a contradiction. Thus $\pv(\Psi_{\ver}^{\a\times\b})\le e^{-c\a}$
completing the induction step.
\end{proof}
As preparation for the proof of item \ref{enu:ergodic_tetrachotomy},
we make a definition, and a claim, that will also be used later on.
Define the event\marginpar{$G$}
\begin{equation}
G^{\a\times\b}\coloneqq\left\{ \sigma\in\Omega:\substack{\rect{(N+1)\a\times(N-1)\b}{(0,0)}\text{ is not divided}\\
\text{by vertical sticks of both parities}
}
\right\} .\label{eq:G_def}
\end{equation}

\begin{claim}
\label{claim:same_parity_perco}There is a universal $\lambda_{0}(\a)$
such that for $\lambda>\lambda_{0}(\a)$, 
\[
\pv(\sg{\a\times\b}{\es{G^{\a\times\b}})<e^{-c\a}}.
\]
\end{claim}
\begin{proof}
Denote $R=\rect{K\times L}{(0,0)}$ with $K=(N+1)\a,L=(N-1)\b$. Let
$E$ be the event that each row of faces in $R$ has a vacant face.
Note that $\Omega\setminus G^{\a\times\b}\subset E$.

For a face $f$ in $R$, Corollary \ref{cor:single_vacancy_Z} says
that $\zedi[f][\text{\ensuremath{f} is vacant}]\le\lambda^{-1/4}$.
Lemma \ref{lem:chessboard_recursive}, and Proposition \ref{prop:seminorm}
imply together that for a row of faces $S=\rect{K\times1}{(0,\ell)}$
we have $\zedi[S][\text{\ensuremath{S} has a vacant face}]\le K\lambda^{-\frac{1}{4}}$.
Again by Lemma \ref{lem:chessboard_recursive} we get $\zedi[][E]\le\left(K\lambda^{-\frac{1}{4}}\right)^{L}.$

By item \ref{enu:partition_chessboard} of Lemma \ref{lem:grid_change},
\begin{align*}
\pv(\sg{\a\times\b}{\es{G^{\a\times\b}}}) & \le\sqrt[(N+1)(N-1)]{\zedi[R][\Omega\setminus G^{\a\times\b}]}\le\sqrt[(N+1)(N-1)]{\zedi[R][E]}.\\
 & \le\left((N+1)\a\lambda^{-1/4}\right)^{\frac{(N-1)\b}{(N+1)(N-1)}}\le e^{-c\a}
\end{align*}
Where the last inequality holds by choosing $\lambda_{0}(\a)$ such
that for $\lambda>\lambda_{0}$, it holds that $(N+1)\a\lambda^{-1/4}<e^{-1}$,
and $\frac{\b}{N+1}\ge\frac{c_{\ref{lem:main}}\lambda^{1/2}}{N(N+1)}\ge c\a$.
\end{proof}
We continue with the proof of the lemma.
\begin{proof}
[Proof of Item \ref{enu:ergodic_tetrachotomy}]By item \ref{enu:ergodic_dichotomy_thin},
we may assume that one of $\Psi_{\ver}^{\a\times\b}$ and $\Psi_{\hor}^{\b\times\a}$
is $e^{-c\a}$-strongly-percolating. We prove for the case that $\Psi_{\ver}^{\a\times\b}$
percolates. The other case is similar. Thus
\begin{equation}
\pv(\Psi_{\ver}^{\a\times\b})=\pv(\sg{\a\times\b}{\es{D_{\ver}^{\a\times\b}}})\le e^{-c\a}.\label{eq:psi_ver_pv}
\end{equation}

By Lemma \ref{lem:perc_intersection}, and by (\ref{eq:psi_ver_pv})
and Claim \ref{claim:same_parity_perco},
\[
\pv(\sg{\a\times\b}{\es{D_{\ver}^{\a\times\b}\cap G^{\a\times\b}}})\le2\sqrt[2]{\max\{\pv(\sg{\a\times\b}{\es{G^{\a\times\b}}}),\pv(\Psi_{\ver}^{\a\times\b})\}}\le2e^{-c\a/2}
\]
holds when $\lambda>\lambda_{0}$ and $\a_{0}\le\a\le\b$, for $c$
and $\a_{0}$ of item \ref{enu:ergodic_dichotomy_thin} and $\lambda_{0}$
of Claim \ref{claim:same_parity_perco}.

At this point we fix $c$ and $\a_{0}$ to their final values. We
may decrease $c$ and choose $\a_{0}$ sufficiently large depending
on $c$ such that under the assumptions of the current item
\[
\pv(\sg{\a\times\b}{\es{D_{\ver}^{\a\times\b}\cap G^{\a\times\b}}})\le e^{-c\a}<\eps_{0}.
\]

Note that the conditions of Proposition \ref{prop:peierls_two} are
satisfied for
\begin{equation}
K=\a,L=\b,E=D_{(\ver,0)}^{\a\times\b}\cap G^{\a\times\b},F=D_{(\ver,1)}^{\a\times\b}\cap G^{\a\times\b}.\label{eq:propagation_2}
\end{equation}
Thus
\[
\min\{\pv(\sg{\a\times\b}{\es{D_{(\ver,0)}^{\a\times\b}\cap G^{\a\times\b}}}),\pv(\sg{\a\times\b}{\es{D_{(\ver,1)}^{\a\times\b}\cap G^{\a\times\b}}})\}\le e^{c\a}.
\]
Assume WLOG that $\pv(\sg{K\times L}{\es{D_{(\ver,0)}^{\a\times\b}\cap G^{\a\times\b}}})\le e^{-c\a}$
(the other case is similar). Then in particular $\Psi_{(\ver,0)}^{\a\times\b}$
is $e^{-c\a}$-strongly-percolating. In addition, as $e^{-c\a}<\eps_{0}$
and 
\[
(D_{(\ver,0)}^{\a\times\b}\cap G^{\a\times\b})\cap(D_{(\ver,1)}^{\a\times\b}\cup D_{(\hor,0)}^{\a\times\b}\cup D_{(\hor,1)}^{\a\times\b})=\emptyset,
\]
it holds almost surely that the other three sets have only finite
$\boxtimes$-components.
\end{proof}

\subsection{Proof of Proposition \ref{prop:isometries and phase}}

By the assumptions of the Proposition, we may require that $\a$ is
sufficiently large, and $\lambda$ is large as a function of $\a$.
Thus Theorem \ref{thm:ergodic_tetrachotomy}, the definition of $\phase$
and Claim \ref{claim:same_parity_perco} apply.
\begin{proof}
[Proof of items \ref{enu:ph_ref_diag} and \ref{enu:ph_ref_vert}]
Immediate from the definitions of $\Psi_{(\ver,0)}^{\a\times\b}$,$\Psi_{(\ver,1)}^{\a\times\b}$,$\Psi_{(\hor,0)}^{\b\times\a}$,
and $\Psi_{(\hor,1)}^{\b\times\a}$.
\end{proof}
\begin{proof}
[Proof of item \ref{enu:ph_trans_x}]It suffices to \emph{rule out}
the following possibilities:
\begin{align}
\phase(\mu) & =\phase(\eta_{(1,0)}\mu)=(\ver,i)\label{eq:phase_poss_11}\\
\phase(\mu) & =(\hor,i),\,\phase(\eta_{(1,0)}\mu)=(\ver,j)\label{eq:phase_poss_12}\\
\phase(\mu) & =(\ver,i),\,\phase(\eta_{(1,0)}\mu)=(\hor,j)\label{eq:phase_poss_13}
\end{align}
for any $i,j\in\{0,1\}$.

Assume by contradiction possibility (\ref{eq:phase_poss_11}). Applying
Theorem \ref{thm:ergodic_tetrachotomy} to $\mu$ and $\eta_{(1,0)}\mu$,
and Claim \ref{claim:same_parity_perco} to $\mu$, shows that
\[
\max\left\{ \pv_{\mu}(\sg{\a\times\b}{\es{D_{(\ver,0)}^{\a\times\b}}}),\pv_{\mu}(\sg{\a\times\b}{\es{\eta_{(1,0)}D_{(\ver,0)}^{\a\times\b}}}),\pv_{\mu}(\sg{\a\times\b}{\es{G^{\a\times\b}}})\right\} <e^{-c\a}
\]
for some universal $c>0$. Applying Lemma \ref{lem:two_point_peierls}
(with $d=0$) thus gives
\[
\max\left\{ \mu(\Omega\setminus D_{(\ver,0)}^{\a\times\b}),\mu(\Omega\setminus\eta_{(1,0)}D_{(\ver,0)}^{\a\times\b}),\mu(\Omega\setminus G^{\a\times\b})\right\} <\frac{e^{-c\a}}{\eps_{0}}.
\]
Taking $\a$ large enough so that $e^{-c\a}/\eps_{0}<1/3$ we conclude
that
\[
\mu(D_{(\ver,0)}^{\a\times\b}\cap\eta_{(1,0)}D_{(\ver,0)}^{\a\times\b}\cap G^{\a\times\b})>0.
\]
However this is a contradiction since the event on the LHS is empty.

Now assume by contradiction possibility (\ref{eq:phase_poss_12}).
Then by Theorem \ref{thm:ergodic_tetrachotomy},
\[
\max\left\{ \pv_{\mu}(\sg{\a\times\b}{\es{D_{\hor}^{\b\times\a}}}),\pv_{\mu}(\sg{\b\times\a}{\es{\eta_{(1,0)}D_{\ver}^{\a\times\b}}})\right\} <e^{-c\a}
\]
which as before leads to 
\[
\mu(D_{\hor}^{\b\times\a}\cap\eta_{(1,0)}D_{\ver}^{\a\times\b})>0.
\]
The event on the LHS is empty when $\a\le\b$ and $0\le1<N\a+1\le N\b$,
since then each horizontal stick that divides $\rect{N\b\times N\a}{(0,0)}$
crosses each vertical stick that divides $\rect{N\a\times N\b}{(1,0)}$.
This holds when $\lambda$ is sufficiently large as a function of
$\a$, thus we have a contradiction.

Possibility (\ref{eq:phase_poss_13}) leads similarly to a contradiction,
considering the two events $\eta_{(\a,0)}D_{\ver}^{\a\times\b},\eta_{(1,0)}D_{\hor}^{\b\times\a}$
instead of $D_{\hor}^{\b\times\a},\eta_{(1,0)}D_{\ver}^{\a\times\b}$.
\end{proof}
\begin{proof}
[Proof of item \ref{enu:ph_trans_y}]It suffices to rule out the
following possibilities:
\begin{align}
\phase(\mu) & =(\ver,i),\,\phase(\eta_{(0,1)}\mu)=(\ver,1-i)\label{eq:phase_poss_21}\\
\phase(\mu) & =(\ver,i),\,\phase(\eta_{(0,1)}\mu)=(\hor,j)\label{eq:phase_poss_22}\\
\phase(\mu) & =(\hor,i),\,\phase(\eta_{(0,1)}\mu)=(\ver,j)\label{eq:phase_poss_23}
\end{align}
for any $i,j\in\{0,1\}$. Possibility (\ref{eq:phase_poss_21}) leads
to a contradiction in a manner similar to the previous ones, by noting
that
\[
D_{(\ver,i)}^{\a\times\b}\cap\eta_{(0,1)}D_{(\ver,1-i)}^{\a\times\b}\cap\eta_{(0,1)}G^{\a\times\b}=\emptyset.
\]
and each of the events in this intersection has high probability.
Possibilities (\ref{eq:phase_poss_22}) and (\ref{eq:phase_poss_23})
are impossible, as switching the $x$ and $y$ axis leads respectively
to possibilities (\ref{eq:phase_poss_12}) and (\ref{eq:phase_poss_13}).
\end{proof}

\section{Characterization of the invariant Gibbs Measures and decay of correlations\label{sec:decay}}

Throughout this section and Section \ref{sec:Columnar-order}, fix
$\a\in2\nat$ to be large enough for the following arguments (its
value is a large universal constant). Also fix $\lambda_{0}$ to be
a threshold depending on $\a$ and chosen sufficiently large for the
following arguments, and assume $\lambda>\lambda_{0}$. Lastly, we
continue to use the length scale $\b$ defined in (\ref{eq:b def})
and introduce a third (and final) length scale\marginpar{$\c$}
\[
\c=\c(\lambda):=\lfloor\sqrt{\a}\rfloor\b.
\]
We will use the $\phase$ notation introduced after Theorem \ref{thm:ergodic_tetrachotomy}. 

In this section we establish significant parts of our main results.
We prove item \ref{item:Invariance-and-extremality} of Theorem \ref{thm:main}
and Theorem \ref{thm:main2}. We explicitly state these results (together
with some byproducts) in the following lemma.
\begin{lem}
\label{lem:unique Gibbs measure}Let $\lambda>\lambda_{0}$. For each
$\pi\in\mathcal{P}$, there is a unique $\b!\Z^{2}$-ergodic Gibbs
measure, denoted $\mu_{\pi}$, with $\phase(\mu_{\pi})=\pi$. In addition,
\begin{enumerate}
\item \label{enu:unique_extremal}$\mu_{\pi}$ is extremal for each $\pi\in\mathcal{P}$.
\item \label{enu:unique_invariance}$\mu_{\pi}$ is $2\Z\times\Z$-invariant
when $\pi\in\{(\ver,0),(\ver,1)\}$ and $\Z\times2\Z$-invariant when
$\pi\in\{(\hor,0),(\hor,1)\}$.
\item \label{enu:unique_relations}$\mu_{(\ver,1)}$ is created by translating
$\mu_{(\ver,0)}$ by one lattice space in the horizontal direction.
The measures $\mu_{(\hor,0)}$ and $\mu_{(\hor,1)}$ are formed from
$\mu_{(\ver,0)}$ and $\mu_{(\ver,1)}$, respectively, by switching
the $x$ and $y$ axes.
\item \label{enu:unique_decompostion}Every periodic Gibbs measure is a
convex combination of $(\mu_{\pi})_{\pi\in\mathcal{P}}.$
\end{enumerate}
\end{lem}
In addition, we establish the quantitative decay of correlations estimate
corresponding to the first term in the minimum in item \ref{item:Decay-of-correlations}
of Theorem \ref{thm:main}.

Following these facts, the tasks remaining to complete the proofs
of our main results are to prove that $\mu_{(\ver,0)}$ satisfies
item \ref{item:Columnar-order} (columnar order) of Theorem \ref{thm:main}
and to refine the quantitative decay of correlations estimate to include
the second term in the minimum in item \ref{item:Decay-of-correlations}
of Theorem \ref{thm:main}. These tasks will be taken up in Section
\ref{sec:Columnar-order}.

\subsection{Disagreement percolation}

The proofs of our main results are based on the concept of disagreement
percolation, as introduced by van den Berg \cite{cmp/1104252313}
and further studied in the context of the hard-core model by van den
Berg and Steif \cite{VANDENBERG1994179}. The following theorem states
the results that will be used. For two configurations $\sigma,\sigma'\in\Omega$,
denote their \textbf{disagreement set} by\marginpar{$\Delta_{\sigma,\sigma'}$}
\[
\Delta_{\sigma,\sigma'}:=\{v\in\Z^{2}\colon\sigma(v)\neq\sigma'(v)\}.
\]
Define a \textbf{path of disagreement} as a $\boxtimes$-path in $\Delta_{\sigma,\sigma'}$.
The motivation for considering the $\boxtimes$ connectivity in particular
is that our model is a random Markov field with respect to the graph
$(\V,\E_{\boxtimes})$. 
\begin{thm}
\label{thm:disagreement percolation}Let $\mu,\mu'$ be Gibbs measures.
Let $\sigma,\sigma'$ be independent samples from $\mu,\mu'$, respectively.
Suppose that
\begin{equation}
\mathbb{P}(\text{\ensuremath{\Delta_{\sigma,\sigma'}} has an infinite \ensuremath{\boxtimes}-connected component})=0.\label{eq:no infinite disagreement path}
\end{equation}
Then
\begin{enumerate}
\item \label{item:equal_extremal}$\mu=\mu'$ and $\mu$ is extremal.
\item \label{item:decay of correlations diagreement perc}Let $f,g:\Omega\to[-1,1]$.
Suppose that $f$ is $A$-local and $g$ is $B$-local for $A,B\subset\V$
(locality is defined in (\ref{eq:local definition})). Then
\begin{equation}
\cov(f(\sigma),g(\sigma))\le2\mathbb{P}(\text{a \ensuremath{\boxtimes}-path in \ensuremath{\Delta_{\sigma,\sigma'}} intersects \ensuremath{A} and \ensuremath{B}})\label{eq:covariance bound from disagreements}
\end{equation}
with $\cov(\cdot,\cdot)$ denoting the covariance between two random
variables.
\end{enumerate}
\end{thm}
The equality of the measures under the assumption (\ref{eq:no infinite disagreement path})
is proved in \cite[Theorem 1]{cmp/1104252313}. Extremality also follows,
as one may apply the equality clause to the measures in the extremal
decomposition of $\mu$ (for extremal decomposition, see \cite[Theorem (7.26)]{Georgii2011}).
The covariance bound (\ref{eq:covariance bound from disagreements})
is an extension of \cite[Theorem 2.4]{VANDENBERG1994179}. For completeness,
a self-contained proof is provided in subsection \ref{subsec:Disagreement-percolation-proofs}.
While we state the theorem for the specific hard-core model studied
here, we remark that the disagreement percolation method applies to
general Markov random fields, defined on general graphs.

The following lemma gives a quantitative bound on the size of disagreement
components, and in particular shows that disagreement components do
not percolate. The lemma is proved in subsection \ref{subsec:disag_bound}.
After its statement we proceed to derive the main results of the current
section.
\begin{lem}
\label{lem:decay_1}Let $\lambda>\lambda_{0}$. There exist universal
$C,c>0$ such that the following holds. Let $\mu,\mu'$ be $\b!\Z^{2}$-ergodic
Gibbs measures with $\phase(\mu)=\phase(\mu')=(\ver,0)$. Let $\sigma,\sigma'$
be independent samples from $\mu,\mu'$, respectively. Then for each
$A\subset\Z^{2}$ and $B\subset\Z^{2}$,
\[
\mathbb{P}(\text{a \ensuremath{\boxtimes}-path in \ensuremath{\Delta_{\sigma,\sigma'}} intersects \ensuremath{A} and \ensuremath{B}})\le\sum_{u\in A}\sup_{v\in B}\alpha_{1}(u,v)
\]
where for $u=(x_{1},y_{1})$ and $v=(x_{2},y_{2})\in\Z^{2}$,
\[
\alpha_{1}(u,v):=C\exp\left(-c|x_{2}-x_{1}|-c\frac{|y_{2}-y_{1}|}{\sqrt{\lambda}}\right).
\]

In particular, there are no infinite disagreement components in the
sense that (\ref{eq:no infinite disagreement path}) holds.
\end{lem}
Applying Theorem \ref{thm:disagreement percolation} to the conclusions
of the lemma above, yields the following proposition, which is the
key to the proof Lemma \ref{lem:unique Gibbs measure}.
\begin{prop}
\label{prop:verzero_unique}Let $\lambda>\lambda_{0}$. Then there
is a unique $\b!\Z^{2}$-ergodic Gibbs measure $\mu$ with $\phase(\mu)=(\ver,0)$,
and $\mu$ is extremal.
\end{prop}
\begin{proof}
Let $\lambda>\lambda_{0}$. Let $\mu,\mu'$ be $\b!\Z^{2}$-ergodic
Gibbs measures with $(\ver,0)=\phase(\mu)=\phase(\mu')$, and let
$\sigma,\sigma'$ be independent samples from $\mu,\mu'$ respectively.
By Lemma \ref{lem:decay_1}, the condition (\ref{eq:no infinite disagreement path})
holds, thus by item \ref{item:equal_extremal} of Theorem \ref{thm:disagreement percolation},
$\mu=\mu'$, and $\mu$ is extremal.
\end{proof}
Finally, Lemma \ref{lem:decay_1}, put together with item \ref{item:decay of correlations diagreement perc}
of Theorem \ref{thm:disagreement percolation}, also yields the quantitative
decay of correlations estimate corresponding to the first term in
the minimum in item \ref{item:Decay-of-correlations} of Theorem \ref{thm:main}.
This is used later in subsection \ref{subsec:correlations}, where
a complete proof of item \ref{item:Decay-of-correlations} is given.

\subsection{Proof of Lemma \ref{lem:unique Gibbs measure}}

Here we prove the main result of the current section. The proof relies
only on Propositions \ref{prop:verzero_unique} and \ref{prop:isometries and phase}.
\begin{proof}
[Proof of Lemma~\ref{lem:unique Gibbs measure}]Let $\lambda>\lambda_{0}$.
We require that $\a$ and $\lambda_{0}$ are sufficiently large so
that $\phase$ is defined for every $\b!\Z^{2}$-ergodic Gibbs measure,
as explained immediately after Theorem \ref{thm:ergodic_tetrachotomy}.
Proposition \ref{prop:verzero_unique} justifies the notation $\mu_{\pi}$
and proves that $\mu_{\pi}$ is extremal, for the case of $\pi=(\ver,0)$.

Let $\mu$ be a be $\b!\Z^{2}$-ergodic Gibbs measure with $\phase(\mu)=(\ver,1)$.
By item \ref{enu:ph_trans_x} of Proposition \ref{prop:isometries and phase},
$\phase(\eta_{(-1,0)}\mu)=(\ver,0)$. Thus $\mu=\eta_{(1,0)}\mu_{(\ver,0)}$,
showing the uniqueness of $\mu$.

Let $i\in\{0,1\}$. Let $\mu$ be a be $\b!\Z^{2}$-ergodic Gibbs
measure with $\phase(\mu)=(\hor,i)$. By item \ref{enu:ph_ref_diag}
of Proposition \ref{prop:isometries and phase}, for $\tau$ defined
by $\tau(x,y)=(y,x)$, it holds that $\phase(\tau^{-1}\mu)=(\ver,i)$.
Thus $\mu=\tau\mu_{(\ver,i)}$, showing the uniqueness of $\mu$.

The above arguments show the uniqueness for each $\pi\in\mathcal{P}$,
justify the notation $\mu_{\pi}$ and prove items \ref{enu:unique_extremal}
and \ref{enu:unique_relations}. Item \ref{enu:unique_invariance}
then follows from items \ref{enu:ph_trans_x} and \ref{enu:ph_trans_y}
of Proposition \ref{prop:isometries and phase} together with the
uniqueness just shown.

The uniqueness results above show that $(\mu_{\pi})_{\pi\in\mathcal{P}}$
are the only $\b!\Z^{2}$-ergodic Gibbs measures. We now show how
this implies item \ref{enu:unique_decompostion}.

We will use the fact that a Gibbs measure invariant with respect to
a full-rank lattice $\L$ has a unique decomposition as a mixture
of $\L$-ergodic Gibbs measures. This theorem is stated and proved
in \cite[Theorem (14.17)]{Georgii2011}. More formally, the theorem
says that for each $\L$-invariant Gibbs measure $\mu$, there is
a unique measure $w_{\mu}$ on the space of all $\L$-ergodic measures
on $\Omega$ such that $\mu=\int v\,dw_{\mu}(v)$, and the unique
measure $w_{\mu}$ is supported on the set of $\L$-ergodic Gibbs
measures.

Let $\L$ be a full-rank lattice such that $\L\subset\b!\Z^{2}$.
We claim that every $\L$-ergodic Gibbs measure is one of $(\mu_{\pi})_{\pi\in\mathcal{P}}$.
Indeed let $\mu$ be an $\L$-ergodic Gibbs measure. Define $\nu$
to be the average of all the shifts of $\mu$ by elements of $\b!\Z^{2}/\mathcal{L}$.
Then $\nu$ is $\b!\Z^{2}$-invariant and thus by the decomposition
theorem, $\nu$ is a linear combination of $(\mu_{\pi})_{\pi\in\mathcal{P}}$.
This decomposition is also the unique decomposition of $\nu$ as a
mixture of $\L$-ergodic measures, since $(\mu_{\pi})_{\pi\in\mathcal{P}}$
are extremal, and in particular $\L$-ergodic. But as $\nu$ is defined
as an average of $\L$-ergodic measures, namely the shifts of $\mu$,
it follows that each shift of $\mu$ and in particular $\mu$ itself
must be one of $(\mu_{\pi})_{\pi\in\mathcal{P}}$.

Let $\mu$ be a periodic measure, invariant with respect to a full-rank
lattice $\mathcal{L}$. Assume WLOG that $\L\subset\b!\Z^{2}$. Then
by the claim of the previous paragraph that $(\mu_{\pi})_{\pi\in\mathcal{P}}$
are the only $\L$-ergodic Gibbs measures, the decomposition theorem
implies item \ref{enu:unique_decompostion}.
\end{proof}

\subsection{\label{subsec:disag_bound}Tail bounds for the connectivity of disagreement
components}

In this subsection we prove Lemma \ref{lem:decay_1}. Recall the variables
and assumptions introduced in the beginning of the section. Let $\lambda>\lambda_{0}$
and let $\mu,\mu'$ be $\b!\Z^{2}$-ergodic Gibbs measures satisfying
$\phase(\mu)=\phase(\mu')=(\ver,0)$. Let $\sigma,\sigma'$ be independent
samples from $\mu,\mu'$, respectively.

Heuristically, to prove Lemma \ref{lem:decay_1} one needs to show
that long disagreement paths are rare. To this end we will define
``sealed rectangles'' and ``semi-sealed rectangles''. A rectangle
$R$ will be defined to be semi-sealed in $\sigma$, if $\sigma$
satisfies, in the vicinity of $R$, a set of conditions which are
typical of a configuration drawn from a $(\ver,0)$ Gibbs measure.
The rectangle $R$ is said to be sealed in $(\sigma,\sigma')$ if
it satisfies these conditions for both $\sigma$ and $\sigma'$.

The conditions are designed in such a way that the assumption that
$R$ is sealed ensures that a disagreement path starting in $R$ can
only reach points in the vicinity of $R$. Therefore a long path of
disagreement will imply a long sequence of neighboring non-sealed
rectangles, which will be shown to be unlikely by a Peierls argument.

\subsubsection{Semi-sealed rectangles}

Here we consider only $\sigma$, the same considerations apply also
to $\sigma'$. We say that $\rect{N\a\times N\c}{(0,0)}$ is \textbf{semi-sealed}
if the event $\Sigma\subset\Omega$ \marginpar{$\Sigma$}holds, where
$\Sigma=\Sigma_{1}\cap\Sigma_{2}$ and $\Sigma_{1},\Sigma_{2}$ are
defined below. More generally, for $(x,y)\in\Z^{2}$, we say that
$\rect{N\a\times N\c}{(N\a x,N\c y)}$ is semi-sealed if $\eta_{(N\a x,N\c y)}\Sigma$
holds. Our goal in this subsubsection is to prove ``strong-percolation
of semi-sealed rectangles'' in the sense of Lemma \ref{lem:semi-sealed}
below. We first prove the Lemma assuming Proposition \ref{prop:seal_bounds},
and then prove the Proposition.

Consider events $\Sigma_{0},\Sigma_{1},\Sigma_{2}$ defined as follows:
\begin{itemize}
\item $\sigma\in\Sigma_{0}$ \marginpar{$\Sigma_{0}$}iff every $N\a\times1$
rectangle contained in 
\[
\rect{N\a\times3N\c}{(-N\a,-N\c)}\cup\rect{N\a\times3N\c}{(+N\a,-N\c)}
\]
 intersects the interior of a tile with even horizontal parity.
\item $\sigma\in\Sigma_{1}$ \marginpar{$\Sigma_{1}$}iff all tiles of $\sigma$
with center in $\rect{N\a\times3N\c}{(0,-N\c)}$ have even horizontal
parity.
\item $\sigma\in\Sigma_{2}$ \marginpar{$\Sigma_{2}$}iff every $1\times N\c$
rectangle contained in 
\[
\rect{N\a\times N\c}{(0,-N\c)}\cup\rect{N\a\times N\c}{(0,+N\c)}
\]
contains a vacant face of $\sigma$.
\end{itemize}
Recall that $\a$ and $\lambda_{0}$ were introduced at the beginning
of the section.
\begin{lem}
\label{lem:semi-sealed}For every $\eps>0$ we may choose $\a$ sufficiently
large, and $\lambda_{0}$ sufficiently large as a function of $\a$,
such that $\sg{N\a\times N\c}{\es{\Sigma}(\sigma)}$ is $\eps$-strongly-percolating.
\end{lem}
\begin{proof}
Let $\delta>0$. Recalling the definitions of $\b$ and $\c$, we
check that for every sufficiently large $\a$ there is a sufficiently
large $\lambda_{0}$ such that all the bounds of Proposition \ref{prop:seal_bounds}
are less than $\delta$.

For each $\eta\in\{\eta_{v}:v\in\a\Z\times\b\Z\}$, and every $\b!\Z^{2}$-ergodic
Gibbs measure $\mu$, $\phase(\mu)=\phase(\eta\mu)$. Thus the bounds
of Proposition \ref{prop:seal_bounds} hold also for translations
of the relevant events by elements of $\a\Z\times\b\Z$. Thus we may
apply item \ref{enu:partition} of Lemma \ref{lem:grid_change} three
times, with $k=N\a,l=N\b$ and respectively with:
\begin{alignat*}{3}
K & =3N\a, & \quad L & =3N\b, & \quad E & =\Sigma_{0},\\
K & =3N\a, & L & =5N\b, & E & =\Sigma_{1}\cup\Sigma_{0}^{c},\\
K & =N\a, & L & =3N\b, & E & =\Sigma_{2},
\end{alignat*}
to obtain respectively
\begin{align*}
\pv_{\mu}(\sg{N\a\times N\c}{\es{\Sigma_{0}}}) & \le\sqrt[9]{9\sqrt[9]{\delta}},\\
\pv_{\mu}(\sg{N\a\times N\c}{\es{\Sigma_{1}\cup\Sigma_{0}^{c}}}) & \le\sqrt[15]{15\sqrt[15]{\delta}},\\
\pv_{\mu}(\sg{N\a\times N\c}{\es{\Sigma_{2}}}) & \le\sqrt[3]{3\sqrt[3]{\delta}}.
\end{align*}

Note that $\Sigma=\Sigma_{1}\cap\Sigma_{2}\supset\Sigma_{0}\cap(\Sigma_{1}\cup\Sigma_{0}^{c})\cap\Sigma_{2}$,
thus by Lemma \ref{lem:perc_intersection}, 
\begin{alignat*}{2}
\pv_{\mu}(\sg{N\a\times N\c}{\es{\Sigma})}\le3(\max\{\, & \pv_{\mu}(\sg{N\a\times N\c}{\es{\Sigma_{0}}}),\\
 & \pv_{\mu}(\sg{N\a\times N\c}{\es{\Sigma_{1}\cup\Sigma_{0}^{c}}}),\\
 & \pv_{\mu}(\sg{N\a\times N\c}{\es{\Sigma_{2}}}) & \})^{1/3}.
\end{alignat*}
Given $\eps$, we are done by taking $\delta$ sufficiently small.
\end{proof}
\begin{prop}
\label{prop:seal_bounds}There is a universal $c>0$ such that
\begin{enumerate}
\item \label{enu:sig0_bound}${\displaystyle \pv_{\mu}(\sg{3N\a\times3N\c}{\es{\Sigma_{0}})}\le\frac{9N^{2}\c}{\b}e^{-c\a}}$
\item \label{enu:sig1_bound}${\displaystyle \pv_{\mu}(\sg{3N\a\times5N\c}{\es{\Sigma_{1}\cup\Sigma_{0}^{c}})}\le(3N\c+1)(6N\a)^{4}\lambda^{-1}}$
\item \label{enu:sig2_bound}${\displaystyle \pv_{\mu}(\sg{N\a\times3N\c}{\es{\Sigma_{2}})}\}\le2N\a e^{-c\lambda^{-1/2}N\c}}$
\end{enumerate}
\end{prop}
\begin{proof}
[Proof of item \ref{enu:sig0_bound}]We shall apply item \ref{enu:lump}
of Lemma \ref{lem:grid_change}, for
\begin{align*}
K & =3N\a,k=\a,L=3N\c,l=\b\\
E & =D_{(\ver,0)}^{\a\times\b}.
\end{align*}
We let $H$ be as defined in the lemma. The lemma yields that
\begin{equation}
\pv_{\mu}(\sg{3N\a\times3N\c}{\bigcap_{\eta\in H}\es{\eta E}})\le\frac{9N^{2}\c}{\b}\pv_{\mu}(\sg{\a\times\b}{\es E}).\label{eq:bp_1_1}
\end{equation}
As $\Psi_{(\ver,0)}^{\a\times\b}(\sigma)=\sg{\a\times\b}{\es E}$
, Theorem \ref{thm:ergodic_tetrachotomy} yields
\begin{equation}
\pv_{\mu}(\sg{\a\times\b}{\es E})\le e^{-c_{\ref{thm:ergodic_tetrachotomy}}\a}.\label{eq:bp_1_2}
\end{equation}
Note that the application of the Theorem above is the only place in
the proof of the current proposition where we used the assumptions
on $\a$ and $\lambda$ being sufficiently large and the assumption
that $\phase(\mu)=(\ver,0)$. It is also the only place in the current
subsection where we directly use the assumption that $\phase(\mu)=(\ver,0)$.

If $\sigma\in\bigcap_{\eta\in H}\eta E$ then in particular for each
$0\le i<\bigcap_{i=0}^{3\c/\b-1}$ it holds that $\sigma\in\eta_{(-N\a,-N\c+iN\b)}E$,
so $R_{i}\coloneqq\rect{N\a\times N\b}{(-N\a,-N\c+iN\b)}$ is divided
by a $(\ver,0)$ stick. Each $N\a\times1$ rectangle in $\rect{N\a\times3N\c}{(-N\a,-N\c)}$
is contained in some $R_{i}$, and thus intersects a $(\ver,0)$ stick
of $\sigma$, and thus also intersects the interior of a tile in $\sigma$
with even horizontal parity. A similar argument holds for $N\a\times1$
rectangles contained in $\rect{N\a\times3N\c}{(+N\a,-N\c)}$. Thus
\begin{equation}
\bigcap_{\eta\in H}\eta E\subset\Sigma_{0}.\label{eq:bp_1_3}
\end{equation}
and the claim follows by (\ref{eq:bp_1_1}), (\ref{eq:bp_1_2}), and
(\ref{eq:bp_1_3}).
\end{proof}
\begin{proof}
[Proof of item \ref{enu:sig1_bound}]Denote $R=\rect{3N\a\times5N\c}{(-N\a,-2N\c)}$.
By Corollary \ref{cor:complement of eps rare}, 
\[
\pv_{\mu}(\sg{3N\a\times5N\c}{\es{\Sigma_{1}\cup\Sigma_{0}^{c}}})\le\zedi[R][\Sigma_{0}\setminus\Sigma_{1}]
\]
thus it suffices to bound the RHS.

Assume $\sigma\in\Sigma_{0}\setminus\Sigma_{1}$, then since $\sigma\notin\Sigma_{1}$,
the rectangle $\rect{N\a\times3N\c}{(0,-N\c)}$ contains the center
of a tile with odd horizontal parity. Equivalently, there is a point
$(x_{0},y_{0})\in[0,N\a]\times[-N\c,2N\c]$ with $x_{0}$ even and
$\sigma(x_{0},y_{0})=1$. Consider the four rectangles:
\begin{align*}
[-N\a,x_{0}]\times[y_{0}-1,y_{0}],\,[x_{0},2N\a]\times[y_{0}-1,y_{0}],\\{}
[-N\a,x_{0}]\times[y_{0},y_{0}+1],\,[x_{0},2N\a]\times[y_{0},y_{0}+1].
\end{align*}
By $\sigma\in\Sigma_{0}$, each of them intersects the interior of
a tile of $\sigma$ that has even horizontal parity. Each of them
also intersects $T_{(x_{0},y_{0})}$, which has odd horizontal parity.
Thus each of the four rectangles contains a vacant face of $\sigma$.
Thus we see that whenever $\sigma\in\Sigma_{0}\setminus\Sigma_{1}$,
the rectangle $R$ contains a $3N\a\times2$ rectangle that contains
$4$ vacancies of $\sigma$. This corresponds to at most $(3N\c+1)(6N\a)^{4}$
sets of $4$ faces in $R$, such that for each $\sigma\in\Sigma_{0}\setminus\Sigma_{1}$,
one of those sets has all of its faces vacant.

For $4$ given faces in $R$, the event that they are all vacant has
chessboard norm at most $\lambda^{-1}$ by Corollary \ref{cor:single_vacancy_Z}
and Proposition \ref{lem:chessboard_recursive}. The bound on $\zedi[R][\Sigma_{0}\setminus\Sigma_{1}]$
follows from the subadditivity and positivity of the chessboard seminorm.
\end{proof}
\begin{proof}
[Proof of item \ref{enu:sig2_bound}]For $i\in\Z$, $0\le i<N\a$,
$j\in\{-N\c,N\c\}$ denote $R_{ij}=\rect{1\times N\c}{(i,j)}$ and
let $E_{ij}$ be the event that $R_{ij}$ contains no vacant face.
By Corollary \ref{cor:single_vacancy_Z} and Proposition \ref{lem:chessboard_recursive},
\[
\zedi[R_{ij}][E_{ij}]\le\left(1-c_{\ref{cor:single_vacancy_Z}}\lambda^{-1/2}\right)^{N\c}\le e^{-c_{\ref{cor:single_vacancy_Z}}\lambda^{-1/2}N\c}.
\]

Denote $R=\rect{N\a\times3N\c}{(0,-N\c)}$. By Corollary \ref{cor:complement of eps rare},
Proposition \ref{lem:chessboard_recursive}, and subadditivity and
positivity of the chessboard seminorm,
\[
\pv_{\mu}(\sg{N\a\times3N\c}{\es{\Sigma_{2}}})\le\zedi[][\Sigma_{2}^{c}]=\zedi[][\bigcup_{\substack{i,j}
}E_{ij}]\le\sum_{i,j}\zedi[R_{i}][E_{ij}]\le2N\a e^{-c_{\ref{cor:single_vacancy_Z}}\lambda^{-1/2}N\c}.\qedhere
\]
\end{proof}

\subsubsection{Bounding the disagreement components}

We say that a rectangle $\rect{N\a\times N\c}{(N\a x,N\c y)}$ is
\textbf{sealed} in $(\sigma,\sigma')$, if it is semi-sealed in both
$\sigma$ and $\sigma'$. The following deterministic statement shows
that large $\boxtimes$-components of $\Delta_{\sigma,\sigma'}$ are
disjoint from the union of the rectangles that are sealed in $(\sigma,\sigma')$. 
\begin{prop}
\label{prop:sealing}Let $\sigma,\sigma'\in\Omega$. Suppose that
$S=\rect{N\a\times N\c}{(N\a x_{0},N\c y_{0})}$ is sealed in $(\sigma,\sigma')$.
Let $v=(x_{1},y_{1})\in S$. If $v\in\Delta_{\sigma,\sigma'}$ then
the $\boxtimes$-component of $v$ in $\Delta_{\sigma,\sigma'}$ is
contained in $\{(x_{1},y)\in\V:N\c(y_{0}-1)<y<N\c(y_{0}+2)\}$. 
\end{prop}
\begin{proof}
It suffices to prove for the case of $x_{0}=y_{0}=0$. The assumption
that $S$ is sealed means that $\sigma,\sigma'\in\Sigma$. By $\sigma,\sigma'\in\Sigma_{1}$,
for each $(x,y)$ with $0\le x\le N\a$, $-N\c\le y\le2N\c$ and even
$x$, it holds that $\sigma(x,y)=\sigma'(x,y)=0$. In particular each
point in $S\cap\Delta_{\sigma,\sigma'}$ must have an odd first coordinate.
Thus $x_{1}$ is odd.

By $\sigma\in\Sigma_{2}$, the rectangle $\rect{1\times N\c}{(x_{1},N\c)}$
contains a vacant face, $\rect{1\times1}{(x_{1},y)}$, of $\sigma$.
Thus $\sigma(x_{1},y)=\sigma(x_{1},y+1)=0$. Since either $\sigma'(x_{1},y)=0$
or $\sigma'(x_{1},y+1)=0$, and $N\c\le y<2N\c$, there is a point
$(x_{1},y_{3})\notin\Delta_{\sigma,\sigma'}$ with $N\c\le y_{3}\le2N\c$
(with either $y_{3}=y$ or $y_{3}=y+1$). 

By a similar argument, there is a point $(x_{1},y_{4})\notin\Delta_{\sigma,\sigma'}$
with $-N\c\le y_{4}\le0$. By the first paragraph, $\Delta_{\sigma,\sigma'}\cap\{(x,y):x\in\{x_{1}-1,x_{1}+1\},\,y_{4}\le y\le y_{3}\}=\emptyset$. 

Thus the $\boxtimes$-component of $v$ in $\Delta_{\sigma,\sigma'}$
is contained in $B\coloneqq\{(x_{1},y)\in\V:y_{4}<y<y_{3}\}$ since
we have shown that every point outside of $B$ and $\boxtimes$-adjacent
to a point in $B$ is not in $\Delta_{\sigma,\sigma'}$. As $B\subset\{(x_{1},y)\in\V:-N\c<y<2N\c\}$,
the proof is complete.
\end{proof}
Denote\marginpar{$\Pi$}
\[
\Pi=\Pi(\sigma,\sigma')\coloneqq\xre{N\a\times N\c}{\es{\Sigma}(\sigma)}\,\cap\,\xre{N\a\times N\c}{\es{\Sigma}(\sigma')}.
\]
This random set represents the set of sealed rectangles, as $\rect{N\a\times N\c}{(N\a x,N\c y)}$
is sealed iff $(x,y)\in\Pi$. Fix some $\eps$ with $0<\eps<\eps_{0}$.
By Lemma \ref{lem:semi-sealed}, we may choose $\lambda_{0}$ sufficiently
large so that $\pv(\xre{N\a\times N\c}{\es{\Sigma}(\sigma)})<(\eps/2)^{2}$.
Thus by lemma \ref{lem:perc_intersection}, we have
\begin{equation}
\pv(\Pi)\le2\sqrt[2]{\max\{\pv(\xre{N\a\times N\c}{\es{\Sigma}(\sigma)}),\pv(\xre{N\a\times N\c}{\es{\Sigma}(\sigma')})}<\eps.\label{eq:sealed_bound}
\end{equation}

We proceed to prove Lemma \ref{lem:decay_1}.
\begin{proof}
[Proof of Lemma \ref{lem:decay_1}]Define $f:\V\to\V$ by $f(x,y)=\left(\left\lfloor \frac{x}{N\a}\right\rfloor ,\left\lfloor \frac{y}{N\c}\right\rfloor \right)$.
Let $u=(x_{u},y_{u}),v=(x_{v},y_{v})$ be in $\V$. Suppose that $u-v\notin\{0\}\times[-4N\c,4N\c]$
and that a $\boxtimes$-path $P$ in $\Delta_{\sigma,\sigma'}$ connects
$u$ to $v$.

We claim that this implies that $f(u)$ and $f(v)$ are connected
by a $\boxtimes$-path of points $(x,y)$ satisfying $(x,y)\notin\Pi$.
Each point $w$ on $P$ is connected by a $\boxtimes$-path in $\Delta_{\sigma,\sigma'}$,
to a point $w'$ such that $w-w'\notin\{0\}\times[-2N\c,2N\c]$. By
Proposition \ref{prop:sealing}, this means that each point in $P$
is contained in a non-sealed rectangle, i.e., a rectangle of the form
$\rect{N\a\times N\c}{(N\a x,N\c y)}$ where $(x,y)\notin\Pi$. The
claim follows, as $\{f(w):w\in P\}$ contains the required path.

Fix a point $u\in A$. Denote $\alpha=\sup_{v\in B}\alpha_{1}(u,v)$
and $d'=\inf_{v\in B}\left\Vert f(u)-f(v)\right\Vert _{\infty}$.
Then $d'\ge\frac{1}{2}\inf_{v\in B}\left(\frac{|x_{u}-x_{v}|}{N\a}+\frac{|y_{u}-x_{v}|}{N\c}\right)-1$.
Choosing $c,C$ appropriately, it holds that 
\begin{equation}
(\eps/\eps_{0})^{d'+1}\le\sup_{v\in B}\alpha_{1}(u,v)\label{eq:quant_proof_d_alpha}
\end{equation}
and $\alpha_{1}(u,v)\ge1$ whenever $u-v\in\{0\}\times[-4N\c,4N\c]$.
To prove the lemma, by a union bound it suffices to show that
\[
\mathbb{P}(\text{a \ensuremath{\boxtimes}-path in \ensuremath{\Delta_{\sigma,\sigma'}} intersects \ensuremath{\{u\}} and \ensuremath{B}})\le\sup_{v\in B}\alpha_{1}(u,v).
\]
In the case that there is $v\in B$ with $u-v\in\{0\}\times[-4N\c,4N\c]$,
there is nothing to prove as the RHS is at least $1$. Otherwise,
by the claim, if a $\boxtimes$-path in $\Delta_{\sigma,\sigma'}$
connects $u$ to some $v\in B$, then $f(u)$ is connected to $f(v)$
by a $\boxtimes$-path disjoint from $\Pi$. By (\ref{eq:sealed_bound})
and Lemma \ref{lem:two_point_peierls}, the probability that this
holds for some $v\in B$ is at most $(\eps/\eps_{0})^{d'+1}$, and
by (\ref{eq:quant_proof_d_alpha}) the proof is complete.
\end{proof}

\subsection{Disagreement percolation - proofs\label{subsec:Disagreement-percolation-proofs}}

In this section we provide a proof of Theorem \ref{thm:disagreement percolation}.
The proof of the theorem is based on the following lemma.
\begin{lem}
\label{lem:van-den-berg_involution}Let $A\subset\V$ be a finite
set. For $\sigma,\sigma'\in\Omega$, define $C_{A}(\sigma,\sigma')$
to be be the set of points which are connected to a point in $A$
by a path of disagreement (``the cluster of disagreement of $A$'').
Let $m:\Omega^{2}\to\Omega^{2}$ be defined by $m(\sigma,\sigma')=(\omega,\omega')$
where 
\begin{equation}
\omega(v),\omega'(v)=\begin{cases}
\sigma'(v),\sigma(v) & \text{\ensuremath{C_{A}(\sigma,\sigma')} is finite and \ensuremath{v\in C_{A}(\sigma,\sigma')} }\\
\sigma(v),\sigma'(v) & \text{o/w}
\end{cases}.\label{eq:berg_exchange}
\end{equation}
Let $(\sigma,\sigma')$ be sampled from $\Omega^{2}$ with the measure
$\mu\times\mu'$. Then $(\sigma,\sigma')$ has the same distribution
as $m(\sigma,\sigma')$.
\end{lem}
\begin{proof}
Denote by $C'_{A}(\sigma,\sigma')$ the ``exterior $\boxtimes$-boundary
of $C_{A}(\sigma,\sigma')$''. Precisely, it is the set of vertices
that are in $A$ or $\boxtimes$-adjacent to a vertex in $C_{A}(\sigma,\sigma')$,
but not in $C_{A}(\sigma,\sigma')$.

Consider the family $\mathcal{F}$ of events $E\subset\Omega^{2}$
consisting of
\begin{enumerate}
\item events contained in the event that $C_{A}(\sigma,\sigma')$ is infinite,
and
\item events of the form $E=\{(\sigma,\sigma'):\sigma\rstr_{D}=\rho\rstr_{D},\sigma'\rstr_{D}=\rho'\rstr_{D}\}$
where $\rho,\rho'\in\Omega$, and $D$ is finite and $C_{A}(\rho,\rho')\cup C'_{A}(\rho,\rho')\subset D$.
\end{enumerate}
The family $\mathcal{F}$ is a $\pi$-system. To see that it generates
the sigma algebra of $\Omega^{2}$, consider for an event $E$ its
partition according to the possibilities for $C_{A}(\sigma,\sigma')$
being finite, and the possibility that it is infinite. This gives
a \emph{countable} partition of $E$, and each part is easily seen
to be in the sigma algebra generated by $\mathcal{F}$. Thus it remains
to show for an event $E\in\mathcal{F}$ that 
\begin{equation}
\P((\sigma,\sigma')\in E)=\P((\sigma,\sigma')\in m^{-1}(E)).\label{eq:exchange_preserve}
\end{equation}

For the case of the first item, this is since $m^{-1}(E)=E$. For
the case of the second item, fix $\rho,\rho',D$ and the corresponding
event $E$, as in the second item. Denote $C=C_{A}(\rho,\rho')$ and
$C'=C'_{A}(\rho,\rho')$. Then $C'\subset D\setminus C$ and $\rho\rstr_{C'}=\rho'\rstr_{C'}$.
Thus by the domain Markov property, 
\begin{equation}
\begin{aligned}\P(\sigma\rstr_{C}=\rho\rstr_{C}\,\big|\,\sigma\rstr_{D\backslash C}=\rho\rstr_{D\backslash C}) & =\P(\sigma'\rstr_{C}=\rho\rstr_{C}\,\big|\,\sigma'\rstr_{D\backslash C}=\rho'\rstr_{D\backslash C}),\\
\P(\sigma'\rstr_{C}=\rho'\rstr_{C}\,\big|\,\sigma'\rstr_{D\backslash C}=\rho'\rstr_{D\backslash C}) & =\P(\sigma\rstr_{C}=\rho'\rstr_{C}\,\big|\,\sigma\rstr_{D\backslash C}=\rho\rstr_{D\backslash C}).
\end{aligned}
\label{eq:exchange_conditional}
\end{equation}

Since $C'\subset D$, it holds that $C_{A}(\sigma,\sigma')=C$ for
all $(\sigma,\sigma')\in E$. Thus 
\[
m^{-1}(E)=\{(\sigma,\sigma'):\substack{\sigma\rstr_{D\backslash C}=\rho\rstr_{D\backslash C},\\
\sigma'\rstr_{D\backslash C}=\rho'\rstr_{D\backslash C},
}
\enskip\substack{\sigma\rstr_{C}=\rho'\rstr_{C},\\
\sigma'\rstr_{C}=\rho\rstr_{C}
}
\}.
\]
Writing $E$ in an analogous way, we show (\ref{eq:exchange_preserve})
by expressing each side as a product of four terms and then using
(\ref{eq:exchange_conditional}).
\end{proof}
\begin{proof}
[Proof of Theorem \ref{thm:disagreement percolation}]Let $\mu,\mu'$
and $\sigma,\sigma'$ be as in the theorem and assume (\ref{eq:no infinite disagreement path}).
Define $(\omega,\omega')=m(\sigma,\sigma')$ as in Lemma \ref{lem:van-den-berg_involution}.

To show that $\mu=\mu'$, it suffices to prove $\mu(f)=\mu'(f)$ for
every function $f:\Omega\to\R$ which is $A$-local for some \emph{finite}
set $A\subset\V$. Define $C_{A}=C_{A}(\sigma,\sigma')$ as in Lemma
\ref{lem:van-den-berg_involution}. By the assumption (\ref{eq:no infinite disagreement path}),
$C_{A}$ is almost surely finite, thus (\ref{eq:berg_exchange}) gives
that almost surely $\omega\rstr_{A}=\sigma'\rstr_{A}$ and thus almost
surely $f(\omega)=f(\sigma')$. In addition, by Lemma \ref{lem:van-den-berg_involution},
$\sigma$ and $\omega$ have the same distribution, and thus
\[
\E(f(\sigma))=\E(f(\omega))=\E(f(\sigma')).
\]

Thus we have shown $\mu=\mu'$. We first prove item \ref{item:decay of correlations diagreement perc}
and then conclude from it the extremality of $\mu$. Let $f,g,A,B$
be as in item \ref{item:decay of correlations diagreement perc}.
As $f,g$ may be approximated (in $L^{2}$) by functions depending
on restrictions to finite sets, and as replacing $A,B$ with finite
subsets only decreases the RHS of (\ref{eq:covariance bound from disagreements}),
we may assume WLOG that $A$ and $B$ are finite. Again define $C_{A}=C_{A}(\sigma,\sigma')$
as in Lemma \ref{lem:van-den-berg_involution}. Let $E$ be the event
that a $\boxtimes$-path in $\Delta_{\sigma,\sigma'}$ intersects
both $A$ and $B$. The event $E^{c}$ is the event that $C_{A}$
is disjoint from $B$. Since $C_{A}$ is almost surely finite, (\ref{eq:berg_exchange})
gives that $\omega\rstr_{B}=\sigma\rstr_{B}$ and $g(\omega)=g(\sigma)$
hold almost surely on $E^{c}$. As before, $f(\omega)=f(\sigma')$
almost surely. Thus $\mathbb{P}\left(f(\omega)\cdot g(\omega)\neq f(\sigma')\cdot g(\sigma)\right)\le\mathbb{P}\left(E\right)$.
By Lemma \ref{lem:van-den-berg_involution}, since $f,g$ are bounded
between $-1$ and $1$, and by the independence of $\sigma,\sigma'$,
we have
\[
\E\left(f(\sigma)g(\sigma)\right)=\E\left(f(\omega)g(\omega)\right)\le\E\left(f(\sigma')g(\sigma)\right)+2\mathbb{P}(E)=\E(f(\sigma))\E(g(\sigma))+2\mathbb{P}(E).
\]

Finally, to show that $\mu$ is extremal, it suffices to show that
the covariance is $0$ between every bounded $f,\tilde{g}$ where
$f$ is $A$-local for a finite set $A$, and $\tilde{g}$ is measurable
with respect to the tail sigma algebra. This may be seen by approximating
$\tilde{g}$ by a function $g$ which is $B$-local where $B$ is
the complement of a large box around the origin. The assumption (\ref{eq:no infinite disagreement path})
shows that $\mathbb{P}\left(E\right)\to0$ as the box grows to infinity.
\end{proof}

\section{Columnar order\label{sec:Columnar-order}}

The measure $\mu_{(\ver,0)}$ referred to in Theorem \ref{thm:main}
has been defined in the previous section (in Lemma \ref{lem:unique Gibbs measure})
for $\lambda>\lambda_{0}$. In this section we prove some additional
properties of $\mu_{(\ver,0)}$, completing the proof of the theorem.
We increase $\lambda_{0}$ to be sufficiently large for the arguments
in this section.

\subsection{Offset tiles are rare}

The measure $\mu_{(\ver,0)}$ is characterized by columns of tiles
with even horizontal parity, i.e. tiles whose center has an odd first
coordinate. Here we bound the probability that a tile with odd horizontal
parity (``an offset tile'') appears in a given position. In subsection
\ref{subsec:tile_probabilities} we show that the bound is sharp up
to a multiplicative constant.
\begin{thm}
\label{thm:columnar_order}There is $C>0$ such that for $\lambda>\lambda_{0}$,
every $(x,y)\in\V$ with even $x$ satisfies
\[
\mu_{(\ver,0)}(\sigma(x,y)=1)\le C\lambda^{-1}.
\]
\end{thm}
\begin{proof}
Fix $\lambda$ sufficiently large for the following computations and
fix $(x_{T},y_{T})\in\V$ with even $x_{T}$. We denote $\mu=\mu_{(\ver,0)}$
and aim to show for some universal $C>0$ that $\mu(\sigma(x_{T},y_{T})=1)\le C\lambda^{-1}$.
For $\sigma\in\Omega$ define
\begin{align*}
X_{-\downarrow}(\sigma) & =\max\{x\in\Z:\text{\ensuremath{x<x_{T}} and \ensuremath{\rect{1\times1}{(x-1,y_{T}-1)}} is a vacant face in \ensuremath{\sigma}}\},\\
X_{-\uparrow}(\sigma) & =\max\{x\in\Z:\text{\ensuremath{x<x_{T}} and \ensuremath{\rect{1\times1}{(x-1,y_{T})}} is a vacant face in \ensuremath{\sigma}}\},\\
X_{+\downarrow}(\sigma) & =\min\{x\in\Z:\text{\ensuremath{x>x_{T}} and \ensuremath{\rect{1\times1}{(x,y_{T}-1)}} is a vacant face in \ensuremath{\sigma}}\},\\
X_{+\uparrow}(\sigma) & =\min\{x\in\Z:\text{\ensuremath{x>x_{T}} and \ensuremath{\rect{1\times1}{(x,y_{T})}} is a vacant face in \ensuremath{\sigma}}\}.
\end{align*}
These variables are almost surely finite.

Fix some integers $x_{-\downarrow},x_{-\uparrow}<x_{T}$ and $x_{+\downarrow},x_{+\uparrow}>x_{T}$.
Define an event
\[
J\coloneqq\left\{ \sigma\in\Omega:\substack{\sigma(x_{T},y_{T})=1\text{ and }\\
(X_{-\downarrow}(\sigma),X_{-\uparrow}(\sigma),X_{+\downarrow}(\sigma),X_{+\uparrow}(\sigma))=(x_{-\downarrow},x_{-\uparrow},x_{+\downarrow},x_{+\uparrow})
}
\right\} ,
\]
and denote
\begin{align*}
x_{-} & \coloneqq\min\{x_{-\downarrow},x_{-\uparrow}\},\\
x_{+} & \coloneqq\max\{x_{+\downarrow},x_{+\uparrow}\}.
\end{align*}
We claim that for some universal constants $c,C>0$, it holds that
\begin{equation}
\mu(J)\le C\lambda^{-1}e^{-c(x_{+}-x_{-})}.\label{eq:four_vac_bound}
\end{equation}
The theorem follows by summing over the possible values of $x_{-\downarrow},x_{-\uparrow},x_{+\downarrow},x_{+\uparrow}$.
We now prove (\ref{eq:four_vac_bound}).

Consider the segment $s=[x_{-},x_{+}]\times\{y_{T}\}$. For $\sigma\in J$,
no $(\ver,0)$ stick intersects with $s$. This implies that whenever
$s$ divides $\rect{\a N\times\b N}{(\a x,\b y)}$, it holds that
$(x,y)\notin\Psi_{(\ver,0)}^{\a\times\b}$. Thus it holds that the
set
\[
A_{0}=\left\{ \left(x,\left\lfloor \frac{y_{T}}{\b}\right\rfloor \right):x\in\Z,x_{0}<\a x<\a x+N\a<x_{1}\right\} 
\]
is disjoint from $\Psi_{(\ver,0)}^{\a\times\b}(\sigma)$ for each
$\sigma\in J$. Note that $\#A_{0}\ge(x_{1}-x_{0})/\a-N-1$.

By Corollary \ref{cor:single_vacancy_Z}, $\mu(J)\le\lambda^{-1}$.
Fix some $\eps_{1}>0$ satisfying $\eps_{1}<\eps_{0}$ and define
$\tilde{\Omega}$ to be some arbitrary event satisfying $J\subset\tilde{\Omega}$
and $\mu(\tilde{\Omega})=\lambda^{-1}\eps_{1}^{-4}$. Define a measure
$\tilde{\mu}$ to be $\mu$ conditioned on $\tilde{\Omega}$.

We define a random set 
\[
\Theta(\sigma)\coloneqq\begin{cases}
\emptyset & \sigma\in J\\
\V & \text{o/w}
\end{cases}.
\]
We claim that for sufficiently large $\lambda$, the following holds:
\begin{equation}
\pv_{\tilde{\mu}}(\Theta\cup\Psi_{(\ver,0)}^{\a\times\b})\le\eps_{1}.\label{eq:four_vac_tilde_perc}
\end{equation}
Given this claim, Lemma \ref{lem:two_point_peierls} implies that
\[
\tilde{\mu}(J)\le\tilde{\mu}(A_{0}\cap(\Theta\cup\Psi_{(\ver,0)}^{\a\times\b})=\emptyset)\le(\eps_{1}/\eps_{0})^{\#A_{0}},
\]
and (\ref{eq:four_vac_bound}) follows by taking into account the
measure of $\tilde{\Omega}$ and the size of $A_{0}$. Thus it remains
to prove (\ref{eq:four_vac_tilde_perc}).

We claim that in the current situation, item \ref{enu:partition_chessboard}
of Lemma \ref{lem:grid_change} holds (for every $k,l,K,L,R,E,r,H$
satisfying its conditions) when we replace the conclusion (\ref{eq:chess_pv})
with
\begin{equation}
\pv_{\tilde{\mu}}(\Theta\cup\sg{k\times l}{\es E)\le\sqrt[r]{\max\{\lambda^{-1/4},{\zedi[][E^{c}]}\}}}.\label{eq:conditional_cheass_pv}
\end{equation}
The proof is similar, except that $\eps:=\max\{\lambda^{-1/4},\zedi[][E^{c}]\}$
and that (\ref{eq:chessboard_subgrid}) is proved differently: we
let $\eta\in H$ and show that $\sg{K\times L}{\es{E^{c}}}$ is $\eps$-rare
for the measure $\eta\tilde{\mu}$. Indeed let $A\subset\V$ be a
non-empty finite set. Then one may check using the chessboard estimate
that $\eta\mu(J\cap\{A\subset\sg{K\times L}{\es{E^{c}}}\})\le\lambda^{-1}(\zedi)^{\#A-4}$.
Thus $\tilde{\mu}(A\subset\sg{K\times L}{\es{E^{c}}})\le\eps^{\#A}$.

Recall $G^{\a\times\b}$ from (\ref{eq:G_def}).

We show that for sufficiently large $\lambda$,
\begin{equation}
\pv_{\tilde{\mu}}(\Theta\cup(\Psi^{\a\times\b}\cap\sg{\a\times\b}{\es{G^{\a\times\b}}}))<\eps_{1}.\label{eq:tilde_basic_perc}
\end{equation}

Assume that $\lambda_{0}$ is such that $\lambda^{-1/4}$ is smaller
than the bound of (\ref{eq:strong_perc_eps_1}). Then we obtain $\pv_{\tilde{\mu}}(\Theta\cup\Psi^{\a\times\b})\le e^{-c_{\ref{lem:chotomies}}\a}$,
in the same way that (\ref{eq:dev_pv}) is obtained in the proof of
Lemma \ref{lem:chotomies}, except that instead of using Lemma \ref{lem:grid_change},
we use (\ref{eq:conditional_cheass_pv}).

Similarly, we obtain $\pv_{\tilde{\mu}}(\Theta\cup\sg{\a\times\b}{\es{G^{\a\times\b}}})\le e^{-c_{\ref{claim:same_parity_perco}}\a}$
as in Claim \ref{claim:same_parity_perco} except that again in the
proof, instead of Lemma \ref{lem:grid_change}, we use (\ref{eq:conditional_cheass_pv}).

Combining the bounds of the last two paragraphs, Lemma \ref{lem:perc_intersection}
gives (\ref{eq:tilde_basic_perc}) when taking $\a$ to be large enough.
By (\ref{eq:tilde_basic_perc}), there is an $\eps_{1}$-rare set
$B$ (with respect to $\tilde{\mu}$) for which $I$, the unique infinite
$\square$-component of $\V\setminus B$, is contained in $\Theta\cup(\Psi^{\a\times\b}\cap\sg{\a\times\b}{\es{G^{\a\times\b}}})$.
Thus $I$ is $\eps_{1}$-strongly-percolating and almost surely connected
with respect to $\tilde{\mu}$.

On the event $J$ (up to measure $0$) the following holds: $I$ is
connected and $I\subset\Psi^{\a\times\b}\cap\sg{\a\times\b}{\es{G^{\a\times\b}}}$.
Both $I$ and $\Psi_{(\ver,0)}^{\a\times\b}$ satisfy that the $\boxtimes$-components
of their complements are almost surely finite. Thus $\Psi_{(\ver,0)}^{\a\times\b}\cap I\neq\emptyset$,
and since $I\subset\Psi^{\a\times\b}$, Lemma \ref{lem:stick_collision}
implies that $I\subset\Psi_{\ver}^{\a\times\b}$. The fact that the
conditions of Proposition \ref{prop:peierls_two} are satisfied for
(\ref{eq:propagation_2}) implies that there is no edge $uv\in\E_{\square}$
with $u\in\Psi_{(\ver,0)}^{\a\times\b}\cap\sg{\a\times\b}{\es{G^{\a\times\b}}}$,
and $v\in\Psi_{(\ver,1)}^{\a\times\b}\cap\sg{\a\times\b}{\es{G^{\a\times\b}}}$.
Thus $I\subset\Psi_{(\ver,0)}^{\a\times\b}$.

Therefore with respect to $\tilde{\mu}$ the set $I$ is $\eps_{1}$-strongly-percolating
and almost surely contained in $\Theta\cup\Psi_{(\ver,0)}^{\a\times\b}$.
This completes the proof of (\ref{eq:four_vac_tilde_perc}).
\end{proof}

\subsection{Correlations\label{subsec:correlations}}

In this subsection we prove item \ref{item:Decay-of-correlations}
of Theorem \ref{thm:main}.
\begin{lem}
\label{lem:decay2}There exists a universal $C>0$ such that the following
holds for $\lambda>\lambda_{0}$. Let $\sigma,\sigma'$ be independently
sampled from $\mu_{(\ver,0)}$. Then for each $A\subset\Z^{2}$ and
$B\subset\Z^{2}$,
\[
\mathbb{P}(\text{a \ensuremath{\boxtimes}-path in \ensuremath{\Delta_{\sigma,\sigma'}} intersects \ensuremath{A} and \ensuremath{B}})\le\sum_{u\in A}\sup_{v\in B}\alpha_{2}(u,v)
\]
where for $u=(x_{1},y_{1})$ and $v=(x_{2},y_{2})\in\Z^{2}$,
\[
\alpha_{2}(u,v):=\left(\frac{C\log\lambda}{\sqrt{\lambda}}\right)^{\mathbf{1}_{x_{1}\neq x_{2}}}.
\]
\end{lem}
\begin{proof}
Let $\lambda>\lambda_{0}$. Let $\sigma,\sigma'$ be independently
sampled from $\mu_{(\ver,0)}$. Fix $u=(x_{1},y_{1})$. Let $d\in\nat$.
Let $E$ be the event a $\boxtimes$-path in $\Delta_{\sigma,\sigma'}$,
connects $u$ to a point outside of $\{(x_{1},y)\in\V:|y-y_{1}|<d\}$.
It suffices to show that 
\begin{equation}
\mathbb{P}(E)\le\frac{C\log\lambda}{\sqrt{\lambda}}.\label{eq:decay_2_main}
\end{equation}
If $x_{1}$ is even, then by Theorem \ref{thm:columnar_order} there
is probability of at most $2C_{\ref{thm:columnar_order}}\lambda^{-1}$
that $u\in\Delta_{\sigma,\sigma'}$ thus (\ref{eq:decay_2_main})
follows immediately. Assume that $x_{1}$ is odd. Let $E_{1}$ be
the event that $\Delta_{\sigma,\sigma'}$ intersects $\{(x,y)\in\V:|y-y_{1}|\le d,|x-x_{1}|=1\}$.
Then by Theorem \ref{thm:columnar_order}, $\mathbb{P}(E_{1})\le4(2d+1)C_{\ref{thm:columnar_order}}\lambda^{-1}$.
Let $E_{2}$ be the event that $u$ is connected by a $\boxtimes$-path
to a point in $B=\{(x_{1},y)\in\V:|y-y_{1}|=d\}$. By Lemma \ref{lem:decay_1}
(taking $A=\{u\}$ and $B$ as above), $\mathbb{P}(E_{2})\le C_{\ref{lem:decay_1}}\exp\left(-c_{\ref{lem:decay_1}}\frac{d}{\sqrt{\lambda}}\right)$.
As $E\subset E_{1}\cup E_{2}$, for $d=\left\lfloor \frac{\sqrt{\lambda}\log\lambda}{c_{\ref{lem:decay_1}}}\right\rfloor $
and sufficiently large $\lambda_{0}$, (\ref{eq:decay_2_main}) is
satisfied.
\end{proof}
\begin{cor}
\label{cor:decay_final}Let $\lambda>\lambda_{0}$. Then $\mu_{(\ver,0)}$
satisfies item \ref{item:Decay-of-correlations} of Theorem \ref{thm:main}.
\end{cor}
\begin{proof}
Let $\lambda>\lambda_{0}$. Let $\sigma,\sigma'$ be independently
sampled from $\mu_{(\ver,0)}$. Let $u\in\V$ and $B$ be a finite
subset of $\V$. Define $B_{1}=\{v\in B:\alpha_{1}(u,v)<\alpha_{2}(u,v)\}$
and $B_{2}=B\setminus B_{1}$. We apply Lemma \ref{lem:decay_1} to
$\{u\}$ and $B_{1}$, and Lemma \ref{lem:decay2} to $\{u\}$ and
$B_{2}$. By a union bound, this shows that
\begin{align*}
\mathbb{P}(\text{a \ensuremath{\boxtimes}-path in \ensuremath{\Delta_{\sigma,\sigma'}} intersects \ensuremath{\{u\}} and \ensuremath{B}}) & \le\sup_{v\in B_{1}}(\alpha_{1}(u,v))+\sup_{v\in B_{2}}(\alpha_{2}(u,v))\\
 & \le2\sup_{v\in B}(\min\{\alpha_{1}(u,v),\alpha_{2}(u,v)\})\\
 & \le\frac{1}{2}\sup_{v\in B}(\alpha(u,v))
\end{align*}

where for the last inequality we require $C_{\ref{thm:main}}\ge4\max\{C_{\ref{lem:decay_1}},C_{\ref{lem:decay2}}\}$
and $c_{\ref{thm:main}}\le c_{\ref{lem:decay_1}}$. We then use a
union bound over $u\in A$ to show that the RHS of (\ref{eq:covariance bound from disagreements})
is at most $\sum_{v\in A}\sup_{v\in B}(\alpha(u,v))$.

We finish the proof using Theorem \ref{thm:disagreement percolation},
recalling that (\ref{eq:no infinite disagreement path}) holds by
Lemma \ref{lem:decay_1}.
\end{proof}
\begin{rem}
\label{Rem:sharpness of second term in correlation decay}The correlation
decay estimate (\ref{eq:decay of correlations}) of Theorem (\ref{thm:main})
shows, in particular, that $\cov_{\mu_{(\ver,0)}}(\sigma(1,0),\sigma(3,0))\le\tfrac{C\log\lambda}{\sqrt{\lambda}}$.
It is natural to ask how sharp is this bound. We believe that $\cov_{\mu_{(\ver,0)}}(\sigma(1,0),\sigma(3,0))$
is of the order $\lambda^{-1/2}$ so that our bound has the correct
power of $\lambda$ but adds an unnecessary logarithmic term. Indeed,
van den Berg--Steif \cite[Theorem 2.4]{VANDENBERG1994179} give a
precise formula in terms of disagreement paths for such covariances
and we believe that in our setup the terms in this formula are dominated
by the disagreement paths that start at $(1,0)$, go vertically to
distance of order $\sqrt{\lambda},$ move horizontally to the column
of $(3,0)$ and then move vertically to $(3,0)$. Such disagreement
paths should occur with probability of order $\lambda^{-1/2}$.
\end{rem}

\subsection{Probability for a tile at a given position\label{subsec:tile_probabilities}}

In this subsection we give for each point in $\Z^{2}$ an estimate
for the probability (with respect to $\mu_{(\ver,0)}$) that a tile
is centered at it.
\begin{thm}
Let $\lambda>\lambda_{0}$. Then $\mu_{(\ver,0)}$ satisfies item
\ref{item:Columnar-order} of Theorem \ref{thm:main}.
\end{thm}
\begin{proof}
Fix $(x,y)$ where $x$ is odd. Denote:
\begin{align*}
E_{1} & =\{\sigma(x,y)+\sigma(x,y+1)=1\}\\
E_{2} & =\{\sigma(x-1,y)+\sigma(x-1,y+1)+\sigma(x+1,y)+\sigma(x+1,y+1)\ge1\}\\
E_{3} & =\{\text{the faces \ensuremath{\rect{1\times1}{(x-1,y)}} and \ensuremath{\rect{1\times1}{(x,y)}} are vacant}\}
\end{align*}
and note that $\{E_{1},E_{2},E_{3}\}$ is a partition of $\Omega$.
By Theorem \ref{thm:columnar_order}, $\mu_{\ver,0}(E_{2})\le4C_{\ref{thm:columnar_order}}\lambda^{-1}$.
By Corollary \ref{cor:single_vacancy_Z}, $\mu_{\ver,0}(E_{3})\le\lambda^{-1/2}$,
and $\mu_{\ver,0}(E_{1})\le\left(1-c\lambda^{-1/2}\right)^{2}$. Thus
$\mu_{\ver,0}(E_{1})=1-\Theta(\lambda^{-1/2})$. Since $\mu_{\ver,0}$
is $2\Z\times\Z$ translation invariant, $\mu_{\ver,0}(E_{1})=2\mu_{\ver,0}(\sigma(x,y)=1)$
and the theorem follows for the case of odd $x$.

Now fix $(x,y)$ where $x$ is even. By Theorem \ref{thm:columnar_order},
it remains to show that $\mu_{\ver,0}(\sigma(x,y)=1)=\Omega(\lambda^{-1})$
(note that here the symbol $\Omega$ represents the asymptotic notation
rather than the set of configurations). Now denote 
\begin{align*}
E & =\{\sigma(x,y)=1\}\\
E_{1} & =\{\sigma(x-1,y)=1\}\\
E_{2} & =\{\sigma(x+1,y)=1\}
\end{align*}
By the previous case, it holds that $\mu_{\ver,0}(E_{1})=\mu_{\ver,0}(E_{2})=\Omega(1)$.
By Corollary \ref{cor:decay_final}, $\mathrm{Cov}(E_{1},E_{2})=o(1)$.
Thus $\mu_{\ver,0}(E_{1}\cap E_{2})=\Omega(1)$. By a local surgery
(remove one of the two tiles and slide the other), it follows that
$\mu_{\ver,0}(E)\ge\lambda^{-1}\mu_{\ver,0}(E_{1}\cap E_{2})=\Omega(\lambda^{-1})$.\pagebreak{}
\end{proof}

\part{Concluding remarks\label{part:Discussion}}

\section{Discussion and open questions}

In this section we discuss some of the predictions, open questions
and research directions related to this work.

\subsection{The $2\times2$ hard squares model\label{subsec:remarks on 2x2 hard squares}}

\textbf{Intermediate fugacity and critical behavior:} This work establishes
that the $2\times2$ hard-square model exhibits columnar order in
the high-fugacity regime. As discussed in the introduction, classical
results imply that the model is disordered with a unique Gibbs measure
in the low-fugacity regime. What happens at intermediate fugacities?
The physics literature predicts a single transition point from the
disordered to the columnar phase, with the transition being continuous
and belonging to the Ashkin--Teller universality class \cite{ramolaColumnarOrderAshkinTeller2015}
(at a point close to the Ising universality class \cite[Figure 5]{ramola2012thesis}).
These predictions have not been mathematically justified. 

\textbf{Boundary conditions: }Our work characterizes the periodic
Gibbs measures of the $2\times2$ hard-square model. However, we do
not prove that any specific sequence of finite volumes and boundary
conditions converge in the infinite-volume limit.

A related question is whether non-periodic Gibbs measures exist for
the model. We expect the answer is negative, as in other two-dimensional
models \cite{russo1979infinite,higuchi1979absence,aizenman1980translation,dobrushin1985problem,georgii2000percolation,coquille2014gibbs,glazman2021structure}.

\textbf{Decay of correlations: }Theorem \ref{thm:main} gives an upper
bound on the exponential rate of correlation decay in $\mu_{(\ver,0)}$
which is anisotropic. Specifically, the correlation length in the
horizontal direction is at most a universal constant while in the
vertical direction it is at most $C\sqrt{\lambda}$. On a mesoscopic
scale (for distances $1\ll d\le\sqrt{\lambda})$ it is clear from
our results that correlations are indeed anisotropic. However, we
do not establish lower bounds for the correlations as the distances
grow without bound (i.e., when $\lambda$ is fixed and $d\to\infty$).
A natural question is whether in the limit of large distances, the
exponential rate of correlation decay is indeed highly anisotropic
as our bound suggests. 

The same question may be asked for models where a proof of nematic
order was given, such as those listed in subsubsection \ref{subsec:disscusion_LC}.
We mention in particular the result of Jauslin--Lieb, where the proven
correlation bounds \cite[equation (19)]{jauslin2018nematic} are very
similar in form to those of the present work.

\subsection{\label{subsec:Cubes-and-rods}Cubes and rods on $\protect\Z^{d}$}

We briefly discuss related models in which the $2\times2$ hard squares
are replaced by cubes and rods on $\Z^{d}$.

\textbf{Cubes:} For $k\times k\times\cdots\times k$ cubes on $\Z^{d}$
we expect the high-fugacity regime to behave similarly to our results
for the $2\times2$ hard-square model. In particular, we conjecture
that there are exactly $dk^{d-1}$ extremal and periodic Gibbs measures
(where $d$ accounts for the possible orientations of columns and
$k^{d-1}$ accounts for translations perpendicular to the columns).
Elements of our approach may well be relevant to proving such a result,
at least when $d=2$. However, even in two dimensions our analysis
does not apply as is due to the absence of reflection positivity when
$k\ge3$. In higher dimensions, the case $k=2$ may be more accessible
as reflection positivity is again available. 

Interestingly, a recent physics study \cite{vigneshwar2019phase}
(see also \cite{Vigneshwar2019thesis}) predicts that the $2\times2\times2$
hard-cube model undergoes \emph{three} phase transitions as the fugacity
increases. One of the predicted phases is a sublattice phase, at intermediate
fugacity, where cubes preferentially occupy one of the eight sublattices.

As a possible complication, we point out that tilings of cubes in
high dimensions present new phenomena. For instance, refuting a conjecture
of Keller \cite{Keller+1930+231+248}, it has been shown \cite{mackeyCubeTilingDimension2002,lagariasKellersCubetilingConjecture1992}
that for $d\ge8$ one may tile $\R^{d}$ with unit cubes in a way
that no two cubes share a complete $(d-1)$-dimensional face (while
this is not possible for $d\le7$ \cite{perronUeberLueckenloseAusfuellung1940,perronUeberLueckenloseAusfuellung1940a,brakensiekResolutionKellersConjecture2022}).
Moreover, these tilings may be chosen so that all cube centers lie
in $\frac{1}{2}\Z^{d}$ (equivalently, a tiling with this feature
is possible using $2\times2\times\cdots\times2$ cubes with centers
in $\Z^{d}$) and the tiling is $2\Z^{d}$-translation invariant.

We also note a connection between cube packings and a famous problem
in information theory. The Shannon capacity of a graph $G$ is defined
as $\lim_{d\to\infty}(\alpha(G^{\boxtimes d}))^{1/d}$ where $\alpha(H)$
is the size of the largest independent set in the graph $H$ and $G^{\boxtimes d}$
stands for the strong product of $G$ with itself $d$ times \cite{shannonZeroErrorCapacity1956}.
The Shannon capacity remains unknown even for fairly simple graphs.
In particular, the Shannon capacity of the cycle $C_{k}$ with $k$
odd has not been determined for $k\ge$7 (see, e.g., \cite{bohmanLimitTheoremShannon2005}).
The connection is that $\alpha(C_{k}^{d})$ equals the maximal number
of $2\times2\times\cdots\times2$ cubes with centers having integer
coordinates that may be packed in the torus $(\R/k\Z)^{d}$.

\textbf{Rods in two dimensions:} The random packing of $1\times k$
and $k\times1$ tiles on $\Z^{2}$ has been studied extensively in
the physics literature \cite{Ghosh2007_rods2transitions,fernandez2008_rods_tri,kundu2013_rods,Vogel2020_rods2D,Shah2021_lattice_rods}
(see also the literature reviews in the theses \cite{kundu2015thesis,Nath2016thesis}).
The following behavior is predicted: For $k\le6$, the model is disordered
for all fugacities. For $k\ge7$, the model exhibits two phase transitions.
At low fugacity the model is disordered (low-density disordered, LDD),
at intermediate fugacities the model has a nematic phase, and at high
fugacities the model is again disordered (high-density disordered,~HDD).

The case $k=2$ is the monomer--dimer model. As discussed in subsubsection
\ref{subsec:monomer_dimer}, it was shown to have a unique Gibbs measure
for all fugacities. As mentioned in subsubsection \ref{subsec:disscusion_LC},
Disertori--Giuliani \cite{disertori2013nematic} rigorously established
the nematic phase at an \emph{intermediate} range of fugacities for
large values of $k$.

The properties of the HDD phase are unclear, though simulations clearly
demonstrate that horizontal and vertical rods appear with equal density
(unlike in the nematic phase). It would be interesting to improve
our understanding of this phase and a starting point may be the study
of the fully-packed regime (the limit $\lambda=\infty$). Is there
a unique maximal-entropy Gibbs measure in this case? Estimates of
the entropy-per-site are provided by \cite{gagunashvili_rods1979,DharRajesh2021_lattice_rods}.

\textbf{Rods in three dimensions: }The predicted phase diagram for
$1\times1\times k$ rods (and their lattice rotations) on $\Z^{3}$
is less complete; see \cite{vigneshwar2017different,Shah2021_lattice_rods,DharRajesh2021_lattice_rods}
for recent results.

\subsection{A simplified lattice model with nematic order}

In developing the technique of this paper the following simplified
spin model proved handy for pointing out the essential features. We
describe this model for its intrinsic interest and with the hope that
it may lend similar help to the study of some of the models described
above.

\textbf{Oriented monomer model:} Configurations are functions $\sigma:\Z^{d}\to\{0,e_{1},\dots,e_{d}\}$
with $e_{i}$ being the $i$th vector of the standard basis of $\R^{d}$.
The state $0$ represents a vacancy while each state $e_{i}$ may
be thought of as a ``monomer oriented in the $i$th coordinate direction''.
Oriented monomers of equal orientations which are adjacent in the
direction of their orientation are thought to join together to form
rods (an oriented monomer which is not joined in this way is thought
of as a rod of length 1). A configuration may thus be imagined as
a packing of rods. We wish to study the model in which the probability
of a configuration is proportional to $t^{2N(\sigma)}$ with $N(\sigma)$
representing the number of rods (i.e., the fugacity $\lambda=t^{2}$).
Equivalently, the weight of a configuration assigns weight $t$ to
each end of a rod. An essential simplification available for this
model is that the probability measure may be represented by nearest-neighbor
interactions, via the Hamiltonian:
\[
H(\sigma)=\sum_{uv\in\E_{\square}}\left|<u-v,\sigma(u)-\sigma(v)>\right|
\]

with $<\cdot,\cdot>$ denoting the standard inner product and $\E_{\square}$
denoting the edge set of $\Z^{d}$. It is straightforward that $H(\sigma)=2N(\sigma)$.
We then define the probability of a configuration $\sigma$ to be
proportional to $e^{-\beta H(\sigma)}=t^{H(\sigma)}$ with $t=e^{-\beta}$.
This way of writing the model shows that it is reflection positive
(for reflections through planes of vertices) in all dimensions.

We are interested in the low-fugacity / low-temperature regime of
the model. There, since rod ends are disfavored, the rods that appear
tend to be long and orientational symmetry breaking may occur. Indeed,
we expect this regime to exhibit a nematic phase, with exactly $d$
extremal and periodic Gibbs measures, with each measure characterized
by an orientation in which monomers appear abundantly while monomers
of other orientations are rare. Moreover, in the $i$th such measure
(where $e_{j}$ for $j\neq i$ are rare) we expect a typical configuration
to resemble a perturbation of a union of one-dimensional systems in
the $i$th direction in which only the states $0$ and $e_{i}$ are
allowed. In particular, the density of $e_{i}$ should be approximately
$1/2$ since such one-dimensional systems are invariant to swapping
the two allowed states ($0$ and $e_{i}$). We are able to prove these
properties when $d=2$ using the techniques of the current work (thinking
of the sticks of the $2\times2$ hard-square model as the rods of
the oriented monomer model). 

We point out connections between the oriented monomer model and the
existing literature. First, if one removes the possibility of vacancies
then, in two dimensions, one recovers the ``exactly-solvable'' version
of the model of Ioffe--Velenik--Zahradník \cite{ioffe2006entropy}
which has an exact mapping to the Ising model. Second, there is also
similarity with the two-dimensional interacting dimer model studied
by Heilmann--Lieb \cite[Model I]{heilmann1979lattice} and Jauslin--Lieb
\cite{jauslin2018nematic}: The models become identical when a specific
relation between the dimer activity and dimer interaction energy is
imposed and, further, the interacting dimers are replaced by interacting
oriented monomers. Lastly, the oriented monomer model with a fixed
number of vacancies is equivalent to a model studied in \cite{lopezPhaseDiagramSelfassembled2010}.

As mentioned, we hope that the oriented monomer model may be handy
in understanding orientational order in higher dimensions as we believe
that it represents some of the essential difficulties in those problems,
while cutting down on some technicalities. Specifically, it resembles
the lattice rod models of subsection \ref{subsec:Cubes-and-rods}
but has the advantage of having reflection positivity and the further
advantage that the nematic phase is expected at a perturbative regime
(low fugacity) rather than at intermediate fugacity. In addition,
it may be of help in analyzing the lattice hard cubes packing model.
As mentioned, this was indeed the case for us when studying the hard-square
model.

\subsection{Packing Euclidean disks on the lattice with the sliding phenomenon\label{subsec:remarks on sliding}}

Continuing a discussion from subsubsection \ref{subsec:background_lattice_balls},
we consider hard-core models of Euclidean disks of fixed diameter
$D$ with centers restricted to lie on a planar lattice. There is
a finite list of diameters for which the maximal-density packings
exhibit a ``sliding instability''. For these diameters, it is not
known whether there are multiple Gibbs measures at high fugacity (except
for the one case resolved by the current work) and our goal in this
section is to speculate on this question.

\subsubsection{Basic definitions}

The \textbf{base lattice} $\mathbb{W}$ is either the square ($\Z^{2}$),
the triangular ($\mathbb{A}_{2}$) or the hexagonal/honeycomb ($\mathbb{H}_{2}$)
lattice, normalized so that nearest-neighbor points are at (Euclidean)
distance $1$. Configurations are packings of Euclidean disks of diameter
$D$ with disjoint interiors and centers on the base lattice. We restrict
to values of $D$ which are attainable as the Euclidean distance between
points in the base lattice (for $\Z^{2}$, all values of the form
$D^{2}=a^{2}+b^{2}$ and for $\mathbb{A}_{2}$ and $\mathbb{H}_{2}$
all values of the form $D^{2}=a^{2}+b^{2}+ab$). The model is known
in the physics literature as the $k$-nearest-neighbor ($k$-NN) hard-core
lattice gas, with $k$ being the number of distinct positive lattice
distances smaller than $D$ (see Table \ref{tab:square lattice correspondence}
and Table \ref{tab:hexagonal lattice correspondence}).

\begin{table}[t]
$\mathbb{Z}^{2}:$$\quad$%
\begin{tabular}{|c|c|c|c|c|c|c|c|c|c|c|c|c|c|c|c|c|c|c|c|c|c|c|}
\hline 
{\tiny k} & {\tiny 1} & {\tiny 2\cellcolor{yellow}} & {\tiny 3} & {\tiny 4\cellcolor{yellow}} & {\tiny 5\cellcolor{yellow}} & {\tiny 6} & {\tiny 7} & {\tiny 8} & {\tiny 9} & {\tiny 10\cellcolor{yellow}} & {\tiny 11\cellcolor{yellow}} & {\tiny 12} & {\tiny 13} & {\tiny 14\cellcolor{yellow}} & $\ldots$ & {\tiny 21\cellcolor{yellow}} & $\ldots$ & {\tiny 31\cellcolor{yellow}} & $\ldots$ & {\tiny 34\cellcolor{yellow}} & $\ldots$ & {\tiny 39\cellcolor{yellow}}\tabularnewline
\hline 
{\tiny$D^{2}$} & {\tiny 2} & {\tiny 4\cellcolor{yellow}} & {\tiny 5} & {\tiny 8\cellcolor{yellow}} & {\tiny 9\cellcolor{yellow}} & {\tiny 10} & {\tiny 13} & {\tiny 16} & {\tiny 17} & {\tiny 18\cellcolor{yellow}} & {\tiny 20\cellcolor{yellow}} & {\tiny 25} & {\tiny 26} & {\tiny 29\cellcolor{yellow}} & $\ldots$ & {\tiny 45\cellcolor{yellow}} & $\ldots$ & {\tiny 72\cellcolor{yellow}} & $\ldots$ & {\tiny 80\cellcolor{yellow}} & $\ldots$ & {\tiny 90\cellcolor{yellow}}\tabularnewline
\hline 
\end{tabular}

\caption{\label{tab:square lattice correspondence}The correspondence between
the $k$-NN and $D^{2}$ notation for $\protect\Z^{2}$. The first
few cases of sliding are highlighted.}
\end{table}
\begin{table}[t]
$\mathbb{H}_{2}:$$\quad$%
\begin{tabular}{|c|c|c|c|c|c|c|c|c|c|c|c|c|c|c|c|c|}
\hline 
{\tiny k} & {\tiny 1} & {\tiny 2\cellcolor{yellow}} & {\tiny 3\cellcolor{yellow}} & {\tiny 4} & {\tiny 5} & {\tiny 6} & {\tiny 7} & {\tiny 8} & {\tiny 9} & {\tiny 10} & {\tiny 11} & {\tiny 12} & {\tiny 13\cellcolor{yellow}} & {\tiny 14} & $\ldots$ & {\tiny 46\cellcolor{yellow}}\tabularnewline
\hline 
{\tiny$D^{2}$} & {\tiny 3} & {\tiny 4\cellcolor{yellow}} & {\tiny 7\cellcolor{yellow}} & {\tiny 9} & {\tiny 12} & {\tiny 13} & {\tiny 16} & {\tiny 19} & {\tiny 21} & {\tiny 25} & {\tiny 27} & {\tiny 28} & {\tiny 31\cellcolor{yellow}} & {\tiny 36} & $\ldots$ & {\tiny 133\cellcolor{yellow}}\tabularnewline
\hline 
\end{tabular}

\caption{\label{tab:hexagonal lattice correspondence}The correspondence between
the $k$-NN and $D^{2}$ notation for $\mathbb{A}_{2}$ and $\mathbb{H}_{2}$.
All the cases of sliding in $\mathbb{H}_{2}$ are highlighted.}
\end{table}

A maximal-density \emph{periodic} packing is called a \textbf{periodic
ground state} (PGS). Following \cite{MSS2020_short,MSS2018_trihex,MSS2019_square},
the \textbf{sliding phenomenon} is defined to occur for the pair $(\mathbb{W},D)$
if and only if there are infinitely many PGSs (this definition suffices
for our planar setting; see \cite{mazel2021kepler} for a study of
$\Z^{3}$).

\subsubsection{\label{subsec:PGS_is_perfect}Geometric characterization of periodic
ground states}

We now introduce key notions from the approach used by Mazel--Stuhl--Suhov
(MSS) \cite{MSS2020_short,MSS2018_trihex,MSS2019_square} to give
a geometric description of the set of PGSs. This approach allows to
determine the cases where sliding occurs, and provides further information
regarding the ground states of these cases, which should be important
in understanding the high-fugacity behavior (cf. the discussion of
fully-packed configurations in subsection \ref{subsec:discuss_results}).
The notions described here are used in the following subsubsections
to comment on the sliding cases. For simplicity, we restrict to the
case $\mathbb{W}=\Z^{2}$; the treatment of the other cases follows
similar ideas, and we indicate some of the differences in subsubsection
\ref{subsec:sliding_hex}.

MSS make the following definitions: A $\Z^{2}$-triangle is a triangle
with vertices on $\Z^{2}$. A $\Z^{2}$-triangle with side lengths
$\ge D$ and angles $\le90^{\circ}$ is called an \textbf{M-triangle}
if it has minimal area among such $\Z^{2}$-triangles. Given a configuration,
the triangles forming the Delauney triangulation of the set of disk
centers are called \textbf{C-triangles} (of the configuration). A
configuration is said to be \textbf{perfect} if all of its C-triangles
are M-triangles.
\begin{thm}
[\cite{MSS2019_square}]\label{thm:PGS_is_perfect}A periodic configuration
is a PGS iff it is perfect.
\end{thm}
The theorem follows from the following claims: (i) Every triangulation
by M-triangles is the Delauney triangulation of a configuration with
density $1/S(D)$, with $S(D)$ denoting twice the area of an M-triangle.
(ii) A perfect configuration exists, since an M-triangle may be extended
to a triangulation consisting of translations and $180^{\circ}$ rotations
of it. (iii) All configurations have density at most $1/S(D)$ and
a non-perfect periodic configuration has density that is strictly
less than $1/S(D)$. The last claim is nontrivial and is established
in \cite[Lemmas 3.5, 3.6]{MSS2019_square}.

The theorem provides a handy tool to study PGSs for a given exclusion
diameter $D$. As a first step, one should understand the $M$-triangles
for that $D$ (a task which may be carried out by a computer search).
Then, one may study the ways in which these triangles may be assembled
together to form periodic triangulations.

A necessary condition for sliding to occur for a given exclusion diameter
$D$, is the existence of two distinct M-triangles with a common edge,
termed the \textbf{sliding base}, both having their third vertex on
the same side of the common edge. See Figure \ref{fig:sliding_base}
for the case $D^{2}=29$. 

\subsubsection{Sliding on $\protect\Z^{2}$}

The list of sliding cases on $\Z^{2}$ was confirmed by \cite{MSS2019_square,krachun2020_MSS_sliding_list}
(a partial list is also in \cite{nath_sliding2016}). The first 9
cases are highlighted on Table \ref{tab:square lattice correspondence}.
We remind that the first among these cases ($D^{2}=4$) is the subject
of the current work. At this point we refer the reader to \cite[subsection 2.2]{MSS2019_square}
for the list of sliding cases on $\Z^{2}$, a visualization of sliding
bases in some specific cases, and some discussion of the resulting
ground states.
\begin{figure}[H]
\hfill{}(i)~\includegraphics[height=0.15\textheight]{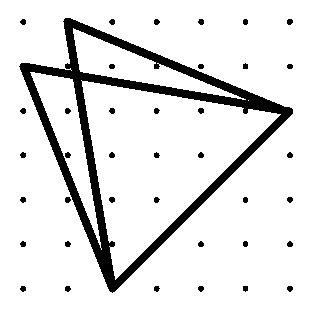}\hfill{}(ii)~\includegraphics[height=0.15\textheight]{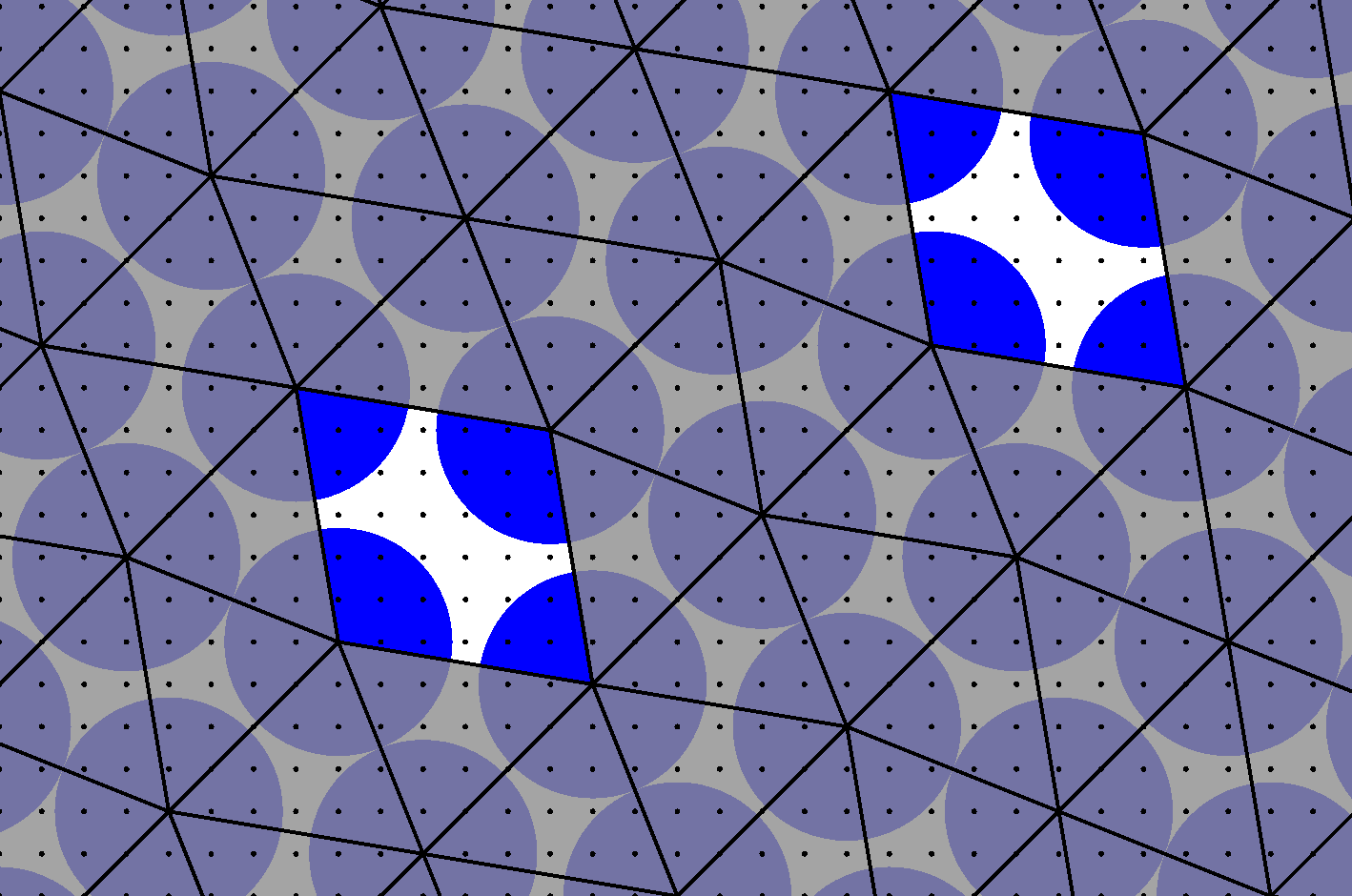}\hfill{}

\caption{\label{fig:sliding_base}(i) Two M-triangles for $D^{2}=29=5^{2}+2^{2}$,
sharing their sliding base.\protect \\
(ii) A configuration for $D^{2}=29$. The C-triangles which are M-triangles
are displayed in transparent grey color.}

\end{figure}

Monte Carlo simulations \cite{Nath2014_kNN} carried out for $D^{2}\le20$
indicate that the high-fugacity phase is columnar in all of the sliding
cases (interestingly, two phase transitions are predicted for $D^{2}=8,18,20$).
Here, we point to an extra feature present only in the sliding cases
with $D^{2}>20$ which leads us to believe that the lattice's $90^{\circ}$
rotational symmetry is broken in the high-fugacity phase (leading
to multiple Gibbs measures).

\begin{figure}[H]
\includegraphics[width=1\textwidth]{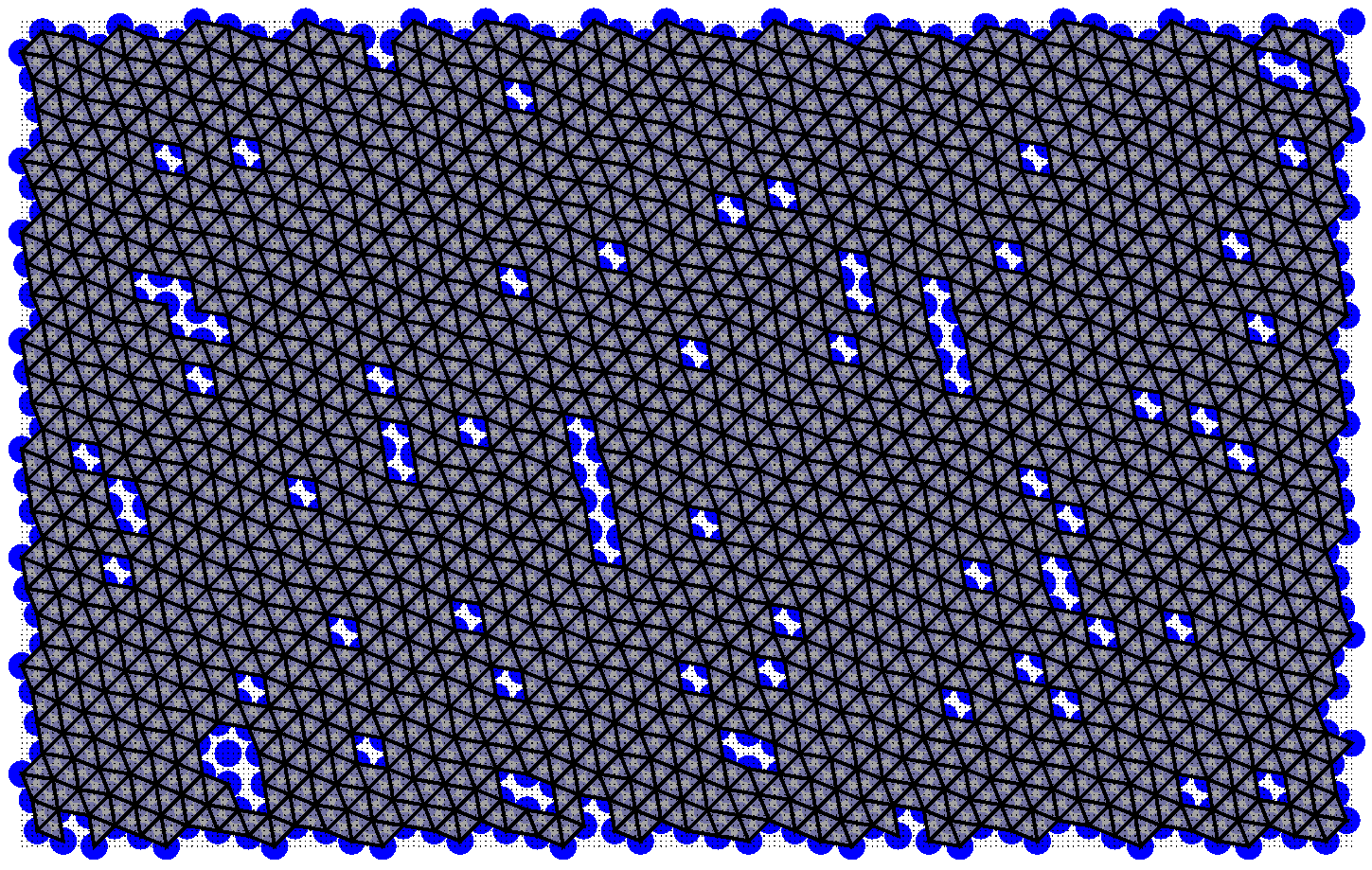}

\caption{\label{fig:sim_29}Simulation for $D^{2}=29$ with fugacity $\lambda=2000$
on a torus of dimensions $259\times161$. In line with the proposed
heuristic, all the M-triangles have their sliding base parallel to
the line $x=y$, and the disks are arranged in columns parallel to
this line.}
\end{figure}

In all sliding cases except for the first five cases (i.e., when $D^{2}>20$),
the following property holds: each M-triangle has at most one sliding
base. Consequently, two internally disjoint M-triangles that share
an edge either do not have sliding bases, or have their unique sliding
bases parallel to each other (See Figure \ref{fig:sliding_base}).
This leads to the following heuristic argument supporting the multiplicity
of Gibbs measures in high fugacity: Consider a configuration sampled
from a high-fugacity Gibbs measure. It is natural to believe that
most $C$-triangles are $M$-triangles, leading to the existence of
a unique infinite connected component of $M$-triangles. If this is
the case, by the property above, either all the M-triangles in the
unique infinite component have no sliding bases, or all of them have
their sliding bases oriented parallel to the same line. We believe
that the second possibility is entropically favored (see Figure \ref{fig:sim_29}).
The orientation of the line to which the sliding bases are parallel
is thus a (tail measurable and translation invariant) observable which
may be used to distinguish different Gibbs measures (e.g., a high-fugacity
ergodic Gibbs measure will be singular with respect to its $90^{\circ}$
rotation).

In the cases where the above heuristic applies, we conjecture typical
configurations to display the following order. All C-triangles but
a rare set, are M-triangles with a sliding base parallel to a shared
direction. Thus centers are arranged in columns parallel to the shared
direction. The columns are occasionally interrupted by gaps (analogous
to the double vacancies in the 1D systems of the $2\times2$ hard
square model). In contrast to the case studied in this work (where
correlations between neighboring columns are small), the columns interact
more strongly since shifting a column parallel to its direction does
not in general result in a valid configuration.

\subsubsection{\label{subsec:sliding_hex}Sliding on $\mathbb{\mathbb{H}}_{2}$}

On the triangular lattice $\mathbb{A}_{2}$, due to its symmetry,
there is always an equilateral lattice triangle with side length $D$.
By arguments similar to those described in subsubsection \ref{subsec:PGS_is_perfect},
it follows that for the disk models on $\mathbb{A}_{2}$ the sliding
phenomenon never occurs.

For the case of $\mathbb{\mathbb{H}}_{2}$, MSS analyze the PGSs using
an approach similar to the one described above for $\Z^{2}$, with
some modification. The notion of M-triangles is replaced by that of
\textbf{MRA-triangles,} defined in \cite[subsection 4.2]{MSS2018_trihex}
using a notion of ``redistributed area''. In the equivalent of Theorem
\ref{thm:PGS_is_perfect} for the case of $\mathbb{\mathbb{H}}_{2}$,
the direct implication that every PGS is perfect still holds; however
the reverse implication does not, since MRA-triangles do not necessarily
all have the same area. It also happens that there are finitely many
values of $D$ where there are PGSs that are not lattices in the sense
of an additive group, but still no sliding occurs.

On $\mathbb{\mathbb{H}}_{2}$, there are exactly four cases of sliding
\cite{MSS2018_trihex}: $D^{2}=4,7,31,133$. We refer the reader to
\cite[Section 8]{MSS2018_trihex} for a discussion and visualization
of the resulting perfect configurations.

For the cases $D^{2}=31,133$ we note that all MRA-triangles are of
the same area, implying that every perfect configuration has maximal
density. While the heuristic presented in the previous subsubsection
does not apply to these cases, as there exist (equilateral) MRA-triangles
for which all sides are sliding bases, the specific geometry of these
cases still leads us to conjecture that columnar order arises at high
fugacity, with three possible orientations.

\begin{figure}[t]
\centering{}\hspace*{\fill}(i)~\includegraphics[height=0.2\textwidth]{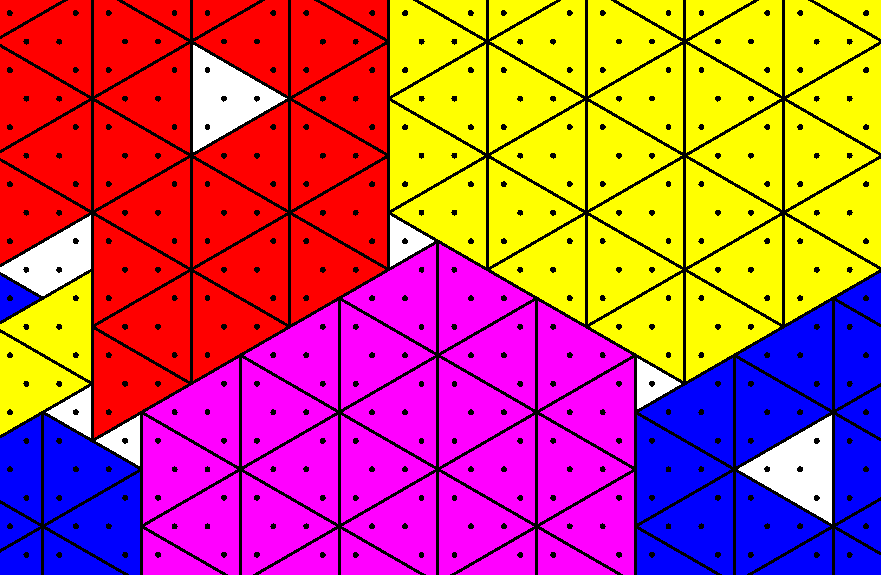}\hspace*{\fill}(ii)~\includegraphics[height=0.2\textwidth]{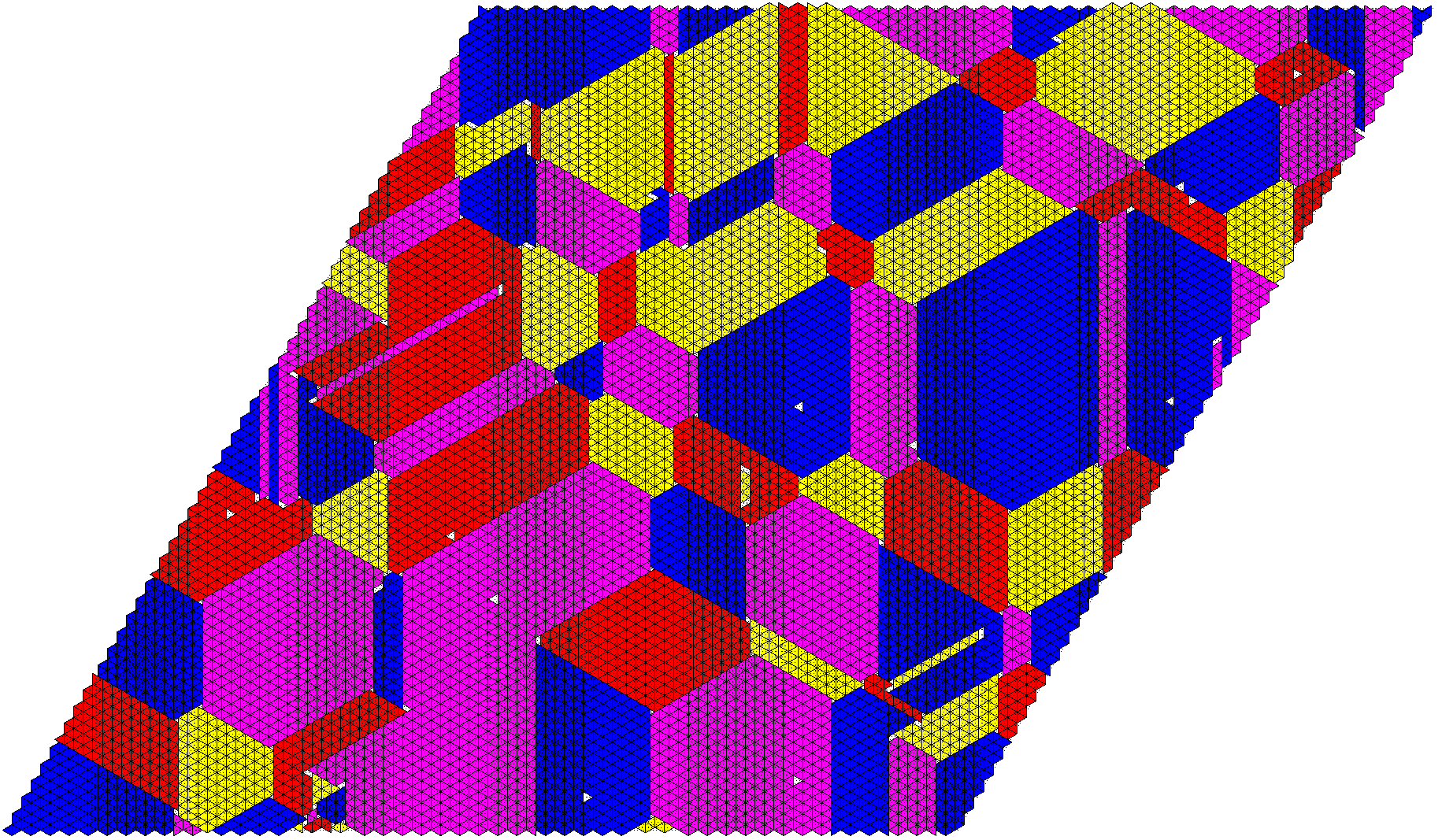}\hspace*{\fill}(iii)~\includegraphics[height=0.2\textwidth]{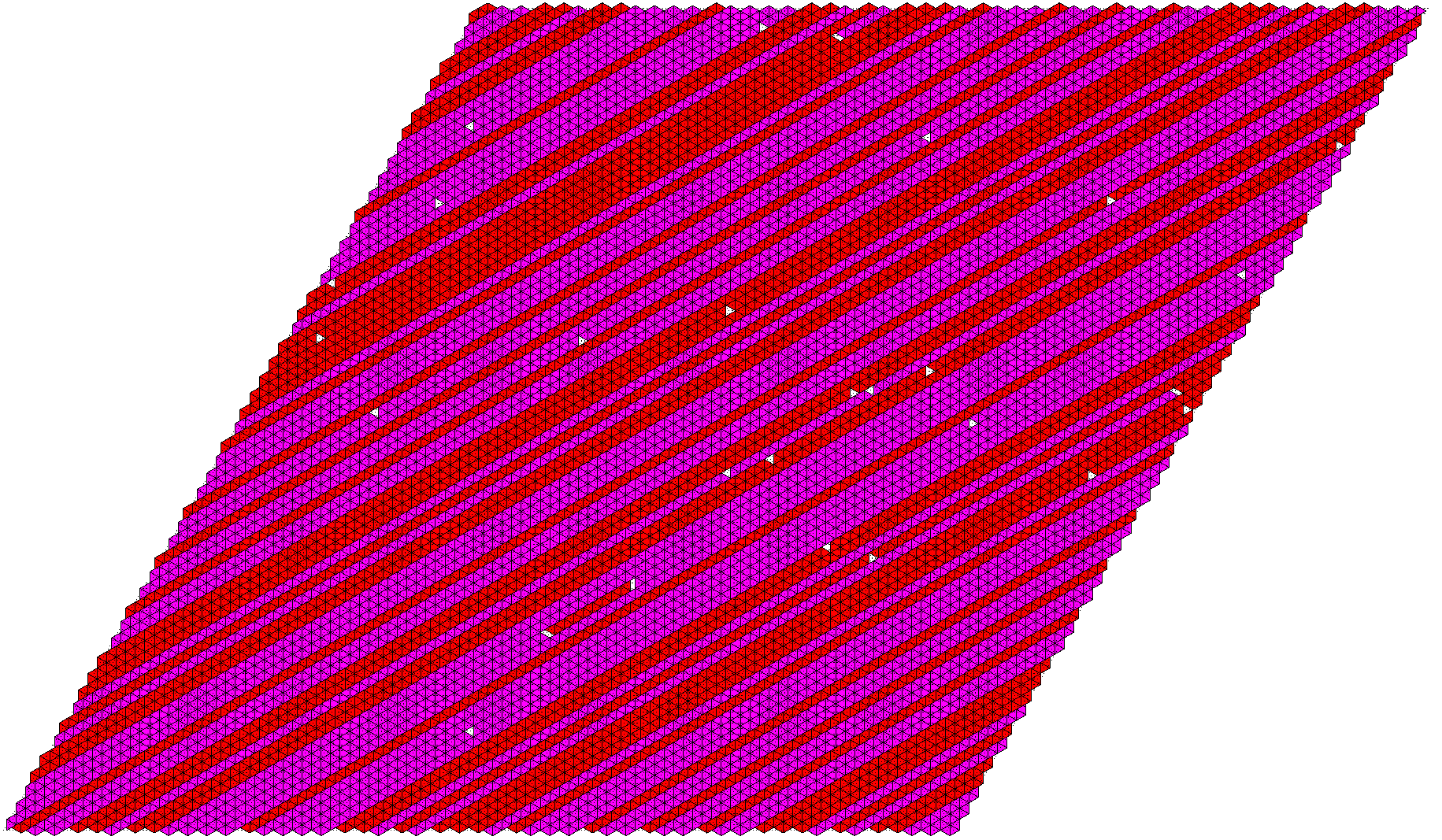}\hspace*{\fill}\caption{\label{fig:hex_D4}Images of the $D^{2}=4$ model on the hexagonal
lattice. The model is represented using triangular tiles, consisting
of four lattice points.\protect \\
(i) A closeup image of a configuration. The black dots are elements
of $\mathbb{H}_{2}$.\protect \\
(ii) and (iii) MCMC simulations with $\lambda=700$ (see description
in text).}
\end{figure}

For the case $D^{2}=4$, configurations may be equivalently represented
as a packing of equilateral triangles of side length $2$, with vertices
restricted to the $\mathbb{A}_{2}$ lattice (in this equivalence,
one rescales $\mathbb{\mathbb{H}}_{2}$ to be dual to $\mathbb{A}_{2}$).
This is illustrated in Figure \ref{fig:hex_D4}(i), where tiles are
painted in four colors corresponding to the parities of each tile's
center when expressed in the basis $1,e^{2\pi i/3}$. MCMC simulations
with local moves at fugacity $\lambda=700$ did not converge and led
to different results depending on the starting position. Figure \ref{fig:hex_D4}(ii)
depicts the result starting from an empty configuration, in which
the domains of uniform color vaguely resemble the faces of a randomly-deformed
hexagonal lattice. Figure \ref{fig:hex_D4}(iii) depicts the result
starting from a fully-packed configuration with tiles of a single
color, in which columnar order is exhibited. Thewes--Fernandes \cite[Section B]{Fernandes2020_kNN_hexa}
consider this model in the physical literature, predict a columnar
high-fugacity phase and further discuss the intermediate fugacity
regime.

\begin{figure}[t]
\centering{}\hspace*{\fill}(i)~\includegraphics[height=0.3\textwidth]{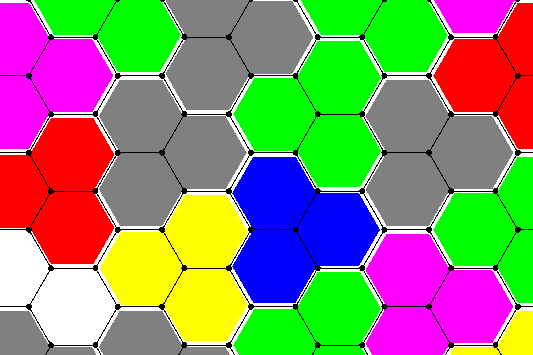}\hspace*{\fill}(ii)~\includegraphics[height=0.3\textwidth]{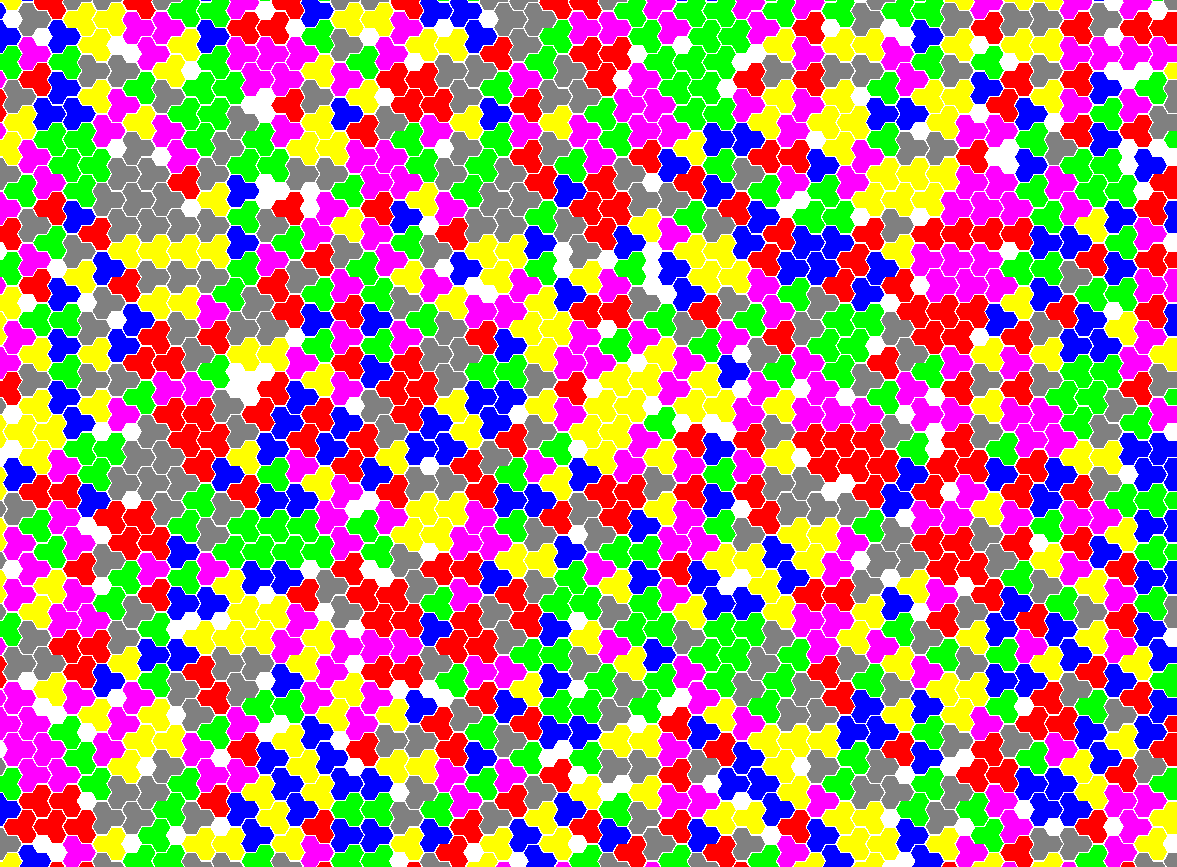}\hspace*{\fill}\caption{\label{fig:hex_D7}(i) A configuration for $D^{2}=7$, viewed as a
packing of \textquotedblleft trimers\textquotedblright{} with centers
on $\mathbb{\mathbb{H}}_{2}$.\protect \\
(ii) A portion of a result of an MCMC simulation with $\lambda=600$.}
\end{figure}

For the case $D^{2}=7$, configurations may equivalently be represented
as packings of ``trimers'', where each trimer is a union of three
pairwise neighboring faces of $\mathbb{H}_{2}$, see Figure \ref{fig:hex_D7}.
For the fully-packed version of this model an exact solution was found
by Verberkmoes--Nienhuis \cite{TriangularTrimers_short1999,TriangularTrimers_long2001}
(see also Propp \cite{propp2022trimer} for related enumeration problems).
The case $D^{2}=7$ is discussed by Thewes--Fernandes \cite[Section C]{Fernandes2020_kNN_hexa}
where, interestingly, it is predicted that the model is disordered
at all finite fugacities.

\bibliographystyle{plain}
\bibliography{biblio}

@article{turan1984succinct,
  title={On the succinct representation of graphs},
  author={Tur{\'a}n, Gy{\"o}rgy},
  journal={Discrete Applied Mathematics},
  volume={8},
  number={3},
  pages={289--294},
  year={1984},
  publisher={North-Holland}
}

@book{Georgii2011,
author = {Hans-Otto Georgii},
doi = {doi:10.1515/9783110250329},
url = {https://doi.org/10.1515/9783110250329},
title = {Gibbs Measures and Phase Transitions},
year = {2011},
publisher = {De Gruyter},
ISBN = {978-3-11-025032-9}
}

@book{friedli_velenik_2017,
place={Cambridge},
title={Statistical Mechanics of Lattice Systems: A Concrete Mathematical Introduction},
DOI={10.1017/9781316882603},
ISBN={978-1-107-18482-4},
publisher={Cambridge University Press},
author={Friedli, Sacha and Velenik, Yvan},
year={2017}
}

@InCollection{biskup2009reflection,
  author    = {Biskup, Marek},
  title     = {Reflection positivity and phase transitions in lattice spin models},
  booktitle = {Methods of contemporary mathematical statistical physics},
  publisher = {Springer},
  year      = {2009},
  pages     = {1--86},
}

@Article{cmp/1104252313,
  author    = {J. van den Berg},
  title     = {{A uniqueness condition for {G}ibbs measures, with application to the $2$-dimensional {I}sing antiferromagnet}},
  journal   = {Communications in Mathematical Physics},
  year      = {1993},
  volume    = {152},
  number    = {1},
  pages     = {161 -- 166},
  doi       = {cmp/1104252313},
  publisher = {Springer},
  url       = {https://doi.org/},
}

@Article{VANDENBERG1994179,
  author   = {J. {van den Berg} and J.E. Steif},
  title    = {Percolation and the hard-core lattice gas model},
  journal  = {Stochastic Processes and their Applications},
  year     = {1994},
  volume   = {49},
  number   = {2},
  pages    = {179--197},
  issn     = {0304-4149},
  doi      = {https://doi.org/10.1016/0304-4149(94)90132-5},
  url      = {https://www.sciencedirect.com/science/article/pii/0304414994901325},
}

@article{biskup2006forbidden,
  title={Forbidden gap argument for phase transitions proved by means of chessboard estimates},
  author={Biskup, Marek and Koteck{\'y}, Roman},
  journal={Communications in mathematical physics},
  volume={264},
  number={3},
  pages={631--656},
  year={2006},
  publisher={Springer}
}

@article{timar2013boundary,
  title={Boundary-connectivity via graph theory},
  author={Tim{\'a}r, {\'A}d{\'a}m},
  journal={Proceedings of the American Mathematical Society},
  volume={141},
  number={2},
  pages={475--480},
  year={2013}
}

@Article{PeledSpinka2019,
  author  = {Ron Peled and Yinon Spinka},
  journal = {Sojourns in Probability Theory and Statistical Physics - I},
  title   = {Lectures on the Spin and Loop {O}(n) Models},
  year    = {2019},
}

@article{shlosman1986method,
  title={The method of reflection positivity in the mathematical theory of first-order phase transitions},
  author={Shlosman, SB},
  journal={Russian Mathematical Surveys},
  volume={41},
  number={3},
  pages={83--134},
  year={1986}
}

@article{dobrushin1968description,
  title={Description of a random field by means of conditional probabilities and conditions for its regularities},
  author={Dobrushin, RL},
  journal={Theo. Probab. Appl.},
  volume={13},
  pages={197--224},
  year={1968}
}

@article{heilmann1972theory,
  title={Theory of Monomer-Dimer Systems},
  author={Heilmann, Ole J and Lieb, Elliott H},
  journal={Commun. math. Phys},
  volume={25},
  pages={190--232},
  year={1972}
}

@article{heilmann1970monomers,
  title={Monomers and dimers},
  author={Heilmann, Ole J and Lieb, Elliott H},
  journal={Physical Review Letters},
  volume={24},
  number={25},
  pages={1412},
  year={1970},
  publisher={APS}
}

@article{van1999absence,
  title={On the Absence of Phase Transition in the Monomer-Dimer Model},
  author={van den Berg, J},
  journal={Perplexing Problems in Probability: Festschrift in Honor of Harry Kesten},
  volume={44},
  pages={185--195},
  year={1999},
  publisher={Springer Science \& Business Media}
}

@article{disertori2013nematic,
  title={The nematic phase of a system of long hard rods},
  author={Disertori, Margherita and Giuliani, Alessandro},
  journal={Communications in Mathematical Physics},
  volume={323},
  number={1},
  pages={143--175},
  year={2013},
  publisher={Springer}
}

@article{ruelle1971existence,
  title={Existence of a phase transition in a continuous classical system},
  author={Ruelle, David},
  journal={Physical Review Letters},
  volume={27},
  number={16},
  pages={1040},
  year={1971},
  publisher={APS}
}

@incollection{heilmann1979lattice,
  title={Lattice models for liquid crystals},
  author={Heilmann, Ole J and Lieb, Elliott H},
  booktitle={Statistical Mechanics},
  pages={299--313},
  year={1979},
  publisher={Springer}
}

@article{jauslin2018nematic,
  title={Nematic liquid crystal phase in a system of interacting dimers and monomers},
  author={Jauslin, Ian and Lieb, Elliott H},
  journal={Communications in Mathematical Physics},
  volume={363},
  number={3},
  pages={955--1002},
  year={2018},
  publisher={Springer}
}

@article{alberici2016cluster,
  title={A cluster expansion approach to the {H}eilmann--{L}ieb liquid crystal model},
  author={Alberici, Diego},
  journal={Journal of Statistical Physics},
  volume={162},
  number={3},
  pages={761--791},
  year={2016},
  publisher={Springer}
}

@article{ioffe2006entropy,
  title={Entropy-Driven Phase Transition in a Polydisperse Hard-Rods Lattice System},
  author={Ioffe, D and Velenik, Y and Zahradn{\'\i}k, M},
  journal={Journal of Statistical Physics},
  volume={122},
  number={4},
  pages={761--786},
  year={2006}
}

@article{galvin2004phase,
  title={On phase transition in the hard-core model on $\mathbb{Z}^d$},
  author={Galvin, David and Kahn, Jeff},
  journal={Combinatorics, Probability and Computing},
  volume={13},
  number={2},
  pages={137--164},
  year={2004},
  publisher={Cambridge University Press}
}

@inproceedings{peled2014odd,
  title={Odd cutsets and the hard-core model on $\mathbb{Z}^d$},
  author={Peled, Ron and Samotij, Wojciech},
  booktitle={Annales de l'IHP Probabilit{\'e}s et statistiques},
  volume={50},
  number={3},
  pages={975--998},
  year={2014}
}

@article{ziff1992spanning,
  title={Spanning probability in 2{D} percolation},
  author={Ziff, Robert M},
  journal={Physical review letters},
  volume={69},
  number={18},
  pages={2670},
  year={1992},
  publisher={APS}
}

@article{malarz2005square,
  title={Square-lattice site percolation at increasing ranges of neighbor bonds},
  author={Malarz, Krzysztof and Galam, Serge},
  journal={Physical Review E},
  volume={71},
  number={1},
  pages={016125},
  year={2005},
  publisher={APS}
}

@article{dobrushin1968problem,
  title={The problem of uniqueness of a {G}ibbsian random field and the problem of phase transitions},
  author={Dobrushin, Roland L’vovich},
  journal={Functional Analysis and its Applications},
  volume={2},
  number={4},
  pages={302--312},
  year={1968},
  publisher={Springer}
}

@article{disertori2020plate,
  title={Plate-nematic phase in three dimensions},
  author={Disertori, Margherita and Giuliani, Alessandro and Jauslin, Ian},
  journal={Communications in Mathematical Physics},
  volume={373},
  number={1},
  pages={327--356},
  year={2020},
  publisher={Springer}
}

@article{onsager1949effects,
  title={The effects of shape on the interaction of colloidal particles},
  author={Onsager, Lars},
  journal={Annals of the New York Academy of Sciences},
  volume={51},
  number={4},
  pages={627--659},
  year={1949},
  publisher={Blackwell Publishing Ltd Oxford, UK}
}

@inproceedings{Nath2016thesis,
  title={Phase behaviour and ordering in hard core lattice gas models [HBNI Th109, Ph.D. thesis]},
  author={Nath, Trisha},
  year={2016},
  organization={IMSc},
}

@inproceedings{Vigneshwar2019thesis,
  title={Entropy driven phase transition in hard core lattice gas models in three dimensions [HBNI Th170]},
  author={Vigneshwar, N},
  year={2019},
  organization={IMSc}
}

@article{DharRajesh2021_lattice_rods,
archivePrefix = {arXiv},
arxivId = {2012.07223},
author = {Dhar, Deepak and Rajesh, R.},
doi = {10.1103/PhysRevE.103.042130},
eprint = {2012.07223},
issn = {24700053},
journal = {Physical Review E},
number = {4},
pages = {1--12},
pmid = {34005993},
publisher = {American Physical Society},
title = {{Entropy of fully packed hard rigid rods on $d$-dimensional hypercubic lattices}},
volume = {103},
year = {2021}
}

@article{Vogel2020_rods2D,
author = {Vogel, E. E. and Saravia, G. and Ramirez-Pastor, A. J. and Pasinetti, Marcelo},
doi = {10.1103/PhysRevE.101.022104},
issn = {24700053},
journal = {Physical Review E},
number = {2},
pages = {1--12},
pmid = {32168581},
publisher = {American Physical Society},
title = {{Alternative characterization of the nematic transition in deposition of rods on two-dimensional lattices}},
volume = {101},
year = {2020}
}

@article{Shah2021_lattice_rods,
archivePrefix = {arXiv},
arxivId = {2109.07881},
author = {Shah, Aagam and Dhar, Deepak and Rajesh, R.},
eprint = {2109.07881},
pages = {1--14},
title = {{The phase transition from nematic to high-density disordered phase in a system of hard rods on a lattice}},
url = {http://arxiv.org/abs/2109.07881},
year = {2021}
}

@article{Nath2014_kNN,
archivePrefix = {arXiv},
arxivId = {1404.6902},
author = {Nath, Trisha and Rajesh, R.},
doi = {10.1103/PhysRevE.90.012120},
eprint = {1404.6902},
issn = {15502376},
journal = {Physical Review E - Statistical, Nonlinear, and Soft Matter Physics},
number = {1},
pages = {1--18},
pmid = {25122264},
title = {{Multiple phase transitions in extended hard-core lattice gas models in two dimensions}},
volume = {90},
year = {2014}
}

@article{Fernandes2007_kNN,
archivePrefix = {arXiv},
arxivId = {arXiv:cond-mat/0612372v1},
author = {Fernandes, Heitor C. Marques and Arenzon, Jeferson J. and Levin, Yan},
doi = {10.1063/1.2539141},
eprint = {0612372v1},
issn = {00219606},
journal = {Journal of Chemical Physics},
number = {11},
primaryClass = {arXiv:cond-mat},
title = {{Monte Carlo simulations of two-dimensional hard core lattice gases}},
volume = {126},
year = {2007}
}

@article{Ghosh2007_rods2transitions,
author = {Ghosh, A. and Dhar, D.},
doi = {10.1209/0295-5075/78/20003},
issn = {02955075},
journal = {Epl},
number = {2},
title = {{On the orientational ordering of long rods on a lattice}},
volume = {78},
year = {2007}
}

@article{MSS2018_trihex,
archivePrefix = {arXiv},
arxivId = {1803.04041},
author = {Mazel, A. and Stuhl, I. and Suhov, Y.},
eprint = {1803.04041},
pages = {1--54},
title = {{High-density hard-core model on triangular and hexagonal lattices}},
url = {http://arxiv.org/abs/1803.04041},
year = {2018}
}

@article{MSS2019_square,
archivePrefix = {arXiv},
arxivId = {1909.11648},
author = {Mazel, A. and Stuhl, I. and Suhov, Y.},
eprint = {1909.11648},
issn = {23318422},
journal = {arXiv},
title = {{High-density hard-core model on $\mathbb{Z}^2$ and norm equations in ring $\mathbb{Z} [{\sqrt[6]{-1}}]$}},
url = {http://arxiv.org/abs/1909.11648},
year = {2019}
}

@article{MSS2020_short,
archivePrefix = {arXiv},
arxivId = {2011.14156},
author = {Mazel, A. and Stuhl, I. and Suhov, Y.},
eprint = {2011.14156},
pages = {1--15},
title = {{The hard-core model on planar lattices: the disk-packing problem and high-density {G}ibbs distributions}},
url = {http://arxiv.org/abs/2011.14156},
year = {2020}
}

@article{krachun2020_MSS_sliding_list,
  title={Extreme {G}ibbs measures for high-density hard-core model on $\mathbb{Z}^2$},
  author={Krachun, Dmitry},
  journal={arXiv:1912.07566},
  year={2019}
}

@article{Fernandes2020_kNN_hexa,
  title = {Phase Transitions in Hard-Core Lattice Gases on the Honeycomb Lattice.},
  author = {Thewes, Filipe C. and Fernandes, Heitor C.M. M.},
  year = {2020},
  month = jun,
  journal = {Physical review. E},
  volume = {101},
  number = {6-1},
  eprint = {2002.04603},
  eprinttype = {arxiv},
  pages = {062138},
  publisher = {{American Physical Society}},
  issn = {2470-0053},
  doi = {10.1103/PhysRevE.101.062138},
  archiveprefix = {arXiv},
  arxivid = {2002.04603},
  pmid = {32688552}
}

@article{pirogov1975phase,
  title={Phase diagrams of classical lattice systems},
  author={Pirogov, Sergey Anatol'evich and Sinai, Yakov Grigor'evich},
  journal={Teoreticheskaya i Matematicheskaya Fizika},
  volume={25},
  number={3},
  pages={358--369},
  year={1975},
  publisher={Russian Academy of Sciences, Steklov Mathematical Institute of Russian~…}
}

@article{pirogov1976phase,
  title={Phase diagrams of classical lattice systems continuation},
  author={Pirogov, Sergey Anatol'evich and Sinai, Yakov Grigor'evich},
  journal={Teoreticheskaya i Matematicheskaya Fizika},
  volume={26},
  number={1},
  pages={61--76},
  year={1976},
  publisher={Russian Academy of Sciences, Steklov Mathematical Institute of Russian~…}
}

@article{nisbet1974hard,
  title={Hard-core lattice gases with residual degrees of freedom at close packing},
  author={Nisbet, RM and Farquhar, IE},
  journal={Physica},
  volume={73},
  number={2},
  pages={351--367},
  year={1974},
  publisher={Elsevier}
}

@article{bellemans1967phase,
  title={Phase Transitions in Two-Dimensional Lattice Gases of Hard-Square Molecules},
  author={Bellemans, Andr{\'e} and Nigam, RK},
  journal={The Journal of Chemical Physics},
  volume={46},
  number={8},
  pages={2922--2935},
  year={1967},
  publisher={American Institute of Physics}
}

@article{mazel2021kepler,
  title={Kepler's conjecture and phase transitions in the high-density hard-core model on $\mathbb{Z}^3$},
  author={Mazel, A and Stuhl, I and Suhov, Y},
  journal={arXiv preprint arXiv:2112.14250},
  year={2021}
}

@article{peled2020long,
  title={Long-range order in discrete spin systems},
  author={Peled, Ron and Spinka, Yinon},
  journal={arXiv preprint arXiv:2010.03177},
  year={2020}
}

@article{brightwell1999nonmonotonic,
  title={Nonmonotonic behavior in hard-core and {W}idom--{R}owlinson models},
  author={Brightwell, Graham R and H{\"a}ggstr{\"o}m, Olle and Winkler, Peter},
  journal={Journal of statistical physics},
  volume={94},
  number={3},
  pages={415--435},
  year={1999},
  publisher={Springer}
}

@book{gennesPhysicsLiquidCrystals1993,
  title = {The {{Physics}} of {{Liquid Crystals}}},
  author = {de Gennes, P. G. and Prost, J.},
  year = {1993},
  publisher = {{Clarendon Press}},
  isbn = {978-0-19-851785-6},
  langid = {english}
}

@article{ramolaColumnarOrderAshkinTeller2015,
  title = {Columnar {{Order}} and {{Ashkin-Teller Criticality}} in {{Mixtures}} of {{Hard Squares}} and {{Dimers}}},
  author = {Ramola, Kabir and Damle, Kedar and Dhar, Deepak},
  year = {2015},
  month = may,
  journal = {Physical Review Letters},
  volume = {114},
  number = {19},
  pages = {190601},
  publisher = {{American Physical Society}},
  doi = {10.1103/PhysRevLett.114.190601}
}

@phdthesis{ramola2012thesis,
  title = {Onset of Order in Lattice Systems: {{Kitaev}} Model and Hard Squares [Ph.D. thesis]},
  author = {Ramola, Kabir},
  year = {2012}
}

@phdthesis{kundu2015thesis,
  title = {Phase Transitions in Systems of Hard Anisotropic Particles on Lattices [HBNI Th82, Ph.D. thesis]},
  author = {Kundu, Joyjit},
  year = {2015},
  school = {Homi Bhabha National Institute}
}

@article{gagunashvili_rods1979,
  title = {Close Packing of Rectilinear Polymers on a Square Lattice},
  author = {Gagunashvili, N. D. and Priezzhev, V. B.},
  year = {1979},
  month = jun,
  journal = {Theoretical and Mathematical Physics},
  volume = {39},
  number = {3},
  pages = {507--510},
  issn = {1573-9333},
  doi = {10.1007/BF01017997},
  langid = {english}
}

@article{nath_sliding2016,
  title = {The High Density Phase of The $k$-{NN} Hard Core Lattice Gas Model},
  author = {Nath, Trisha and Rajesh, R.},
  year = {2016},
  month = jul,
  journal = {Journal of Statistical Mechanics: Theory and Experiment},
  volume = {2016},
  number = {7},
  pages = {073203},
  publisher = {{IOP Publishing}},
  issn = {1742-5468},
  doi = {10.1088/1742-5468/2016/07/073203},
  langid = {english}
}

@article{kundu2013_rods,
  title = {Nematic-Disordered Phase Transition in Systems of Long Rigid Rods on Two-Dimensional Lattices},
  author = {Kundu, Joyjit and Rajesh, R. and Dhar, Deepak and Stilck, J{\"u}rgen F.},
  year = {2013},
  month = mar,
  journal = {Physical Review E},
  volume = {87},
  number = {3},
  pages = {032103},
  publisher = {{American Physical Society}},
  doi = {10.1103/PhysRevE.87.032103}
}

@article{fernandez2008_rods_tri,
  title = {Critical Behavior of Long Straight Rigid Rods on Two-Dimensional Lattices: {{Theory}} and {{Monte Carlo}} Simulations},
  shorttitle = {Critical Behavior of Long Straight Rigid Rods on Two-Dimensional Lattices},
  author = {{Matoz-Fernandez}, D. A. and Linares, D. H. and {Ramirez-Pastor}, A. J.},
  year = {2008},
  month = jun,
  journal = {The Journal of Chemical Physics},
  volume = {128},
  number = {21},
  pages = {214902},
  publisher = {{American Institute of Physics}},
  issn = {0021-9606},
  doi = {10.1063/1.2927877}
}

@article{coquille2014gibbs,
  title={On the {G}ibbs states of the noncritical {P}otts model on $\mathbb{Z}^2$},
  author={Coquille, Loren and Duminil-Copin, Hugo and Ioffe, Dmitry and Velenik, Yvan},
  journal={Probability Theory and Related Fields},
  volume={158},
  number={1},
  pages={477--512},
  year={2014},
  publisher={Springer}
}

@article{glazman2021structure,
  title={Structure of {G}ibbs measures for planar {FK}-percolation and {P}otts models},
  author={Glazman, Alexander and Manolescu, Ioan},
  journal={arXiv preprint arXiv:2106.02403},
  year={2021}
}

@article{aizenman1980translation,
  title={Translation invariance and instability of phase coexistence in the two dimensional {I}sing system},
  author={Aizenman, Michael},
  journal={Communications in Mathematical Physics},
  volume={73},
  number={1},
  pages={83--94},
  year={1980},
  publisher={Springer}
}

@inproceedings{higuchi1979absence,
  title={On the absence of non translation invariant {G}ibbs states for the two dimensional {I}sing model},
  author={Higuchi, Y},
  booktitle={Colloquia Math. Sociatatis Janos Bolyai},
  volume={27},
  pages={517--534},
  year={1979},
  organization={Random fields}
}

@article{georgii2000percolation,
  title={Percolation and number of phases in the two-dimensional {I}sing model},
  author={Georgii, Hans-Otto and Higuchi, Yasunari},
  journal={Journal of Mathematical Physics},
  volume={41},
  number={3},
  pages={1153--1169},
  year={2000},
  publisher={American Institute of Physics}
}

@article{russo1979infinite,
  title={The infinite cluster method in the two-dimensional {I}sing model},
  author={Russo, Lucio},
  journal={Communications in Mathematical Physics},
  volume={67},
  number={3},
  pages={251--266},
  year={1979},
  publisher={Springer}
}

@article{dobrushin1985problem,
  title={The problem of translation invariance of {G}ibbs states at low temperatures},
  author={Dobrushin, RL and Shlosman, SB},
  journal={Mathematical physics reviews},
  volume={5},
  pages={53--195},
  year={1985}
}

@article{vigneshwar2019phase,
  title={Phase diagram of a system of hard cubes on the cubic lattice},
  author={Vigneshwar, N and Mandal, Dipanjan and Damle, Kedar and Dhar, Deepak and Rajesh, R},
  journal={Physical Review E},
  volume={99},
  number={5},
  pages={052129},
  year={2019},
  publisher={APS}
}

@article{vigneshwar2017different,
  title={Different phases of a system of hard rods on three dimensional cubic lattice},
  author={Vigneshwar, N and Dhar, Deepak and Rajesh, R},
  journal={Journal of Statistical Mechanics: Theory and Experiment},
  volume={2017},
  number={11},
  pages={113304},
  year={2017},
  publisher={IOP Publishing}
}

@article{lopezPhaseDiagramSelfassembled2010,
  title = {Phase Diagram of Self-Assembled Rigid Rods on Two-Dimensional Lattices: {{Theory}} and {{Monte Carlo}} Simulations},
  shorttitle = {Phase Diagram of Self-Assembled Rigid Rods on Two-Dimensional Lattices},
  author = {L{\'o}pez, L. G. and Linares, D. H. and {Ramirez-Pastor}, A. J. and Cannas, S. A.},
  year = {2010},
  month = oct,
  journal = {The Journal of Chemical Physics},
  volume = {133},
  number = {13},
  pages = {134706},
  publisher = {{American Institute of Physics}},
  issn = {0021-9606},
  doi = {10.1063/1.3496482}
}

@article{TriangularTrimers_short1999,
  title = {Triangular {{Trimers}} on the {{Triangular Lattice}}: {{An Exact Solution}}},
  shorttitle = {Triangular {{Trimers}} on the {{Triangular Lattice}}},
  author = {Verberkmoes, Alain and Nienhuis, Bernard},
  year = {1999},
  month = nov,
  journal = {Physical Review Letters},
  volume = {83},
  number = {20},
  pages = {3986--3989},
  publisher = {{American Physical Society}},
  doi = {10.1103/PhysRevLett.83.3986}
}

@article{TriangularTrimers_long2001,
  title = {Bethe Ansatz Solution of Triangular Trimers on the Triangular Lattice},
  author = {Verberkmoes, Alain and Nienhuis, Bernard},
  year = {2001},
  month = may,
  journal = {Physical Review E},
  volume = {63},
  number = {6},
  pages = {066122},
  publisher = {{American Physical Society}},
  doi = {10.1103/PhysRevE.63.066122}
}

@article{jauslinHighFugacityExpansionLee2018,
  title = {High-Fugacity Expansion, {L}ee--{Y}ang Zeros, and Order-Disorder Transitions in Hard-Core Lattice Systems},
  author = {Jauslin, Ian and Lebowitz, Joel L.},
  year = {2018},
  month = dec,
  journal = {Communications in Mathematical Physics},
  volume = {364},
  number = {2},
  pages = {655--682},
  issn = {1432-0916},
  doi = {10.1007/s00220-018-3269-7},
  langid = {english}
}

@article{jauslinCrystallineOrderingLarge2018,
  title = {Crystalline Ordering and Large Fugacity Expansion for Hard-Core Lattice Particles},
  author = {Jauslin, Ian and Lebowitz, Joel L.},
  year = {2018},
  month = apr,
  journal = {The Journal of Physical Chemistry B},
  volume = {122},
  number = {13},
  pages = {3266--3271},
  publisher = {{American Chemical Society}},
  issn = {1520-6106},
  doi = {10.1021/acs.jpcb.7b08977}
}

@article{angelescuLatticeModelLiquid1982,
  title = {A Lattice Model of Liquid Crystals with Matrix Order Parameter},
  author = {Angelescu, N. and Zagrebnov, V. A.},
  year = {1982},
  month = nov,
  journal = {Journal of Physics A: Mathematical and General},
  volume = {15},
  number = {11},
  pages = {L639--L643},
  publisher = {{IOP Publishing}},
  issn = {0305-4470},
  doi = {10.1088/0305-4470/15/11/012},
  langid = {english}
}

@article{zagrebnovLongrangeOrderLatticegas1996,
  title = {Long-Range Order in a Lattice-Gas Model of Nematic Liquid Crystals},
  author = {Zagrebnov, V. A.},
  year = {1996},
  month = nov,
  journal = {Physica A: Statistical Mechanics and its Applications},
  series = {The {{Nature}} of {{Crystalline States}}},
  volume = {232},
  number = {3},
  pages = {737--746},
  issn = {0378-4371},
  doi = {10.1016/0378-4371(96)00181-1},
  langid = {english}
}

@article{biskupOrbitalOrderingTransitionMetal2005,
  title = {Orbital {{Ordering}} in {{Transition-Metal Compounds}}: {{I}}. {{The}} 120-{{Degree Model}}},
  shorttitle = {Orbital {{Ordering}} in {{Transition-Metal Compounds}}},
  author = {Biskup, M. and Chayes, L. and Nussinov, Z.},
  year = {2005},
  month = apr,
  journal = {Communications in Mathematical Physics},
  volume = {255},
  number = {2},
  pages = {253--292},
  issn = {1432-0916},
  doi = {10.1007/s00220-004-1272-7},
  langid = {english}
}

@article{biskupOrderDisorderOrder2004,
  title = {Order by {{Disorder}}, without {{Order}}, in a {{Two-Dimensional Spin System}} with {{O}}(2) {{Symmetry}}},
  author = {Biskup, Marek and Chayes, Lincoln and Kivelson, Steven A.},
  year = {2004},
  month = dec,
  journal = {Annales Henri Poincar\'e},
  volume = {5},
  number = {6},
  pages = {1181--1205},
  issn = {1424-0661},
  doi = {10.1007/s00023-004-0196-2},
  langid = {english}
}

@article{abraham1980interacting,
  title={Interacting dimers on the simple cubic lattice as a model for liquid crystals},
  author={Abraham, Douglas B and Heilmann, Ole J},
  journal={Journal of Physics A: Mathematical and General},
  volume={13},
  number={3},
  pages={1051},
  year={1980},
  publisher={IOP Publishing}
}

@article{bricmont1984structure,
  title={The structure of {G}ibbs states and phase coexistence for non-symmetric continuum {W}idom {R}owlinson models},
  author={Bricmont, Jean and Kuroda, Koji and Lebowitz, Joel L},
  journal={Zeitschrift f{\"u}r Wahrscheinlichkeitstheorie und Verwandte Gebiete},
  volume={67},
  number={2},
  pages={121--138},
  year={1984},
  publisher={Springer}
}

@article{nussinov2004orbital,
  title={Orbital order in classical models of transition-metal compounds},
  author={Nussinov, Zohar and Biskup, Marek and Chayes, Lincoln and van den Brink, Jeroen},
  journal={EPL (Europhysics Letters)},
  volume={67},
  number={6},
  pages={990},
  year={2004},
  publisher={IOP Publishing}
}

@article{propp2022trimer,
  title={Trimer covers in the triangular grid: twenty mostly open problems},
  author={Propp, James},
  journal={arXiv preprint arXiv:2206.06472},
  year={2022}
}

@article{bohmanLimitTheoremShannon2005,
  title = {A Limit Theorem for the {{Shannon}} Capacities of Odd Cycles. {{II}}},
  author = {Bohman, Tom},
  year = {2005},
  journal = {Proceedings of the American Mathematical Society},
  volume = {133},
  number = {2},
  pages = {537--543},
  issn = {0002-9939, 1088-6826},
  doi = {10.1090/S0002-9939-04-07470-2},
  urldate = {2024-10-17},
  langid = {english}
}

@article{frohlichPhaseTransitionsReflection1978,
  title = {Phase Transitions and Reflection Positivity. {{I}}. {{General}} Theory and Long Range Lattice Models},
  author = {Fr{\"o}hlich, J{\"u}rg and Israel, Robert and Lieb, Elliot H. and Simon, Barry},
  year = {1978},
  month = aug,
  journal = {Communications in Mathematical Physics},
  volume = {62},
  number = {1},
  pages = {1--34},
  issn = {1432-0916},
  doi = {10.1007/BF01940327},
  urldate = {2024-07-03},
  langid = {english}
}

@article{frohlichPhaseTransitionsReflection1980,
  title = {Phase Transitions and Reflection Positivity. {{II}}. {{Lattice}} Systems with Short-Range and {{Coulomb}} Interactions},
  author = {Fr{\"o}hlich, J{\"u}rg and Israel, Robert B. and Lieb, Elliott H. and Simon, Barry},
  year = {1980},
  month = mar,
  journal = {Journal of Statistical Physics},
  volume = {22},
  number = {3},
  pages = {297--347},
  issn = {1572-9613},
  doi = {10.1007/BF01014646},
  urldate = {2022-11-08},
  langid = {english}
}

@online{hadas2022chessboard,
  title = {A minimal introduction to the Chessboard Estimate},
  author = {Hadas, Daniel},
  year = 2022,
  note= {\url{http://daniel.hadaso.net/chessboard.pdf}},
}

@article{mackeyCubeTilingDimension2002,
  title = {A {{Cube Tiling}} of {{Dimension Eight}} with {{No Facesharing}}},
  author = {{Mackey}},
  year = {2002},
  month = aug,
  journal = {Discrete \& Computational Geometry},
  volume = {28},
  number = {2},
  pages = {275--279},
  issn = {1432-0444},
  doi = {10.1007/s00454-002-2801-9},
  urldate = {2024-10-17},
  langid = {english}
}

@article{lagariasKellersCubetilingConjecture1992,
  title = {Keller's Cube-Tiling Conjecture Is False in High Dimensions},
  author = {Lagarias, Jeffrey C. and Shor, Peter W.},
  year = {1992},
  journal = {Bulletin of the American Mathematical Society},
  volume = {27},
  number = {2},
  pages = {279--283},
  issn = {0273-0979, 1088-9485},
  doi = {10.1090/S0273-0979-1992-00318-X},
  urldate = {2024-10-17},
  langid = {english}
}

@article{brakensiekResolutionKellersConjecture2022,
  title = {The {{Resolution}} of {{Keller}}'s {{Conjecture}}},
  author = {Brakensiek, Joshua and Heule, Marijn and Mackey, John and Narv{\'a}ez, David},
  year = {2022},
  month = aug,
  journal = {Journal of Automated Reasoning},
  volume = {66},
  number = {3},
  pages = {277--300},
  issn = {1573-0670},
  doi = {10.1007/s10817-022-09623-5},
  urldate = {2024-10-17},
  langid = {english}
}

@article{shannonZeroErrorCapacity1956,
  title = {The Zero Error Capacity of a Noisy Channel},
  author = {Shannon, C.},
  year = {1956},
  month = sep,
  journal = {IRE Transactions on Information Theory},
  volume = {2},
  number = {3},
  pages = {8--19},
  issn = {2168-2712},
  doi = {10.1109/TIT.1956.1056798},
  urldate = {2024-10-17}
}

@article{perronUeberLueckenloseAusfuellung1940,
  title = {{{\"U}ber l{\"u}ckenlose Ausf{\"u}llung desn-dimensionalen Raumes durch kongruente W{\"u}rfel}},
  author = {Perron, Oskar},
  year = {1940},
  month = dec,
  journal = {Mathematische Zeitschrift},
  volume = {46},
  number = {1},
  pages = {1--26},
  issn = {1432-1823},
  doi = {10.1007/BF01181421},
  urldate = {2024-10-17},
  langid = {ngerman}
}

@article{perronUeberLueckenloseAusfuellung1940a,
  title = {{{\"U}ber l{\"u}ckenlose Ausf{\"u}llung desn-dimensionalen Raumes durch kongruente W{\"u}rfel. II}},
  author = {Perron, Oskar},
  year = {1940},
  month = dec,
  journal = {Mathematische Zeitschrift},
  volume = {46},
  number = {1},
  pages = {161--180},
  issn = {1432-1823},
  doi = {10.1007/BF01181436},
  urldate = {2024-10-17},
  langid = {ngerman}
}

@article{Keller+1930+231+248,
url = {https://doi.org/10.1515/crll.1930.163.231},
title = {Über die lückenlose Erfüllung des Raumes mit Würfeln.},
title = {},
author = {Ott-Heinrich Keller},
pages = {231--248},
volume = {1930},
number = {163},
journal = {Journal für die reine und angewandte Mathematik},
doi = {doi:10.1515/crll.1930.163.231},
year = {1930},
lastchecked = {2024-10-17}
}

\end{document}